Thesis

presented to the

National Institute of Applied Sciences (INSA) of Rennes

in partial fulfillment of the thesis requirements for the degree of

Doctor

in Electronics and Telecommunications

# Cross-layer Optimization for Next Generation Wi-Fi

by

Getachew REDIETEAB

Thesis defended on October 5th 2012 with the following examination committee:

Reviewers

| | |
|---|---|
| Jean-Marie Gorce | Professor, INSA of Lyon |
| Christophe Le Martret | Thales Expert HDR, THALES Communications & Security |

Examiners

| | |
|---|---|
| Michel Terré | Professor, CNAM Paris |
| Erik Ström | Professor, CHALMERS University Göteborg |
| Laurent Cariou | Doctor-engineer, ORANGE LABS Rennes (supervisor) |
| Philippe Christin | Engineer, ORANGE LABS Rennes (supervisor) |
| Jean-François Hélard | Professor, INSA of Rennes (thesis supervisor) |

Thèse

présentée devant

l'Institut National des Sciences Appliquées (INSA) de Rennes

pour obtenir le grade de

Docteur de l'INSA de Rennes

Spécialité : Electronique et Télécommunications

# Optimisation cross-layer pour des futures générations de réseaux Wi-Fi

par

Getachew REDIETEAB

Thèse soutenue le 05 octobre 2012 devant le jury composé de :

Rapporteurs

| | |
|---|---|
| Jean-Marie Gorce | Professeur d'Université, INSA de Lyon |
| Christophe Le Martret | Expert Thales HDR, THALES Communications & Security |

Examinateurs

| | |
|---|---|
| Michel Terré | Professeur d'Université, CNAM Paris |
| Erik Ström | Professeur d'Université, CHALMERS University Göteborg |
| Laurent Cariou | Docteur-ingénieur, ORANGE LABS Rennes (co-encadrant) |
| Philippe Christin | Ingénieur, ORANGE LABS Rennes (co-encadrant) |
| Jean-François Hélard | Professeur d'Université, INSA de Rennes (directeur de thèse) |

# Abstract

IEEE's 802.11 standard, commonly known as wireless fidelity (Wi-Fi), has become de facto the reference in wireless local access networks. From the initial 1997 specification to the undergoing IEEE 802.11ac standardization, a leap in throughput has been observed with every new generation. The expectations for next generations on issues like throughput, range, reliability, and power consumption are even higher. This is quite a challenge considering all the work already done. Cross-layer optimization of physical (PHY) and medium access control (MAC) layers can be an interesting exploration path for further enhancement. IEEE 802.11ac and IEEE 802.11ah have already started using this type of optimization.

During this thesis we have studied cross-layer optimization techniques, with a focus on the IEEE 802.11ac standard. A new multichannel aggregation scheme involving cross-knowledge between PHY and MAC layers has been proposed to improve performance in collision-prone environments. While testing this solution, we have shown that some functionalities or phenomena directly involved PHY and MAC layers. An accurate modeling of both PHY and MAC mechanisms is thus needed to have a realistic characterization of such functionalities and phenomena. A cross-layer simulator, compliant with IEEE 802.11n/ac specifications, has thus been implemented. To the best of our knowledge, this is the first simulator incorporating detailed PHY and MAC functionalities for the IEEE 802.11ac standard.

The multiple-user multiple-input, multiple-output (MU-MIMO) technique, which is one of the main innovations of the IEEE 802.11ac, needs both PHY and MAC layer considerations. We have thus used the implemented cross-layer simulator to evaluate the performance of MU-MIMO and compared it with the single-user MIMO (SU-MIMO). The aim of these studies was to evaluate the 'real' gains of MU-MIMO solutions (accounting for PHY+MAC) over SU-MIMO solutions, and not the generally accepted ones (accounting for MAC only or PHY only). The impact of the channel sounding interval has particularly been studied.

Finally, we have proposed an ultra short PHY layer version of acknowledgment frames for overhead reduction in machine to machine IEEE 802.11ah communications. Thus, the different cross-layer optimization principles, which have been studied and modeled through the PHY+MAC simulator for IEEE 802.11ac systems, can be applied to future generations of Wi-Fi systems.

# Résumé Etendu

Les normes IEEE 802.11, également connues sous l'appellation commerciale Wi-Fi, sont devenues une référence en termes de réseaux locaux sans fil. Depuis la première spécification (1997) jusqu'à la norme IEEE 802.11ac (en cours de normalisation), une augmentation en débit conséquente a été obtenue avec chaque nouvelle génération. Les attentes, en termes de débit, de portée, de fiabilité et de consommation énergétique, pour les prochaines générations sont encore plus élevées. Il y a donc un réel défi technologique à relever. Une approche cross-layer (ou inter-couche) permettant d'optimiser conjointement la couche physique (PHY) et la couche d'accès au canal (MAC, pour Medium Access Control) pourrait donc offrir des pistes d'amélioration intéressantes. C'est le thème que l'on propose d'aborder dans cette thèse. D'ailleurs, les normes IEEE 802.11ac et IEEE 802.11ah intègrent déjà certaines techniques relevant d'une telle d'optimisation.

## Réseaux locaux sans fil et techniques associées

### Techniques de la couche physique

Les réseaux locaux sans fil sont mis en place dans des environnements présentant, en général, plusieurs réflecteurs (plafond, vitres, chaises, etc.). Un récepteur recevra donc autant de répliques, déphasées en temps, du signal transmis qu'il y a de réflecteurs. C'est la propagation multitrajets. Celle-ci induit une sélectivité en fréquence et une interférence inter-symbole. La technique de modulation multiporteuse (OFDM pour Orthogonal Frequency Division Multiplexing) permet de compenser ces effets. La subdivision en sous-porteuses implique alors une étape de synchronisation préalable. Une phase d'estimation des coefficients du canal, suivie d'une phase de transmission d'information de signalisation, ont également lieu afin de pouvoir décoder le signal transmis.

Les systèmes multi-antennaires (MIMO, pour Multiple-Input, Multiple-Output) permettent de bénéficier de la diversité spatiale, grâce à une redondance du signal reçu, ou du multiplexage spatial, transmettant ainsi des flux indépendants de manière simultanée. Dans les deux cas, le canal entre chaque paire d'antennes doit être estimé. Ces estimées peuvent être retournées à l'émetteur pour que celui-ci puisse adapter le signal transmis au canal en appliquant un précodage approprié ; c'est le beamforming. Les performances (en termes de débit ou de portée) peuvent alors être améliorées, au prix d'un retour régulier des coefficients du canal. Ce précodage peut aussi être utilisé pour transmettre, de manière simultanée, plusieurs flux indépendants vers plusieurs utilisateurs en profitant de la différence entre leurs canaux. C'est la transmission multiutilisateur (MU-MIMO, pour Multiple-User Multiple-Input, Multiple-Output) permettant de multiplier les débits, à condition que l'interférence inter-utilisateur (CTI, pour CrossTalk Interference) soit minimisée. Pour ce faire, des techniques de précodage linéaire peu complexes, tels que l'inversion de canal ou la diagonalisation par blocs, peuvent être utilisés.

## Techniques de la couche MAC

Les mécanismes de la couche MAC permettent, entre autres, le partage du canal entre plusieurs utilisateurs. Ainsi, ils doivent tenir compte des spécificités du canal. Celui-ci est fortement atténuateur. De plus, le signal transmis étant diffusé, les stations peuvent « s'entendre », peu importe le réseau auquel elles appartiennent, à condition qu'elles soient à portée les unes des autres. Ceci implique que des stations peuvent être cachées les unes des autres.

Plusieurs mécanismes élémentaires sont alors mis en place. Premièrement, toute transmission est précédée d'une période de contention d'accès au canal au cours de laquelle le canal est « écouté » afin d'éviter les collisions. Dans ce même but, le temps d'attente est aléatoire d'une station à l'autre. Deuxièmement, les trames reçues doivent donc être acquittées par le destinataire (dont l'adresse MAC est informée dans la trame), faute de quoi les trames devront être retransmises. Troisièmement, le temps d'occupation canal restant est indiqué dans la trame envoyée afin de protéger les trames à suivre, qui y sont directement associées (par exemple les trames d'acquittement). D'ailleurs, les conditions canal évoluant au cours du temps, le débit de transmission peut être adapté de manière dynamique grâce à un algorithme de sélection de débit PHY.

Des mécanismes avancés sont aussi définis. La qualité de service est introduite en définissant les catégories d'accès. Chaque catégorie est attribuée une priorité différente dans le processus d'accès au canal. Certaines catégories sont autorisées à occuper le canal pour plus d'une transmission à la fois, à condition de ne pas dépasser un temps total prédéfini à chaque accès. Suite à cela, l'acquittement par bloc de trames a été introduit pour gagner en efficacité, puis utilisé pour acquitter des paquets agrégés. L'agrégation consiste à mutualiser les phases de contention et entêtes de mise en trame pour une meilleure efficacité. Ainsi, en sortie de la couche MAC, les paquets MAC (MPDU, pour MAC

Protocol Data Unit) sont agrégés en un A-MPDU (Aggregate MPDU).

## Normes IEEE 802.11 basées sur l'OFDM

La première norme IEEE 802.11 à implémenter l'OFDM est la norme IEEE 802.11a, définie pour la bande 5 GHz. La norme IEEE 802.11g popularise cette technique en la translatant dans la bande 2.4 GHz. Ces deux normes permettent d'atteindre des débits couche PHY de 54 Mbit/s en utilisant une bande passante de 20 MHz (le débit couche MAC étant en moyenne de 20 Mbit/s). Différentes valeurs de temps d'accès au canal sont aussi définies par la couche MAC.

Avec l'arrivée de nouvelles applications nécessitant plus de débit, la norme IEEE 802.11n a été définie pour les bandes 2.4 GHz et 5 GHz. En effet des débits PHY jusqu'à 600 Mbit/s peuvent être atteints, grâce au multiplexage spatial du MIMO, à l'augmentation de la bande passante (40 MHz) et à l'utilisation d'un intervalle de garde plus court. La couche MAC gagne aussi en efficacité : l'agrégation MAC et l'acquittement par bloc sont utilisés. L'objectif de 100 Mbit/s de débit couche MAC peut donc être ainsi atteint, et même dépassé. La qualité de service est également incluse.

La norme IEEE 802.11ac se définit dans le prolongement de la norme 802.11n. La bande passante est encore augmentée (jusqu'à 160 MHz) et le MU-MIMO est introduit pour plus de débit. Les débits PHY allant à 6,93 Gbit/s par utilisateur devront permettre d'atteindre l'objectif fixé de 1 Gbit/s de débit MAC multiutilisateur. La taille maximale d'agrégation est augmentée dans ce but.

La dernière norme, à ce jour, est la norme IEEE 802.11ah. Elle reprend une grande partie des spécifications 802.11ac en les adaptant aux communications M2M (Machine to Machine), comme les réseaux de capteurs, pour les bandes sous 1 GHz. L'accent est mis sur l'augmentation du nombre de stations, la robustesse et l'économie d'énergie.

## Solutions cross-layer dans les réseaux locaux sans fil

Il y a un intérêt grandissant pour la conception cross-layer dans le monde des réseaux sans fil car celle-ci permet une meilleure adaptation aux variations du canal. En effet, la structure actuelle s'est beaucoup inspirée du monde filaire. Le canal évoluant peu dans ce dernier, les techniques héritées doivent donc être repensées sans pour autant casser le modèle en couches. Ce dernier est utile pour isoler les fonctions et leur implémentation. Par conséquent, les solutions cross-layer doivent être conçues en tenant compte de leur impact sur d'autres couches, d'une part, et des possibilités d'évolution du système sur le long terme, d'autre part. Notre vision de l'optimisation cross-layer porte donc plus sur les possibilités d'évolution des normes actuelles que sur l'allocation de ressources (autre vision de l'optimisation cross-layer).

Une telle optimisation peut se voir dans l'utilisation de l'identifiant de groupe pour économiser de l'énergie dans les systèmes IEEE 802.11ac activant le MU-MIMO. Un autre exemple d'implémentation est l'utilisation d'un entête MAC robuste en ECMA-368.

## Plateformes de simulation pour les réseaux locaux sans fil

Dans le monde des réseaux locaux sans fil, les performances PHY ou MAC peuvent être évaluées en se basant sur des simulations liens ou systèmes. Les simulateurs liens (ou PHY) permettent de modéliser le canal et les chaînes de transmission/réception de manière fine. Toutefois les couches supérieures, dont la couche MAC, sont simplifiées. A l'inverse, une modélisation précise des mécanismes d'accès au canal et de gestion des files d'attentes peut être effectuée par le biais de simulateurs systèmes (ou MAC), où les spécificités de la couche PHY sont abstraites.

Cependant, certains phénomènes impliquant les deux couches, une modélisation précise et réaliste des mécanismes de ces deux couches serait très avantageuse. Il existe en effet, plusieurs efforts faits dans ce sens, mais aucun, à notre connaissance, ne porte sur les normes IEEE 802.11n et IEEE 802.11ac.

# La norme IEEE 802.11ac

Servant de base pour les travaux de cette thèse, cette norme est présentée plus en détail ici.

## Couches PHY et MAC

En IEEE 802.11ac, les transmissions de données peuvent se faire en utilisant juste un canal de 20 MHz de bande passante ou en concaténant plusieurs pour former des canaux de 40 MHz, 80 MHz ou 160 MHz. Ce dernier canal peut être composé de deux canaux de 80 MHz non-contigus. Une augmentation en débit conséquente peut donc être obtenue par le biais de la concaténation de canaux. Un autre moyen de monter en débit est d'augmenter le nombre de flux spatiaux à huit.

Les stations 802.11ac, fonctionnant dans la bande 5 GHz, doivent coexister avec les stations IEEE 802.11a/n. La première partie de l'entête PHY, illustrée dans la Figure 1, est donc héritée. L'autre partie est spécifique aux fonctionnalités 802.11ac. Elle est d'ailleurs précodée avec les données lorsque le beamforming est utilisé.

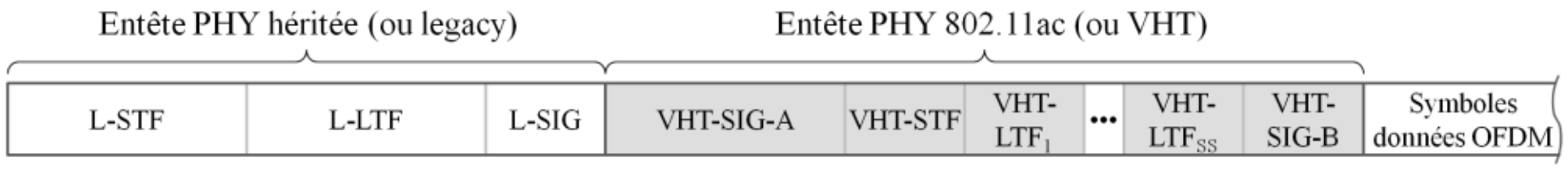

**Figure 1  Mise en trame PHY en IEEE 802.11ac**

La couche MAC, quant à elle, étend les règles antérieures. L'accès au canal pour des transmissions multicanaux comprend maintenant les transmissions à 80 MHz et 160 MHz. Les mécanismes de protection sont aussi améliorés. Enfin, le protocole de retour de coefficients canal devient unique.

## Transmission multi-antennaire vers plusieurs utilisateurs (MU-MIMO)

Le MU-MIMO, qui est la nouveauté du 802.11ac, implique des modifications PHY et MAC. Transmettre des données indépendantes vers plusieurs utilisateurs implique des modulations et précodages indépendantes ; plusieurs chaînes de traitement sont donc nécessaires. De plus la deuxième partie de l'entête et les données sont précodées différemment. Les procédures d'acquittement et de sondage de canal sont aussi adaptées à plusieurs utilisateurs. D'ailleurs, en raison du précodage simultané, le MU-MIMO requiert que les estimées canal soient bien à jour. Le sondage canal doit donc être effectué plus fréquemment qu'en SU-MIMO (Single-User MIMO).

# Agrégation cross-layer pour transmissions multicanaux

L'augmentation en débit MAC de la norme IEEE 802.11n n'aurait pu être obtenue sans l'agrégation A-MPDU. Mais cette augmentation s'appuie aussi sur une montée en débit PHY. Cette dernière s'obtient, en partie, par la concaténation de canaux. Nous avons vu plus haut que quatre, voire huit, canaux peuvent être concaténés en IEEE 802.11ac pour une plus grande efficacité au niveau de la couche PHY. Toutefois, une telle concaténation augmente les risques de collision. L'A-MPDU étant étalée sur toute la bande, une collision sur un des canaux peut en corrompre toutes les MPDU. Une illustration de cette agrégation, nommée 'verticale' en raison de l'étalement des MPDU sur les canaux, est donnée dans la Figure 2 (a). Un tel agencement n'a été optimisé que d'un point de vue PHY, afin de profiter au mieux de la diversité fréquentielle. Or, plus le nombre de canaux augmente, plus le pourcentage de collision, $P_{collision}$, augmente.

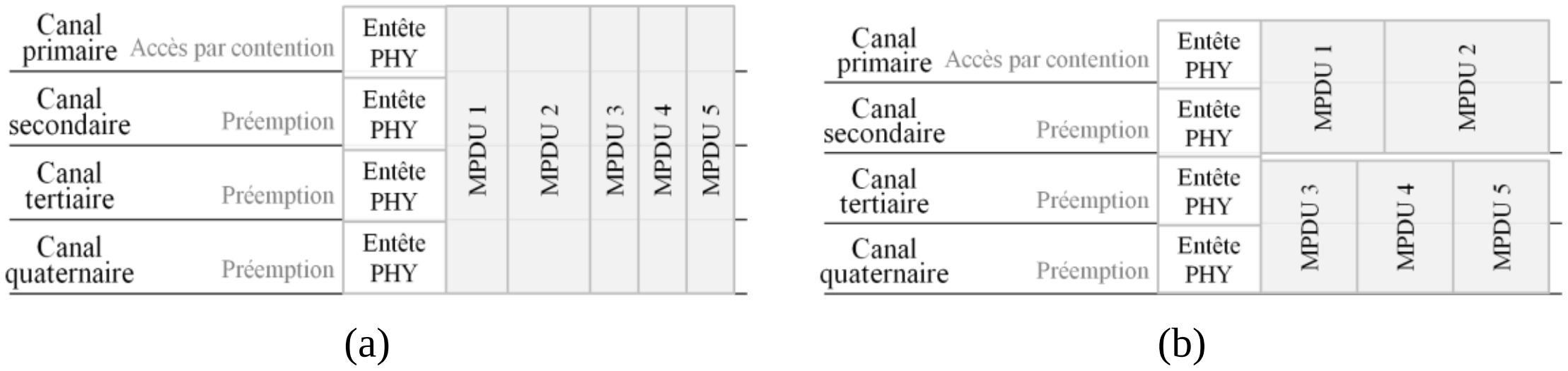

**Figure 2 Exemples d'agrégation 'verticale' (a) et d'agrégation 'horizontale' (b)**

D'ailleurs, dans des endroits densément peuplés, le problème de superposition de réseaux locaux sans fil grandit et il devient très difficile de trouver des bandes de 80 MHz, et encore moins de 160 MHz. La concaténation de canaux non-contigus devient donc une solution intéressante. Pour ceci, il faut utiliser autant de chaînes radiofréquences (comprenant entre autres, un convertisseur analogique numérique, CAN) qu'il y a de blocs de canaux non-contigus.

## Technique cross-layer proposée : l'agrégation 'horizontale'

Au vu d'un tel contexte, nous proposons d'étaler les MPDU sur moins de canaux afin de réduire la probabilité de collision. Autrement dit, l'agrégat initial pourrait être segmenté en $B$ sous-agrégats, et la bande de transmission en $B$ blocs de canaux, de telle manière à avoir $B$ transmissions parallèles. La probabilité de collision sur chaque MPDU est alors divisée par un facteur $B$. La Figure 2 (b) illustre ceci pour $B=2$. La différence avec la technique classique est que les MPDU sont agencés horizontalement, expliquant l'appellation agrégation 'horizontale' donnée. Le type d'agrégation devra toutefois être signalé dans l'entête PHY, de telle manière à ce que la couche PHY du récepteur puisse traiter les blocs de données de manière indépendante.

## Evaluation des performances

L'impact d'une collision dépend de la puissance de l'interférent l'ayant engendrée, et donc du rapport signal sur interférent (RSI). Les résultats obtenus avec un algorithme de sélection de débit sont données dans la Figure 3. Une augmentation du débit MAC allant jusqu'à 86% avec $B=2$ et 107% avec $B=4$ peut être observée, relativement au débit MAC de l'agrégation verticale. De plus, l'agrégation horizontale est plus intéressante pour des RSI en dessous 9 dB, voire 12 dB.

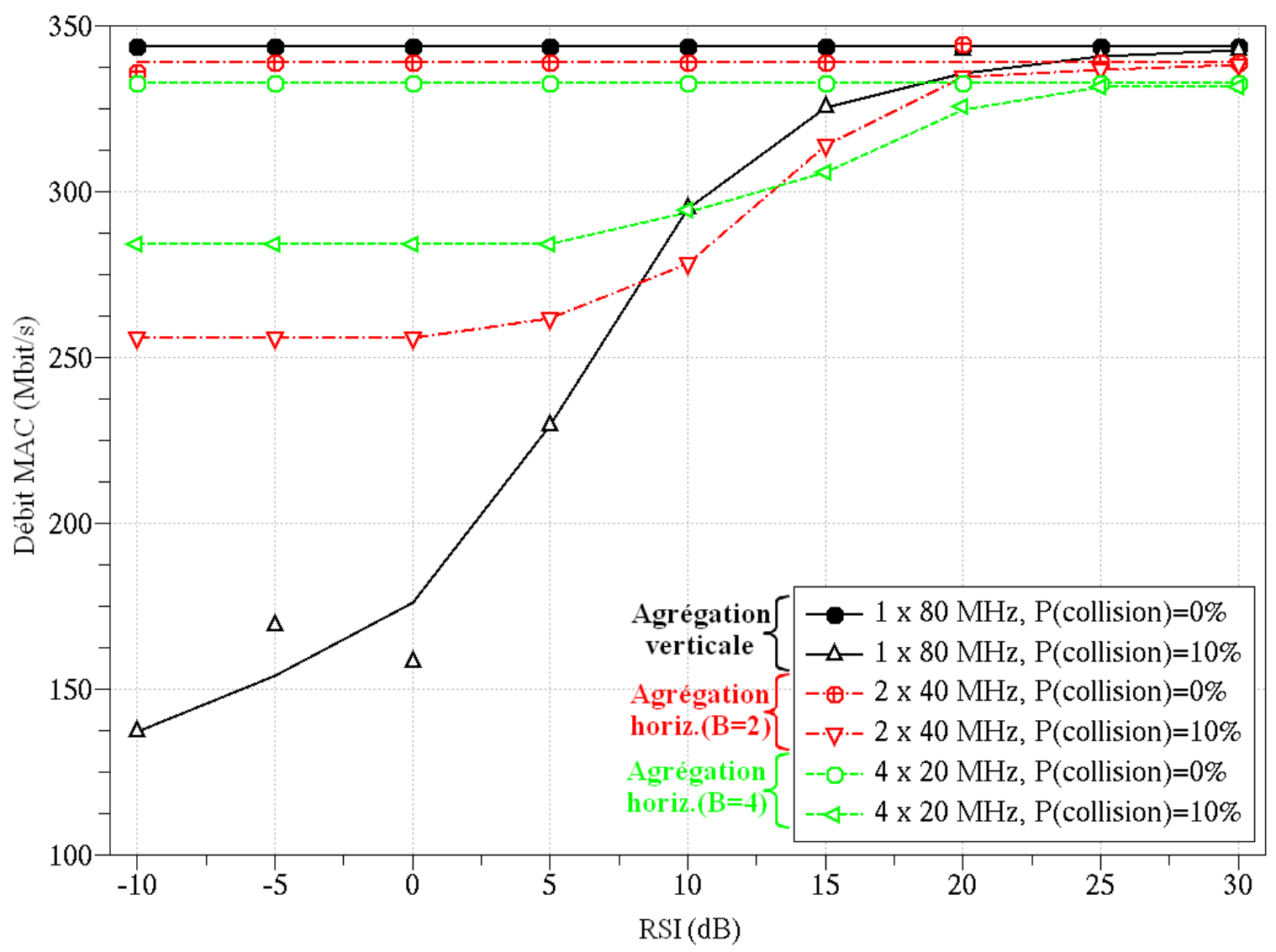

**Figure 3 Débit MAC en fonction du RSI pour différents schémas d'agrégation avec $P_{collision}$ = 0% et 10%**

Ainsi, l'agrégation horizontale permet de réduire l'impact des collisions. Le nombre de blocs devrait être choisi en tenant compte des performances désirées et de la complexité d'implémentation. De ce point de vue, l'agrégation $B=2$ offre le meilleur compromis dans le cas illustré.

# Simulateur PHY+MAC pour systèmes IEEE 802.11n/ac

## Architecture générale

La plateforme de simulation pour systèmes IEEE 802.11n/ac est composée de deux entités, PHY et MAC, séparées. Celles-ci interagissent de manière ponctuelle pour échanger des informations et données qu'elles ont traitées au préalable. L'entité PHY, type Ptolemy II, est composée d'une chaîne de transmission 802.11n/ac, d'un canal TGn/TGac et d'une chaîne de réception adaptée. L'évolution du canal dans le temps ainsi que l'influence des techniques de transmission utilisées peuvent être finement modélisées. L'entité MAC, quant à elle basée sur la structure de ns-3, modélise les mécanismes de contention, de qualité de service, mais aussi les implications du MU-MIMO. Elle est d'ailleurs conforme aux spécifications des normes 802.11n/ac.

## Validation du simulateur

La Figure 4 permet de comparer les performances du simulateur implémenté (noté XLS, pour Cross-Layer Simulator) à celles de ns-2 en utilisant différentes tailles d'agrégation, pour un AP et une station. Les résultats obtenus sont bien similaires.

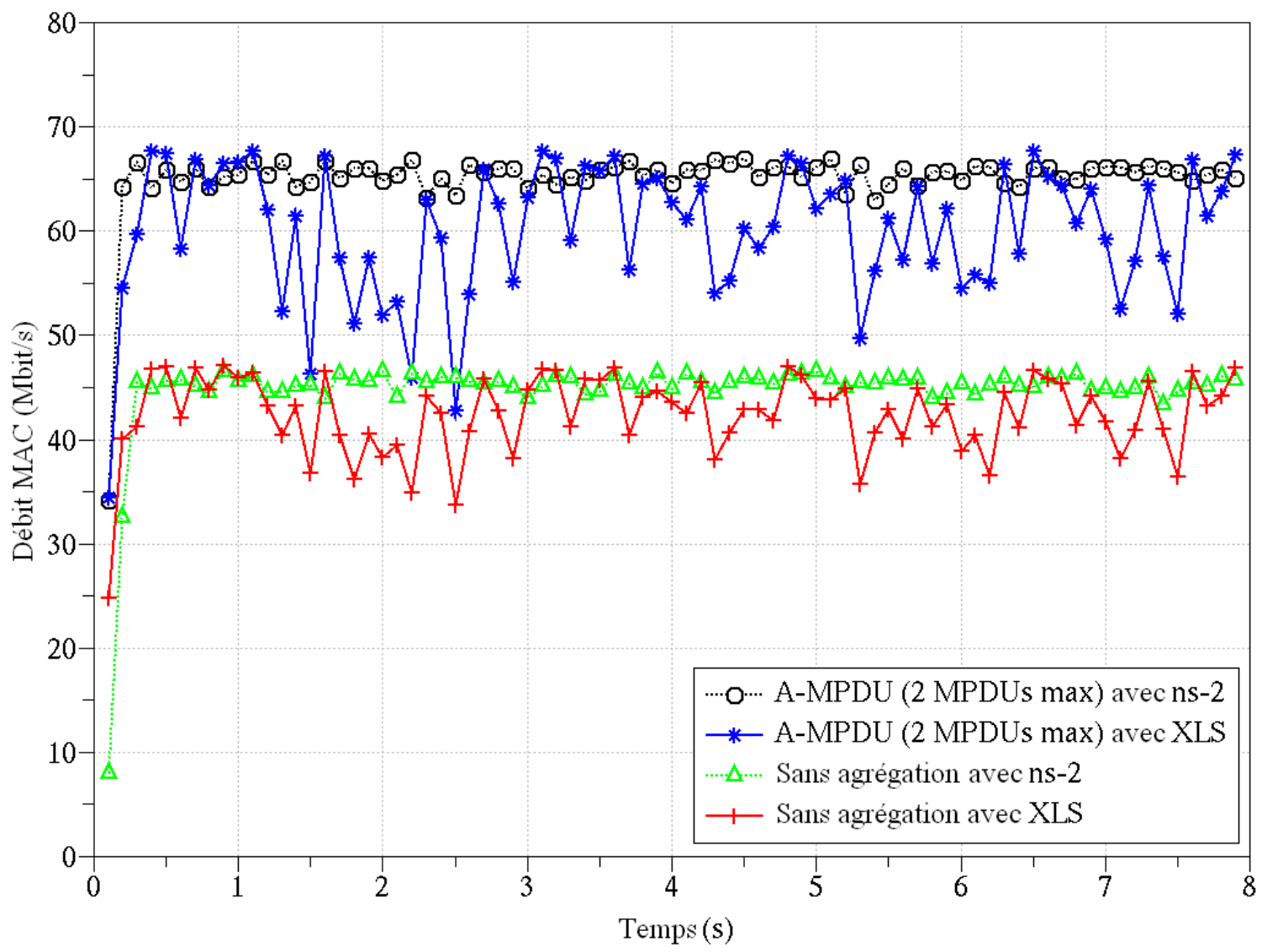

**Figure 4 Débit MAC en fonction du temps avec et sans agrégation en utilisant le simulateur de référence ns-2 et le simulateur cross-layer implémenté**

Il est intéressant de noter que les variations de débit avec le XLS sont dues à l'adaptation du débit PHY aux conditions du canal par l'algorithme de sélection de débit. Une telle précision est un avantage pour des techniques (telles le MU-MIMO) dont les performances sont liées à l'état du canal.

## Apport d'un tel simulateur dans la modélisation d'une configuration avec stations cachées

Lorsque deux stations sont cachées, l'une vis-à-vis de l'autre mais toutes deux à portée de l'AP, les trames qu'elles transmettent à ce dernier peuvent entrer en collision. La Figure 5 donne le débit MAC montant total lorsqu'une station, émettant des trames courtes, s'éloigne de l'AP et de l'autre station toutes les 2 secondes. Tant que les deux stations sont à portée (pendant les 2 premières secondes), les performances entre ns-2 et XLS sont similaires. Toutefois en configuration de stations cachées (les 4 secondes qui suivent), le simulateur implémenté donne de meilleurs résultats. En effet, ns-2 considère que les trames ayant subi une collision sont toutes perdues, les chaînes de réception étant abstraites. Le XLS permet lui d'évaluer si le récepteur arrive à récupérer le signal malgré la collision, modélisant mieux la réalité.

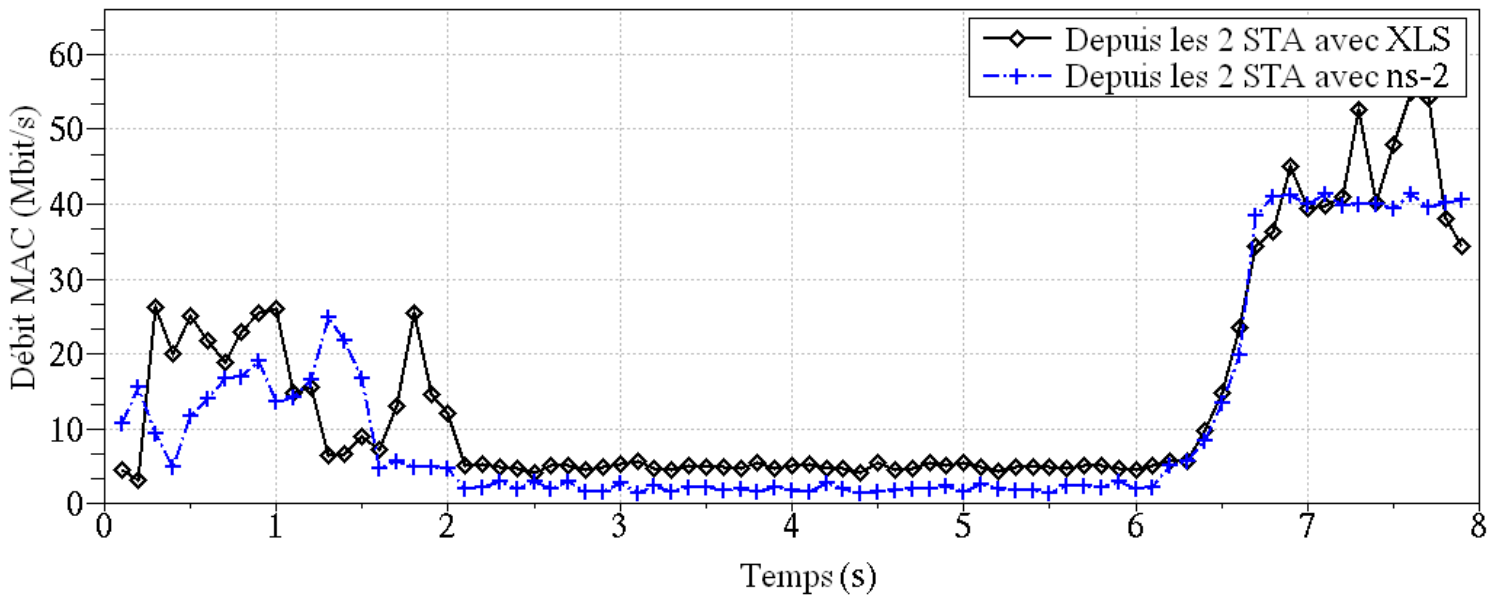

**Figure 5 Débit MAC total en fonction du temps en configuration de stations cachées en utilisant le simulateur de référence et le simulateur cross-layer implémenté**

Le simulateur PHY+MAC implémenté est donc une plateforme intéressante pour modéliser des phénomènes et pour tester des solutions cross-layer impliquant la couche PHY et la couche MAC. Les études sur le MU-MIMO pourront profiter de cet aspect du simulateur.

# Comparaison PHY+MAC des performances du SU-MIMO et du MU-MIMO en IEEE 802.11ac

## Pourquoi une analyse cross-layer ?

La plupart des études comparant le SU-MIMO et le MU-MIMO tiennent compte soit seulement de la couche PHY, soit uniquement de la couche MAC. Dans le premier cas, les gains du SDMA sont démontrés sans tenir compte des mécanismes de contention et de sondage. Or, le MU-MIMO et le SU-MIMO avec beamforming impliquent une connaissance du canal à l'émetteur. Le sondage est donc un aspect important et doit être pris en compte. Dans le deuxième cas, l'interférence qu'il pourrait y avoir entre les différents utilisateurs (ou CTI) est ignorée. Or, en MU-MIMO, les performances sont très liées à l'état du canal et au précodage utilisé. Il faut donc modéliser les évolutions du canal, et le vieillissement des estimées, de manière précise.

## Evaluation des performances

Ainsi, pour tenir compte des implications PHY et MAC du MU-MIMO, nous utilisons le simulateur cross-layer IEEE 802.11ac exposé plus haut. Ceci nous permet de comparer les deux techniques. L'importance du CTI est évaluée en considérant le MU-MIMO comme deux transmissions isolées et donc utilisant des précodeurs indépendants. La Figure 6 illustre le débit MAC vers une station, parmi deux stations servies par le point d'accès (AP, pour Access Point). Le MU-MIMO permet bien de doubler le débit par rapport au SU-MIMO. De plus, l'inclusion du CTI n'a pas d'impact sur le débit. Des simulations MAC auraient donné des résultats similaires pour ce canal.

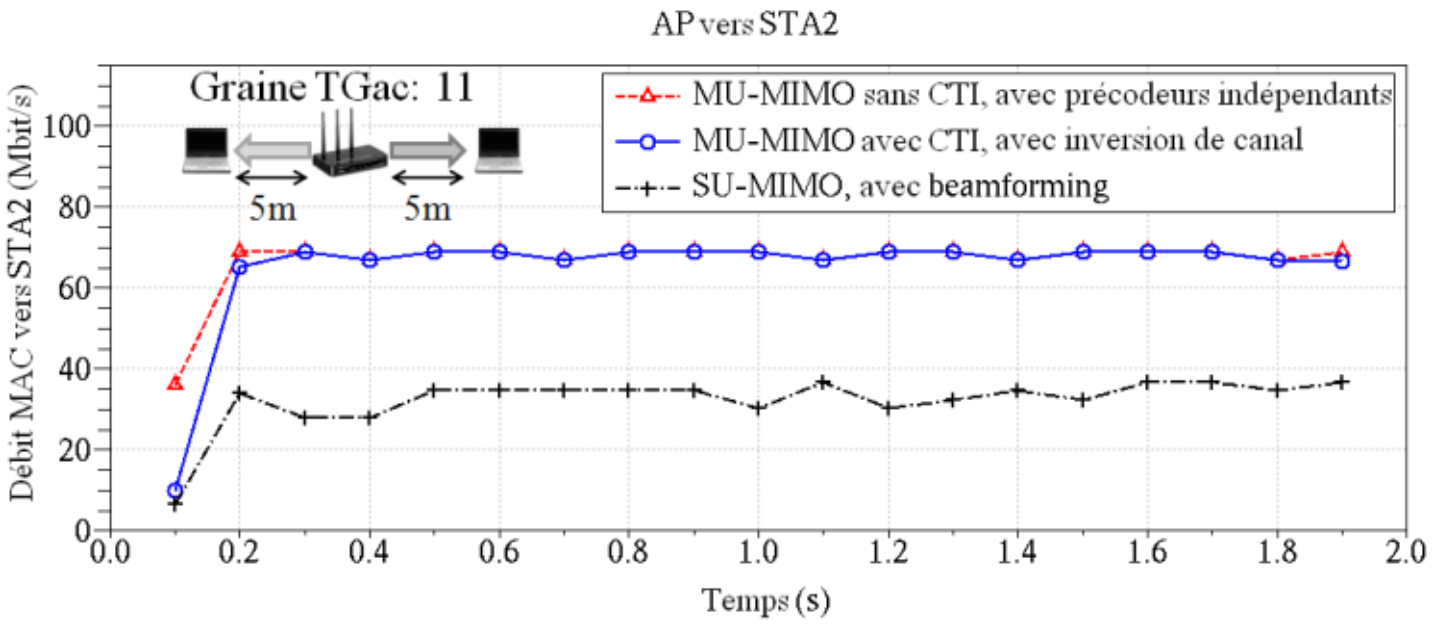

**Figure 6  Débit MAC vers la station 2 en fonction du temps pour les schémas étudiés pour des canaux descendants favorables**

Toutefois, si les canaux vers les deux stations sont corrélés, les performances du MU-MIMO changent comme le montrent les courbes de la Figure 7. Les débits atteints par le SU-MIMO et le MU-MIMO avec précodeurs indépendants indiquent que les canaux sont aussi bons qu'avant. Par contre leur corrélation impacte beaucoup le MU-MIMO, lorsque le CTI est pris en compte. Ce dernier peut être, par moment, moins intéressant que le SU-MIMO. Ceci n'a pu être observé que parce que les couches PHY et MAC ont été finement modélisées.

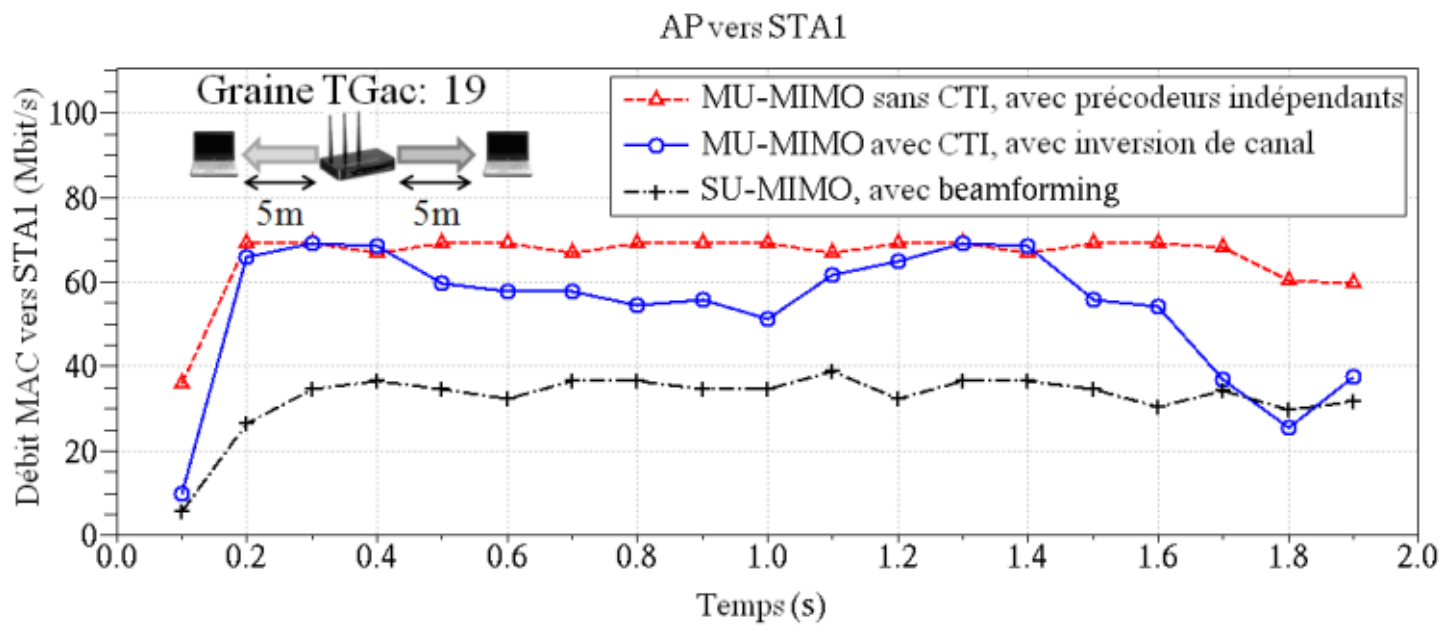

**Figure 7 Débit MAC vers la station 1 en fonction du temps pour les schémas étudiés pour des canaux descendants corrélés**

Si les deux stations sont placées plus loin de l'AP, les performances du MU-MIMO empirent comme l'illustre la Figure 8. L'avantage d'utiliser de la transmission multiutilisateur est perdu lorsque les canaux sont corrélés et que les signaux reçus sont faibles.

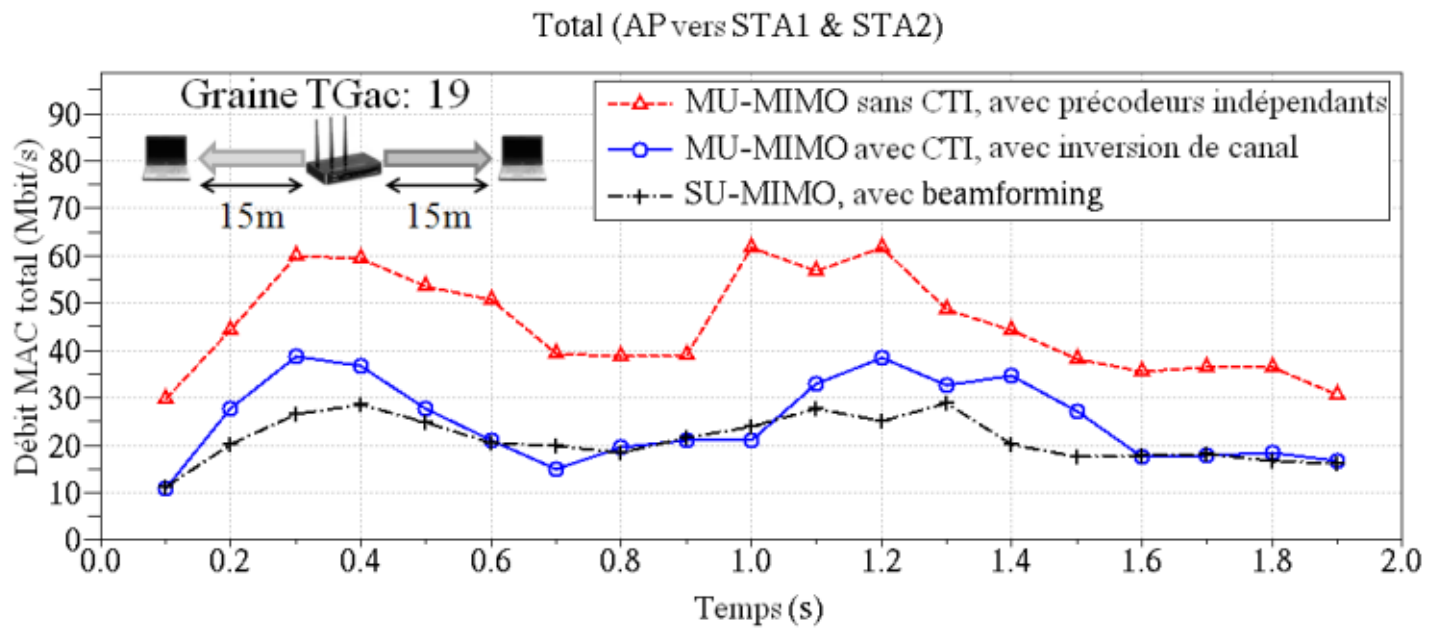

**Figure 8 Débit MAC total en fonction du temps pour les schémas étudiés pour des canaux descendants corrélés et des stations éloignées**

Nous pouvons donc voir que les performances du MU-MIMO sont grandement conditionnées par l'état du canal. Un algorithme de basculement entre MU-MIMO et SU-MIMO pourrait être envisagé.

# Analyse PHY+MAC de la période de sondage canal pour le MU-MIMO en IEEE 802.11ac

Nous avons vu que le MU-MIMO nécessite une connaissance précise des conditions des canaux vers les stations concernées. La procédure de sondage des canaux en question doit se faire de manière régulière. Aucune indication, quant à la fréquence, n'est donnée dans la norme. Néanmoins, des études montrent qu'elle doit se faire plus fréquemment que les 100 ms pris habituellement pour le SU-MIMO.

## Evaluation des performances d'un point de vue MAC

Dans un premier temps, nous allons déterminer l'intervalle à utiliser d'après une analyse MAC. La couche PHY y est abstraite par des tables de qualités remplies au préalable. Les débits MAC totaux obtenus pour différents intervalles de sondage avec le MU-MIMO et le SU-MIMO sont donnés dans la Figure 9. Ainsi, les performances peuvent être améliorées en prenant un intervalle de 40 ms (voire plus). Ce gain, quoique minime pour deux stations, devient plus important lorsqu'il y a plusieurs groupes et que la contention entre en jeu.

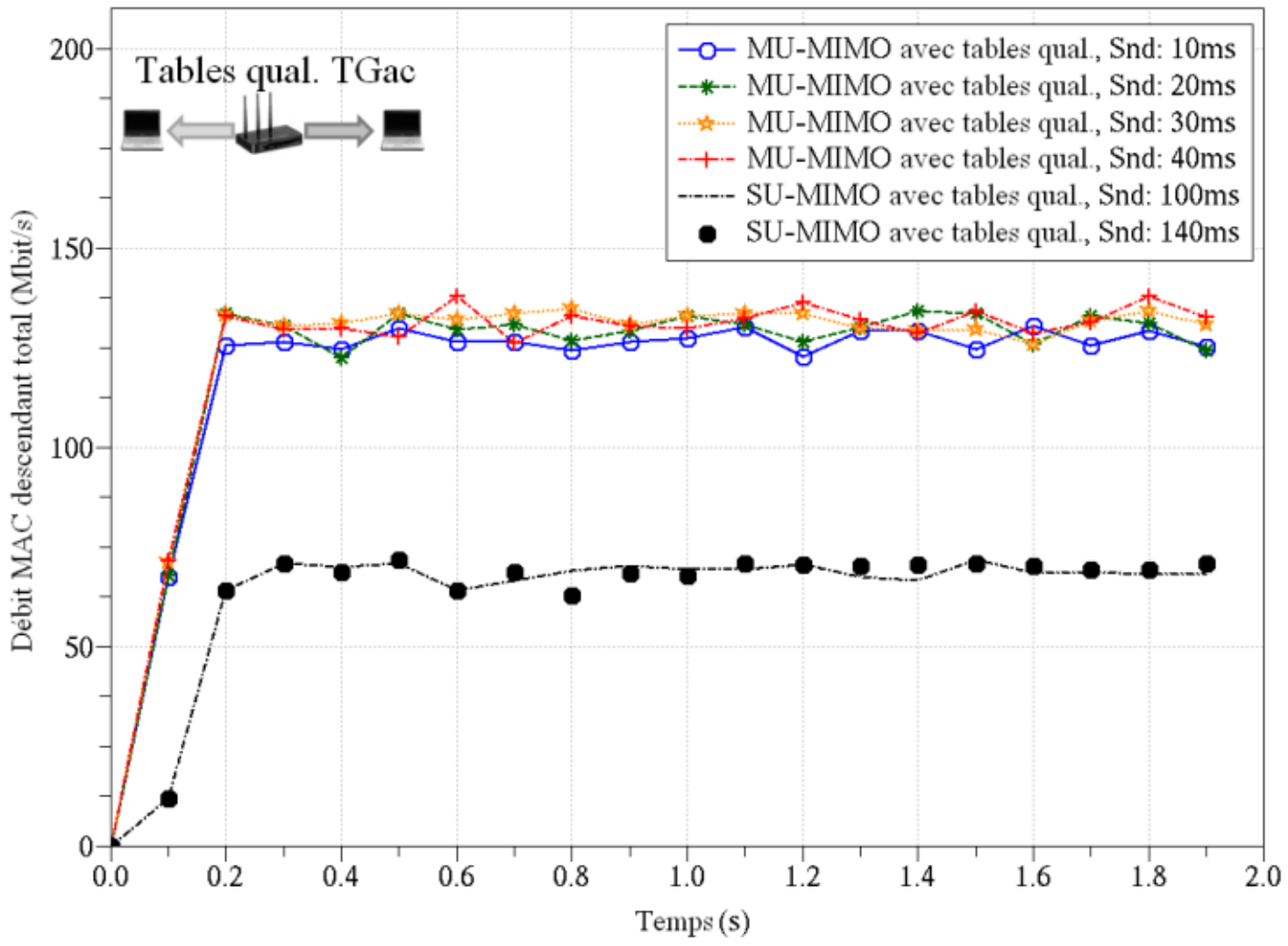

**Figure 9 Débit MAC total en fonction du temps pour différents intervalles de sondage. Les résultats sont obtenus en utilisant un simulateur MAC**

## Evaluation des performances d'un point de vue PHY+MAC

Si maintenant le vieillissement des estimées et le CTI sont pris en compte, les résultats donnés dans la Figure 10 sont obtenus. Les débits pour le SU-MIMO correspondent bien à ceux obtenus avec l'analyse MAC. Toutefois, pour le MU-MIMO, l'écart entre l'intervalle le plus court (10 ms) et le plus long (40 ms) est non seulement inversé, par rapport à ci-dessus, mais multiplié par plus d'un facteur 5 par moments.

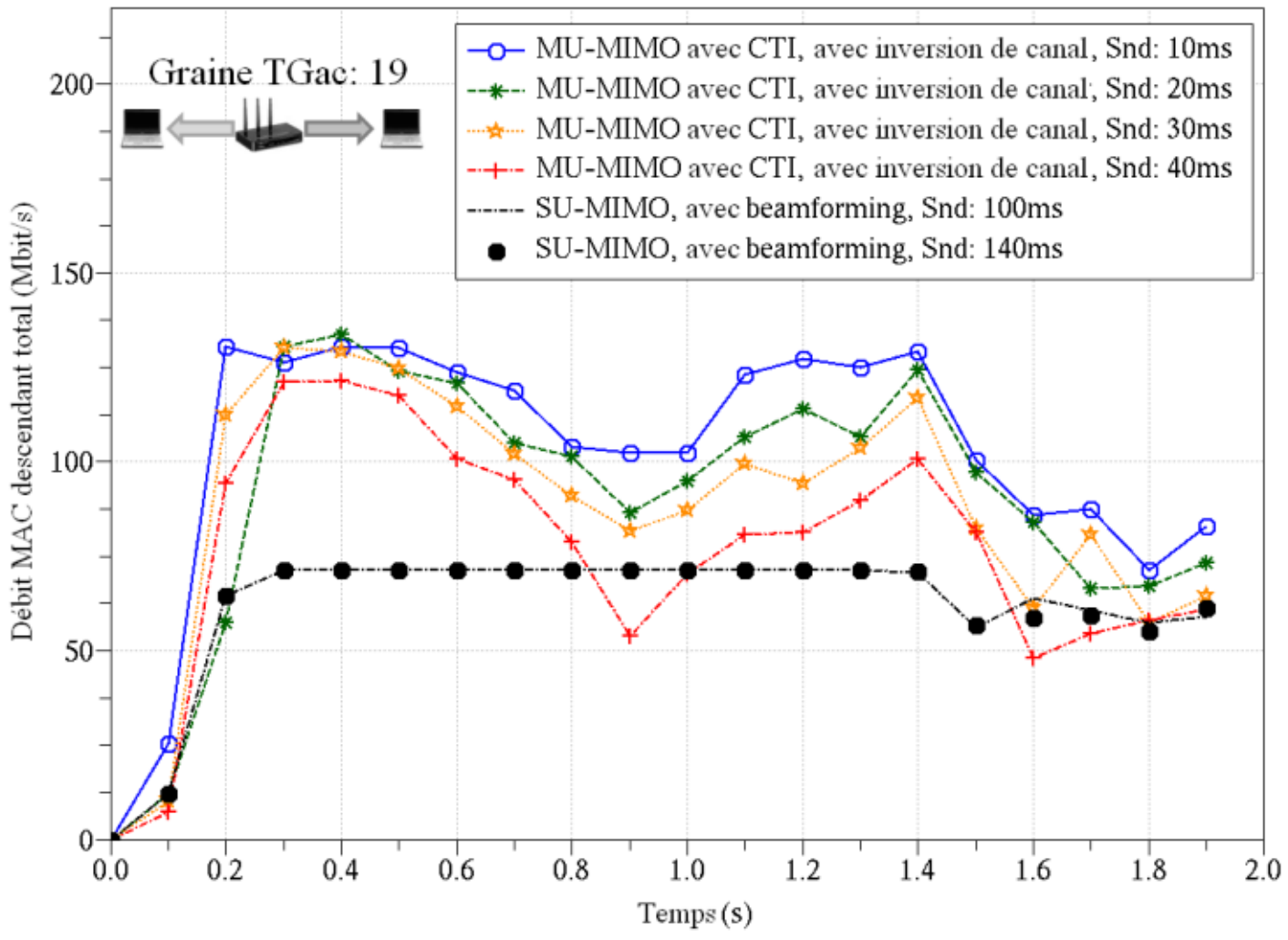

**Figure 10  Débit MAC total en fonction du temps pour différents intervalles de sondage. Les résultats sont obtenus en utilisant un simulateur PHY+MAC**

Prendre en compte les paramètres PHY conduit à des recommandations opposées : il vaut mieux faire un sondage fréquent pour gagner en débit. Ce constat est encore plus flagrant lorsque le nombre de stations est augmenté à trois. La puissance reçue par chaque station étant plus faible lorsque le MU-MIMO est activé, la connaissance du canal doit être encore plus précise qu'avec deux stations. A l'inverse, les performances sont grandement améliorées par l'utilisation de la diversité spatiale. Les antennes supplémentaires permettent en effet d'améliorer la réception et donc de réduire la fréquence de sondage.

# Optimisation cross-layer de réduction de consommation pour les futurs réseaux locaux sans fil

## La norme IEEE 802.11ah

La norme IEEE 802.11ah utilise la bande sous 1 GHz pour les communications M2M. Des applications, telles que le réseau de distribution d'électricité intelligent, les réseaux de capteurs, l'automatisation de processus industriels et le médical, sont visées. Ainsi les défis à relever sont les suivants : le grand nombre de stations associées, la robustesse des signaux pour une grande portée et la réduction des consommations.

Cette norme capitalise sur les avancées et le cadre posé par la norme IEEE 802.11ac (même schémas de modulation, même nombre de sous-porteuses) tout en y apportant des optimisations. Ainsi plusieurs trames MAC ont été raccourcies, dont l'acquittement (ACK, pour acknowledgment). Cet ACK court cross-layer permet d'acquitter en utilisant uniquement l'entête PHY. Le risque d'avoir un faux positif est minimisé par l'insertion d'un identifiant dans le champ de signalisation.

## Acquittement ultra-court pour la norme IEEE 802.11ah

Nous proposons un format d'ACK qui permet de réduire le surcoût associé tout en assurant une robustesse aux faux positifs. La trame d'ACK ultra-courte proposée ne consiste qu'en un champ de synchronisation (cf Figure 11 où elle est comparée à la trame d'acquittement classique). Or, la séquence utilisée est très robuste. Le risque de non détection de l'acquittement est donc réduit.

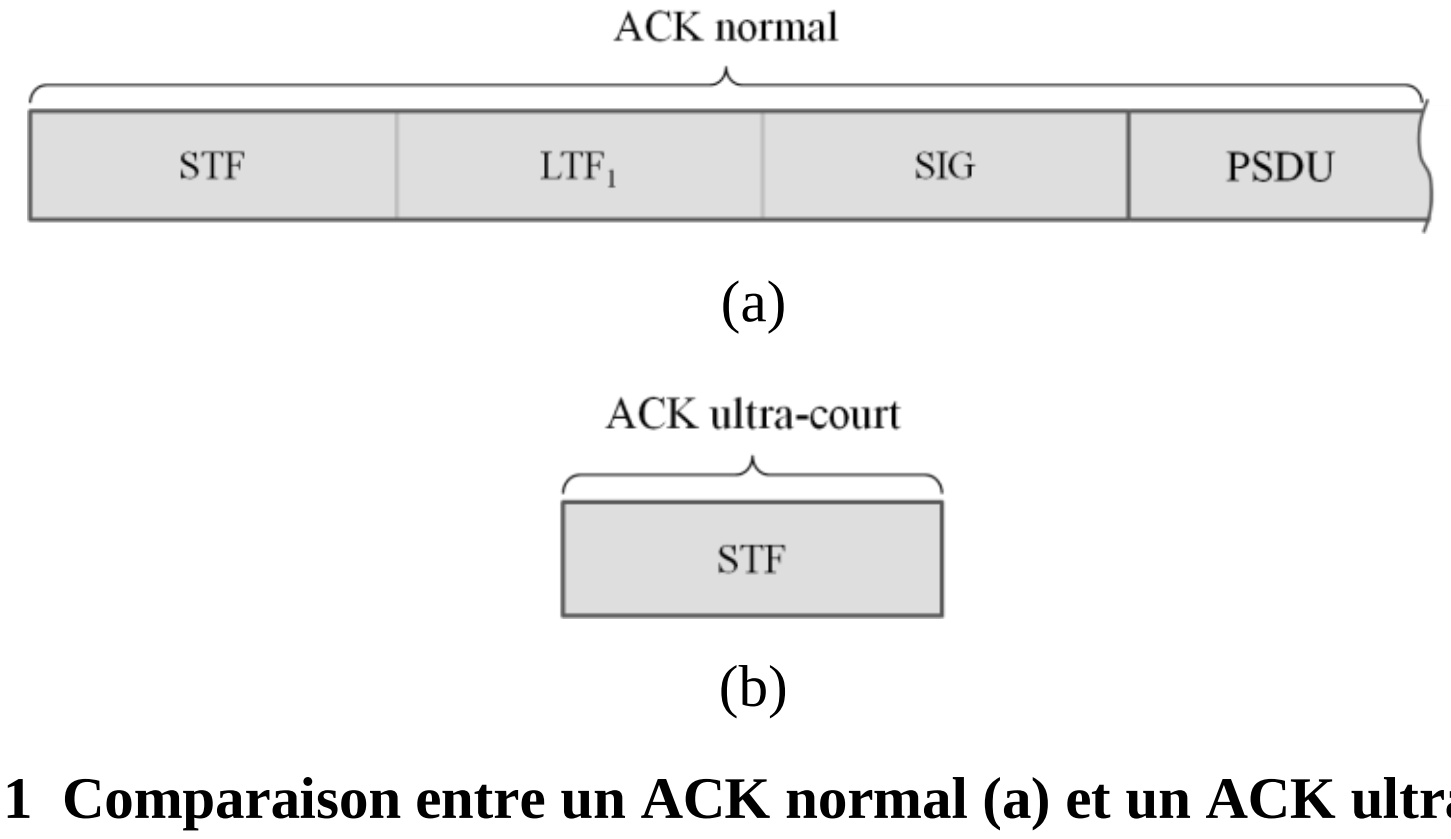

(a)

(b)

**Figure 11 Comparaison entre un ACK normal (a) et un ACK ultra-court (b)**

Pour minimiser le risque de faux positifs, nous proposons d'utiliser le retournement temporel. La station recevant la trame enregistre la réponse du canal et utilise sa version retournée pour mettre en forme l'ACK ultra-court. Ce signal sera donc concentré au niveau de l'émetteur de la trame.

Les Figure 12 (a) et Figure 12 (b) illustrent la capacité en termes d'utilisateurs d'un réseau sans fil, sans collision, pour l'automatisation de processus industriels et les réseaux de capteurs (respectivement). Dans le premier cas, une augmentation de 72 % et 32% par rapport à l'ACK normal et à l'ACK court (respectivement) peut être observée. Dans le deuxième cas, utiliser l'ACK ultra-court au lieu de l'ACK court permet de rajouter 9300 stations supplémentaires. Ce nombre double si l'ACK normal est pris comme référence.

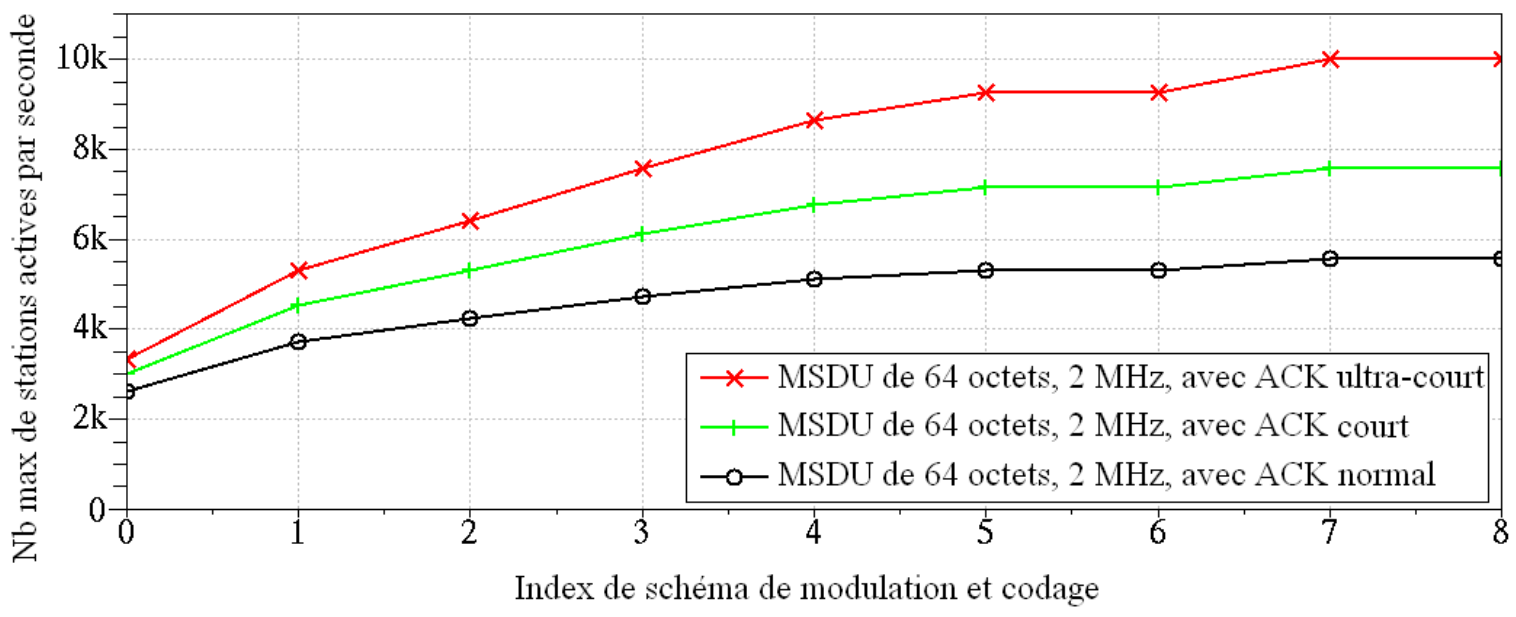

(a)

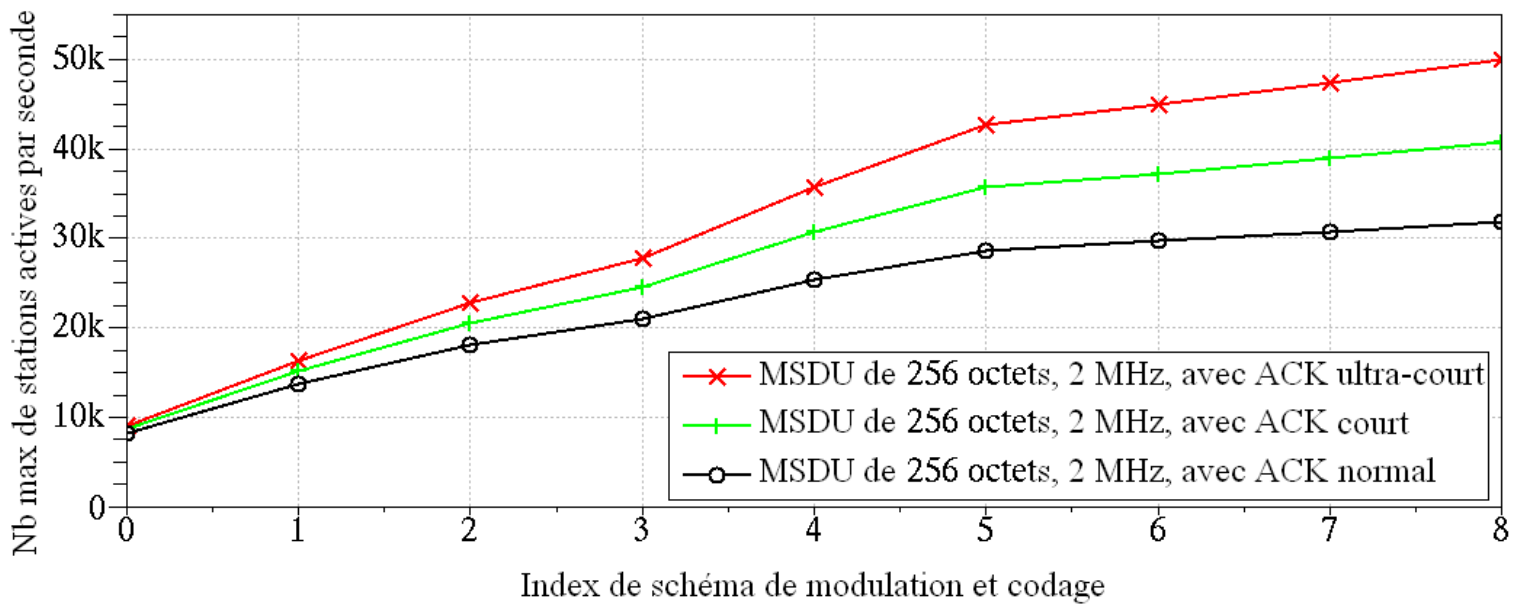

(b)

**Figure 12 Nombre maximum de stations actives par seconde en fonction du schéma de modulation pour l'automatisation de processus industriels, 64 octets, (a) et les réseaux de capteurs, 256 octets (b)**

La solution d'ACK ultra-courte offre donc des performances intéressantes pour le 802.11ah.

# Conclusions et perspectives

Durant cette thèse, nous avons étudié et proposé des techniques d'optimisation cross-layer principalement dans le cadre de la norme IEEE 802.11ac. Une nouvelle technique d'agrégation pour transmissions multicanaux, où la couche MAC tient compte de la couche PHY et inversement, a été proposée. Elle permet d'améliorer les performances en cas de collisions.

Lors de l'évaluation des performances de cette solution, nous avons mis en évidence que certaines fonctionnalités (ou phénomènes) impliquent autant la couche PHY que la couche MAC. Une évaluation réaliste de telles fonctionnalités (ou phénomènes) requiert donc une modélisation fine des mécanismes de ces deux couches. Un simulateur cross-layer conforme à la norme IEEE 802.11ac a donc été développé. C'est le premier simulateur, à notre connaissance, modélisant finement les techniques PHY et MAC de cette norme.

La technique phare du 802.11ac, le MU-MIMO, implique la couche PHY et la couche MAC. Nous avons donc utilisé le simulateur développé pour évaluer les performances 'réelles' de solutions MU-MIMO (c'est-à-dire tenant compte des couches PHY+MAC, et non juste de la PHY ou de la MAC) pour les comparer à celles de solutions SU-MIMO. L'impact de la périodicité du sondage canal sur le MU-MIMO a aussi été particulièrement étudié.

Enfin, une nouvelle structure de trame d'acquittement ultra-courte permettant d'améliorer la capacité des réseaux IEEE 802.11ah a été proposée. En conclusion, les différents principes d'optimisation cross-layer étudiés et modélisés par le simulateur PHY+MAC dans le cadre de la norme IEEE 802.11ac peuvent être appliqués aux différentes générations des futurs systèmes Wi-Fi.

# Acknowledgments

I would firstly like to thank my industrial supervisors Laurent and Philippe. They have taught me much. I would also like to thank them for implicating me in their normalization work. I have greatly appreciated working with them. A special thanks to Philippe, with whom we shared the same office, for his availability. I would also like to thank Pr. Hélard for his supervision. His experience and professionalism have been decisive in the orientations and the writing of this thesis.

My thanks also go to all the members of the WASA/CREM and WASA/WIDE teams at Orange Labs. I have much appreciated the different discussions we had, at lunch or on coffee breaks, with Marie, Marie-Hélène, Sophie, Isabelle, Hélène, Chrisitian, Jean-Philippe, Pierre S., Lin, Jean-Claude, Jean-Christophe, Yann, Pascal, Abdel, Denis, Jean-François, and Alain. The time I have spent with the other PhD students − Pierre A., Jean, Marco, Akl − were also very enjoyable. Thank you all.

I would like to thank Pr. Gorce, Dr. Le Martret, Pr. Terré, and Pr. Ström for having taken part in my examination committee. Their comments have been very constructive and encouraging. A special thanks to Pr. Ström for having come all the way up from Göteborg to attend the defense. I would also like to thank Marie for her state of the art on SDMA techniques, Hélène and Laurent their document on IEEE 802.11ac, and Nicolas for his documentation on rate adaptation algorithm and for the framework of the simulator. I would also like to thank Laurent, Philippe, Pr. Hélard, Yann, and Marco for having proof read this thesis.

Last but not least, I would like to thank my wife, Caroline, for being a source of encouragement. She has been a real support, especially during the writing of this thesis. I am also grateful that my family – Abbaby, Emmamy, Simret, Tigist, Gash Behaïlu – and friends – Gérard, Daniel, Nathan, Timothée, Alexis, Raphaël – were there for me. I could not have made it without them.

# Table of Contents

# Nomenclature

## Acronyms

The acronyms used throughout the thesis are summarized here. The meaning of an acronym is usually indicated once, upon first occurrence, at the beginning of each chapter.

| | | |
|---|---|---|
| **A** | AC | Access category |
| | AC_BE | Best effort AC |
| | AC_BK | Background AC |
| | AC_VI | Video AC |
| | AC_VO | Voice AC |
| | ACK | Acknowledgment |
| | ACK ID | ACK identifier |
| | ADC | Analog to digital converter |
| | AID | Association identifier |
| | AIFS | Arbitration inter-frame spacing |
| | A-MPDU | Aggregate MPDU |
| | A-MPDU_2 | A-MPDU of up to 2 MPDUs |
| | A-MPDU_8 | A-MPDU of up to 8 MPDUs |
| | A-MPDU_20 | A-MPDU of up to 20 MPDUs |
| | AMRR | Adaptive multi-rate retry algorithm |
| | A-MSDU | Aggregate MSDU |
| | AP | Access point |
| | AWGN | Additive white Gaussian noise |
| **B** | BA | Block ACK |
| | BAR | BA request |
| | BCC | Block convolutional coding |
| | BD | Block diagonalization |
| | BER | Bit error rate |
| | BF | Beamforming |
| | BO | Backoff |
| | BPSK | Binary phase-shift keying |
| | BSS | Basic service set |
| | BSSID | BSS identifier |
| **C** | CCA | Clear channel assessment |
| | CRC | Cyclic redundancy check |
| | CSD | Cyclic shift diversity |
| | CSI | Channel state information |
| | CSMA | Carrier sense multiple access |
| | CSMA/CA | CSMA with collision avoidance |
| | CTI | Crosstalk interference |
| | CTS | Clear to send |
| | CW | Contention window |
| **D** | DCF | Distributed coordination function |
| | DIFS | DCF inter-frame spacing |
| **E** | EDCA | Enhanced distributed channel access |
| **F** | FCS | Frame check sequence |
| | FDM | Frequency division multiplexing |
| | FEC | Forward error correction |
| | FIFO | First-in, first-out |
| | FFT | Fast Fourier transform |
| **G** | GI | Guard interval |
| **H** | HCF | Hybrid contention function |
| | HCS | Header check sequence |
| | HDMI | High-definition multimedia interface |
| | HT | High throughput |
| | HT-LTF | High throughput LTF |
| | HT-SIG | High throughput SIG |
| | HT-STF | High throughput STF |
| **I** | ID | Identifier |
| | iFFT | Inverse FFT |
| | IFS | Inter-frame spacing |
| | I/O | In/out |
| | ISM | Industrial, scientific, and medical band |
| **L** | LAN | Local access network |
| | LDPC | Low-density parity check |
| | LTF | Long training field |
| | LUT | Lookup table |
| **M** | M2M | Machine to machine |
| | MAC | Medium access control |
| | MCS | Modulation and coding scheme |
| | MCS0 Rep2 | MCS 0 with 2 block-wise repetition |
| | MIMO | Multiple-input, multiple-output |
| | MMSE | Minimum mean square error |
| | MPDU | MAC protocol data unit |
| | MSDU | MAC service data unit |
| | MU | Multiple-user |
| | MU-MIMO | Multiple-user MIMO |
| **N** | NAV | Network allocation vector |
| | NDP | Null data packet |
| | NDPA | NDP announcement |
| | ns-2 | Network simulator 2 |
| | ns-3 | Network simulator 3 |
| **O** | OFDM | Orthogonal frequency division multiplexing |
| **P** | PCF | Point coordination function |
| | PDU | Protocol data unit |
| | PER | Packet error rate |
| | PHY | Physical layer |
| | PIFS | PCF inter-frame spacing |
| | PLCP | Physical layer convergence procedure |
| | PMD | Physical medium dependent |
| | PPDU | PHY protocol data unit |
| | PSDU | PHY service data unit |
| **Q** | QAM | Quadrature amplitude modulation |
| | QoS | Quality of service |
| | QPSK | Quadrature phase-shift keying |
| **R** | RTS | Request to send |
| | Rx | Receive (or receiver) |
| **S** | SDMA | Spatial division multiple access |
| | SDU | Service data unit |
| | SIFS | Short inter-frame spacing |
| | SIG | Signaling field |
| | SIR | Signal-to-interference ratio |
| | SISO | Single-input, single-output |
| | SNR | Signal-to-noise ratio |
| | SS | Spatial stream |
| | STBC | Space-time block coding |
| | STF | Short training field |
| | STS | Space-time stream |
| | SU | Single-user |
| | SU-MIMO | Single-user MIMO |
| | SVD | Singular value decomposition |
| **T** | TGac | Very high throughput task group |
| | TGn | High throughput task group |
| | Tx | Transmit (or transmitter) |
| | TxOP | Transmit opportunity |
| **U** | UDP | User datagram protocol |
| **V** | VHT | Very high throughput |
| | VHT-LTF | Very high throughput LTF |
| | VHT-SIG-A | Very high throughput SIG part A |
| | VHT-SIG-B | Very high throughput SIG part B |
| | VHT-STF | Very high throughput STF |
| **W** | Wi-Fi | Wireless fidelity |
| | WLAN | Wireless local access network |
| | WMM | Wi-Fi multimedia |
| **X** | XLS | Cross-layer simulator (or simulation) |
| | XLS-CE | XLS with continuous estimation |
| **Z** | ZF | Zero-forcing |

# Notations

In this thesis notations are generally confined to the chapter in which they are used. The main notations are given here.

| | |
|---|---|
| $n_T$ | Number of transmit antennas |
| $n_R$ | Number of receive antennas |
| $n_R^c$ | Number of receive antennas at station $c$ |
| $N$ | Total number of served stations in MU-MIMO |
| $C$ | Selected set of stations in MU-MIMO transmission |
| $B$ | Number of groups of channels in horizontal aggregation |
| $P_{collision}$ | Probability of having a collision on a 20 MHz channel |
| $r_0, r_1, r_2, r_3$ | 1st, 2nd, 3rd, and 4th data rates for AMRR algorithm (resp.) |
| $c_0, c_1, c_2, c_3$ | 1st, 2nd, 3rd, and 4th retry counts for AMRR algorithm (resp.) |
| $n_T \times n_R$ | System with $n_T$ transmit and $n_R$ receive antennas |
| $1 \times 80$ MHz | Vertical aggregation scheme for 80 MHz channels |
| $2 \times 40$ MHz | Horizontal aggregation scheme with $B = 2$ for 80 MHz channels |
| $4 \times 20$ MHz | Horizontal aggregation scheme with $B = 4$ for 80 MHz channels |
| $\mathbf{H}$ | MIMO channel matrix with $n_R$ lines and $n_T$ columns |
| $\mathbf{x}$ | Vector of $n_T$ transmitted symbols |
| $\mathbf{y}$ | Vector of $n_R$ received symbols |
| $\mathbf{U}$ | Unitary matrix holding the $n_R$ left singular vectors of $\mathbf{H}$ |
| $\mathbf{V}$ | Unitary matrix holding the $n_T$ right singular vectors of $\mathbf{H}$ |
| $\mathbf{W}_c$ | Precoding matrix for station $c$ |

# Introduction

## Context and Motivation

Over the past two decades, wireless technologies have blossomed and have even become standard household and office appliances. The deployment facility and freedom of movement they enable have made them a very attractive solution for setting up local access networks (LAN). The number of shipped wireless LANs (WLAN) chipsets can testify of this “wireless revolution” [1]. Between 2010 and 2014, this number was expected to grow from 366 million units to 2 billion [1].

IEEE’s 802.11 standard, commonly known as wireless fidelity (Wi-Fi), has become de facto the reference in WLANs. Since the initial 1997 specification, the IEEE 802.11 standard has been extended to different ‘flavors’. Designations such as IEEE 802.11b, IEEE 802.11a, IEEE 802.11g, and IEEE 802.11n are nowadays commonly used even by non professionals. With the leap in throughput observed with every new generation of the IEEE 802.11 standard, expectations for the next generation are even higher. Issues like throughput, range, reliability, and power consumption are to be better tackled, while ensuring a certain level of compatibility with existent appliances. This is quite a challenge considering all the work already done!

IEEE 802.11 standards address specifications of physical (PHY) and medium access control (MAC) layer mechanisms. The many technical improvements were naturally brought to each layer in a rather traditional manner. A clear line was drawn between the PHY and MAC layers so as to improve each layer independently of the other. This method has proven to be quite useful and important optimizations have been established. However, optimizing the two layers independently may lead to redundancies and may even be counter productive. Therefore, cross-layer optimization, i.e. taking into account both layers in the scope, can be an interesting exploration path for further enhancements. IEEE 802.11ac and IEEE 802.11ah have already started using this type of optimization.

Cross-layer optimization can be decomposed in different objectives: removing redundancy between PHY and MAC layers, migrating functionalities from one layer to another, or improving interactions between layers, so as to improve performance. It is thus very interesting to see the global picture in order to put optimization effort on the winning horse. Special care should however be taken so as not to end up with “spaghetti design” [2], where all functionalities are intertwined in a chaotic mass.

# Interesting Contributions

During this thesis, we have concentrated on studying cross-layer optimization techniques mainly for the IEEE 802.11ac standard. A new multichannel aggregation scheme involving cross-knowledge between PHY and MAC layers has been proposed to improve performance in collision-prone environments.

While testing this solution, we have shown that some functionalities or phenomena directly involved PHY and MAC layers. An accurate modeling of both PHY techniques and MAC mechanisms is thus needed to have a realistic characterization of such functionalities and phenomena. A cross-layer simulator has thus been implemented and optimized. To the best of our knowledge, this is the first simulator incorporating detailed PHY and MAC functionalities for the IEEE 802.11ac standard.

The multiple-user multiple-input, multiple-output (MU-MIMO) technique, which is one of the main innovations of 802.11ac, needs both PHY and MAC layer considerations. We have thus used the implemented cross-layer simulator to study and evaluate the performance of MU-MIMO and compare it with the single-user MIMO (SU-MIMO) technique. The aim of these studies was to evaluate the ‘real’ gains of MU-MIMO solutions (accounting for PHY+MAC) over SU-MIMO solutions and not the generally accepted ones (accounting for MAC only or PHY only). The impact of the channel sounding interval has particularly been studied.

Cross-layer optimization principles are not limited to 802.11ac. Thus we have also proposed an ultra short PHY layer version of acknowledgment frames for overhead reduction in machine to machine (M2M) IEEE 802.11ah communications. The latter standard is a natural extension of 802.11ac to M2M low power communications.

# Outline

The thesis is organized as follows:

In Chapter 1, we first introduce the fundamentals of WLANs. This chapter quickly gives the background of the PHY and MAC techniques addressed in the orthogonal frequency division multiplexing-based IEEE 802.11 standards. We then present the fundamental functionalities of some of its amendments. Some cross-layer guidelines along with two interesting cross-layer solutions are also given. We finish by presenting the different types of simulation platforms which can be used in PHY/MAC studies of WLANs.

In Chapter 2, we present in detail the new functionalities of the IEEE 802.11ac standard. 802.11ac's PHY and MAC layer functionalities are presented, with a special focus on channelization and MU-MIMO.

In Chapter 3, we propose a new cross-layer multichannel aggregation. MAC protocol data units are aggregated per group of sub-channels so as to mitigate the effects of a collision on one of the used sub-channels. This cross-layer solution is compared to the classical aggregation scheme and performance is evaluated for different collision probabilities. The scaling offset on the analog to digital converter, due to collisions, is also considered.

In Chapter 4, we present the IEEE 802.11n/ac PHY+MAC simulation platform we have implemented and used in the rest of the thesis. The purpose of such an implementation is firstly discussed. The architecture is then given and the obtained model is validated with ns-2 (network simulator 2) simulations. A hidden node scenario is simulated on the cross-layer simulation platform and ns-2. We thus show the interest of using joint PHY+MAC simulations to model complex phenomena in a more realistic manner than just using MAC simulations.

In Chapter 5, we investigate the performance of MU-MIMO and SU-MIMO in the 802.11ac amendment using the cross-layer simulator. This way the relevance of accounting for crosstalk interference in MU-MIMO simulations, while accounting for sounding induced overhead, is evaluated. The analysis is done for single-antenna stations first, so as to discriminate the characteristics of each transmission scheme. It is then extended to a multiple-antenna station scenario. The aim of this section is to compare the 'real' gains of MU-MIMO, over SU-MIMO, with the generally accepted ones.

In Chapter 6, we investigate the channel sounding interval to choose for MU-MIMO in the 802.11ac amendment using again the cross-layer simulator. A compromise between precise channel knowledge, for estimate-error-sensitive precoding purposes, and feedback overhead reduction is needed. We study the impact of this interval on MU-MIMO transmissions for different scenarios and extract some design guidelines.

In Chapter 7, we propose a cross-layer optimization for the IEEE 802.11ah standard. This amendment of the IEEE 802.11 standard addresses M2M communications. Therefore power consumption and channel sharing are important issues. We propose to use an ultra short PHY layer acknowledgment to reduce overhead, thus increasing the maximum number of associated stations, and power consumption. MAC level gains brought by this solution are evaluated.

Finally, we conclude this thesis and give some potential research directions.

# List of Publications

## Patents

- L. Cariou, P. Christin, and G. Redieteab, “Methods for transmitting and receiving data using a plurality of radio channels, transmission devices and corresponding recipient, signal and computer program,” *France Telecom*, FR20100051670 20100308, WO/2011/110779, Sept. 2011.
- P. Christin, L. Cariou, and G. Redieteab, 1250080, *France Telecom*, filed in Feb. 2012.

## Journal Paper

- G. Redieteab, L. Cariou, P. Christin, J.-F. Hélard, and N. Cocaign, “Novel cross-layer simulation platform to include realistic channel modeling in system simulations,” *International Journal on Computer Networks and Communications*, vol. 4, no. 4, 18 pages, Jul. 2012.

## International Conferences

- G. Redieteab, L. Cariou, P. Christin, and J.-F. Hélard, “Cross-layer multichannel aggregation for future WLAN systems,” *IEEE International Conference on Communication Systems*, pp. 740-745, 6 pages, Nov. 2010.
- G. Redieteab, L. Cariou, P. Christin, and J.-F. Hélard, “SU/MU-MIMO in IEEE 802.11ac: PHY+MAC performance comparison for single antenna stations,” *Wireless Telecommunications Symposium*, pp. 1-5, 5 pages, Apr. 2012.
- G. Redieteab, L. Cariou, P. Christin, and J.-F. Hélard, “PHY+MAC channel sounding interval analysis for IEEE 802.11ac MU-MIMO,” *International Symposium on Wireless Communication Systems*, 5 pages, Aug. 2012.

Chapter 1

# Wireless LAN Fundamentals and Techniques

In this chapter we quickly review some fundamental physical (PHY) and medium access control (MAC) layer techniques behind the IEEE 802.11a/g/n/ac standards[1]. A brief overview of the evolution of orthogonal frequency division multiplexing (OFDM) based IEEE 802.11 standards from 802.11a/g to 802.11ac and 802.11ah follows. We then give some cross-layer optimization guidelines and expose two already implemented interesting cross-layer solutions. Simulation results obtained for cross-layer solutions depending on the used platform, we finish this chapter by exposing the different types of simulators which can be used for PHY/MAC studies of wireless local access networks (WLAN).

## 1.1 Physical Layer

The aim of this section is to give some PHY layer material related to the OFDM-based IEEE 802.11 amendments. The PHY layer is split into two sub-layers, the physical layer convergence procedure (PLCP) sub-layer and the physical medium dependent (PMD) sub-layer. The PLCP sub-layer prepares frames for radio transmission. The PMD sub-layer turns the bits into radio waves in the air. We try to expose the principles, rather than the underlying mathematics, behind the techniques of the latter sub-layer. However, some key equations are given so as to better understand precoding techniques. We also expose framing principles of the PLCP sub-layer.

[1] In this thesis, we only address orthogonal frequency division multiplexing based IEEE 802.11 amendments. Thus the 1997 original IEEE 802.11 standard [3] and the 1999 IEEE 802.11b extension [38] are only cited for comparative reasons. In addition, solutions using 60 GHz band are not studied.

### 1.1.1 Orthogonal Frequency Division Multiplexing

In the initial specification of the IEEE 802.11 standard (1997) [3], three PHY layers were published: frequency-hopping spread-spectrum radio PHY, direct-sequence spread-spectrum PHY, and infrared light PHY. OFDM PHY was introduced with the IEEE 802.11a version (1999) [4]. Its multiple assets have made it the building block of following versions, i.e. IEEE 802.11g (2003) [5], IEEE 802.11n (2009) [6], and IEEE 802.11ac (still in draft) [7].

In order to understand why OFDM is well suited to wideband WLAN environments [8] (wideband implying the use of a relatively large bandwidth for transmission), we have to first comprehend the impediments of such environments.

#### 1.1.1.1 Multipath Propagation

An elementary fact is that WLANs are seldom set up in completely empty spaces, way above the ground. Therefore, the direct path, i.e. directly linking the transmitter to the receiver, is not the only transmission path. There are as many reflected paths as there are reflectors, as illustrated in Figure 1.1. The receiver will thus see the combination of many delayed and attenuated replicas of the same transmitted signal. However these experience different propagation conditions. The destructive superposition of the replicas, for a given frequency, can cause serious signal power loss [9].

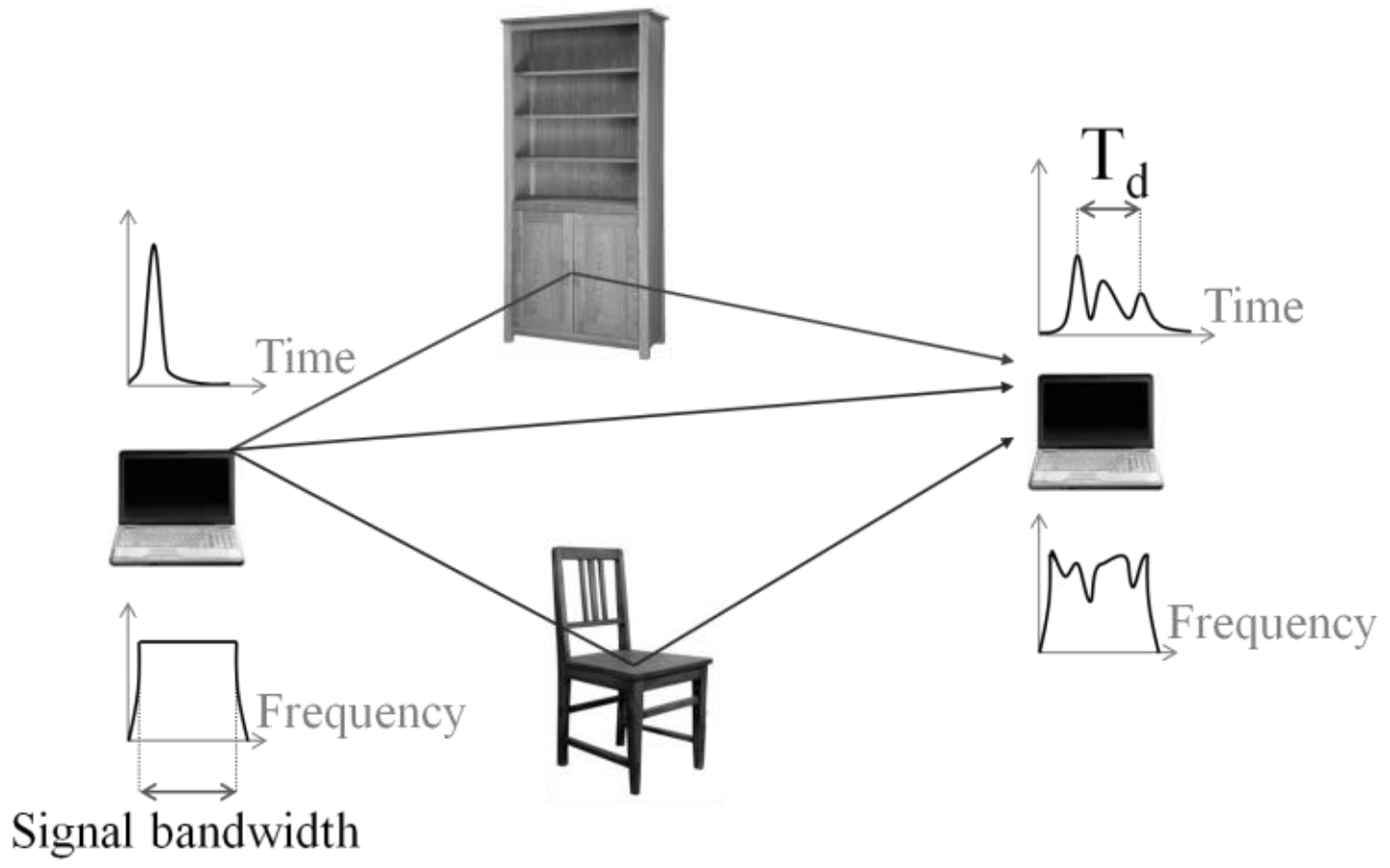

**Figure 1.1 Multipath propagation**

We shall however note that multipath propagation ensures communication and coverage even when there is no direct path [10].

#### 1.1.1.1.1 Frequency Selectivity

The difference in propagation time between the longest and shortest path is a good characterization parameter [11]. This duration is called the delay spread (labeled ‘$T_d$’ in Figure 1.1). Its inverse, the coherence bandwidth, directly gives the range of frequencies over which the channel can be considered ‘flat’ [12], i.e. locally linear. If the signal’s bandwidth is greater than this coherence bandwidth, the channel is considered as ‘frequency selective’. Otherwise the channel is referred to as ‘flat fading’ [11]. Wideband WLAN channels are in the first category. We will note that the environment changes over time also inducing time selectivity.

#### 1.1.1.1.2 Inter-symbol Interference

The delay incurred by multipath propagation smears the received signal. Considering that information is sent through chunks of signals, i.e. symbols, this is especially problematic because successive symbols overlap. Errors are thus introduced in symbol decisions at the receiver. This is known as inter-symbol interference and is a barrier to achieving high data rate transmission [9].

### 1.1.1.2 Frequency Division Multiplexing

Upon reception, the receiver tries to compensate the effects of the channel by applying the good filter. This process is called equalization. If the channel is frequency selective, advanced equalization is required [13]. Multiple tap filtering process greatly increases complexity.

Frequency division multiplexing (FDM) solves this problem by dividing the frequency selective channel into many flat fading sub-channels, as illustrated in Figure 1.2. A single channel tap is sufficient to represent each sub-channel [11], thus simplifying equalization. This is true as long as $\Delta f$, the spacing between subcarriers (the central frequencies of each sub-channel), is smaller than the coherence bandwidth.

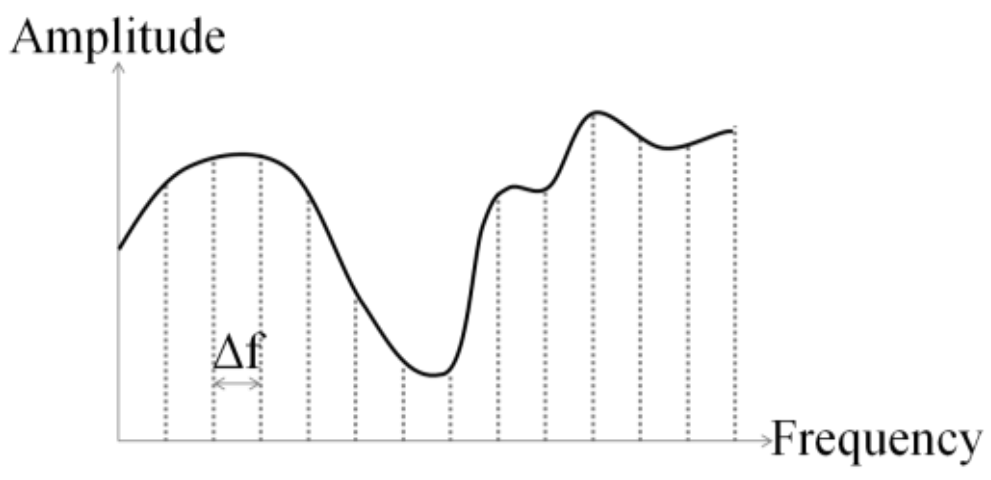

**Figure 1.2 Frequency division to obtain flat fading channels**

The other advantage is that FDM enables coding over sub-channels rather than over the whole bandwidth, as with single carrier modulation (see Figure 1.3 [8]). A high-speed data stream using the whole bandwidth is divided into many low-speed data streams each using a sub-channel, while transmitting the same amount of information. The same 8 coded symbols $c_i$ in Figure 1.3 (a) can be found in Figure 1.3 (b) within the same duration and using the same total bandwidth. Frequency diversity can be obtained by coding across the symbols in different subcarriers, through interleaving [11]. The larger the channel, the less correlated sub-channels tend to be because of frequency selectivity, leading to greater diversity. The lengthening of symbol duration enables to mitigate the impact of inter-symbol interference. FDM systems can thus achieve high data rate transmissions [9].

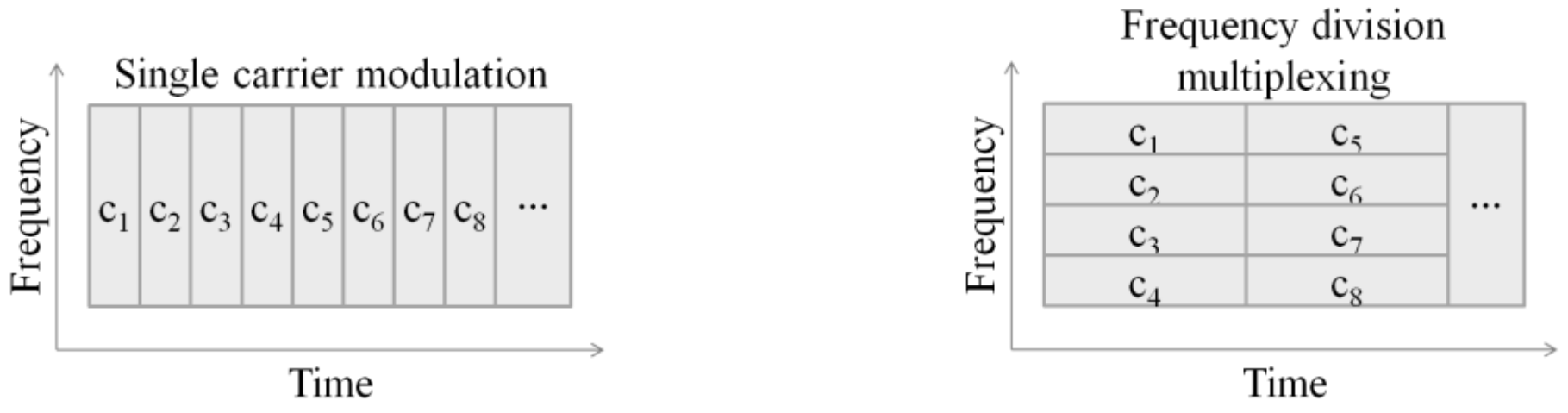

(a) (b)

**Figure 1.3 Comparison between single carrier modulation (a) and FDM (b)**

### 1.1.1.3 Orthogonality

We now come to the first letter of OFDM, orthogonality. In simple FDM, subcarriers are to be placed far apart so as to avoid mutual interference. To avoid wasting bandwidth, thus transmission capacity, OFDM selects sub-channels that overlap without interfering with each other. This double constraint leads to the use of sinc[1] functions for each subcarrier in OFDM, as illustrated in Figure 1.4.

The reader can see that at each subcarrier frequency (dot on the peak), all other subcarriers do not contribute to the overall waveform (dot at zero) [14]. These narrow subcarriers packed tightly, OFDM is bandwidth efficient [8]. This comes at a price though. OFDM is sensitive to frequency shifts, which may cause interference between subcarriers. These frequency shifts are due to environment mobility, also known as Doppler effect, or clock drifts [14]. Phase tracking, through known or pilot symbols, is thus necessary to maintain orthogonality [8].

[1] The normalized sinc function of $x$ is defined as $\mathrm{sinc}(x) = \sin(\pi \cdot x)/(\pi \cdot x)$ for $x \neq 0$ and $\mathrm{sinc}(0) = 1$.

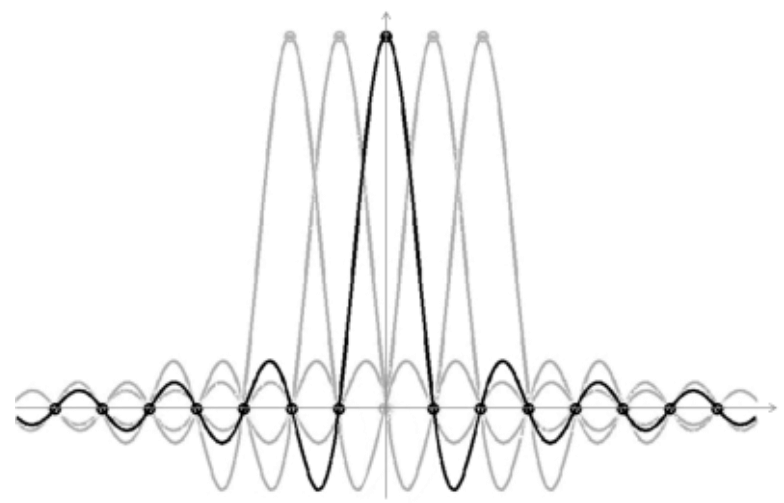

**Figure 1.4  Frequency orthogonality with sinc function**

### 1.1.1.4 Cyclic Prefix Guard Interval

Despite the lengthening of symbol duration through FDM, inter-symbol interference between adjacent OFDM symbols still impairs performance. A guard interval (GI) is used to absorb the impact of inter-symbol interference [8]. GI is simply padding inserted between two OFDM symbols. It should be longer than the delay spread $T_d$.

However ensuring phase continuity between consecutive OFDM symbols is important for maintaining subcarriers orthogonality and avoiding high frequency components. A good solution is to extend the symbol by using a copy of the last portion of the OFDM symbol itself, as illustrated in Figure 1.5 [9].

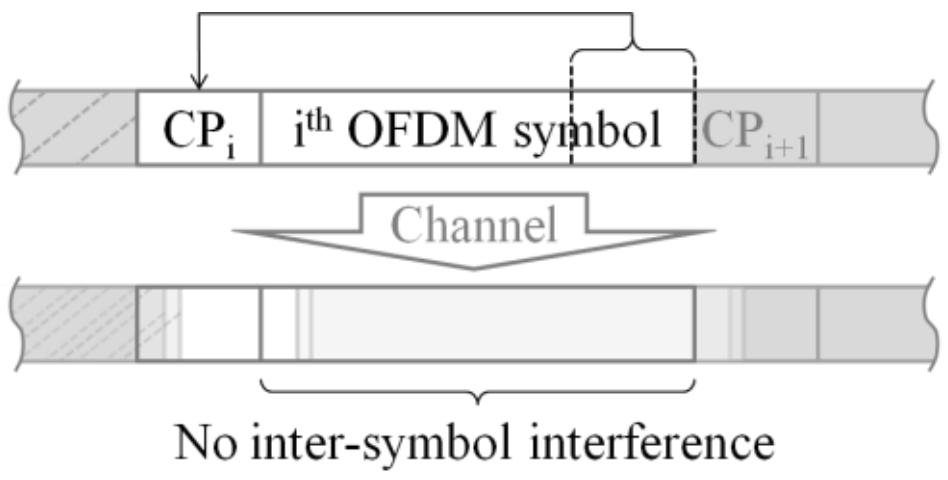

**Figure 1.5  Cyclic prefix guard interval**

This way, even if some subcarriers have more delay than others, the receiver can correctly retrieve all subcarriers, having been cyclically extended. This technique is called cyclic prefix GI [14].

### 1.1.1.5 Synchronization

We have seen that frequency synchronization is an important requirement for OFDM. In addition, locking on to randomly incoming signals and doing time synchronization is crucial for the correct decoding of the rest of the frame. Start-of-frame detection is done through a special sequence, which is known by all stations and is chosen for its good correlation properties. The synchronization field is thus the first field of the PLCP preamble, which itself precedes data transmission. This field is also used for initial frequency offset estimation and initial time synchronization. Finally, automatic gain control can be done through it, so as to calibrate the analog to digital converter for the rest of the frame [8].

#### 1.1.1.6 Channel Estimation

The wireless channel evolves over time. A channel estimation process is thus needed with every incoming frame so as to keep up with these changes. Another special sequence, which is also known by all stations, is used to this aim (once the receive chain is synchronized). The obtained channel matrix estimate is used for equalizing the rest of the frame, this estimate having only one tap per sub-carrier because of OFDM (see section 1.1.1.2). Frequency offset estimation and time synchronization can also be fine-tuned through it [14].

#### 1.1.1.7 In-band Signaling

In-band signaling (i.e. using the same frequency band as data symbols) is contained in the PLCP header[1]. This header follows the PLCP preamble and provides signaling information on data symbols. Indeed the receiver needs to precisely know which modulation and coding are applied by the transmitter on data symbols. The amount of transmitted data is also important. If special options are activated and required for correct decoding, they should be indicated in the PLCP header. If the latter is misinterpreted the data contained in the frame cannot be retrieved [8,14]. This field is thus often 'protected' by an integrity check.

### 1.1.2 Multiple-input, Multiple-output

#### 1.1.2.1 From Single Antenna Systems to Multiple Antenna Systems

Up until 2004, most IEEE 802.11 devices had a single antenna. To increase performance though, some devices had two antennas and selected the 'best' antenna at a given time to improve reception[2].

---

[1] The PLCP preamble and PLCP header are often referred to by just using 'PHY preamble' or 'PHY header' to designate both.

[2] Some even used the two antennas to concentrate energy, using unconventional beamforming techniques. All the while, only one spatial stream could be transmitted.

Still, Wi-Fi devices had a single-input, single-output (SISO) chain [14]. Systems are termed SISO with reference to the transmit (Tx) chain's single input to the environment and the receive (Rx) chain's single output from the environment, as illustrated in Figure 1.6 [8]. Multipath propagation is of course implicitly accounted for.

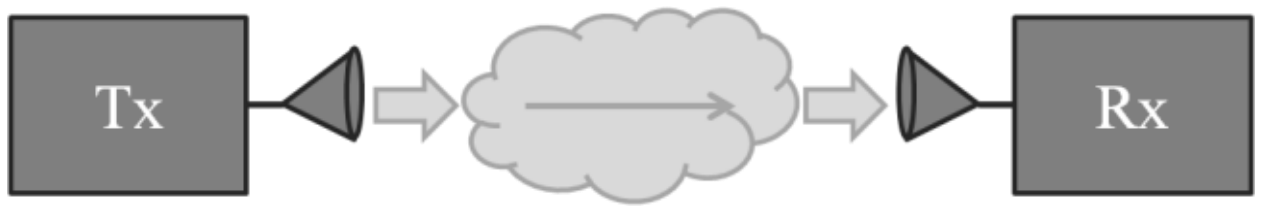

**Figure 1.6 Single-input, single-output system**

With the IEEE 802.11n amendment [6], multiple-input, multiple-output (MIMO) systems were introduced. In these systems, multiple antennas are used both at the transmitter and receiver, as illustrated in Figure 1.7. The global antenna configuration, with $n_T$ transmit antennas and $n_R$ receive antennas, is described with the shorthand '$n_T \times n_R$' notation [14].

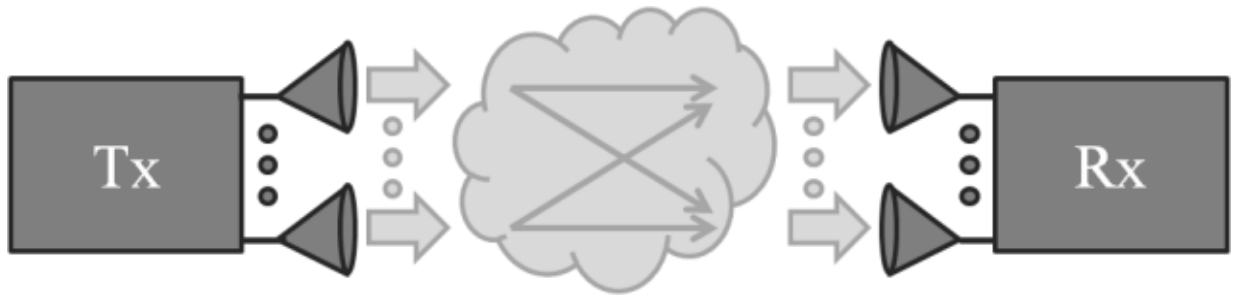

**Figure 1.7 Multiple-input, multiple-output system**

The advantage of using multiple antennas is twofold. The receiver can benefit from spatial diversity, and the transmitter can spatially multiplex independent streams of data.

### 1.1.2.2 Mathematical Model of MIMO Systems

Before detailing these two advantages, it is important to better expose the principle of MIMO. To that sense, a short presentation of the mathematical model can be quite useful. Indeed, techniques such as beamforming (section 1.1.3) and spatial division multiple access (section 1.1.4) are based on such an approach.

A $n_T \times n_R$ MIMO channel can be represented using a matrix $\mathbf{H}$ with $n_R$ lines and $n_T$ columns. The coefficient $h_{ik}$ of this matrix represents the transfer function from the $k^{\text{th}}$ transmitter to the $i^{\text{th}}$ receiver. $\mathbf{x}$ denotes the vector of $n_T$ transmitted symbols, $\mathbf{y}$ denotes the vector of $n_R$ received symbols, and $\mathbf{z}$ denotes the vector of $n_R$ additive white Gaussian noise (AWGN) perceived by the receiver [15]. With this notation, the model for MIMO systems is:

$$\mathbf{y} = \mathbf{H} \cdot \mathbf{x} + \mathbf{z} \tag{1.1}$$

To clarify this compact equation, let us a take a $2 \times 2$ system as an example:

$$\begin{bmatrix} y_1 & y_2 \end{bmatrix} = \begin{bmatrix} h_{11} & h_{12} \\ h_{21} & h_{22} \end{bmatrix} \cdot \begin{bmatrix} x_1 \\ x_2 \end{bmatrix} + \begin{bmatrix} z_1 & z_2 \end{bmatrix} \tag{1.2}$$

$$\begin{cases} y_1 = h_{11} \cdot x_1 + h_{12} \cdot x_2 + z_1 \\ y_2 = h_{21} \cdot x_1 + h_{22} \cdot x_2 + z_2 \end{cases} \tag{1.3}$$

The receiver can estimate the sent $x_k$ symbols by using the received $y_i$ symbols and pre-estimates of the $h_{ik}$ channel taps. One channel tap being used to characterize the response of a transmit-receive antenna pair, this model assumes a flat fading channel. This is true for correctly dimensioned OFDM systems (see section 1.1.1.2). Thus, this MIMO model applies for every subcarrier of the OFDM system. This explains why OFDM and MIMO are often used together.

#### 1.1.2.3 Spatial Diversity

In equation (1.3), we can see that if the same information is sent twice, i.e. $x_1 = x_2$, the receiver has two different versions of the same signal. The two versions will have followed different paths. Decoding is thus improved with this diversity. Techniques such as space-time block coding (STBC) can be used to improve robustness [16,17]. The general model given in equation (1.1) is still applicable. Spatial diversity was the first multiple antenna technique to be used in IEEE 802.11 systems, mostly with multiple antennas used only for receive diversity [14]. With multiple antennas at both the transmitter and receiver, the robustness of the link is greatly improved. More information can thus be sent at once because of this gain in decoding performance [8].

#### 1.1.2.4 Spatial Multiplexing

Spatial multiplexing is rather straightforward from equation (1.1). If the $n_T$ transmitted symbols are all different, the diversity obtained from the difference in antenna position, i.e. the spatial dimension, can be exploited to transmit independent data streams. The maximum data rate is thus literally multiplied by the number of independent data streams, or spatial streams [8]. A MIMO system can support up to $\min(n_T, n_R)$ spatial streams [16].

### 1.1.2.5 Channel State Information

As suggested in section 1.1.2.2, the receiver uses information from the $h_{ik}$ channel taps to estimate the sent symbols. This channel knowledge is referred to as channel state information (CSI). We will note that additional processing is often needed to compensate for inter-stream interference (because $h_{ik,i\neq k} \neq 0$). CSI is normally available at the receiver through a channel estimation process involving the PLCP preamble. CSI can also be returned to the transmitter for improved performance [15,16].

## 1.1.3 Beamforming

The capability to use CSI to perform beamforming was introduced with the IEEE 802.11n amendment [6]. Antenna specific weights are applied to the transmitted[1] signal so as to make the most out of the channel (i.e. by adapting transmission to the current channel state). Reception performance is thus improved, especially at low signal-to-noise ratio (SNR) and when there are more transmit than receive antennas [15,18].

### 1.1.3.1 Singular Value Decomposition

Singular value decomposition (SVD) is the most common way of computing transmitter weights [8]. The SVD of channel matrix $\mathbf{H}$ with $n_R$ lines and $n_T$ columns is as follows:

$$\mathbf{H} = \mathbf{U} \cdot \mathbf{\Sigma} \cdot \mathbf{V}^* \tag{1.4}$$

$\mathbf{U}$ and $\mathbf{V}$ are unitary matrices[2] holding the $n_R$ left singular vectors and $n_T$ right singular vectors, respectively [19]. Each of the $\mathbf{u}_k$ left singular vectors has $n_R$ lines and each of the $\mathbf{v}_i$ right singular vectors has $n_T$ columns. $\mathbf{\Sigma}$ is a diagonal matrix holding the $\min(n_T, n_R)$ singular values $\sigma_s$ [20]. The diagonal values in $\mathbf{\Sigma}$ are non-negative and ordered in decreasing order. These singular values and vectors are defined as follows:

$$\begin{aligned} \mathbf{H} \cdot \mathbf{v}_i &= \sigma_i \cdot \mathbf{u}_i \\ \mathbf{H}^* \cdot \mathbf{u}_k &= \sigma_k \cdot \mathbf{v}_k \end{aligned} \tag{1.5}$$

More details and examples can be found in [20].

[1] In IEEE 802.11n, only transmit beamforming is considered. Thus receive beamforming is not exposed here.

[2] A unitary matrix $\mathbf{A}$ is a matrix satisfying the condition: $\mathbf{A} \cdot \mathbf{A}^* = \mathbf{A}^* \cdot \mathbf{A} = \mathbf{I}$, $\mathbf{I}$ being the identity matrix and $\mathbf{A}^*$ the conjugate of matrix $\mathbf{A}$.

### 1.1.3.2 Beamforming using Singular Value Decomposition

Through SVD, the complex channel matrix $\mathbf{H}$ can be decomposed into a product of unitary matrices and an equivalent diagonal matrix $\mathbf{\Sigma}$.

Using the fact that the matrices $\mathbf{U}$ and $\mathbf{V}$ are unitary, equation (1.4) can be rewritten as:

$$\mathbf{U}^* \cdot \mathbf{H} \cdot \mathbf{V} = \left(\mathbf{U}^* \cdot \mathbf{U}\right) \cdot \mathbf{\Sigma} \cdot \left(\mathbf{V}^* \cdot \mathbf{V}\right) = \mathbf{\Sigma} \tag{1.6}$$

By pre-multiplying (or precoding) by $\mathbf{V}$ and post-multiplying by $\mathbf{U}^*$ the channel matrix can be diagonalized. The equivalent MIMO system, when using SVD based beamforming, is illustrated in Figure 1.8.

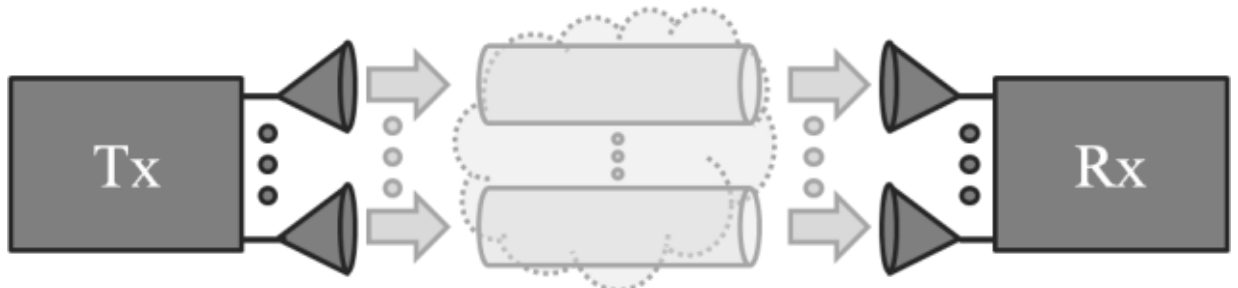

**Figure 1.8 Equivalent MIMO system when using SVD based beamforming**

The advantage of using beamforming is that spatial streams can be separated. Performance is thus much improved [8].

### 1.1.3.3 Channel State Information Feedback

In order to apply the correct vectors $\mathbf{v}_i$ , the transmitter has to possess full knowledge of the channel matrix $\mathbf{H}$ [8]. CSI has to be somewhat regularly returned by the receiver, either by providing channel taps or precoding vectors. All in all, this process is costly, in terms of overhead, because, in Wi-Fi, CSI feedback is done using the same band as the one used for data transfer [15]. Clearly a compromise is needed between having 'fresh' estimates and reducing CSI feedback overhead.

### 1.1.3.4 Consequences on Channel Estimation

Due to this implementation constraint, the transmitter can easily have outdated estimates, whereas the receiver has 'freshly' estimated channel taps. In [21], authors have demonstrated that this difference can result in a severe drop in performance, the decoding vector $\mathbf{U}^*$ not being appropriate. However, it has been shown, in the same paper, that if the transmitter informs which precoding $\mathbf{V}$ vector was used, performance is not degraded and receive process is simplified [8]. This is precisely done by also precoding the part of the PLCP preamble used for channel estimation. The receiver thus uses the channel matrix $\mathbf{H}\cdot\mathbf{V}$ to equalize the beamformed payload.

### 1.1.4 Spatial Division Multiple Access

Spatial diversity between antennas can be further exploited in a multiple-user context by spatial division multiple access (SDMA). SDMA, also called multiple-user MIMO (MU-MIMO) in IEEE 802.11ac [7], enables simultaneous communication with multiple users using the same band [22]. A transmitter can thus simultaneously send multiple independent data streams to different receivers, relying on the differences in their space-time signatures[1]. The equivalent system is illustrated in Figure 1.9, where $N$ receivers are simultaneously served by the transmitter.

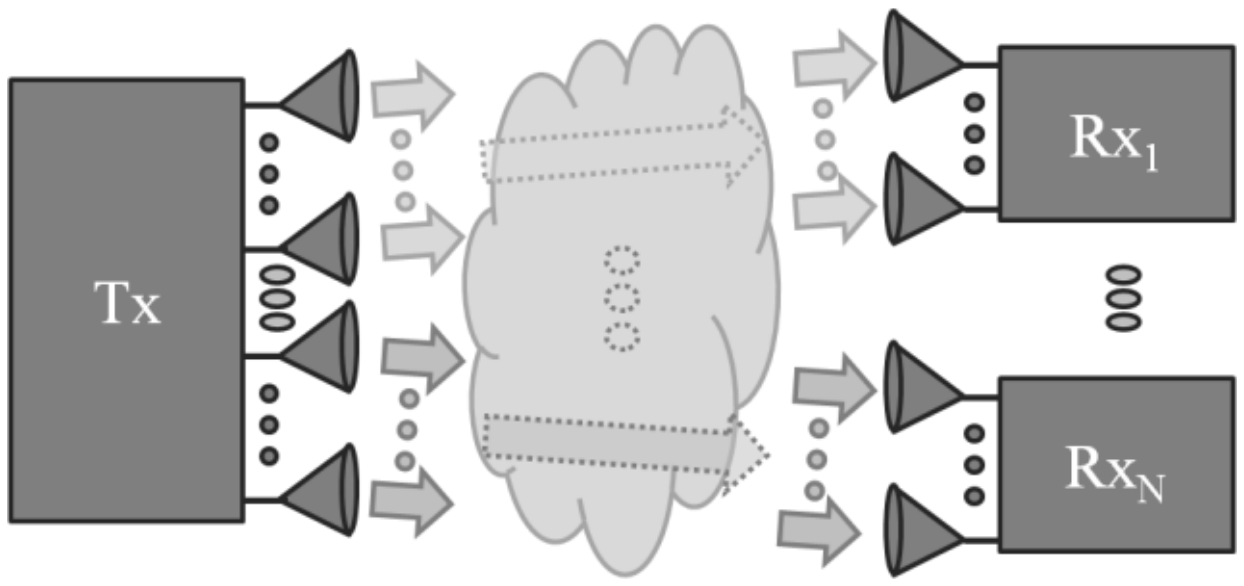

**Figure 1.9 Equivalent MIMO system when using spatial division multiple access**

Just like MIMO, SDMA can be easily combined with OFDM. This enables per-subcarrier processing and simplifies things [23]. Based on this, equation (1.1) can be rewritten for the SDMA context in the following manner:

$$\begin{bmatrix}\mathbf{y}_1\\ \vdots \\ \mathbf{y}_N\end{bmatrix} = \begin{bmatrix}\mathbf{H}_1\\ \vdots \\ \mathbf{H}_N\end{bmatrix}\cdot\mathbf{x} + \begin{bmatrix}\mathbf{z}_1\\ \vdots \\ \mathbf{z}_N\end{bmatrix} \tag{1.7}$$

The $n_T$ symbols of $\mathbf{x}$ are transmitted across the $N$ MIMO channels of dimension $n_T \times n_R^c$, $c$ being the index of the receiver and $n_R^c$ its number of antennas. Each vector of received symbols $\mathbf{y}_c$, with $n_R^c$ lines, is obtained by applying the $n_R^c$ by $n_T$ channel matrix $\mathbf{H}_c$ to $\mathbf{x}$, and accounting for the AWGN noise vector $\mathbf{z}_c$, with $n_R^c$ lines also.

It is to be noted that, in the SDMA context, CSIs of all $N$ channels have to be returned regularly. This is a critical assumption because the transmitter needs to have good knowledge of all channels to reduce inter-user interference as much as possible [15]. This interference is also called crosstalk interference (CTI). Indeed, if special care is not taken in designing the transmitted signal $\mathbf{x}$, symbols intended for one of the users generates interference on other users. This inter-user interference can be mitigated by the use of non-linear or linear precoding techniques.

[1] In IEEE 802.11ac, only downlink SDMA is considered. Uplink SDMA exists but is not exposed here.

#### 1.1.4.1 Non-linear Precoding Techniques

Dirty paper coding is a reference non-linear precoding technique. Its concept was introduced in [24] (also see [25] for a very clear exposition of this concept). It proposes to benefit from the known CTI to transmit information, instead of fighting against this interference. Optimal performance can be achieved this way. But this technique is very difficult to be implemented in practice [12]. There are several sub-optimal and simplified variants of dirty paper coding (see to [12,15] for further details). Other non-linear precoding techniques, such as vector-perturbation [26] and Tomlinson-Harashima precoding [27], also exist. Still the complexity of these non-linear solutions remains relatively important.

#### 1.1.4.2 Linear Precoding Techniques

Linear precoding techniques are quite simple [15,28] yet exhibit dirty paper coding like performance when used with efficient user selection. In these techniques, users are assigned different precoding matrices at the transmitter. The transmitted signal is a linear function expressed as:

$$\mathbf{x} = \sum_{c \in C} \mathbf{W}_c \cdot \mathbf{d}_c \tag{1.8}$$

Where $C$ is the set of selected users, $\mathbf{d}_c$ the vector of data symbols transmitted to user $c$, and $\mathbf{W}_c$ the precoding matrix for user $c$ [12]. Using equations (1.7) and (1.8), the resulting received signal vector for user $c$ is given by:

$$\mathbf{y}_c = \mathbf{H}_c \cdot \mathbf{W}_c \cdot \mathbf{d}_c + \sum_{e \in C, e \neq c} \mathbf{H}_c \cdot \mathbf{W}_e \cdot \mathbf{d}_e + \mathbf{z}_c \tag{1.9}$$

The first term in equation (1.9) represents the useful signal, whereas the second term represents CTI caused by signals precoded for all other users of the SDMA transmission. Linear precoding is thus equivalent to having intelligent beamforming based on CSI from all users and following different optimization criteria. The precoding process can be simplified by intelligently grouping users, i.e. grouping less correlated channels [19].

#### 1.1.4.2.1 Channel Inversion Precoding

A simple way of dealing with CTI is using channel inversion. In this suboptimal technique, also known as zero-forcing (ZF) beamforming, precoders are designed to achieve zero interference between users. In other words, the product $\mathbf{H}_c \cdot \mathbf{W}_e$ is null for all $e \neq c$ of the selected set $C$ [12]. Assuming that the transmitter has at least as many antennas as all selected receivers combined, this can be accomplished by precoding with the pseudo-inverse[1] of the channel matrix $\mathbf{H}$ [19]. Indeed if $\mathbf{x} = \mathbf{H}^\dagger \cdot \mathbf{d}$, then at each receiver this approach results in having:

$$\mathbf{y}_c = \mathbf{d}_c + \mathbf{H}^\dagger \cdot \mathbf{z}_c \tag{1.10}$$

Channel inversion is a good solution for low-noise or high-power situations. In addition, each users' data symbols are perfectly separated. This simplifies Rx processing and power allocation [23]. However, if the channel is not well conditioned[2], the inversion process requires strong signal attenuation at the transmitter, implying low SNR at the receivers [12]. This is due to the stringent requirement that no CTI is tolerated at the receivers [15].

#### 1.1.4.2.2 Minimum Mean Square Error Precoding

The performance of channel inversion can be improved by allowing a limited amount of interference at each receiver [15]. The precoder which minimizes the mean square error between the received and sent symbol, while fulfilling transmit power constraint, is chosen. In this minimum mean square error (MMSE) precoding, the pseudo-inverse is regularized in the following manner:

$$\mathbf{W} = \mathbf{H}^* \left(\mathbf{H} \cdot \mathbf{H}^* + \rho \cdot \mathbf{I}\right)^{-1} \tag{1.11}$$

The regularization factor $\rho$ determines the amount of allowed interference. A tradeoff between Tx power and CTI can be obtained [23]. However, MMSE does not provide separated symbols per receiver. Thus power allocation techniques cannot be performed in a straightforward manner [12].

#### 1.1.4.2.3 Block Diagonalization Precoding

In the case where each receiver has multiple antennas, complete diagonalization of the downlink channel is suboptimal. Indeed each receiver is able to coordinate the processing of its inputs [19]. A solution to this problem is to use block channel inversion, to loosen the constraint. This approach, also called block diagonalization (BD), optimizes the power transfer to a group of antennas belonging to the same receiver, while zero-forcing inter-user interference [15]. Therefore with BD, the received signal vector for user is $c$ in equation (1.9) simplifies to:

$$\mathbf{y}_c = \mathbf{H}_c \cdot \mathbf{W}_c \cdot \mathbf{d}_c + \mathbf{z}_c \tag{1.12}$$

---

[1] The Moore-Penrose pseudo-inverse of matrix $\mathbf{A}$ is $\mathbf{A}^\dagger = \mathbf{A}^* \cdot \left(\mathbf{A} \cdot \mathbf{A}^*\right)^{-1}$.

[2] The channel is considered as ill-conditioned when one of its singular values is very large compared to the others. Transmit ZF implies spending much more energy on the latter than the former to pre-compensate for this difference.

This can be obtained by using a precoder lying in the left null space of the matrix $\tilde{\mathbf{H}}_c$. The latter matrix consists of a remodeling of $\mathbf{H}$ without $\mathbf{H}_c$ [12]. Computation details given in Appendix A, the BD precoder maximizing throughput is:

$$\mathbf{W}_c = \tilde{\mathbf{V}}_c^{(0)} \cdot \overline{\mathbf{V}}_c^{(1)} \tag{1.13}$$

With $\tilde{\mathbf{V}}_c^{(0)}$ holding the $n_T - \operatorname{rank}\left(\tilde{\mathbf{H}}_c\right)$ last left singular vectors of $\tilde{\mathbf{H}}_c$, and thus ensuring that the precoder lies in the null space[1] of $\tilde{\mathbf{H}}_c$. Vector $\overline{\mathbf{V}}_c^{(1)}$ holds the first $\operatorname{rank}\left(\mathbf{H}_c \cdot \tilde{\mathbf{V}}_c^{(0)}\right)$ right singular vectors of $\mathbf{H}_c \cdot \tilde{\mathbf{V}}_c^{(0)}$, maximizing the information rate with zero CTI constraint [19]. We shall note that BD requires that the number of Tx antennas be generally larger than the total number of Rx antennas. However, it offers relatively low computational cost [15].

# 1.2 Medium Access Control Layer

The MAC layer provides, among other things, addressing and channel access control that makes it possible for multiple stations on a network to communicate [8]. The aim of this section is to give some MAC layer material. However, MAC layer mechanisms are specific to the concerned medium. Therefore we focus on the IEEE 802.11 MAC layer and give the background needed to understand the evolutions of OFDM-based IEEE 802.11 amendments exposed in section 1.3, and IEEE 802.11ac specifications detailed in Chapter 2.

## 1.2.1 The Wireless Medium

IEEE 802.11 is often referred to as ‘wireless Ethernet’. Indeed, it does not depart from the IEEE 802.3 standard in any radical way. However, an effort was needed to successfully adapt Ethernet-style networking to radio links [14]. The wireless medium is very different from the wired medium and its specificities justify the used MAC layer mechanisms.

[1] A matrix $\mathbf{A}$ with $n$ columns has $\operatorname{rank}(\mathbf{A})$ independent columns. These vectors span the column space of matrix $\mathbf{A}$. The null space of $\mathbf{A}$, of dimension $n - \operatorname{rank}(\mathbf{A})$, is the set of vectors $\mathbf{x}$, such that $\mathbf{A} \cdot \mathbf{x} = \mathbf{0}$.

#### 1.2.1.1 A Shared Medium

The wireless medium is by definition a shared medium [29]. Stations tuned on the same frequency 'overhear' each others' transmissions. This is due to the natural propagation of radio waves. In addition, wireless implies mobility. The group of stations sharing the same medium is not static and is difficult to control [10]. These two points imply that the medium access control scheme needs to be flexible and fair, with regards to channel access time.

#### 1.2.1.2 Radiofrequency Link Quality

IEEE 802.11 standards use unlicensed frequencies. Therefore one cannot control who transmits over the same medium. These transmissions can create interference [14]. The wireless medium in itself is prone to errors, because of the multipath and changing environment. The important attenuation of signals during transmission should also be considered. A low-latency MAC level error detection and recovery mechanism is thus needed. Another consequence is that stations need to continually adjust the data rate, at which they exchange throughput, according to channel conditions [8]. To obtain reasonable performance despite the important system loss, stations have to transmit at relatively high levels while having high Rx sensitivity. A station cannot currently receive and transmit simultaneously [29].

#### 1.2.1.3 Overlapping Transmissions

As implied above, multiple independent groups of stations can use the same medium to exchange information. These groups, called basic service sets (BSSs), overlap but are not coordinated. Indeed, the Wi-Fi spectrum being unlicensed, a central coordinator allocating channel access time to all its stations, while accounting for neighboring coordinators' allocations, is very difficult to implement. Therefore the use of a distributed channel access suits more Wi-Fi systems, especially in urban areas.

#### 1.2.1.4 Hidden Node Problem

Due to important channel loss, regulations on maximum Tx power, and minimum receiver sensitivity, stations of the same BSS may not be able to detect each others' transmissions. This range restriction causes very frequent collisions because the basic distributed channel access mechanisms of IEEE 802.11 systems are eluded. This phenomenon is the hidden node problem and is illustrated in Figure 1.10 [8,14].

The access point (AP) is the bridge to the distribution system in the infrastructure mode[1] of Wi-Fi. In this mode, the BSS is 'built around' the AP. Being the the central node, it is thus, by definition, in range of all the stations of the BSS. However, two stations can be hidden to each other, like $STA_1$ and $STA_3$ in Figure 1.10. Each would see the medium as idle even if the other is transmitting, thus causing collisions at the AP and $STA_2$.

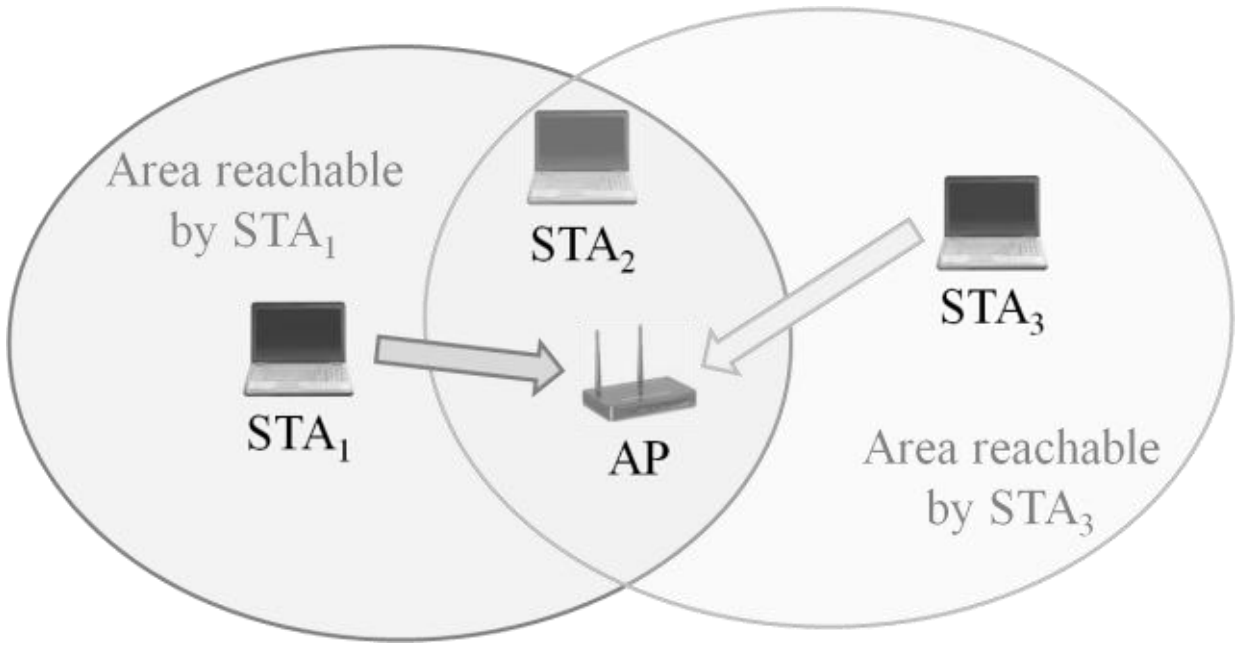

**Figure 1.10 Hidden node problem**

## 1.2.2 Positive Acknowledgment

We have seen that transmission over a wireless medium is error-prone. A low-latency MAC layer retransmission mechanism would be to use positive acknowledgment. In this basic mechanism, all transmitted data frames (except broadcast and multicast frames) should be acknowledged with an acknowledgment (ACK) frame (as illustrated in Figure 1.11).

This 'data+ACK' sequence is thus an atomic operation (i.e. 'all or nothing') [14]. If the station sending the data frame does not receive the ACK frame shortly afterwards, it assumes the data frame has not been received and retransmits it, as soon as possible (as with $Data_2$ in the illustrated example). However, the number of retransmission attempts for a particular data frame is limited. If the retry count exceeds a configured limit, the data frame is discarded by the MAC layer.

The sending station thus retries the transmission of the data as long as the ACK has not been received, even if the data frame had been correctly received and the ACK frame sent. This should normally be rare (collisions excluded) because the ACK frames are modulated more robustly than data frames for greater reliability. The ACK frame being short, the price to pay, in terms of additional

[1] Ad hoc mode is the other mode for building BSSs. All stations have equivalent roles. In the infrastructure mode, all traffic between stations in the BSS has to go through the AP [14].

overhead, should be small despite the used low data rates [8].

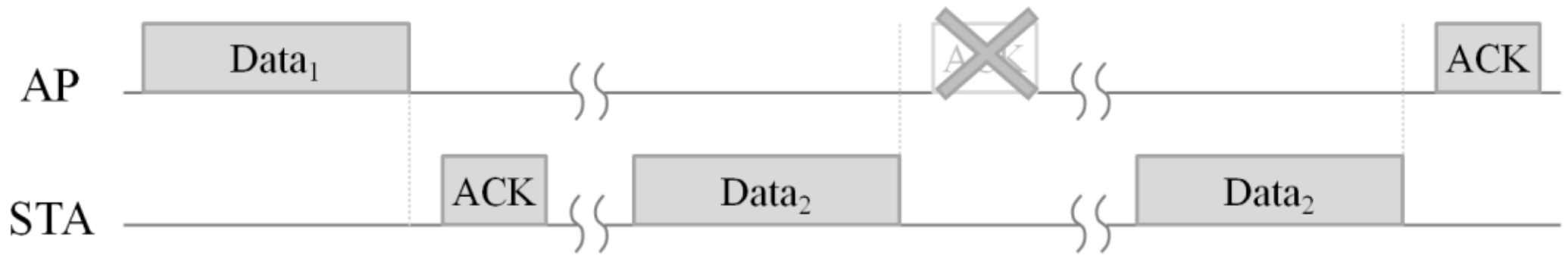

**Figure 1.11 Data+acknowledgment exchange**

## 1.2.3 Carrier Sense Multiple Access with Collision Avoidance

Carrier sense multiple access with collision avoidance (CSMA/CA) is a distributed channel access mechanism used in WLANs. Collision is not completely avoided though. The carrier sense multiple access (CSMA) technique, also used in Ethernet with collision detection, is adapted to the wireless medium by adding positive ACKs, a collision avoidance mechanism used for retransmission, different channel access timings enabling prioritization, and channel access protection mechanisms [29].

### 1.2.3.1 Carrier Sense Multiple Access

In a shared medium, the contention access of CSMA is very interesting. It is flexible both in the number of users and in the supported load. Any station can transmit a frame whenever it needs to, if the channel is idle. A clear channel assessment (CCA) is performed by the PHY layer, before any transmission, by sensing the medium. If a carrier is sensed in the working band, the channel is considered as busy. Transmission is thus deferred until the medium becomes idle again. If the medium remains idle for a fixed duration and an additional random backoff (BO) period, transmission can begin [29]. The random BO period provides the collision avoidance aspect. Since each station having data to send probabilistically chooses a different BO interval, collisions where more than one station begins transmitting at the same time are unlikely in small networks. The station with the smallest BO wins access to the medium and transmits its frame. All other stations freeze their BO counters until the medium becomes idle again [8].

### 1.2.3.2 Retransmission

The random BO interval is bounded by the contention window (CW). The CW takes the initial value

$CW_{min}$. It doubles on each unsuccessful data frame transmission, until it reaches its maximum value $CW_{max}$. The probability that colliding stations draw the same BO again decreases considerably as the CW increases. The CW is reset to $CW_{min}$ when ACK is received or when, the associated maximum retry count being reached, the frame is discarded. Having a longer contention window enables to reduce the probability of repeating collisions and stabilizes CSMA under maximum load [8,14].

#### 1.2.3.3 Virtual Carrier Sensing

In CSMA, physical carrier sensing (or CCA) is used to determine if the medium is available. However, it cannot provide all the necessary information in a hidden node scenario. Virtual carrier sensing was designed to help overcome this problem and is provided by the network allocation vector (NAV). This MAC function is updated by the duration field carried in MAC frames, if the latter have been correctly demodulated. The sending station uses this duration field to indicate for how long it expects the medium to be busy after the last symbol of the frame carrying the field. When the updated NAV reaches 0, the virtual carrier sensing function indicates that the medium is idle [8,14]. Refer to the example given in section 1.2.3.5 for an illustration of virtual carrier sensing.

#### 1.2.3.4 Request to Send / Clear to Send

A station can use the duration field of data frames to set its neighbors' NAVs. Using robust frames (thus covering more neighboring stations) before sending the data frame would offer much better protection from collisions. Request to send (RTS) and clear to send (CTS) frames offer such protection. A RTS frame is sent by the initiator and the station addressed by it responds with a CTS frame. Only then will the initiator send the data frame intended for the addressed station. With these two short and robust frames, stations around the initiator and the replier can update their NAVs [8,14].

#### 1.2.3.5 Frame Exchange Example

An example of a RTS/CTS exchange, summarizing most of the exposed CSMA/CA mechanisms, is illustrated in Figure 1.12. In it, the hidden node scenario illustrated in Figure 1.10 is assumed.

$STA_1$ and $STA_2$, both having frames to transmit, start counting down their drawn BOs when the channel becomes idle. $STA_1$ wins access to the medium and sends a RTS frame to the AP. $STA_2$ sensing a carrier defers its transmission, and updates its NAV after decoding the RTS. The AP replies with a CTS frame, which sets $STA_3$'s NAV. The medium being idle after the data-ACK exchange, $STA_2$ and $STA_3$ contend for the medium, finally won by $STA_2$.

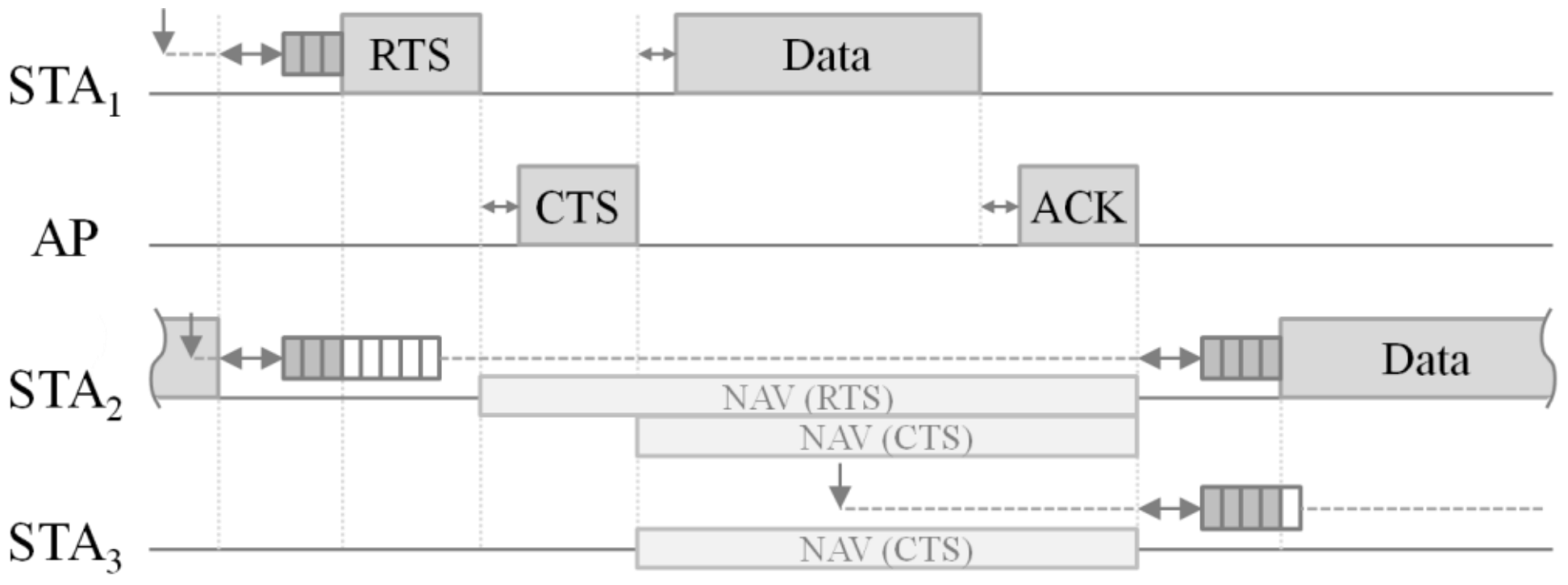

**Figure 1.12 RTS/CTS exchange in a hidden node scenario**

## 1.2.4 Quality of Service

The ability to support quality of service (QoS) requirements of various applications (such as real-time voice and video) in the Wi-Fi MAC layer was introduced with IEEE 802.11e amendment [30]. QoS is enabled through prioritization [8].

### 1.2.4.1 Access Categories

Four access categories (ACs) are defined, when offering QoS. They are termed as background (AC_BK), best effort (AC_BE), video (AC_VI), and voice (AC_VO). A specific set of access parameters is affected for each AC so as to statistically prioritize channel access and offer QoS. Once the AC of a frame is obtained by mapping from user priority, the frame is accordingly stacked in one of the four transmit queues [8]. Frames in these AC-dependent queues internally contend for channel access, and the winning AC gets to send its frame if no other station has started doing so before.

### 1.2.4.2 Transmit Opportunity

A transmit opportunity (TxOP) is a bounded period during which a station may transfer data of a particular AC, as illustrated in Figure 1.13. Once the TxOP has been obtained, through channel access procedures, the station may continue to transmit frames and receive response frames, as long as the total duration does not exceed the TxOP limit set for that AC. This way all stations accessing the network with traffic of the same class will, on average, receive the same amount of air time.

This fairness in resource rather than throughput enables stations using higher data rates to transmit more information than stations with lower data rates [8]. Indeed, the former will send more frames than the latter in an identical duration TxOP.

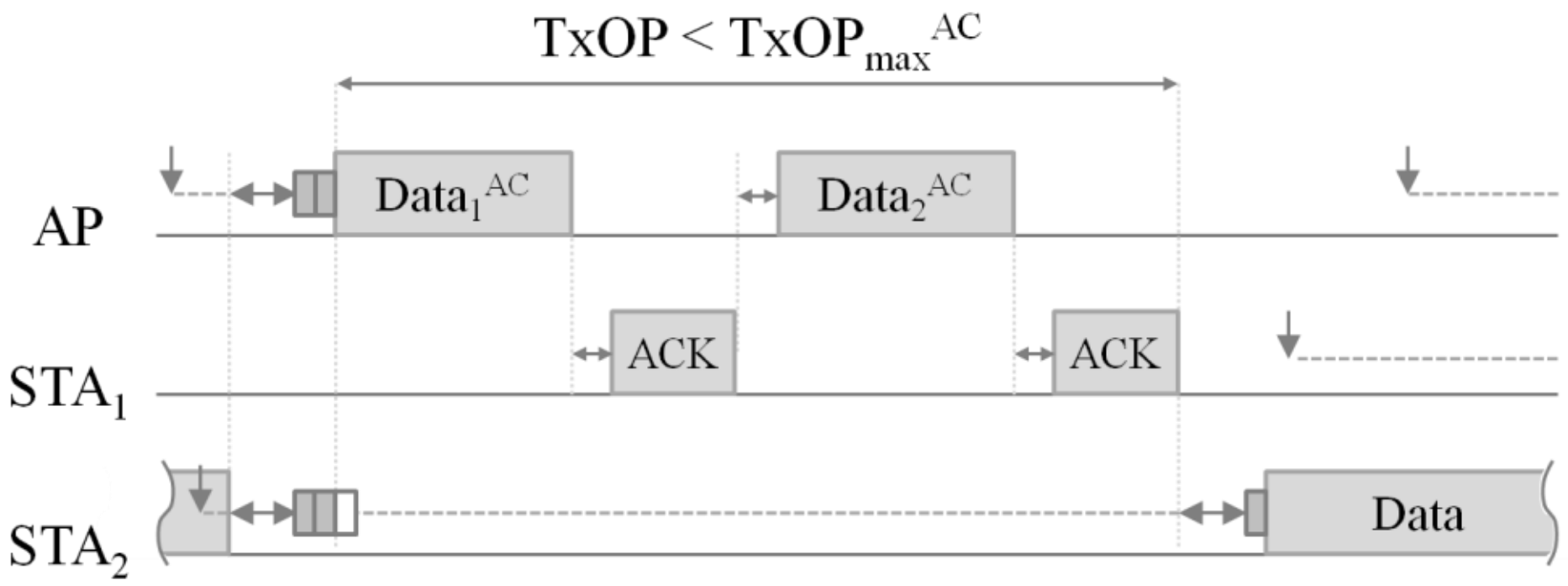

**Figure 1.13  Transmit opportunity**

## 1.2.5 Protocol Layering and Framing

In order to better grasp the MAC add-ons of the IEEE 802.11n standard [6], some basic concepts regarding protocol layering should be introduced. These concepts are common to most standards, by the way. As illustrated in Figure 1.14 [8], both the PHY layer and MAC layer transfer user data as service data units (SDUs). Accordingly the MAC layer uses MAC SDUs (MSDUs) to interact with higher layers and PHY SDUs (PSDUs) with the PHY layer. Protocol data units (PDUs) are used to interact with distant peer layers. MAC PDUs (MPDUs) and PHY PDUs (PPDUs) are thus defined.

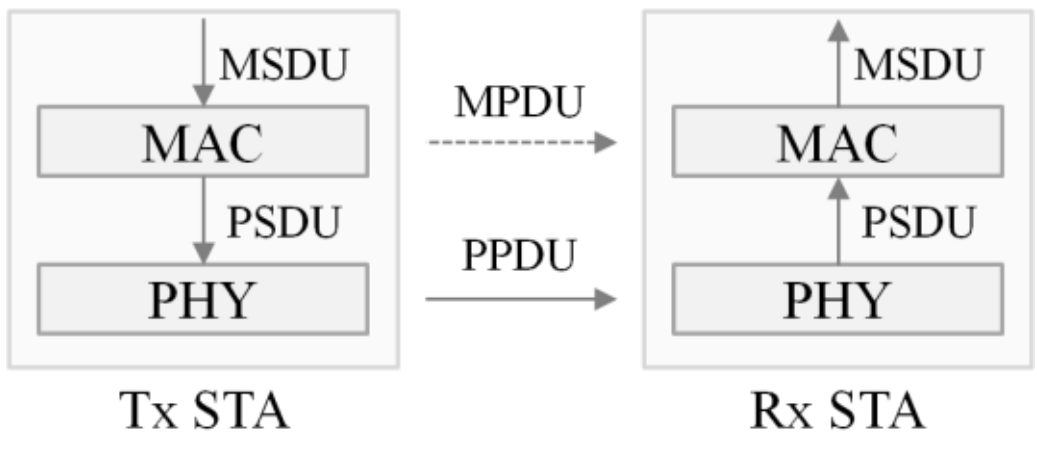

**Figure 1.14  Protocol layering**

We will note that the MSDU (or PSDU) is often referred to as MAC layer (or PHY layer, resp.) payload. Another important fact is that each layer forms a PDU by processing its SDU, prepending its header, and often appending a tail field. This is illustrated in Figure 1.15.

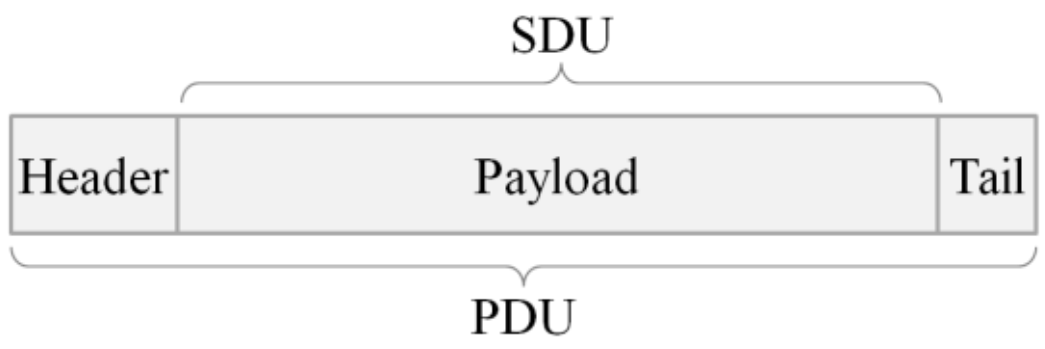

**Figure 1.15 Framing and data units**

For MPDUs, the tail field holds a frame check sequence (FCS), which is a hash of the rest of the frame. The integrity of the whole MAC frame can be verified by the receiving station through this cyclic redundancy check (CRC). Just like the PLCP header in section 1.1.1.7, the MAC header also contains critical information, such as the type of frame, the sending and receiving stations' MAC addresses, and the NAV. That is why this field's integrity is also ensured by the FCS. Considering that address information is in the MAC header, the PHY layer (which considers the latter as part of its payload) processes all frames it can sense on the medium. It is up to the MAC layer to decide whether the receiving station is concerned by the decoded frame or not.

### 1.2.6 Block Acknowledgment

With the introduction of TxOPs, data can be transmitted in bursts. It is thus not efficient to acknowledge each data frame in an independent manner. The block ACK (BA) protocol was introduced with the IEEE 802.11e amendment to improve efficiency by acknowledging blocks of data frames, using a single extended ACK frame [31]. This has two direct implications. The first is that simple data+ACK exchanges are no longer the only atomic exchanges involving data. The use of BA should be signaled in the data frames to be acknowledged. The second is that frame numbers must be tracked for proper acknowledgment and retransmission. A bitmap of correctly received frames is returned in the BA to this aim [8].

### 1.2.7 Aggregation

With the PHY layer improvements of IEEE 802.11n, PHY data rates have increased. However, the contention access and PHY header being overhead with regards to the MAC layer, the latter cannot profit from this improvement [31]. MAC aggregation was thus introduced to increase efficiency. MSDUs can be aggregated and the obtained aggregate MSDU (A-MSDU) is used to form a MPDU. This highly efficient technique is however less robust than legacy framing (i.e. with no MSDU aggregation) because there is only one FCS per MPDU. If the integrity check fails, the whole A-MSDU needs to be retransmitted. On the other hand, aggregate MPDUs (A-MPDUs) are less efficient than A-MSDUs (because delimiters are needed) but much more robust. The validity of each MPDU of the A-MPDU can be established independently and selective retransmission is made possible [8,31]. A-MPDU encapsulation, illustrated in Figure 1.16, is thus more used in practice than A-MSDU encapsulation [32].

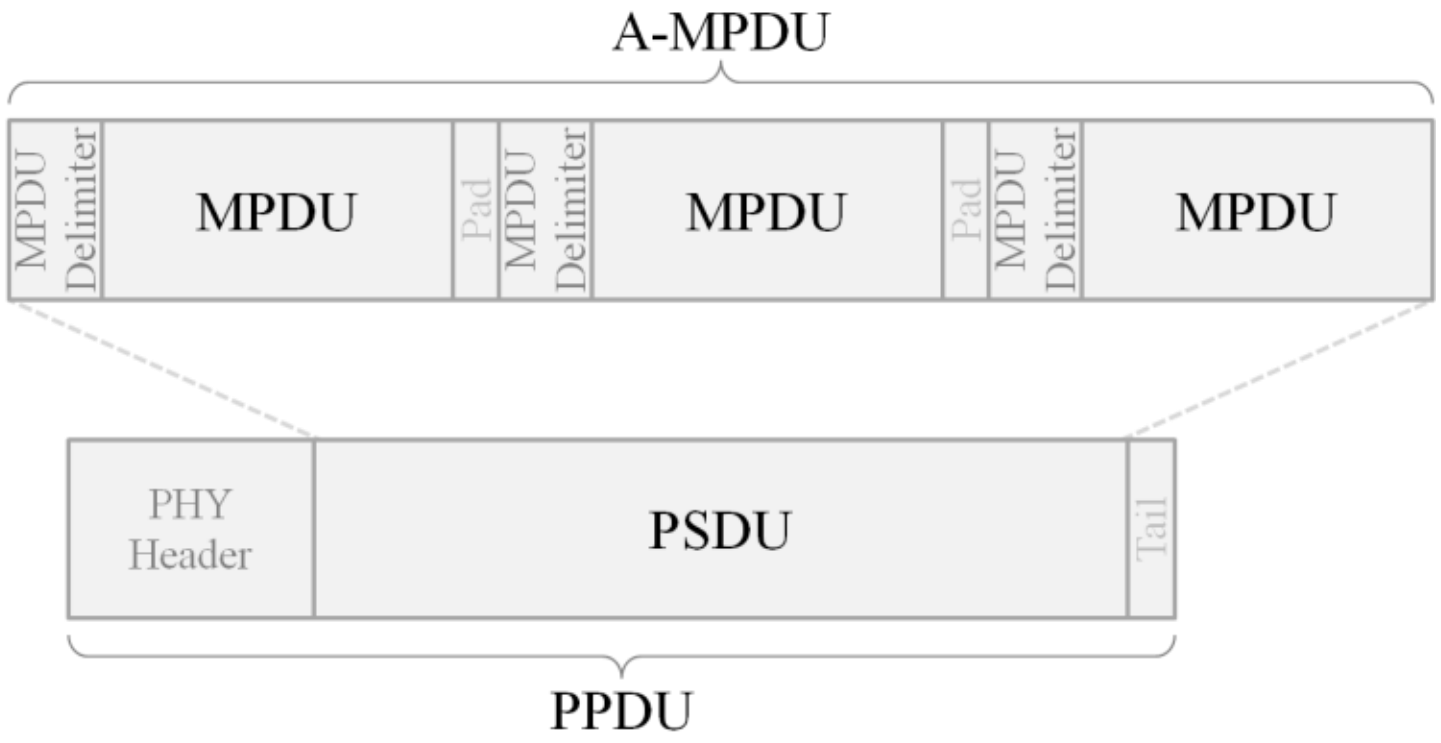

**Figure 1.16 A-MPDU encapsulation**

### 1.2.8 Rate Selection for Data Frames

We have seen in section 1.2.1.2 that channel conditions, between any set of two stations, vary over time. It is thus important to dynamically select the optimal data rate with which to send data towards a particular station. The poorer channel conditions, the more robust the modulation and coding scheme (MCS) should be. But this implies slower transmission. On the other hand, a station can benefit from good channel conditions to transmit data with higher throughput. These rate adaptation algorithms are implementation specific. However, they are generally based on the measured packet error rate (PER) [8] and can be classified either as 'low-latency' or 'high-latency' systems. This latency is the communication latency between the PHY layer and the block that implements the rate control algorithm within MAC layer. Systems using embedded processors have low latency [33].

Low-latency systems allow per-packet adaptation. They are thus more reactive but fail to efficiently handle stable conditions. Sometimes choosing a slightly lower rate can be more profitable because less retransmission is generated. Higher application-level throughputs can be reached this way. On the other hand, high-latency systems require periodic analysis and are thus more stable.

However, they fail to react to short-term channel condition changes. The adaptive multi-rate retry algorithm (AMRR), as detailed in Appendix B, offers short-term adaptation for high-latency systems [33]. It can thus work with both types of systems. That is why we have chosen to use it in our simulations. Other rate adaptation algorithms exist though, some of them being presented in [33,34,35].

We will note that the strategy employed for rate selection has an important impact on MAC throughput. It is thus preferable to compare two simulation scenarios using the same rate adaptation algorithm for a fair evaluation of performance. We will also note that with important changes in PHY or MAC layers (such as aggregation or SDMA), rate adaptation algorithms have to be updated so as to cope with the induced changes in behavior. The purpose of this thesis not being algorithm conception for Wi-Fi systems, we have used AMRR, as exposed in Appendix B, for the reasons stated in the previous paragraph despite its shortcomings.

# 1.3 OFDM-based IEEE 802.11 Standards

Now that the PHY and MAC layer fundamentals have been given, the main functionalities of OFDM amendments of IEEE 802.11 standards can be exposed.

## 1.3.1 IEEE 802.11a/g

The first amendment to introduce OFDM was IEEE 802.11a [4]. Even though it enabled data rates of up to 54 Mbps in the 5GHz band starting 1999, its adoption has been very slow. This was due to the success of IEEE 802.11b devices (11 Mbps data rate max.) which operate in the 2.4 GHz band. New devices wishing to benefit from 802.11a’s higher data rates, while retaining backward compatibility with deployed 802.11b devices, would need to implement two radios (i.e. one for each band) [8]. Consequently the IEEE 802.11g [5] amendment was defined in 2001 to enable OFDM transmission in the 2.4 GHz band. 802.11g devices have the advantage of having backward compatibility with 802.11b devices, whereas 802.11a devices benefit from larger and much less crowded spectrum. Nevertheless, 802.11a and 802.11g amendments are very similar and are often considered as a whole.

### 1.3.1.1 PHY Layer

Most of the changes brought by IEEE 802.11a/g amendments on the original IEEE 802.11 specifications [3] concern the PHY layer. These changes were obtained by adapting OFDM to the WLAN environment. To this aim, 20 MHz-large channels were allocated by regulatory authorities [14]. The GI duration, symbol duration, and total number of subcarriers were fixed accordingly. The main PHY layer parameters of 802.11a/g amendments are detailed in Table 1.1. 802.11a/g devices can transmit data using data rates of up to 54 Mbps, using a specific combination of modulations and coding rates (block convolutional coding, BCC, being used).

**Table 1.1 Main PHY layer characteristics of IEEE 802.11a/g**

| Parameter | Value |
|---|---|
| Carrier frqcy (GHz) | 2.4 (802.11g) |
| | 5 (802.11a) |
| Bandwidth (MHz) | 20 |
| # subcarriers | 48 for data out of 64 |
| Subcarrier spacing | 312.5 kHz |
| Coding | BCC |

| Parameter | Duration (μs) |
|---|---|
| OFDM symbol | 4 |
| GI | 0.8 |
| Short training field | 8 |
| Long training field | 8 |
| Signaling field | 4 |

| Parameter | Possible values |
|---|---|
| Data rates (Mbps) | 6, 9, 12, 18, 24, 36, 48, 54 |
| Modulations[1] | BPSK, QPSK, 16-QAM, 64-QAM |
| Coding rates | 1/2, 2/3, 3/4 |

| Regulatory domain | | France | Europe[2] | United States | Japan |
|---|---|---|---|---|---|
| # of 20 MHz channels | 802.11g | 4 | 13 | 11 | 1 |
| | 802.11a[3] | 19 | 19 | 12 | 7 |

[1] The different types of modulations used in OFDM-based IEEE 802.11 amendments are binary phase-shift keying (BPSK), quadrature phase-shift keying (QPSK), and quadrature amplitude modulation (QAM).
[2] Spain, just like France, has its own regulations, where only 2 channels are allowed [3,38,5].
[3] Europe's regulations have been modified by IEEE 802.11h [36] and Japan's by IEEE 802.11j [37].

### 1.3.1.1.1 PMD Sub-layer

The available spectrum is organized into operating channels. In the 2.4 GHz band, the available 20-MHz channels are closely packed. Only a few of these do not overlap and can be used simultaneously without causing interference (e.g. in Europe, only 3 out of the 13 available channels are non-overlapping [3]). As a result, the 2.4 GHz band has become very crowded with the success of IEEE 802.11b and IEEE 802.11g devices [14]. Regulatory authorities have capitalized on this experience when defining 20-MHz channels in the 5 GHz band. None of them overlap, be it in the United States [4], Europe [36], or Japan [37], facilitating BSS neighboring issues in populated areas.

Each 20-MHz is composed of 52 subcarriers, as illustrated in Figure 1.17. 4 of the subcarriers (numbered -21, -7, 7, and 21) are used as pilot subcarriers (see section 1.1.1.3), while the other 48 subcarriers are used to transmit data. The total number of subcarriers implicated in fast Fourier transform (FFT) operations being 64, the remaining 12 subcarriers are not used, either for signal processing reasons (e.g. to respect transmission mask constraints and thus protect data subcarriers from adjacent channel transmissions, to facilitate filtering, etc.) [14].

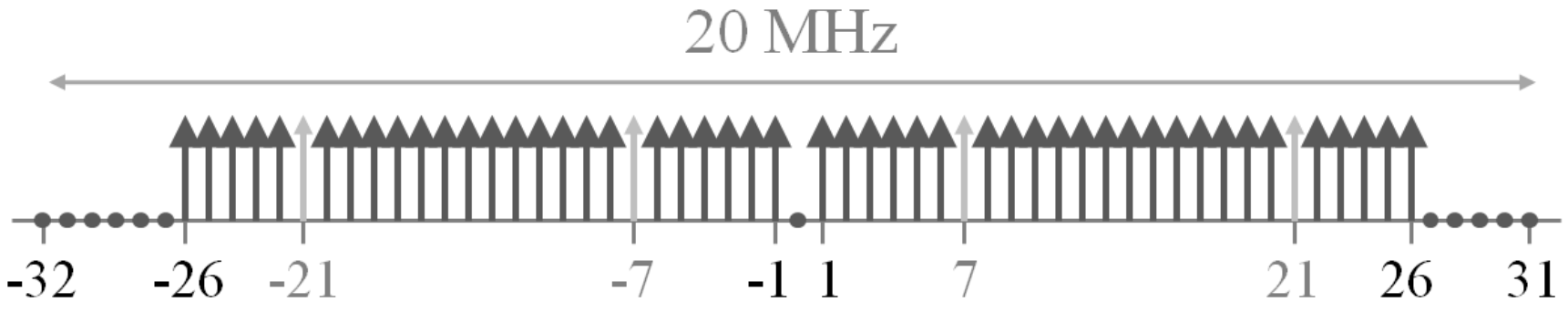

**Figure 1.17 20 MHz subcarrier design for IEEE 802.11a/g**

### 1.3.1.1.2 PLCP Sub-layer

As seen in section 1.1.1.5, synchronization is crucial and should precede any other processing. This is done in IEEE 802.11a/g using the short training field (STF), which is composed of 10 identical 0.8 µs-long training sequences [14] (tagged $s_i$ in Figure 1.18). STF is thus the first field of the PLCP preamble. The STF is followed by the long training field (LTF), which is also 8 µs-long [8]. The LTF is principally used for channel estimation (see section 1.1.1.6) and is composed of two long training sequences. The latter, tagged 'Long $seq_i$' in Figure 1.18, last 3.2 µs each, and are protected by a 1.6 µs GI. These sequences are averaged, making the process more robust to noise and estimation errors. The LTF is the second, and last, field of the IEEE 802.11a/g PLCP preamble. The signaling (SIG) field corresponds to the PLCP header and comes right after the PLCP preamble in the IEEE 802.11 standard as shown in Figure 1.18. As indicated in section 1.1.1.7, this field contains critical in-band signaling. Indeed, a misinterpretation of the information it holds can lead to incorrect decoding of data symbols (even if the latter have no errors). A parity bit is thus appended so as to verify its integrity. 4 µs-long OFDM symbols conveying data follow.

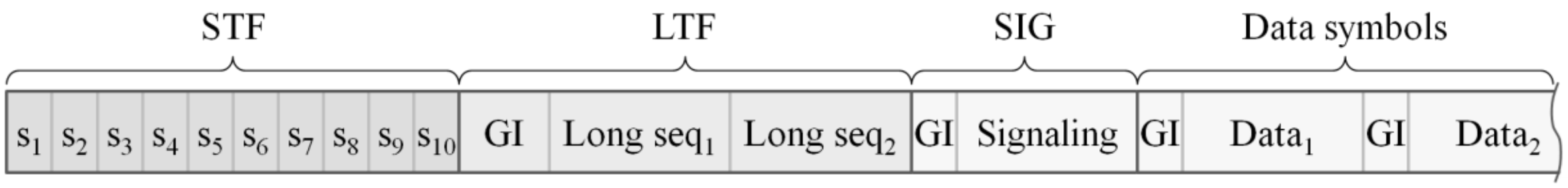

**Figure 1.18 PLCP framing in IEEE 802.11a/g**

### 1.3.1.2 MAC Layer

MAC layers functions of the original IEEE 802.11 standard [3] have not been affected by the OFDM-based IEEE 802.11a/g amendments (except coexistence issues with 802.11b devices in 802.11g). The main MAC layer characteristics given in Table 1.2 are simply inherited from the original 1997 specification.

**Table 1.2 Main MAC layer characteristics of IEEE 802.11a/g**

| Parameter | Max size (octet) |
|---|---|
| MSDU | 2,304 |
| MPDU | 2,346 |
| FCS | 4 (fixed) |

| Parameter | Duration |
|---|---|
| Slot time | 9 µs |
| Short inter-frame spacing | 16 µs |
| Beacon period[1] | 100 ms |

The MPDU maximum size given in Table 1.2 corresponds to the sum of the maximum MAC header size, the maximum frame body size (including maximum MSDU size and encryption related information), and FCS size. The most common 802.11a/g MAC framing[2] is illustrated in Figure 1.19. The first 16 bits of the MAC header, corresponding to the frame control field, identify the type of frame and convey the characteristics of the rest of the MPDU. Indeed it is based on this field that the receiver's MAC layer interprets following bits. The next 2 octets contain the NAV (see section 1.2.3.3). Then comes three 6 octet-long address fields specifying the receiver, transmitter, and source or destination of the MPDU (resp.). In infrastructure mode, all frames go through the AP. Therefore either the receiver or the transmitter address corresponds to the MAC address of the AP (which is also the BSS identifier, or BSSID) depending on whether the frame is uplink or downlink (resp.), with the third address being the destination or source of the frame (resp.).

[1] There is no period given in the standard. This is the commonly set duration [14].

[2] The wireless bridge mode, with 4 address fields, and the ad hoc mode, with 3 address fields (the last one always being the BSSID), are quite rarely used in residential Wi-Fi compared to infrastructure mode.

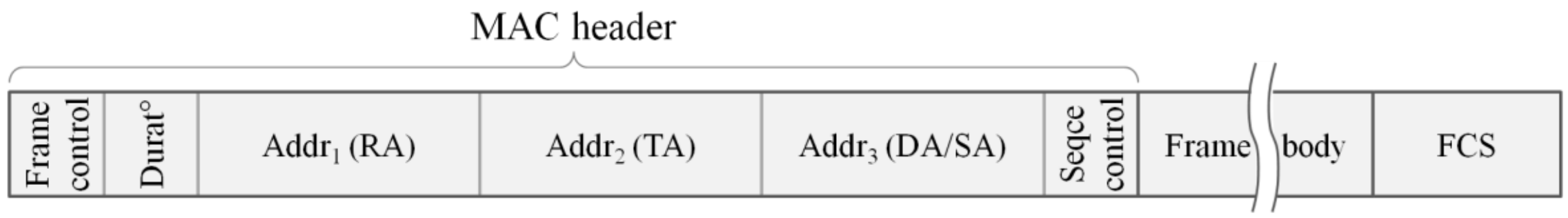

**Figure 1.19 MAC framing in IEEE 802.11a/g**

The last field of the MAC header is the sequence control field. This 2 octet-long field holds the identification number of the MSDU so as to facilitate reassembly (when the original MSDU has been fragmented) and detection of duplicate frames. Depending on the information in the frame control field, the frame body can either contain a data frame, a control frame (e.g. ACK, RTS, CTS), or a management frame (e.g. the beacon frame, which regularly broadcasts BSS parameters).

The specific CSMA/CA mechanism used in the 802.11 MAC is referred to as the distributed coordination function (DCF) [8]. Positive acknowledgment, virtual carrier sensing, RTS/CTS protection, and retransmission mechanisms are all implemented by the DCF. However, there is no differentiation between services.

As with traditional Ethernet, the inter-frame spacing (IFS) plays a large role in coordinating access to the transmission medium and in prioritizing transmissions. The different IFSs of the OFDM-based IEEE 802.11a specification[1] are illustrated in Figure 1.20 [8]. The short IFS (SIFS) is used for high-priority transmissions requiring immediate access to the medium. The data+ACK or RTS+CTS+data+ACK exchanges being atomic, the SIFS is used to seize the medium. This IFS is designed to be as short as possible while still accommodating the latencies incurred in a reasonable implementation (the decoding latency being the longest). Timing for other IFS durations is based on the SIFS and an integral number of time slots. The slot duration is designed to provide enough time

[1] IEEE 802.11g has a slightly different definition of the IFSs so as to ensure IEEE 802.11b compatibility.

for a transmission to be detected by neighboring stations before the next slot boundary.

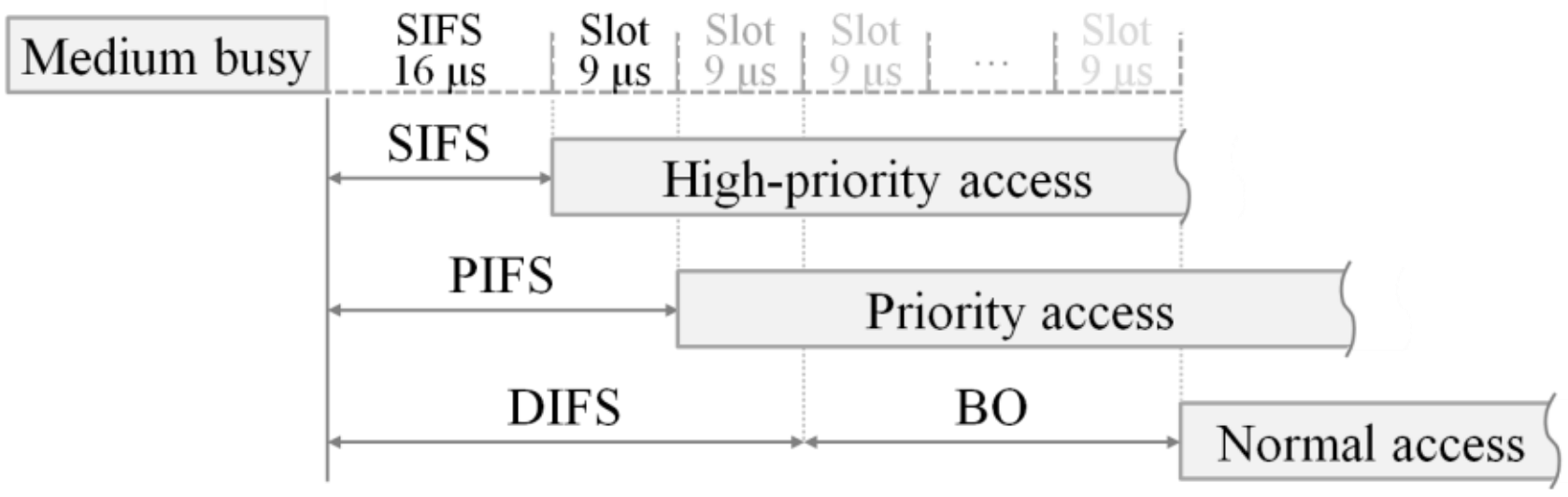

**Figure 1.20 Channel access priorities of IEEE 802.11a with associated timing**

A centralized channel access mechanism, called point coordination function (PCF), was defined in the original IEEE 802.11 specification. This mechanism is seldom used because of BSS overlapping issues (section 1.2.1.3). The PCF IFS (PIFS) defined for this purpose is still used though by the AP (and sometimes stations), to gain priority access to the medium. For all other transmissions, the DCF IFS (DIFS) is used. It is the minimum medium idle time to wait before starting counting down the randomly drawn or frozen BO. We will note that BO duration is a multiple of slot duration and that the CW ranges from 15 to 1,023 slot times [4,38].

## 1.3.2 IEEE 802.11n

The success of IEEE 802.11g expanded the demands on Wi-Fi. The growing expectations of end-users and the requirements in throughput and delay of applications (such as high-definition videos) lead to the definition of the IEEE 802.11n standard [6]. IEEE 802.11n defines mechanisms to provide users some combination of greater throughput, longer range, and increased reliability. This implies having new features on both PHY and MAC layers, some features being mandatory and others optional[1]. A throughput of around 200 Mbps can be obtained (compared to the average 20 Mbps throughput obtained with IEEE 802.11a/g) [39].

### 1.3.2.1 PHY Layer

There are two major PHY layer features brought by the IEEE 802.11n standard. These are the use of MIMO and 40 MHz bandwidth. Advanced coding through low-density parity check (LDPC) codes and the introduction of short GI are other added, but optional, features [39]. Table 1.3 details the main PHY parameters concerned by the evolution, which is tagged high throughput (HT).

[1] The standardization process had been long and painstaking (7 years between the proposal and the final standard) because of optional features. As a result, a lot of industry players released ‘pre-n’ and ‘draft-n’ products during this process. These products are not generally compatible with the final standard [8].

In 802.11n, data rates are indexed as MCSs. There are 8 MCSs per bandwidth and per number of spatial streams (SSs) [6]. These data rates range from 6.5 Mbps (20 MHz, 1 SS, BPSK 1/2, and long GI) to 600 Mbps (40 MHz, 4 SSs, 64-QAM 5/6, and short GI). Still, 802.11n devices have to be interoperable with legacy 802.11a/g devices. Thus legacy data rates have to be supported as well.

**Table 1.3 Main PHY layer characteristics of IEEE 802.11n**

| Parameter | | Value |
|---|---|---|
| Bandwidth (MHz) | | 20, 40 |
| # sub-carriers | 20 MHz | 52 for data out of 64 |
| | 40 MHz | 108 for data out of 128 |
| # spatial streams | | 1 - 4 |
| Coding | | BCC, LDPC |

| Parameter | Duration (μs) |
|---|---|
| GI | 0.8 (long) |
| | 0.4 (short) |
| HT-STF | 4 |
| HT-LTF | 4 each |
| HT-SIG | 8 |

| Parameter | | Possible values |
|---|---|---|
| Data rates (Mbps)[1] | 20 MHz | {6.5, 13, 19.5, 26, 39, 52, 58.5, 65} · #spatial streams |
| | 40 MHz | {13.5, 27, 40.5, 54, 81, 108, 121.5, 135} · #spatial streams |
| Coding rates | | 1/2, 2/3, 3/4, 5/6 |

| Regulatory domain | | Europe | United States | Japan |
|---|---|---|---|---|
| # of 20 MHz channels | 2.4 GHz band | 13 | 11 | 13 |
| | 5 GHz band | 19 | 23 | 22 |

[1] Given values are for long GI only. To determine data rates for short GI multiply tabulated rates by 10/9. This is because OFDM symbol duration decreases from 4 μs to 3.6 μs when shortening the GI. However, coexistence issues arise when using this option. Protective measures should be taken.

#### 1.3.2.1.1 Changes in the PMD Sub-layer

Firstly, multiple Tx/Rx antennas are used to simultaneously transmit up to 4 independently coded and interleaved SSs, through spatial multiplexing (see section 1.1.2.4), thus reaching 600 Mbps. The antennas can also be used to increase range, through spatial diversity (see section 1.1.2.3)[8]. MIMO is thus officially introduced in IEEE 802.11. Secondly, 20 MHz channels can be bonded to form 40 MHz channels. Previously unused border subcarriers on one side of the bonded channels are now around the central carrier frequency. Some can thus be used to convey data. 20 MHz transmissions also benefit from an increase in the number of data subcarriers. Four subcarriers are added. The subcarrier designs for 20 MHz and 40 MHz channels are illustrated in Figure 1.21 (a) and (b) (resp.).

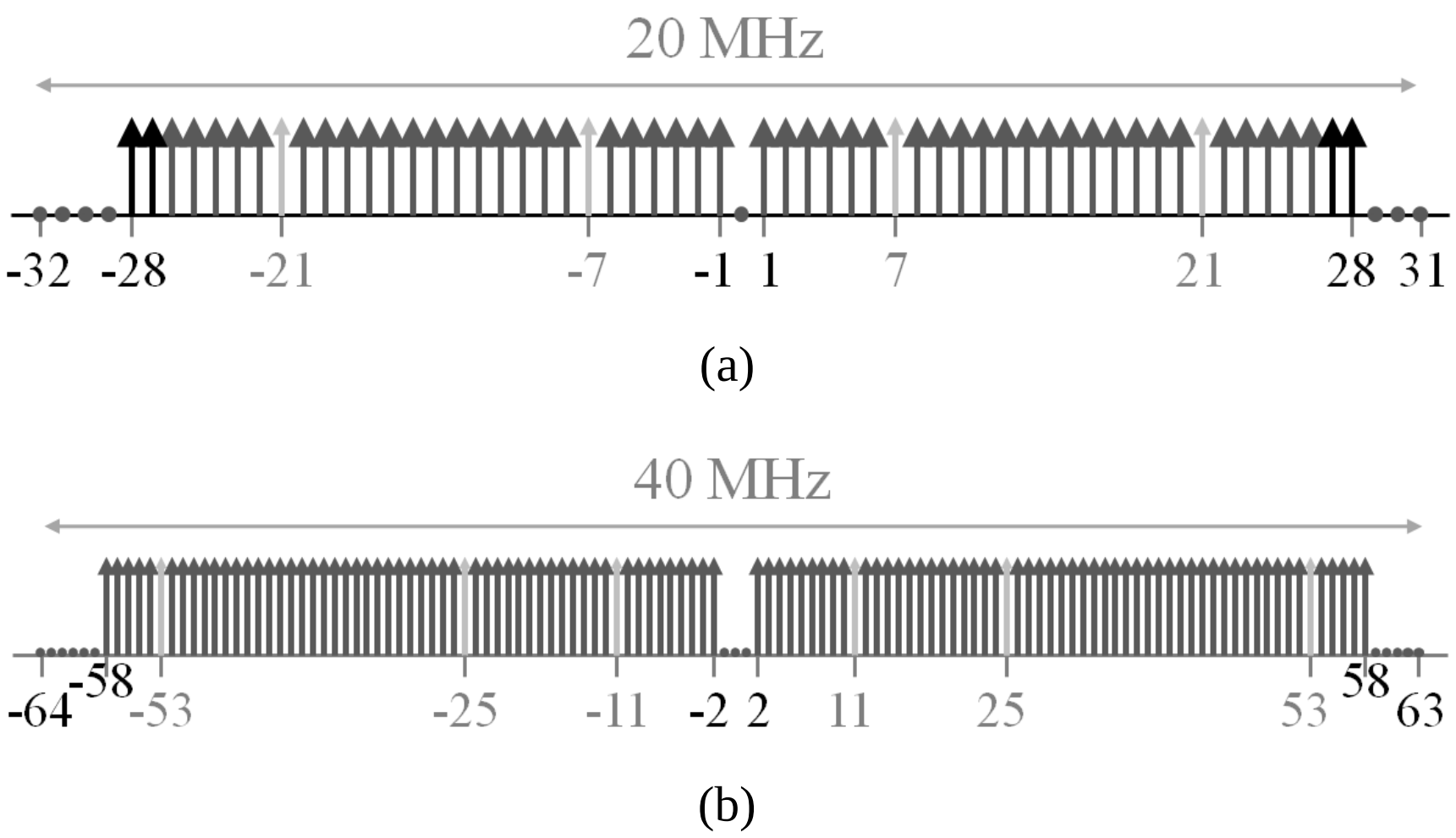

**Figure 1.21 Subcarrier design for IEEE 802.11n with 20 MHz bandwidth (a) and 40 MHz bandwidth (b)**

We will note that channel bonding is facilitated by the fact that regulation authorities have added spectrum in the 5 GHz band. With the resulting additional 20 MHz channels, the odds of having overlapping BSSs when using 40 MHz channels are lowered.

#### 1.3.2.1.2 Changes in the PLCP Sub-layer

When establishing the mathematical model in section 1.1.2.2, we have seen that a channel tap estimate is needed per transmit-receive antenna pair. This cannot be done with the classical LTF. That is why the IEEE 802.11n amendment introduced high throughput LTFs (HT-LTFs). Up to four of these consecutive 4 µs-long symbols are transmitted. One HT-LTF is transmitted per spatial stream[1], so as to efficiently estimate the CSI for each [39].

This is illustrated in Figure 1.22[2]. The HT-LTFs being transmitted using all available antennas, unintentional correlation can be created. To avoid this unintentional beamforming, a different shift (called cyclic shift diversity, CSD) is applied to symbols sent on different antennas. Consequently, automatic gain control has to be redone with a 4 µs-long high throughput STF (HT-STF), before receiving the HT-LTFs [8]. The HT-STF sequence is identical to the one used in L-STF (but shifted over the used antennas) [6]. In addition, the receiver needs to know how many HT-LTFs to expect. This information is contained in the high throughput SIG (HT-SIG) field starting the HT part of the PHY preamble (along with MCS, length, GI, and coding type information). The integrity of this signaling field is checked through an 8-bit CRC and offers much more protection than the single parity bit of the L-SIG.

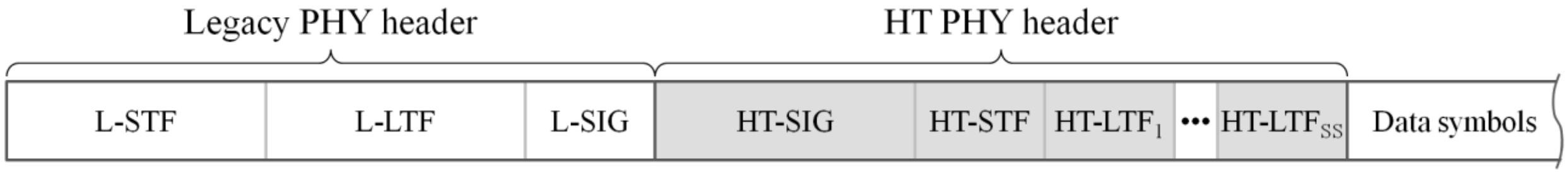

**Figure 1.22 PLCP framing in IEEE 802.11n**

Based on section 1.1.3.4, we will note that HT-LTFs are precoded when using beamforming. The resulting channel can thus be estimated and used for decoding precoded data symbols.

### 1.3.2.2 MAC Layer

Initially, i.e. in 2002, the IEEE 802.11 working group estimated that MAC throughput of over 100 Mbps could be achieved using existing PHY and MAC mechanisms [31]. This was proven to be infeasible considering the important PHY and MAC overhead in IEEE 802.11a/g systems. Consequently, in 2003 the IEEE 802.11n standardization officially started to bring the needed changes. Indeed throughput is the measure of 'useful' information delivered by the system, thus considering the previous overhead [8]. The system cannot benefit from the important increase in data rate offered by the new PHY features if the MAC layer is not also modified. Therefore MAC efficiency, i.e. the ratio between throughput (MAC) and average data rate (PHY), needed to be much improved (see Appendix C for more details and illustrations). Frame aggregation (as explained in section 1.2.7) was introduced for this reason and used with block acknowledgment (see section 1.2.6). The different MAC parameters [6] are detailed in Table 1.4.

[1] When there are 3 SSs, 4 HT-LTFs should be sent for orthogonality reasons [8]. Extension LTFs are not considered in this overview.

[2] This is the common HT-mixed format. The shorter format, HT-greenfield, is optional and is seldom used.

**Table 1.4 Main MAC layer characteristics of IEEE 802.11n**

| Parameter | Max size (octet) |
|---|---|
| A-MSDU | 7,935 |
| MPDU | 7,955 (simple) |
| | 4,095 (in aggregate) |
| A-MPDU | 65,535 |

| Parameter | Value |
|---|---|
| # MDPUs in BA | 64 |
| # access categories | 4 + legacy |
| Sounding for CSI feedback | Explicit |
| | Implicit |

Enhanced distributed channel access (EDCA) is an extension of the basic DCF introduced in the IEEE 802.11e amendment[1] [30]. QoS is supported through prioritization [8]. The MAC headers of 802.11e compliant frames thus introduce a 16 bit QoS control field (as illustrated in Figure 1.23). This field has been modified by the 802.11n amendment [6] to include aggregation information. The latter amendment also added a 4 octet HT control field to provide CSI related information.

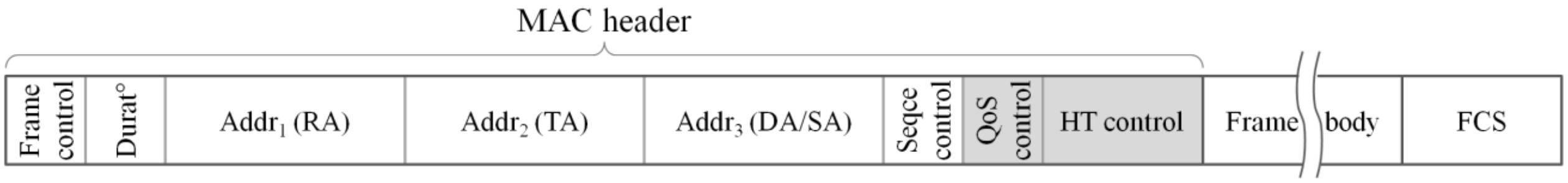

**Figure 1.23 MAC framing in IEEE 802.11n**

In EDCA, channel access timings are defined for every AC (see section 1.2.4.1) so as to implement this prioritization. The arbitration IFS (AIFS) replaces the DIFS, CW range is modified, and TxOPs (see section 1.2.4.2) are limited for every AC. The default EDCA parameters [6] are given in Table 1.5 [8]. These are used as long as the AP does not broadcast a different set of parameters. Usually Wi-Fi multimedia (WMM) parameters [40] are set. The latter are different depending

[1] Actually QoS facility is given by the hybrid coordination function (HCF). EDCA is the contention-based channel access of HCF. The HCF controlled channel access has also been defined but not much used.

whether the device is an AP or a station.

A TxOP limit of zero means that only one frame can be transmitted. Contention access mechanisms have to be used to transmit another frame. Nonetheless, it might seem surprising that voice traffic may occupy such a TxOP considering that voice traffic is composed of small frames (typically 120-octet payload is generated every 20 ms[1]). This is to allow fast retransmissions within the TxOP in case of error (i.e. when ACK has not been received), thus reducing jitter.

**Table 1.5 Default EDCA access parameters**

| AC | CWmin | CWmax | AIFS | TxOP limit |
|---|---|---|---|---|
| Legacy (DCF) | 15 slots | 1,023 slots | SIFS + 2 slots | 0 |
| AC_BK | 15 slots | 1,023 slots | SIFS + 7 slots | 0 |
| AC_BE | 15 slots | 1,023 slots | SIFS + 3 slots | 0 |
| AC_VI | 7 slots | 1,023 slots | SIFS + 2 slots | 3.008 ms |
| AC_VO | 3 slots | 7 slots | SIFS + 2 slots | 1.504 ms |

Whenever beamforming is used (see section 1.1.3), CSI has to be returned to the transmitter so as to perform precoding. This is done through a channel sounding protocol, which is overhead with regards to effective data transmission. In 802.11n, there are two ways of sounding the channel. The beamformer can send a frame containing as many HT-LTFs as there are transmit antennas to trigger an explicit feedback. In this procedure, there are three types of feedback formats (refer to [8] for details). It can also implicitly estimate the channel based on beamformee's PPDUs. In the latter case, however, calibration process is needed to assume channel reciprocity [8].

### 1.3.3 IEEE 802.11ac

New usages, such as compressed video streaming around the house, rapid sync-and-go, and wireless docking, are expected over the coming years. Two projects were started by the IEEE, one for the 5 GHz band, IEEE 802.11ac, and the other for the 60 GHz band, IEEE 802.11ad[2] [41]. The 802.11ac standardization process started in 2008 and should finish by December 2013. In general, 802.11ac could be seen as a lateral extension of 802.11n in which the two basic notions of MIMO and channel bonding are enhanced [42]. More details on 802.11ac mechanisms and techniques are given in Chapter 2.

[1] Considering a G.711 voice over internet protocol codec with real-time transport protocol [94].
[2] IEEE 802.11ad, which includes a single carrier mode in addition to the OFDM mode, is not studied here.

#### 1.3.3.1 PHY Layer

One of the objectives of the IEEE 802.11ac working group is to obtain single link throughput of at least 500 Mbps. This implies at least a five-fold increase in the possible PHY data rates compared to the IEEE 802.11n standard. The consequent changes are listed in Table 1.6. A maximum data rate of 6.93 Gbps (160 MHz, 8 SS, 256-QAM 5/6, and short GI – or $8 \cdot 10/9 \cdot 780$ Mbps ) is thus obtained.

**Table 1.6 Main PHY layer characteristics of IEEE 802.11ac**

| Parameter | | Value |
|---|---|---|
| Bandwidth (MHz) | | 20, 40, 80, 160 |
| # sub-carriers | 20 MHz | 52 for data out of 64 |
| | 40 MHz | 108 for data out of 128 |
| | 80 MHz | 234 for data out of 256 |
| | 160 MHz | 468 for data out of 512 |
| # spatial streams | | 1 - 8 |
| # users per set | | 1 - 4 |

| Parameter | Duration (μs) |
|---|---|
| VHT-STF | 4 |
| VHT-LTF | 4 each |
| VHT-SIG-A | 8 |
| VHT-SIG-B | 4 |

| Parameter | | Possible values |
|---|---|---|
| Data rates (Mbps) | 20 MHz | {6.5, 13, 19.5, 26, 39, 52, 58.5, 65, 78} · #SS |
| | 40 MHz | {13.5, 27, 40.5, 54, 81, 108, 121.5, 135,162, 180} · #SS |
| | 80 MHz | {29.25, 58.5, 87.75, 117, 175.5, 234, 263.25, 292.5, 351, 390} · #SS |
| | 160 MHz | {58.5, 117, 175.5, 234, 351, 468, 526.5, 585, 702, 780} · #SS |
| Modulations | | BPSK, QPSK, 16-QAM, 64-QAM, 256-QAM |

Just as MIMO was the new technique of 802.11n, MU-MIMO is that of 802.11ac. Downlink SDMA (see section 1.1.4) is used. An AP can simultaneously transmit independent groups of streams to multiple stations. It can thus make use of one channel access to transmit 'unicast'[1] data to a group of stations. Up to four independent groups of streams can be transmitted. The antennas available at the AP can therefore be used to increase system efficiency. With MU-MIMO, a total throughput of at least 1 Gbps is expected.

#### 1.3.3.1.1 Changes in the PMD Sub-layer

As an evolution to IEEE 802.11n, IEEE 802.11ac offers enhanced channel bonding. The important increase in data rates obtained with the introduction of 40 MHz bandwidth can thus be pushed a few steps further. In 802.11ac, four or even eight 20 MHz channel can be bonded [41]. The maximum number of spatial streams is also increased to 8. The basic idea is that maximum PHY data rates can be linearly increased with the increase in the number of spatial streams and channel bandwidth [42]. In addition, 256-QAM is used for greater throughput and one 20 MHz channel has been added by the US regulation authorities so as to facilitate channelization (see section 2.1 for more details).

#### 1.3.3.1.2 Changes in the PLCP Sub-layer

IEEE 802.11ac devices are designed to be compatible with legacy IEEE 802.11 devices in the 5 GHz band [41]. Therefore 802.11ac preamble should include L-STF, L-LTF, and L-SIG fields as illustrated in Figure 1.24. The L-SIG field indicates the total length of the very high throughput (VHT) PHY header and PSDU.

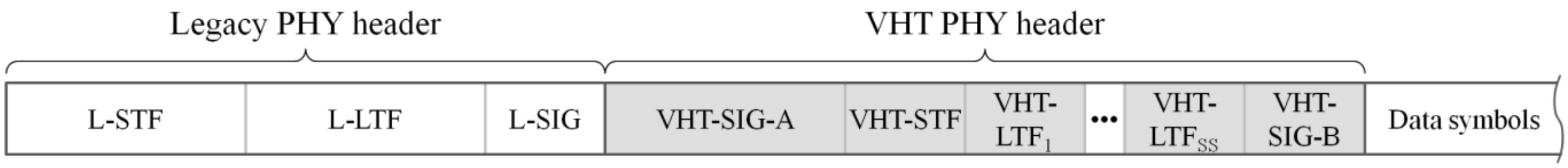

**Figure 1.24 PLCP framing in IEEE 802.11ac**

We have seen in section 1.1.3.4 that when precoding is used, LTFs have to be precoded as well. Different precoding vectors being used per served station in SDMA, the 802.11ac amendment [7] introduced very high throughput LTFs (VHT-LTFs). The same principle is used as with HT-LTFs, except that each VHT-LTF is precoded with a different $\mathbf{w}_c$ vector, so as to enable each receiver $c$ to estimate its precoded channel. Very high throughput STFs (VHT-STFs) are also defined, but are a simple adaptation of HT-LTFs to 802.11ac. In addition, signaling is spread into two fields. The very high throughput SIG A (VHT-SIG-A) field replaces 802.11n HT-SIG and contains common parameters (single-user or multiple-user). The very high throughput SIG B (VHT-SIG-B) contains additional per-user parameters [41]. We shall note that when MU-MIMO is used, all fields after the VHT-SIG-A are precoded. However the latter is not beamformed. It is thus understood by all stations.

[1] Unicast data is meant for a single station in opposition to broadcast (all stations) or multicast (to a group of stations). Here different information is sent to the different stations of the group, thus the use of 'unicast'.

#### 1.3.3.2 MAC Layer

The key MAC enhancements of IEEE 802.11ac address coexistence and medium access with wider channels (refer to section 2.3). The two aggregation techniques introduced by IEEE 802.11n are also slightly modified. Their maximum sizes are increased, as exposed in Table 1.7, so as to have good MAC efficiency at gigabit per second data rates [41]. Indeed with the very high data rates offered by 802.11ac, hybrid aggregation (i.e. combining A-MSDU and A-MPDU) becomes interesting.

**Table 1.7 Main MAC layer characteristics of IEEE 802.11ac**

| Parameter | Max size (octet) |
|---|---|
| A-MSDU | 11,434 |
| MPDU | 11,454 |
| A-MPDU | 1,048,575 |

| Parameter | Value |
|---|---|
| Aggregation types | A-MPDU or hybrid A-MSDU/A-MPDU |
| Acknowledgment | Always BA |

Considering the usage models for 802.11ac, it is much simpler to enable aggregation all the time than constantly switching between A-MPDU and simple MPDU. Consequently only A-MPDUs and BAs are allowed, implying that A-MPDUs containing only 1 MPDU can be used. These A-MPDUs are called single MPDU A-MPDUs. The sounding protocol has also been simplified (see section 2.3.3 for details). We shall also note that the AP can transmit different ACs within one EDCA-driven channel access when using MU-MIMO.

### 1.3.4 IEEE 802.11ah

Current Wi-Fi systems are operating mainly at 2.4/5 GHz and are highly interfered by surrounding networks (especially with channel bonding capabilities). Hence, an alternative and less interfered frequency band will be required in the near future [43]. Using ISM (industrial, scientific, and medical) bands for license-exempt wireless machine to machine communications below 1GHz could help solve this problem. The IEEE 802.11ah standard [44] was thus started in 2010 to enable below 1 GHz transmissions mainly for smart grid and smart utility communications, as well as cellular offload [45]. 802.11ah builds upon the 802.11ac standard. It can thus be easily integrated into current Wi-Fi chip designs. However an important increase in coverage compared to 2.4/5 GHz systems (up to 1 km) is expected [46]. More details on 802.11ah PHY and MAC mechanisms are given in section 7.1.

#### 1.3.4.1 PHY Layer

IEEE 802.11ah indeed explicitly reuses a lot of IEEE 802.11ac's parameters. One major difference (excluding the carrier band) is that bandwidth has been divided by 10 and 1 MHz transmissions are added. The main PHY layer parameters are given in Table 1.8.

**Table 1.8 Main PHY layer characteristics of IEEE 802.11ah**

| Parameter | | Value |
|---|---|---|
| Carrier frqcy (GHz) | | Below 1 GHz |
| Bandwidth (MHz) | | 1, 2, 4, 8, 16 |
| # sub-carriers | 1 MHz | 24 for data out of 32 |
| # sub-carriers | 2/4/8/16 MHz | Same as 20/40/80/160 MHz in 802.11ac (resp.) |

| Parameter | Value |
|---|---|
| Subcarrier spacing | 31.25 kHz |
| Preamble type | Long |
| Preamble type | Short |
| # space-time streams | 1 - 4 |

| Parameter | | Possible values |
|---|---|---|
| Data rates (Mbps) | 1 MHz | 0.375 and {0.65, 1.3, 1.95, 2.6, 3.9, 5.2, 5.85, 6.5} · #SS |
| Data rates (Mbps) | 2/4/8/16 MHz | A $10^{th}$ of 802.11ac's 20/40/80/160 MHz data rates (resp.) |

| Regulatory domain | Europe | United States | Japan[1] |
|---|---|---|---|
| Number of 1 MHz channels | 5 | 26 | 11 |

[1] South Korea, China, and Singapore also have their own regulations [44].

As in the other standards, coexistence with existing networks (e.g. IEEE 802.15.4 [47]) should be taken into account when designing the PLCP. Thus a long preamble has been defined for coexistence with legacy devices when using MIMO with beamforming or MU-MIMO. A short version of the PLCP preamble has also been defined for improved performance when no beamforming is employed, while maintaining coexistence with legacy networks. The major PHY layer challenges that 802.11ah faces concern range and robustness increase.

#### 1.3.4.2 MAC Layer

The MAC header of IEEE 802.11ah will also provide mechanisms that enable co-existence with other systems [43]. On the other hand, some efficiency gains are made by shortening some control/management frames and MAC headers [44]. Indeed, MAC layer changes brought by 802.11ah address the capacity of supporting up to 6000 associated stations (for outdoor smart grid [48]) and the power consumption reduction of each station. In addition, the use of MPDU aggregation becomes mandatory for all MSDUs above 511 octets.

## 1.4 Cross-layer Solutions in WLANs

Over the last few years, there has been increased interest in the idea of cross-layer design for wireless networks [2]. This is motivated by the need to provide better adaptation to variations of the wireless channel [49]. Wireless networks could thus benefit from protocols relying on significant interaction between layers.

### 1.4.1 Cross-layer Optimization Guidelines

The success of the layered architecture for wired networks has had a great impact on network design. By influencing how designers think, it has become the default architecture for designing wireless networks as well [2]. However, there is no notion of link in WLANs. All signals are broadcast and heard by every user within range. The radiated signals may even superpose, thus excluding the Ethernet concept of switching[1] but opening new perspectives. Existing notions should be dispelled and wireless network design should ideally be rethought afresh.

The advantages of cross-layer networking seem worth the effort [50]. Various optimization opportunities do present themselves through cross-layer design. Within the scope of our study, the PHY and MAC layer knowledge of the wireless medium could be shared between these two layers but also with higher ones. An "impedance matching" of the instantaneous radio channel conditions and capacity needs with the traffic and congestion conditions is thus made possible [50]. This aspect is particularly interesting considering that, compared to Ethernet, WLANs suffer from a hostile and quickly changing environment. In addition, taking an architectural shortcut can often lead to performance gain. Functional redundancies can be removed this way. Some functions, which are currently located in one layer, can also be moved to another for greater efficiency. Indeed, there is a strong interconnection among all layers in a wireless network and thus a local layered design approach may not be optimal [49].

[1] A switch allows a receiver to receive just one transmission while shutting out all others [2].

However, once the layering is broken, the luxury of designing a protocol in isolation is lost, and the effect of any single design choice on the whole system needs to be considered. The system can end up having "spaghetti design" [2], where all functionalities are intertwined in a chaotic mass. Indeed some interactions are not easily foreseen and can cause unintended consequences [2]. Cross-layer design can thus potentially work at cross purposes. Special caution should thus be taken when designing cross-layer solutions. A trade-off between performance and architecture is needed. The former has short-term objectives, whereas the latter longer-term objectives. A particular cross-layer suggestion may yield an improvement in throughput or delay performance. But this should be weighed against longer-term considerations, such as ease of evolution and implementation. Cross-layer design proposals must therefore be holistic rather than fragmenting [2].

We will note that cross-layer design is often associated with resource allocation (please look into references given in [49] for more details on these techniques). Issues such as joint design of scheduling, power control adaptive modulation and its influence on CSI are studied. We have chosen to follow the same philosophy when it comes to design principles. However, we have concentrated on the standards themselves and how existing structures could be improved with cross-layer considerations.

### 1.4.2 Group ID for Energy Saving in IEEE 802.11ac

Whenever a frame is sent by a station to another, or frames to a group of stations with MU-MIMO, the sending station puts the MAC address(es) of the destination station(s) in the MAC header(s) of the concerned MPDU(s) (see Figure 1.23 in section 1.3.2.2). Therefore, it is up to the MAC layer of each station to decide whether it is concerned by the received frame(s), as indicated in section 1.2.5. The PHY layer decodes all incoming PPDUs regardless of the destination of the included MPDUs. Important energy consumption due to almost useless processing comes from this layered architecture, when the station is not the destination of received MPDU. Still, virtual carrier sensing (see section 1.2.3.3) benefits from this systematic processing.

Things become more complicated when beamforming is used. This is always the case with MU-MIMO transmissions. An unintended station cannot decode the PSDU because of the precoding. Therefore processing becomes completely useless.

That is why the group identifier (group ID) field was introduced in the VHT-SIG-A of the IEEE 802.11ac amendment [7], as illustrated in Figure 1.25. Through this identifier, stations can determine whether they are part of the multiple-user transmission or not [41]. A station can thus choose not to receive the VHT PSDU [7] and enter power save mode for the duration indicated in the L-SIG [42]. Consequently processing consumption is reduced through this cross-layer design, where a MAC identifier has been added to the PLCP preamble.

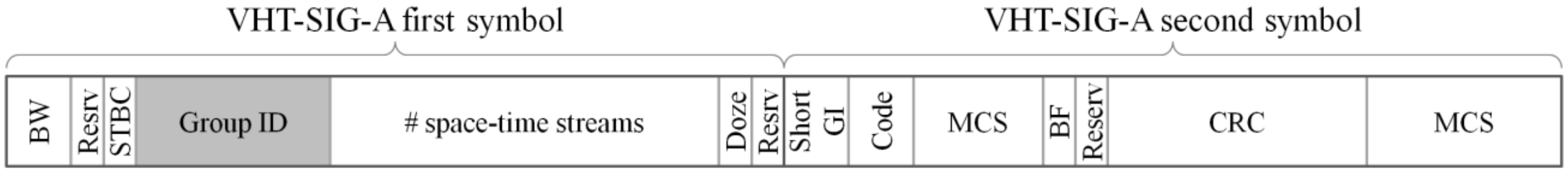

**Figure 1.25 Fields in the VHT-SIG-A (multiple-user transmission)**

A similar procedure is used for single-user transmissions through the partial association ID (partial AID). This abbreviated indication of the intended recipient of the frame (occupying 9 of the 12 bits of the ‘# space-time streams’ field) [7] enables the receiving station to enter power save mode when it ascertains that it is not the intended recipient [42]. The reader can refer to section 2.2 for more details on group ID and partial AID.

### 1.4.3 Robust MAC Header in ECMA-368

In IEEE 802.11 amendments, coexistence issues with legacy devices are important constraints. Therefore cross-layer optimizations are limited and should result in a backward compatible design. In other standards, more freedom can be taken in the design. The ultra wideband ECMA-368 [51] standard is one of those. This high-speed and short-range wireless technology utilizes all or part of the spectrum between 3.1 GHz and 10.6 GHz and supports data rates up to 480 Mbps.

In ECMA-368, the PLCP header contains both PHY and MAC headers as illustrated in Figure 1.26. The integrity of PHY and MAC headers is ensured by the header check sum (HCS). Finally the 3 fields are protected by a Reed-Solomon code. The concatenation of this error-correcting outer code and the classical convolutional inner code[1] improves the robustness of the PLCP header.

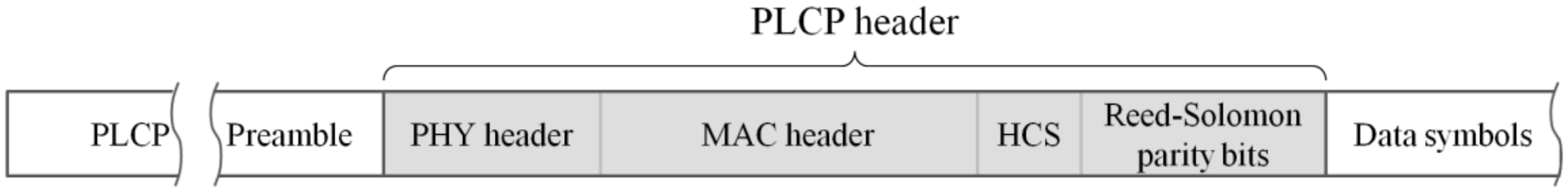

**Figure 1.26 PLCP framing in ECMA-368**

The MAC header being in the PLCP header and well protected from errors, the receiving station can quickly retrieve reliable information from it. The PHY header also benefits from the robustness and error correction capability of the Reed-Solomon coding. Therefore, through this cross-layer design, critical signaling information is well protected, ultimately resulting in increased throughput. The odds of misinterpreting signaling, and thus losing data, are minimized.

In both exposed cross-layer solutions, previously standardized layering has been revisited for improved performance (effective energy consumption, greater robustness, greater throughput), without completely modifying the global architecture. This is precisely the approach we have used for the cross-layer optimization solutions studied in this thesis.

# 1.5 Simulation Platforms for WLANs

In the WLAN context, PHY layer or MAC layer performance evaluation can either use link-level or system-level simulations, depending on the layer to study in depth.

## 1.5.1 Link or PHY-centric Simulators

Elaborated PHY layer simulators, notably including very fine channel models, consider at least one of the following assumptions if not all: full queues [52], oversimplified contention [53], or perfect channel state information feedback [15]. This kind of simulator is based on a 'PHY-centric' study.

[1] Please refer to [96] for details on code concatenation, outer code, and inner code. A concise exposition of these concepts can be found in [97].

Ptolemy II [54] and COSSAP [55] are such simulators. Generally the point of interest is bit error rate (BER) or PER performance for a given SNR. Indeed these PHY-centric simulations concentrate on evaluating a single ‘link’, i.e. Tx chain / channel / Rx chain, between two predefined users.

### 1.5.2 System or MAC-centric Simulators

On the other hand, complex MAC layer simulators allow fine simulation of contention access and queue refilling, which is done according to application layer needs. But the PHY model these ‘MAC-centric’ simulators use is oversimplified considering the complexity of the wireless channel. The latter is often modeled using either graph model [56,57], ON/OFF model [58], or information-theoretic model [16]. Some models even use lookup tables (LUT) to enable a simple link-to-system mapping technique (like PER to SNR correspondence tables [59]). Sometimes more sophisticated techniques are used (BER per block of subcarriers [60]). But in all cases the PHY layer has been simplified. Network simulator 3 (ns-3) [61], or its previous version network simulator 2 (ns-2) [59], and Opnet [62] are such simulators. In MAC-centric simulators the point of interest is the system’s evolution over time. The system, composed of a certain number of users contending for the medium, is modeled without looking into each link.

### 1.5.3 Cross-layer Simulators

Therefore in both types of simulators, i.e. either PHY-centric or MAC-centric, one of the layers has been reduced in complexity, somewhat biasing the behavior of the global system. Creating models that translate actual channel characteristics into system simulations would enable an accurate and fast evaluation of performance [50]. This is crucial when testing cross-layer solutions.

However, building an all-inclusive and exhaustive simulator would be a highly complex solution. The specifications and characteristics of each layer being very different, a common ground might be very difficult to find. All the while, the use of such simulators can be highly profitable, especially when studying phenomena requiring an accurate and realistic modeling of PHY and MAC layer mechanisms.

In [63], authors explain clearly the purpose of using precise PHY layer models in system simulations. Such simulators, referred to as “high-fidelity simulators”, are proven necessary to assess the performance of cross-layer solutions. A generic framework, in which the impact of the wireless channel is finely taken into account in all protocol layers, is thus obtained over the existing

OMNET++ simulator [64]. The MiXiM simulator [65] is based on the same philosophy, though built in a less formalized manner than the previous framework. In both works, the aim is to give a generic framework over an existing simulation tool, ensuring reusability. Nonetheless, the simulator would be less attuned to specific goals than a specifically designed simulator. Considering that this thesis addresses cross-layer optimization for Wi-Fi, the simulator development effort had to focus on the IEEE 802.11 standards.

On the other hand, in [66], authors have finely modeled both PHY and MAC layers of the IEEE 802.11a standard [4], along with a network layer. This simulator, based on ns-3 [61], faithfully accounts for 802.11a specifications (be they PMD, PLCP, or MAC related). Here the aim is to build a detailed PHY layer 802.11a simulator and integrate it in the ns-3 system simulator. However, it does not comprehend either IEEE 802.11ac, or IEEE 802.11n enhancements, thus restricting its use to single antenna Wi-Fi standards. MU-MIMO cannot be studied with such a simulator without important modification.

In [67], system simulations have been done for MU-MIMO transmissions using a fine-grained channel model. However, the considered system is 3GPP LTE-Advanced [68]. Wi-Fi and cellular systems are quite different. In Wi-Fi the access strategy is distributed and the data transmission channel is also used for signaling. An IEEE 802.11 specific simulator has to be used for Wi-Fi MU-MIMO performance analysis studies.

Therefore, there is clearly a need for 802.11ac simulators incorporating not only PHY techniques (MIMO, SDMA) but also MAC techniques (aggregation, MU-MIMO scheduling), while precisely modeling the channel.

# Chapter 2

# IEEE 802.11ac

As exposed in section 1.3.3, the IEEE 802.11ac standard [7] is an emerging very high throughput (VHT) wireless local access network (WLAN) standard that can achieve physical layer (PHY) data rates of up to 7 Gbps for the 5 GHz band [42]. The scope of 802.11ac includes single link throughput supporting at least 500 Mbps, multiple-station throughput of at least 1 Gbps, and backward compatibility and coexistence with legacy 802.11 devices in the 5 GHz band [41]. Consequently, this standard is targeted at higher data rate services such as high-definition television, wireless display (high-definition multimedia interface – HDMI – replacement), wireless docking (wireless connection with peripherals), and rapid sync-and-go (quick upload/download).

In general, 802.11ac could be schematized as a lateral extension of IEEE 802.11n [6] in which the two basic notions of multiple-input, multiple-output (MIMO, see section 1.1.2) and wider channel bandwidth are enhanced [42]. PHY and medium access control (MAC) layer modifications have been introduced with the VHT standard. In addition, channelization has also been modified and multiple-user MIMO (MU-MIMO) introduced for greater efficiency (see section 1.1.4 where the spatial division multiplex access – SDMA – principle behind MU-MIMO has been exposed).

## 2.1 Channelization

In the IEEE 802.11n standard, two 20 MHz channels can be bonded to form a 40 MHz channel bandwidth (see section 1.3.2.1.1). As an evolution to 802.11n, IEEE 802.11ac adds 80 MHz, 160 MHz, and non-contiguous 160 MHz channel bandwidths [41]. The latter is also called 80+80 MHz channel bandwidth and consists in using two non-adjacent 80 MHz channels. The idea of bonding non-adjacent channels is new (see sections 3.1 and 3.2.3.1 for some of its implications). Indeed, in 802.11n, 40 MHz channels comprised two adjacent 20 MHz channels (one primary and the other secondary). These bonded channels do not partially overlap so as to avoid significant in-band interference and resulting coexistence schemes. 802.11ac continues the 802.11n's design philosophy by defining only non-overlapping channels. This is illustrated in Figure 2.1 where 802.11ac channelization, as defined by American and European regulation authorities, is given [41].

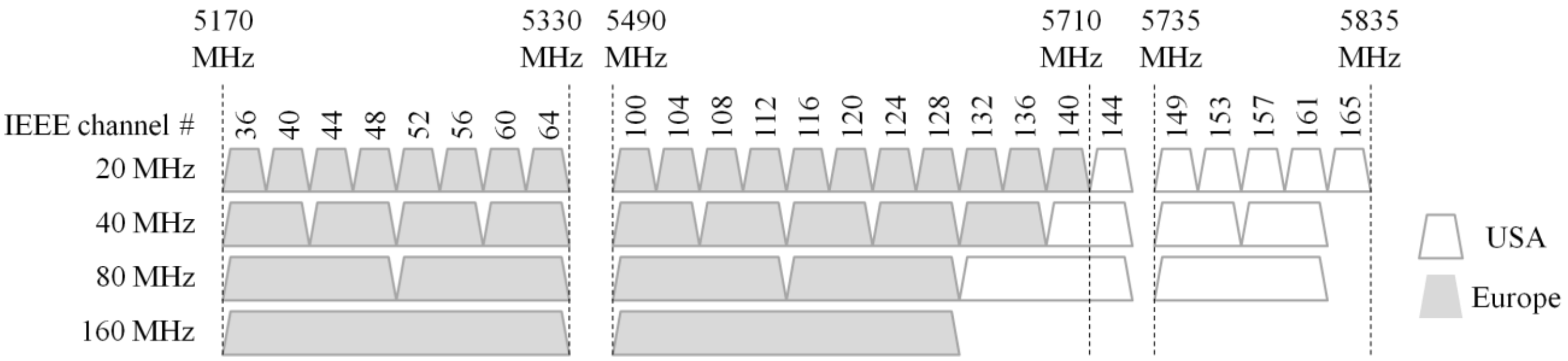

**Figure 2.1 IEEE 802.11ac channelization in USA and Europe**

80 MHz channels are thus formed of two contiguous (or adjacent) 40 MHz bands, in which one of the 20 MHz bands is the primary channel and the rest are secondary channels (actually secondary, tertiary, and quaternary channels) [42]. In North America, channel 144[1] has been added [7] so as to have six 80 MHz channels [41], instead of five if 802.11n's regulations are used. 160 MHz channels are formed of a lower 80 MHz channel and a higher 80 MHz channel which may either be contiguous or non-contiguous [42]. There are only two contiguous 160 MHz channels. This is the reason why non-contiguous 160 MHz channels were introduced, thus enabling many combinations of 80+80 MHz channels [41].

# 2.2 PHY Layer

As seen in section 1.3.3.1, IEEE 802.11ac's design philosophy follows closely that of IEEE 802.11n. Most of 802.11n's mandatory features are maintained. An overview of the mandatory and optional PHY features of 802.11ac is given in Figure 2.2. Contrary to 802.11n, where each station should support up to 2 spatial streams (SSs) and 40 MHz transmissions, only one spatial stream is required in 802.11ac but support for 80 MHz channel bandwidth is added. One reason for such a change is that increasing the number of antennas often results in higher cost. Supporting multiple SSs implies having at least the same number of antennas (and as much reception chains behind these antennas), thus important costs. Consequently a lot of 802.11n devices available on the market could only support 1 SS. In 802.11ac, support for only 1 SS is required so that devices, and especially smartphones, could be labeled as ‘802.11ac compliant’. The 80 MHz mode is made mandatory as the lower cost alternative to the 2 SS and 40 MHz configuration. With 1 SS and 80 MHz transmission, 2 SSs and 40 MHz transmission performance can be achieved and even improved [42].

[1] The central carrier frequency of the corresponding 20 MHz channel can be obtained by adding 5 MHz times the channel number to the channel starting frequency [4] (5 GHz for all channel numbers given in Figure 2.1).

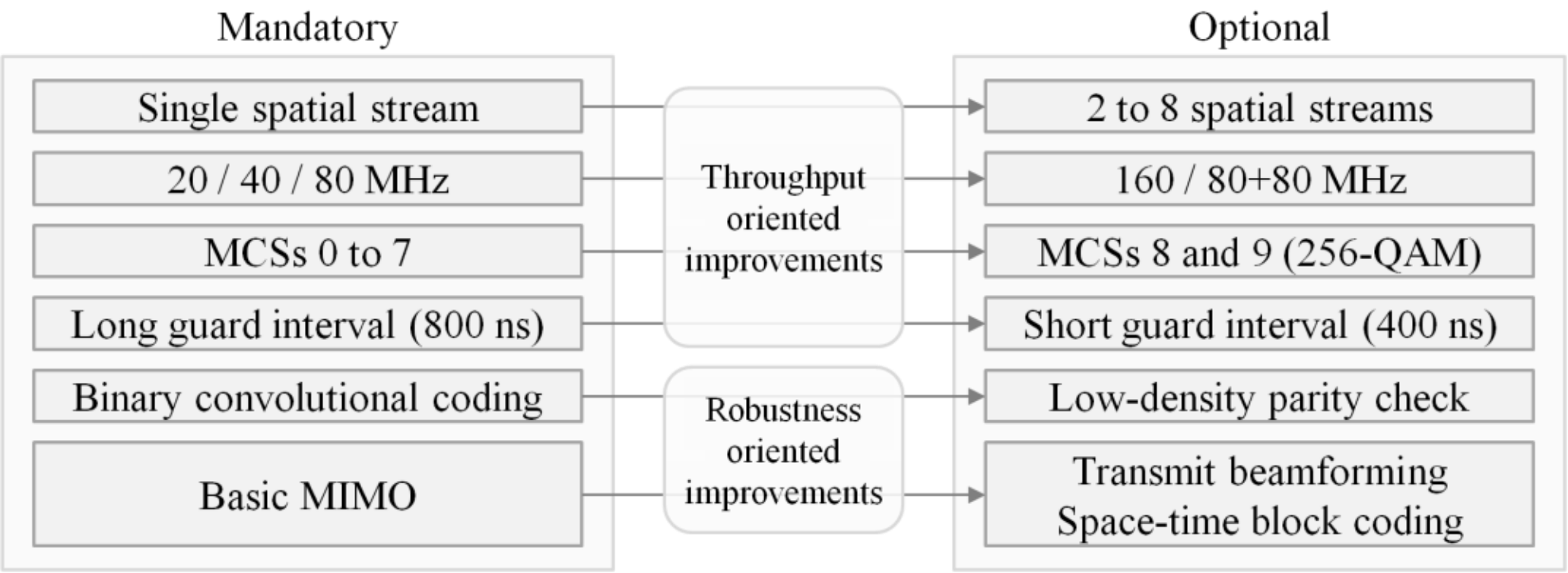

**Figure 2.2 Overview of key mandatory and optional PHY features of IEEE 802.11ac**

## 2.2.1 Subcarrier Design

The 80 MHz channel is indeed composed of 234 data subcarriers, out of a total of 256 subcarriers implicated in fast Fourier transform (FFT), as illustrated in Figure 2.3. This is more than twice the number of data subcarriers in the 40 MHz channel, which has 108 data subcarriers (as listed in Table 1.6).

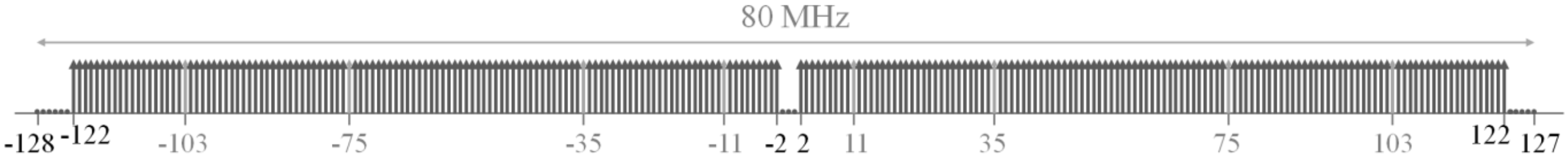

**Figure 2.3 80 MHz subcarrier design for IEEE 802.11ac**

However, the optional 160 MHz subcarrier design is an exact replication of two 80 MHz segments, unifying contiguous and non-contiguous 160 MHz subcarrier design [41]. Another twofold increase in PHY data rates can thus be obtained over the mandatory 80 MHz transmission.

### 2.2.2 Modulation and Coding Schemes

The modulation and coding schemes (MCSs) of IEEE 802.11n, with their 80 MHz extensions, are mandatory in IEEE 802.11ac and listed from 0 to 7 for each bandwidth and number of SSs. MCSs 8 and 9, which use the newly introduced 256-point quadrature amplitude modulation (256-QAM), are optional. All other MCSs, i.e. with 2+ spatial streams or 160 MHz bandwidth, are also optional. An exhaustive list of available very high throughput MCSs can be found in Appendix D.

### 2.2.3 PLCP Framing with Multichannel Transmissions

Despite this enhanced channel bonding, IEEE 802.11ac devices have to coexist and be compatible with legacy 802.11a/n devices in the 5 GHz band. Therefore, basic signaling information, contained in the PHY layer convergence procedure (PLCP) preamble, has to be understandable by all stations within range of the transmitting 802.11ac device. That is why, just like in 802.11n, the fields of the PLCP preamble (as presented in section 1.3.3.1.2) up to, and including, the VHT short training field (VHT-STF) are sent using a 20 MHz waveform signal duplicated over the whole band [41]. Stations within each used channel, be they 20 MHz 802.11ac or legacy stations, can thus retrieve signaling information without having to listen over the whole transmission bandwidth. This is illustrated in Figure 2.4 for a 80 MHz transmission.

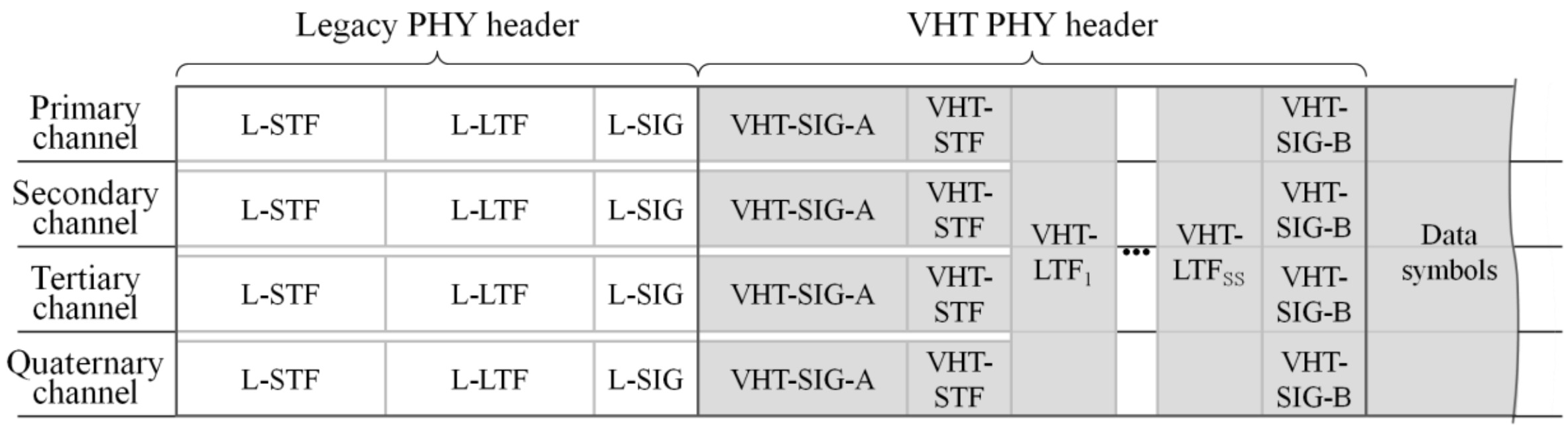

**Figure 2.4 PLCP framing in IEEE 802.11ac for multichannel transmissions (e.g. 80 MHz bandwidth)**

The rest of the fields, especially data symbols, are spread over the whole transmission bandwidth. The VHT long training fields (VHT-LTFs) provide a means for the receiver to estimate the MIMO channel. This way each channel tap of each subcarrier used for data transmission can be correctly estimated. The second signaling field, the VHT signaling B field (VHT-SIG-B), being between data symbols and VHT-LTFs, also occupies the whole bandwidth for ease of implementation. However it roughly contains the same number of bits whatever the bandwidth. Its bits are therefore repeated as many times as necessary and padded, so as to occupy the whole transmission bandwidth for 4 µs [7].

In addition, the first PLCP fields, which are duplicated per 20 MHz channel, are also sent without any precoding (i.e. in an omnidirectional manner) so as to be understandable by all stations. Receiving 802.11ac stations can determine through the first VHT signaling field (VHT-SIG-A) whether they are concerned by the incoming frame. To this aim, the group identifier (group ID) and partial association ID (partial AID), presented in section 1.4.2, are inserted in the VHT-SIG-A field. When stations associate with an access point (AP), they are assigned a 14 bit AID to assist with control and management functions [14]. The 9 bit partial AID is obtained by hashing this AID with the 48 bit basic service set ID (BSSID), thus uniquely identifying a particular station [7]. However, when MU-MIMO is used (see section 2.4 for details), MU-MIMO capable stations need to know whether they are part of the currently served group of client stations. Prior to the multiple-user (MU) transmission, up to 62 different possible groupings are defined by the AP and conveyed to all MU-MIMO capable stations of the basic service set (BSS). Through the 6 bit group ID, stations can also know their order within the group [41].

## 2.2.4 Transmitter Design

The PHY service data unit (PSDU) is processed by the PHY transmission chain to form a PHY protocol data unit (PPDU). The used transmission block diagram for single-user (SU) transmission is given in Figure 2.5 [7]. Transmitter block diagram for MU transmissions is given in 2.4.2.

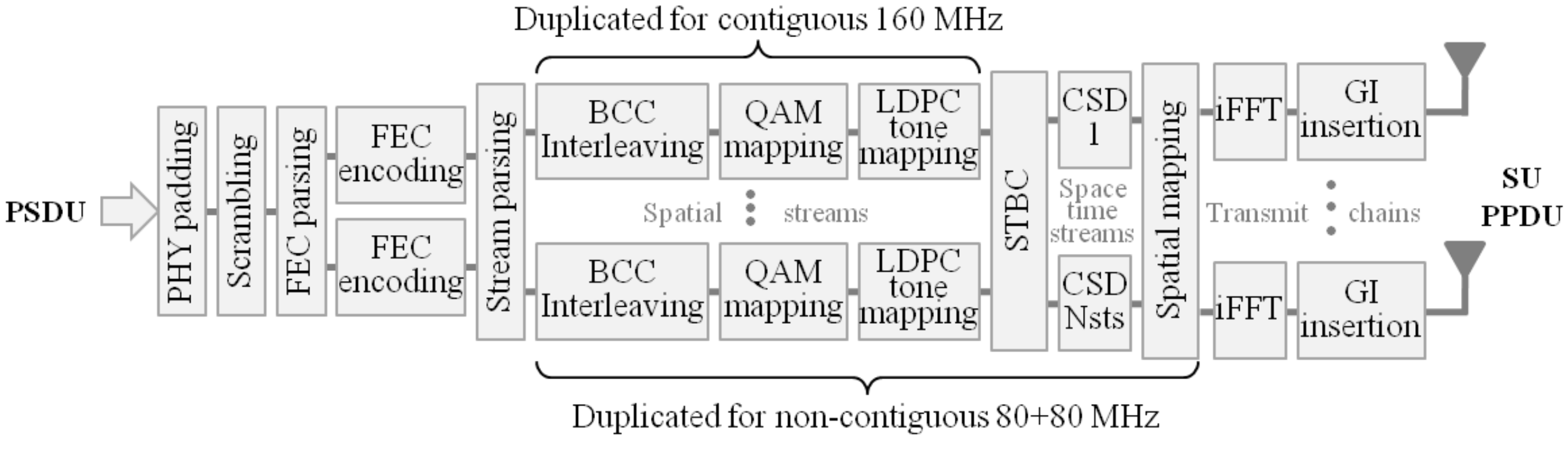

**Figure 2.5 IEEE 802.11ac transmitter block diagram for the data field of single-user PPDUs**

The PSDU bits are firstly padded, to fill all orthogonal frequency division multiplexing (OFDM) symbols, and scrambled. The obtained bits are then encoded using either binary convolutional coding (BCC) or low-density parity check (LDPC). When BCC is used, there may be 1 to 12 forward error correction (FEC) encoders, depending on the maximum data rate. The coded bits are re-arranged into up to 8 spatial streams by the stream parser. Each block of data bits is interleaved and bit-to-symbol mapping is performed for BCC, whereas QAM mapping is performed before tone mapping[1] for LDPC. When space-time block coding (STBC) is used, the STBC block has twice as many outputs than inputs. The outputted space-time streams (STS), to which different cyclic shift diversities (CSDs) are applied to prevent unintentional beamforming (see section 1.3.2.1.2), must not exceed 8. When spatial mapping is used, there may be more transmit chains than space-time streams [7]. The frequency domain symbols are transposed to time domain symbols through inverse FFT (iFFT), prepended with a cyclic prefix guard interval (GI), and windowed, before radio-frequency processing starts and the PPDU is sent over the air.

When 160 MHz channels are used, some blocks are duplicated. For contiguous 160 MHz, the interleaving/tone mapping and constellation mapping functions are duplicated (one block of spatial streams per 80 MHz band). For non-contiguous 160 MHz, STBC, CSD, and spatial mapping functions are included in the duplication [7].

## 2.3 MAC Layer

As exposed in section 1.3.3.2, IEEE 802.11ac builds upon IEEE 802.11n's MAC layer while modifying it to address coexistence and medium access with wider channels (i.e. 80 MHz and 160 MHz) [41]. An overview of mandatory and optional MAC features of 802.11ac is given in Figure 2.6 [42].

Enhancements brought to MAC protocol data unit (MPDU) aggregation and MAC service data unit (MSDU) aggregation are presented in section 1.3.3.2. The new transmit opportunity (TxOP) power save features have been detailed in sections 1.4.2 and 2.2.3. Still, the key feature of 802.11ac is downlink MU-MIMO. MU-MIMO is presented in section 2.4, along with its implications on block acknowledgment (ACK).

[1] LDPC tone mapping is equivalent to block interleaving [7].

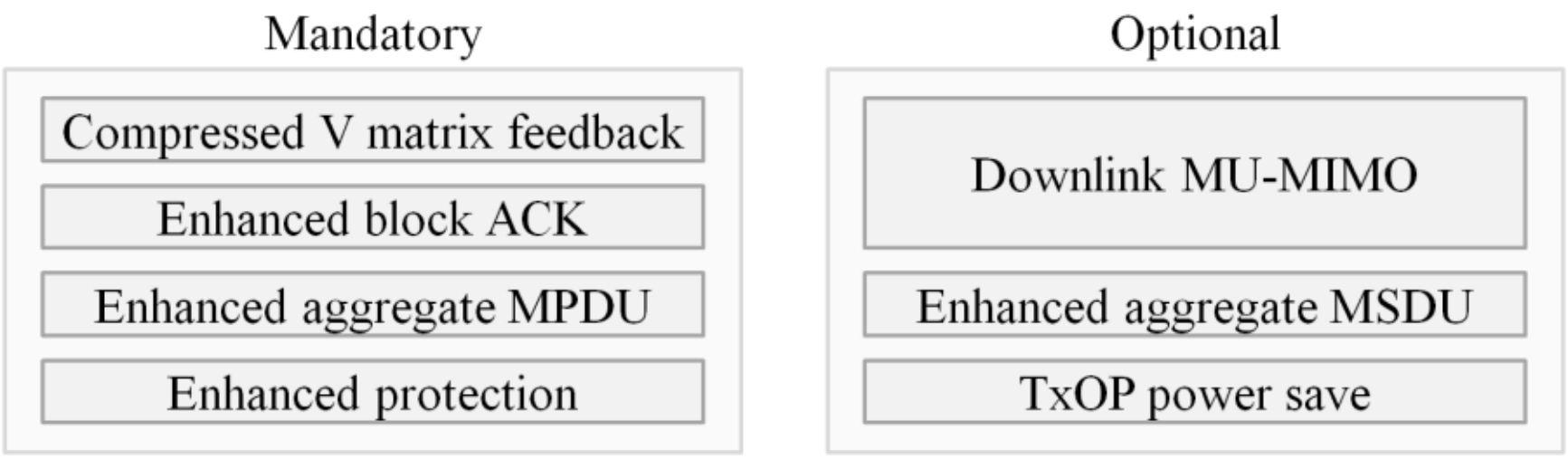

**Figure 2.6  Overview of key mandatory and optional MAC features of IEEE 802.11ac**

### 2.3.1 Multichannel Channel Access

Definition of channel access mechanisms for bonded channels has already been defined in the IEEE 802.11n amendment for 40 MHz transmissions [8]. In IEEE 802.11ac the principle is extended to 80 MHz and 160 MHz transmissions. This channel access mechanism is illustrated in Figure 2.7 for 80 MHz transmissions.

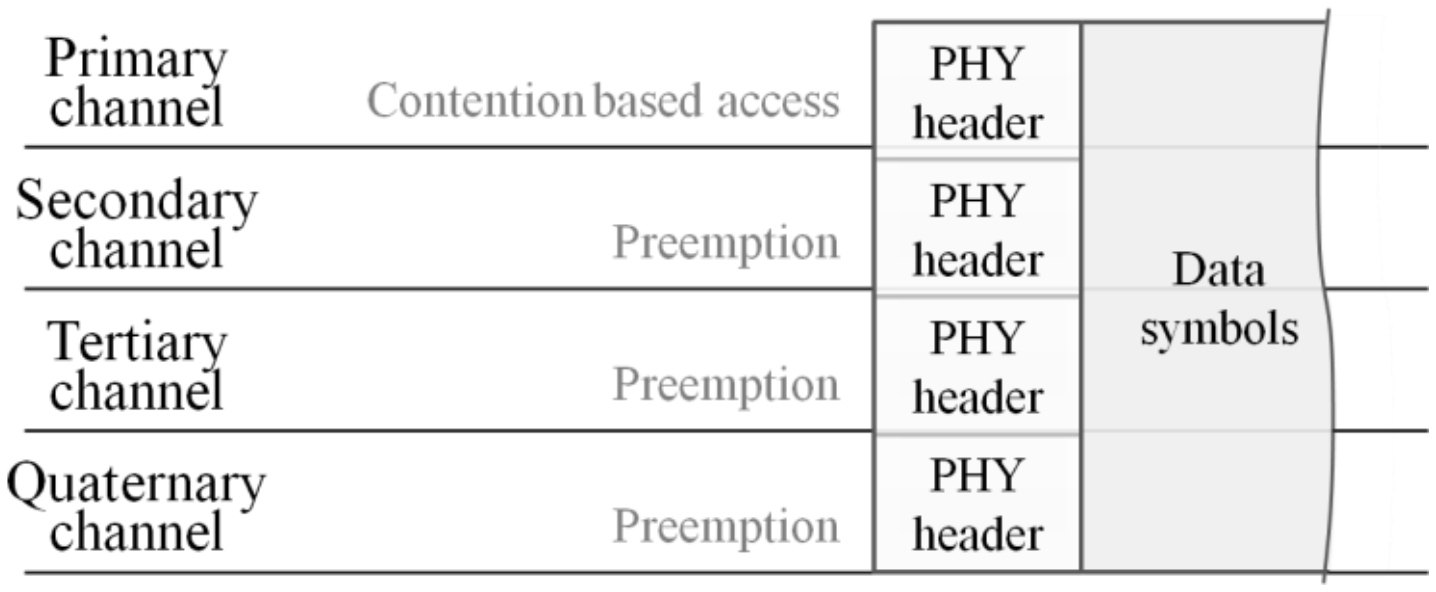

**Figure 2.7  IEEE 802.11ac channel access with channel bonding**

A multichannel station (supporting 80 MHz transmissions in the illustrated example) wins a TxOP through the enhanced distributed channel access (EDCA, see section 1.3.2.2) mechanism on the primary channel. It shall, however, transmit a 80 MHz PPDU only if all other channels have been idle for at least a point coordination function (PCF) inter-frame spacing (PIFS, see section 1.3.1.2). If at least one of the secondary channels has not been idle for a PIFS, then the station must either restart its backoff count, or use the obtained TxOP for 40 MHz or 20 MHz PPDUs [8].

### 2.3.2 Enhanced Protection

With the possibility of bonding up to eight 20 MHz channels in IEEE 802.11ac, it becomes much harder to avoid overlapping between neighboring BSSs despite the important number of available 20 MHz channels (see Table 1.6 in section 1.3.3.1). To address this problem, co-channel operation is improved with three enhancements. Firstly, clear channel assessment (CCA, see section 1.2.3.1) of secondary channels is enhanced so as to improve signal detection. Secondly, the basic request to send and clear to send (RTS and CTS respectively, see section 1.2.3.4) mechanism is modified to improve dynamic channel width operation. The station having sent the RTS frames can send data only on the channels cleared by the CTS frames (with RTS and CTS frames being duplicated on each 20 MHz channel). Thirdly, operating mode notification frames have been introduced to coordinate bandwidth change if interference becomes too frequent [41].

### 2.3.3 Unique Channel Sounding Protocol

A sounding protocol is necessary for beamforming and MU-MIMO, as the beamformer needs to acquire channel state information (CSI) in order to derive the appropriate precoding matrix that could be used to optimize reception at one or more beamformees (see sections 1.1.3 and 1.1.4) [42]. In the IEEE 802.11n standard, the multiplicity of options for the sounding protocol (see section 1.3.2.2) has made things difficult for interoperability when using beamforming (BF) techniques [41]. Consequently, IEEE 802.11ac uses a unique sounding protocol based on the use of null data packets (NDPs) for channel sounding and compressed BF matrices for feedback. A NDP is a PPDU frame with no PSDU, used mainly to estimate the channel. The protocol is illustrated in Figure 2.8 for a single beamformee (see section 2.4.4 for multiple beamformees).

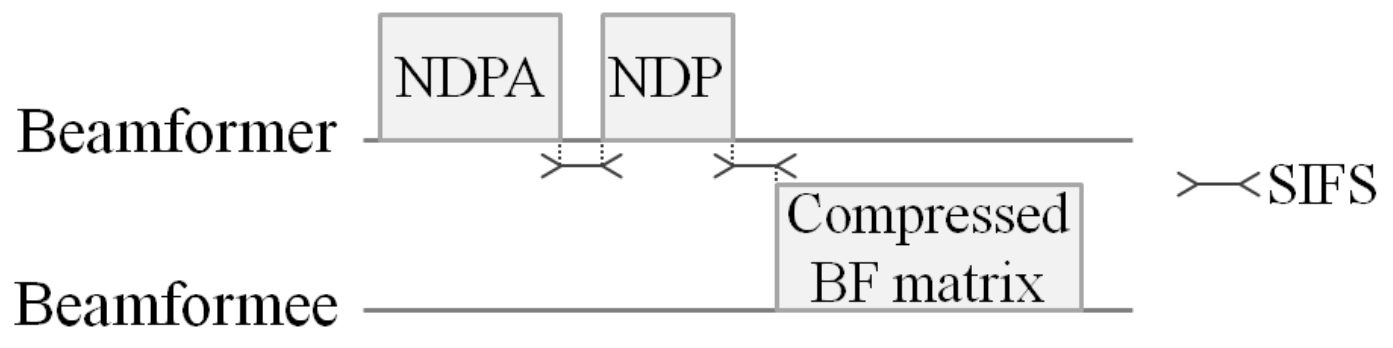

**Figure 2.8 IEEE 802.11ac channel sounding protocol with a single beamformee**

The beamformer announces the beginning of a sounding procedure through a NDP announcement (NDPA) frame [7]. In it the beamformer advertizes the beamformee’s address. The concerned beamformee can thus prepare itself to receive the upcoming NDP frame, and consequently compute its BF matrix using singular value decomposition (see sections 1.1.3.1). The frame exchange is punctuated with short inter-frame spacing (SIFS). Upon reception of the NDP, the beamformee responds with a compressed version of the precoding $\mathbf{V}$ matrix (see section 1.1.3.2).

Compression is used to reduce sounding overhead, while limiting quantization loss. The BF weights are thus expressed in polar coordinates and quantized versions of the $\Psi$ and $\Phi$ angles are sent (refer to [8,7] for more details on how these angles are obtained). Angular information is needed for each transmit-receive antenna pair and for each subcarrier. Grouping of subcarriers may thus be used to further reduce feedback overhead [8].

The duration of the sounding procedure depends on parameters given in Table 2.1. Clearly the parameters having the most impact are the number of spatial streams and antennas, and bandwidth.

**Table 2.1 Parameters conditioning sounding and feedback frames' sizes in IEEE 802.11ac**

| **Frames** / Fields | | Conditioning parameters |
|---|---|---|
| | **NDP** | Beamformee's number of antennas |
| **Compressed BF matrix** | Signal-to-noise ratio info. | Number of spatial streams |
| | Channel matrix element | Bandwidth<br>Subcarrier grouping<br>Beamformee's number of spatial streams<br>Beamformee's number of antennas<br>Number of angle quantization bits (for $\Psi$ and $\Phi$) |
| | MU only information | Bandwidth<br>Subcarrier grouping<br>Beamformee's number of spatial streams |

We shall note that, ideally, the channel remains the same between the time the sounding PPDU is sent and the time the beamformed PPDU is sent [8].

# 2.4 Multiple-user MIMO

MU-MIMO, or downlink SDMA (see section 1.1.4), is one of the important features of IEEE 802.11ac. Through this new technique a cumulated throughput of 1 Gbps can be supported. An AP can simultaneously transmit independent groups of spatial streams to multiple client stations. It can thus make use of one channel access to transmit 'unicast' data to a group of stations. Up to four independent groups of spatial streams can be transmitted. The antennas available at the AP can therefore be used to increase system efficiency [41].

## 2.4.1 PLCP Framing with Multichannel MU-MIMO Transmissions

Using MU-MIMO implies differently precoding the streams to send to each client station. This is illustrated in Figure 2.9, in which the PCLP multichannel framing given in Figure 2.4 is detailed when a MU-MIMO PPDU is sent to 2 client stations. The precoded channels have to be estimated by each of the client stations because different beamforming matrices are used. Thus VHT-LTFs and following fields are all precoded when MU-MIMO is used (see sections 1.3.3.1.2 and 2.2.3).

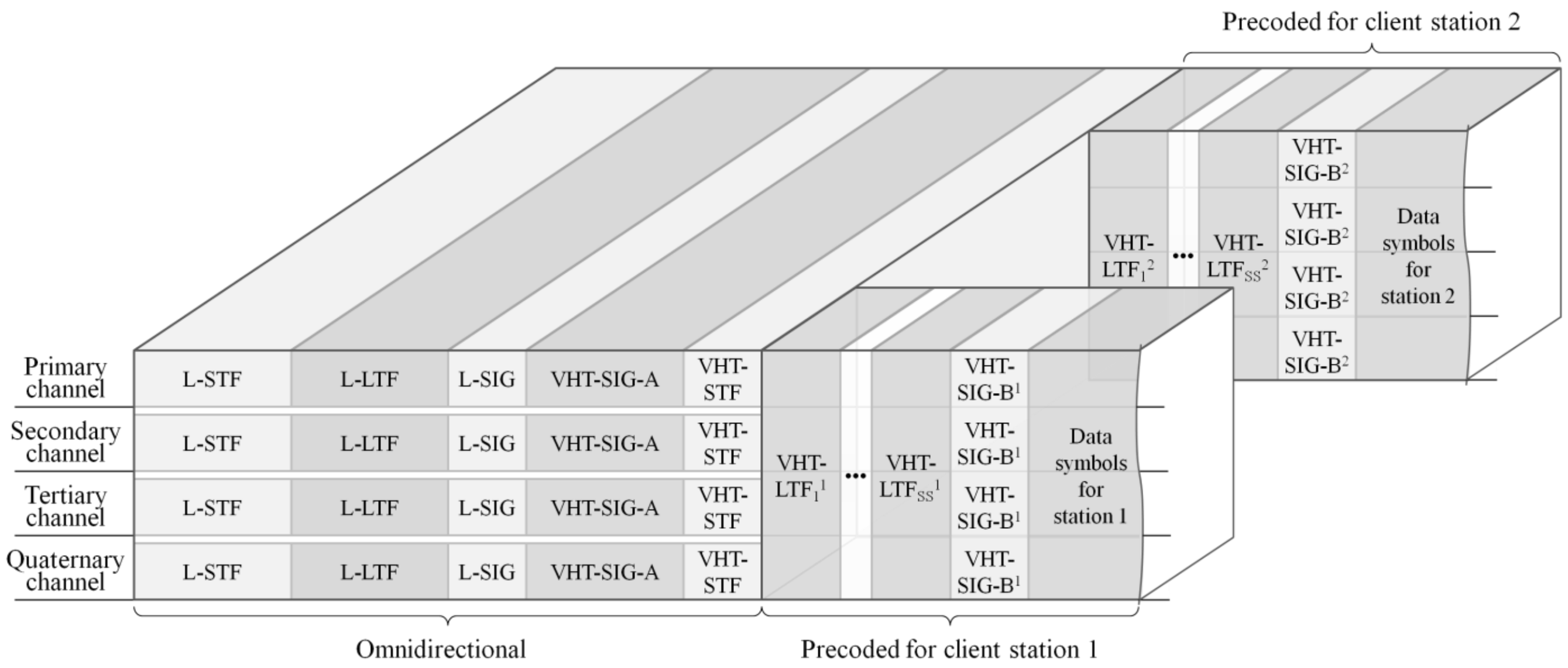

**Figure 2.9 PLCP framing in IEEE 802.11ac for multichannel MU-MIMO transmissions (e.g. 80 MHz bandwidth)**

Precise knowledge of each of the concerned channels is also required at the AP. Consequently, multiple-station channel sounding is accordingly performed in a regular manner (see section 2.4.4), inducing overhead. Another implication is that groups of stations are to be defined by the AP (see section 2.2.3). It is based on the resulting identifier that client stations will retrieve their data from received MU-MIMO PPDUs. Finally, the acknowledgment procedure is also adapted to enable served client stations to acknowledge their received PSDUs (see section 2.4.3). Therefore it can be clearly seen that MU-MIMO has implications on both PHY and MAC layers.

### 2.4.2 Changes in Transmitter Design

When the AP uses MU-MIMO to send independent PSDUs to different clients, it should process the concerned data in a parallel manner. Indeed in single-user MIMO (SU-MIMO), each destination has access to all signals received on its antennas and uses this information in the decoding process. When it comes to MU-MIMO, there is no cooperation between client stations in the receiving process. The transmitter block diagram illustrated in Figure 2.5 should be modified to the transmitter block diagram illustrated in Figure 2.10 so as to send $N$ independently processed PSDUs towards $N$ client stations in a MU PPDU [7].

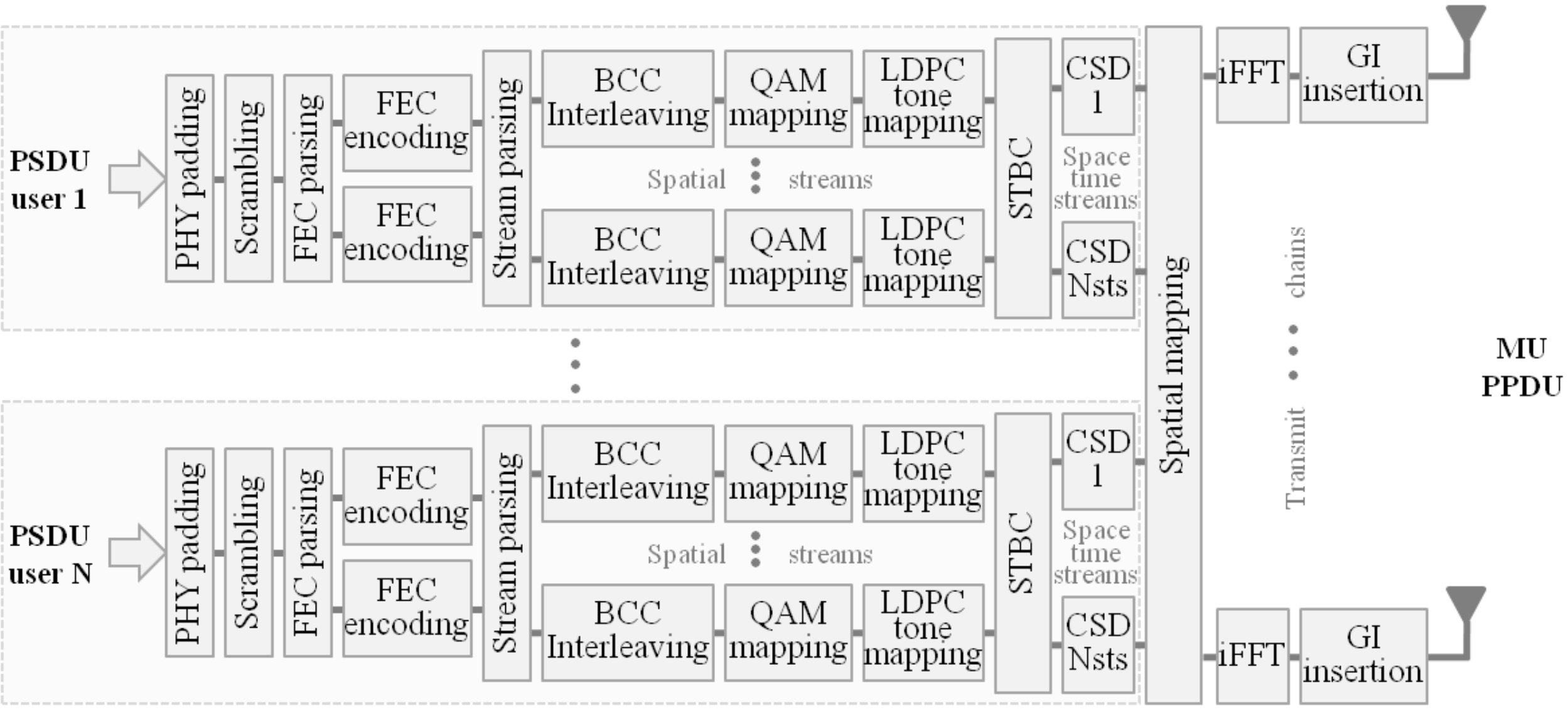

**Figure 2.10 IEEE 802.11ac transmitter block diagram for the data field of multiple-user PPDUs**

The total number of spatial streams shall not exceed the allowed maximum of 8 (see Table 1.6). In addition, for 160 MHz and 80+80 MHz transmissions, each PSDU-processing 'branch' is duplicated following the rules given in section 2.2.4.

### 2.4.3 Multiple-user Acknowledgment

When simultaneously sending PSDUs to client stations, the positive acknowledgment principle (see section 1.2.2) still stands. In addition, MPDU aggregation being mandatory (see section 1.3.3.2), block ACKs (BAs) are used for acknowledging received MPDUs. A sequencing of the BAs sent by clients having successfully decoded at least one of their MPDUs is thus necessary.

Figure 2.11 illustrates the IEEE 802.11ac multiple-user acknowledgment mechanism for 2 client stations. Once the AP has transmitted a MU-MIMO PPDU with PSDUs for both stations, the first client immediately replies with a BA. The AP subsequently polls the second client with a BA request (BAR) and the second client also responds with a BA. The BAR+BA procedure continues until all other clients of the original MU-MIMO transmission are polled [41].

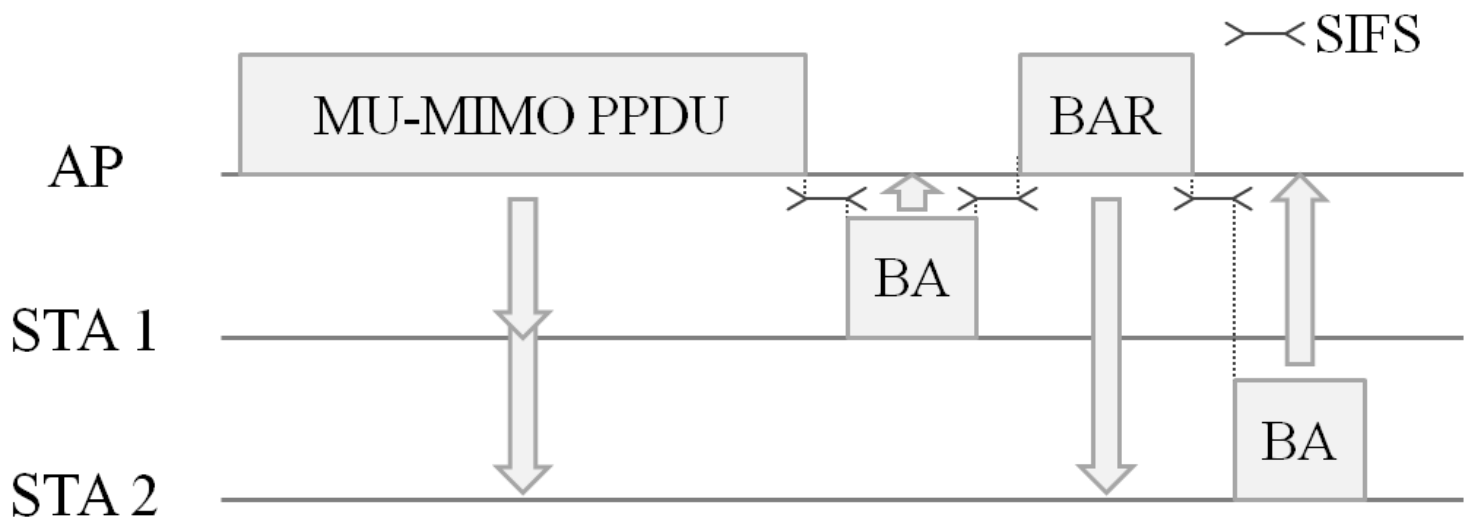

**Figure 2.11 Acknowledgment procedure with MU-MIMO transmissions in IEEE 802.11ac**

This procedure assumes that clients know that they are part of the MU transmission and are aware of their respective orders. This is achieved through the group ID information contained in the VHT-SIG-A of the MU-MIMO PPDU (see section 2.2.3). The VHT-SIG-A also contains a field indicating the number of spatial streams being transmitted to each client (see Figure 1.25 in section 1.4.2).

### 2.4.4 Multiple-user Sounding Protocol

The principle of polling client stations (except the first one) is also used when it comes to the sounding protocol. The resulting MU sounding protocol is illustrated in Figure 2.12 for 2 stations belonging to the same group.

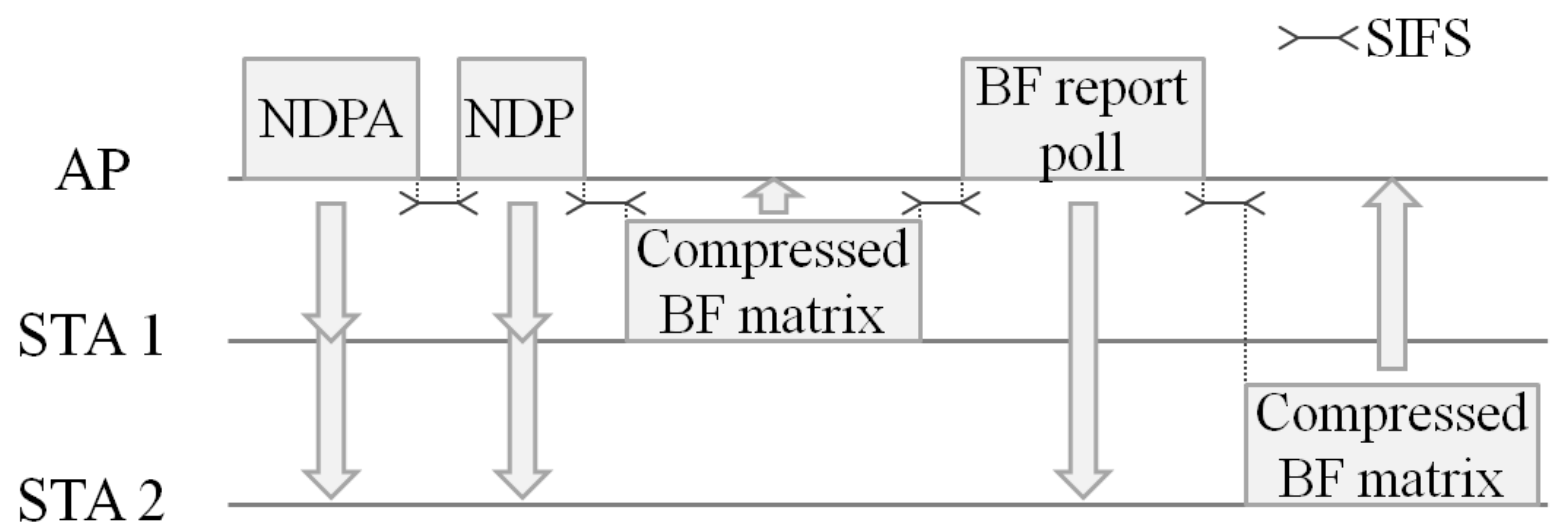

**Figure 2.12 IEEE 802.11ac channel sounding protocol with multiple beamformees**

Indeed it is through the group ID contained in the NDPA that beamformees can prepare themselves for the upcoming NDP frame. The first station, as established by the group ID, follows the SU protocol described in section 2.3.3. However, additional MU information is given (see Table 2.1) to also feedback deviation in signal-to-noise ratio due to precoding [7]. The AP then polls the remaining stations for their respective BF matrices, through BF report poll frames.

With MU-MIMO, antenna weights are much more sensitive to changes in the channel. In the case of SU beamforming, if antenna weights are stale, system performance degrades to the case without beamforming. However, with MU-MIMO, if antenna weights do not accurately match the channel, the streams of one client station introduce interference to the other client stations, leading to important signal degradation [41]. This implies that channel sounding has to be done more frequently when using MU-MIMO than when using SU beamforming. Therefore, in addition to the number of antennas and bandwidth (as seen in section 2.3.3), the number of beamformees (i.e. number of groups and number of clients per group) has an important impact on sounding overhead.

Chapter 3

# Cross-layer Multichannel Aggregation

In the IEEE 802.11n standard [6], one of the major jumps in throughput has been obtained through the use of medium access control (MAC) aggregation (as exposed in section 1.3.2.2). Indeed MAC protocol data unit (MPDU) aggregation, presented in section 1.2.7, was among the head innovations of this standard. Through it, overhead has been considerably reduced by sharing the channel access period and the physical layer (PHY) header among the MPDUs of the aggregate MPDU (A-MPDU). Hence MAC efficiency has been considerably improved (see Appendix C for more details). The increase in PHY data rate, obtained by enlarging bandwidth, can thus result in an increase in throughput. Accordingly, the channel bonding concept has been used in the definition of the 802.11n standard. Two 20 MHz channels (as defined in the IEEE 802.11a/g standards [4,38]) can be bonded [41] (see section 1.3.2.1.1). The multichannel concept is introduced. This concept is pushed a few steps further in the IEEE 802.11ac standard [69] by allowing the use of 4, or even 8, concatenated 20 MHz channels (see sections 1.3.3.1.1 and 2.1). Data is thus spread over more channels. The PHY layer is optimized through the resulting frequency diversity (see section 1.1.1.2).

However, the more the used 20 MHz channels per transmission, the higher the risks of having somebody else transmitting on one of these channels. The overlapping basic service set (BSS) and channel saturation problems become a burden for the MAC layer (see section 1.2.1.3). Channels must be reused and hence collisions appear more frequently. The A-MPDU can indeed be transmitted at very high data rates once access to the whole band has been acquired. But, even though the PHY layer is optimized, a single collision can easily corrupt the whole data. The source station may thus need to retransmit the frame, thus impacting throughput. An exclusive MAC layer optimization will bring about a partitioning of the checksums to minimize sensitivity to collisions (A-MPDU principle).

To the best of our knowledge, system optimization has not focused on MAC layer robustness to collisions in a multichannel context. In [70], authors propose an adaptive aggregation scheme for multichannel transmission. However the MPDUs are spread over all channels for greater MAC efficiency. The issue of collisions has not been addressed. In the solution exposed in this chapter we propose to take into account both PHY and MAC requirements in a realistic environment through a cross-layer multichannel aggregation scheme.

The work presented in this chapter was done during the first steps of the standardization process of the 802.11ac standard. Therefore some assumptions had been taken. 802.11n was taken as a basis and enriched with improvements envisioned by the very high throughput task group (TGac) work group specifications [71] that were available at the time. Most of these specifications are still valid in the current draft [7]. The solution exposed in this chapter can thus be used in 802.11ac systems, but slightly modified for compatibility (especially concerning channelization).

We firstly present the considered system, assuming an increase of transmission bandwidth up to 80 MHz obtained by bonding 4 non-contiguous 20 MHz channels. Then, we present our proposed aggregation solution and discuss its advantages compared to the state-of-the art technique. This is followed by an exposition of simulation scenarios and parameters. Finally, simulation results are presented and discussed.

# 3.1 Explicated Context

## 3.1.1 Multichannel Transmissions

As addressed earlier, the IEEE 802.11ac standard enables the bonding of up to four, or even eight, 20 MHz channels (see section 1.3.3.1.1). Because of the limited number of channels (19 in the 5 GHz band in Europe, as given in Figure 2.1 of section 2.1), channel saturation becomes problematic. Indeed, in densely populated areas the 5 GHz band will surely tend to saturate even with a 20 or 40 MHz bandwidth usage per BSS.

### 3.1.1.1 Non-contiguous Channels

In such a context, the chances of finding an idle 80 MHz bandwidth with contiguous channels become very small. This problem can be partially solved by using non-contiguous channels to form the desired bandwidth (following the principle of 80+80 MHz channels exposed in section 2.1). However it implies duplicating the front-end segments for independent receive filtering and signal forming.

### 3.1.1.2 Front-end Segments

Front-end segments are composed of intermediate frequency filtering and up/down conversion chains, including digital to analog converter and analog to digital converter (ADC) respectively. When using non-contiguous channels, there are as many segments as there are separate blocks of channels. Interference rejection between the blocks of channels is thus simplified. A front-end segment structure in a multichannel receiver using non-contiguous channels is illustrated in Figure 3.1.

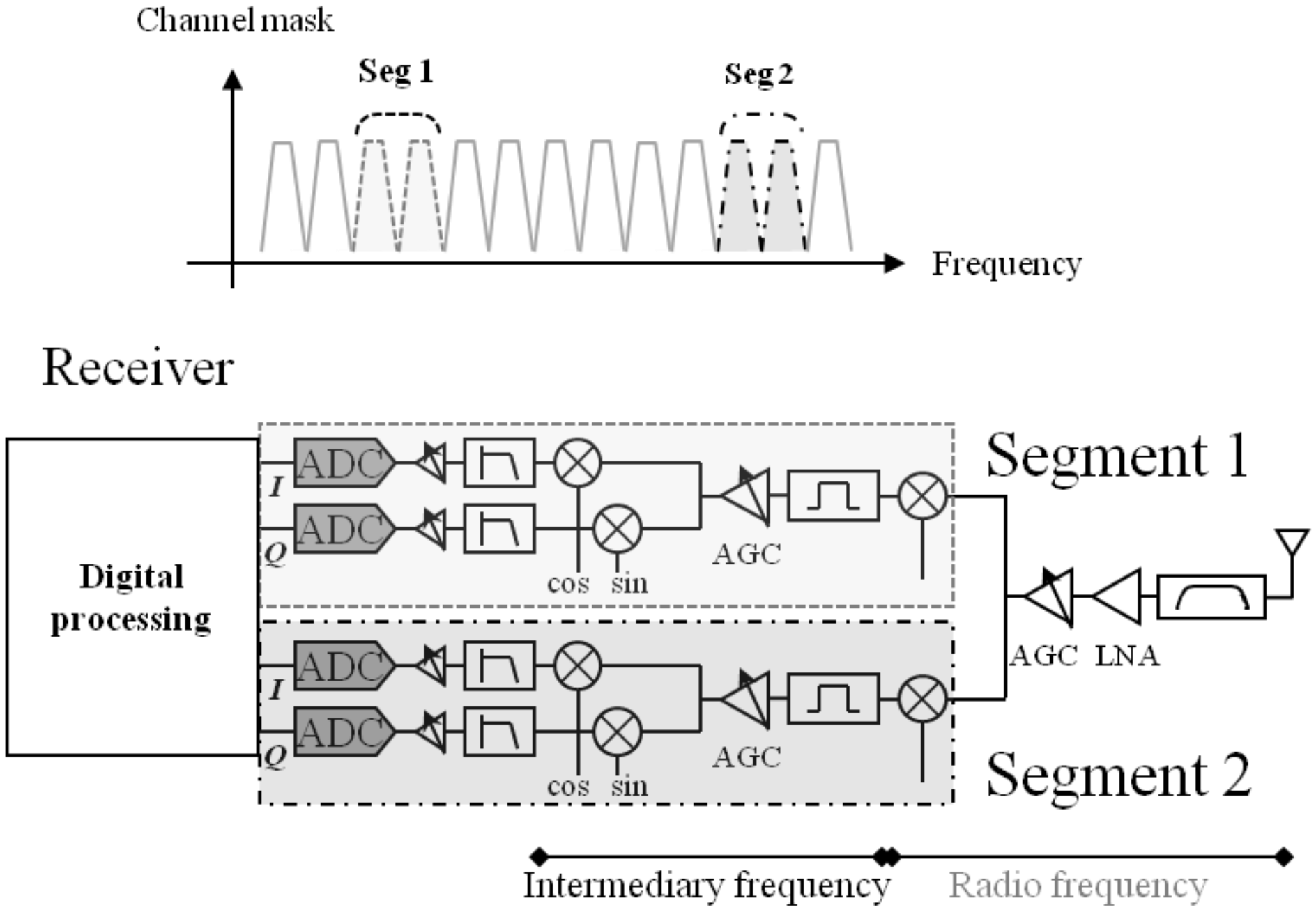

**Figure 3.1 Front-end segment structure in a multichannel receiver using two non-contiguous channels**

The band saturation problem is alleviated, compared to the contiguous case, because channel width can be easily adapted to idle channels.

## 3.1.2 Channel Access Mechanism

Indeed, the IEEE 802.11 channel access mechanism features a random contention-based channel access time with a backoff selection interval which can be exponentially increased so as to minimize the chances of having collisions (see section 1.2.3.1). In the multichannel case, this contention-based mechanism is only applicable for the primary channel. The others are preempted only if they have been idle for a fixed period before the end of the backoff period of the primary channel (see Figure 2.7 in section 2.3.1). If not, the concerned station must either restart its backoff counter or use the obtained transmit opportunity (TxOP, see section 1.2.4.2) to transmit over idle channels.

Even so, collisions cannot be completely avoided. Two stations can draw the same values for their backoff intervals [72]. They can also be out of range of each other but in range of a third station, thus eluding the carrier sense function. This phenomenon is known as the hidden node problem [73] (see section 1.2.1.4 for more details). The probability of occurrence of both situations increases greatly as more 20 MHz channels are used, especially in environments with many overlapping BSSs.

As aggregation in a multichannel context implies spreading the transmitted A-MPDU's data over all the used channels, as illustrated in Figure 3.2, the probability of having a collision in one of the channels can be expressed as follows:

$$P_{collision} = 1 - \left(P_{non\,collision,\ per\,channel}\right)^{number\,of\ channels} \tag{3.1}$$

We consider here that any given frame transmitted over any one of the different channels, is not the duplicate of another (e.g. IEEE 802.11n's 40 MHz duplicate mode [6] not in use) and that the channels have the same probability of collision. This relatively high global collision probability corrupts A-MDPUs greatly. Indeed, the integrity check sequences (which are given per MPDU and are independent of one another, as detailed in section 1.2.7) have also been spread over the used channels. Therefore, even if there is only one collision in one of the channels at a given time, the risks of having a corrupted segment of these sequences, called frame check sequences (FCSs), are very high despite the error-correcting decoding process. All MPDUs could thus automatically be considered as incorrect.

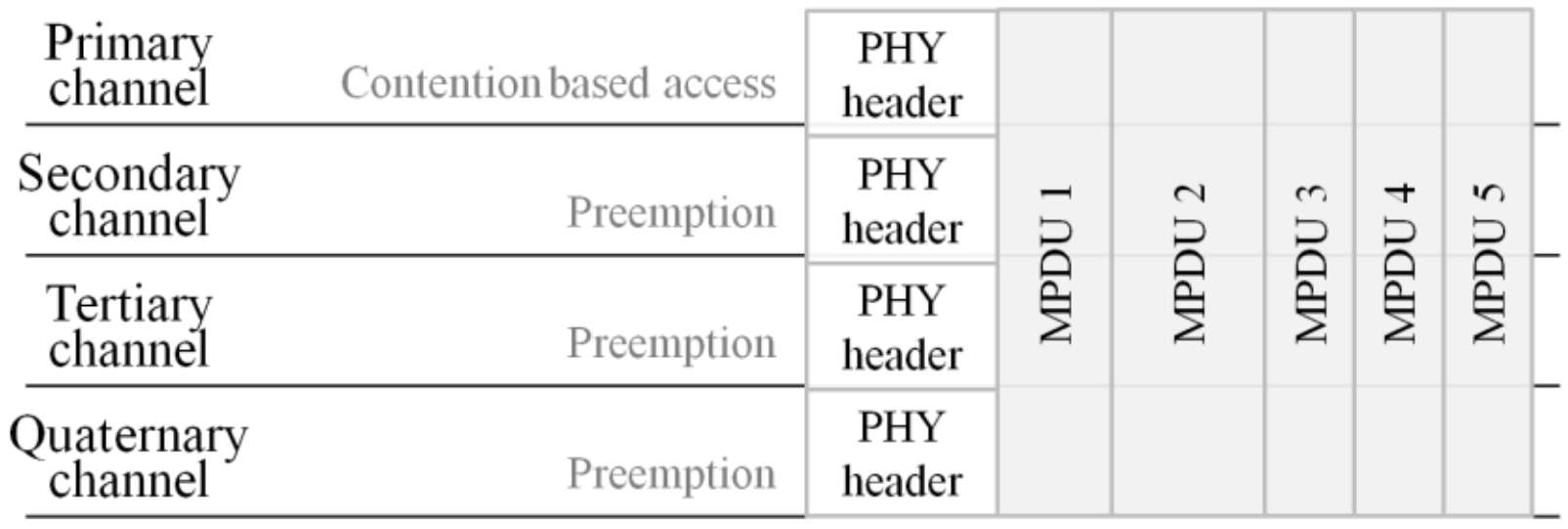

**Figure 3.2 Example of 'vertical' aggregation over 4 channels**

This problematic extension of the classical aggregation to the multichannel case has a structure that we will, from now on, call 'vertical' due to the extension of the payload to all channels (as illustrated in Figure 3.2). We can also clearly see that each MPDU covers all channels. The obtained multichannel system has been optimized for frequency diversity (i.e. from a PHY point of view).

### 3.1.3 Transmit and Receive Chain Structures

The digital part of the transmit and receive chains, illustrated in Figure 3.3, stays more or less the same as in a single channel case [6], except for increased block sizes and header duplication. The IEEE 802.11n standard transmission scheme can be taken as reference[1].

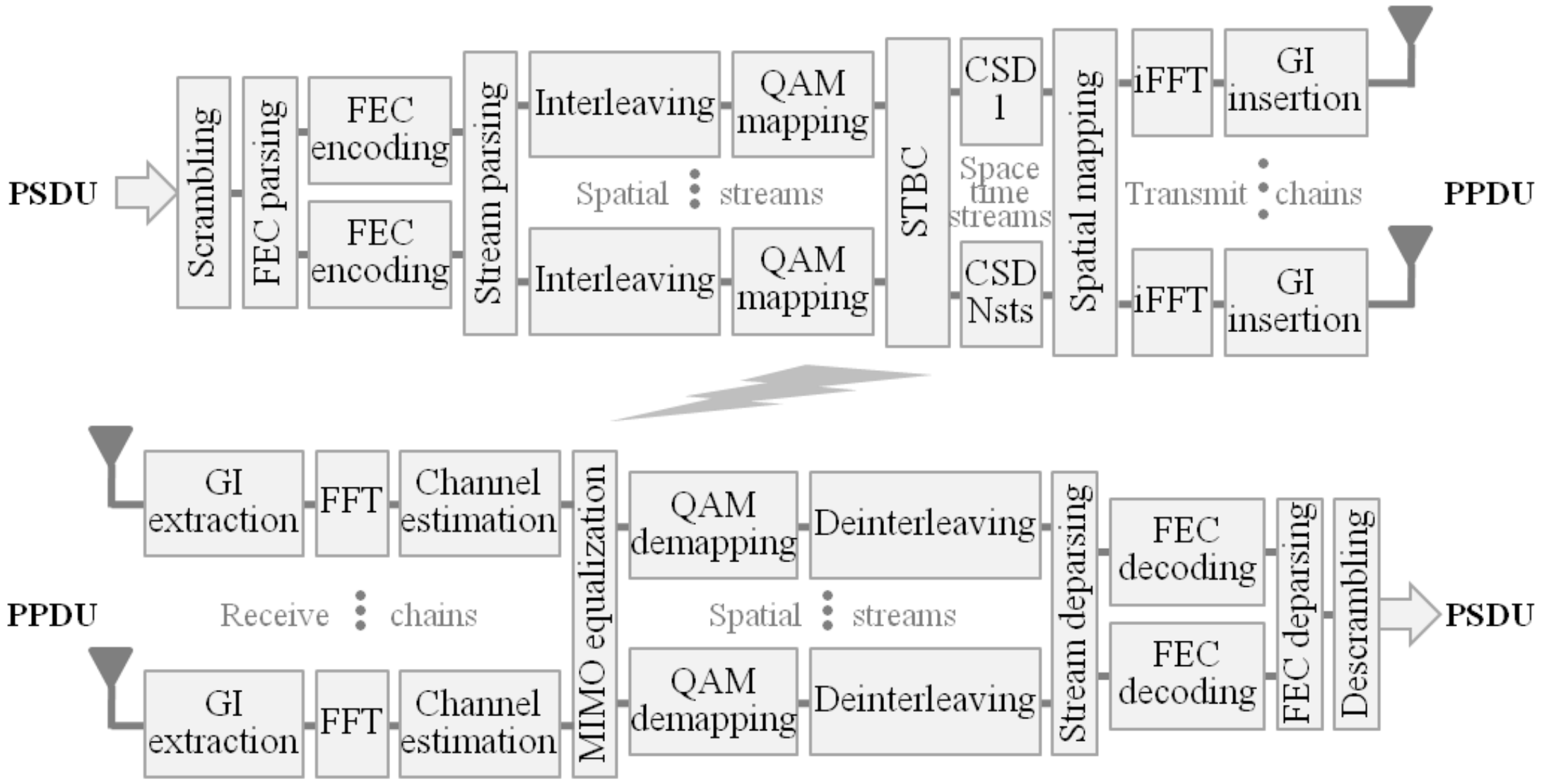

**Figure 3.3 Classical IEEE 802.11n transmission block diagram with adapted reception**

We note that the duplication of some functions in Figure 3.3 (per spatial stream, per space-time stream, and per transmit/receive chain) is only due to the application of a multiple-input, multiple-output (MIMO) scheme.

## 3.2 Proposed Cross-layer Multichannel Aggregation

As seen before, the PHY layer-wise positive aspect of transmitting MPDUs over the whole bandwidth becomes a serious handicap when considering MAC layer-detectable collisions (i.e. through FCSs).

[1] We shall recall that this work was undergone before the first IEEE 802.11ac draft [69].

That is why we propose a type of aggregation that spreads MPDUs over fewer channels to reduce their probability of being affected by a collision on one of the channels (or more).

## 3.2.1 Horizontal Aggregation

### 3.2.1.1 Principle

Another way of seeing this solution is that we partition the initial bandwidth as well as the initial A-MPDU into $B$ subgroups of channels and $B$ separate A-MPDUs (respectively) so as to form $B$ parallel transmissions. The probability of having collisions over each one of these A-MPDUs will thus be reduced as the exponent (number of channels) in (3.1) is divided by $B$ (hypothesis of independent probabilities). An example of this kind of aggregation is given in Figure 3.4 with $B = 2$ .

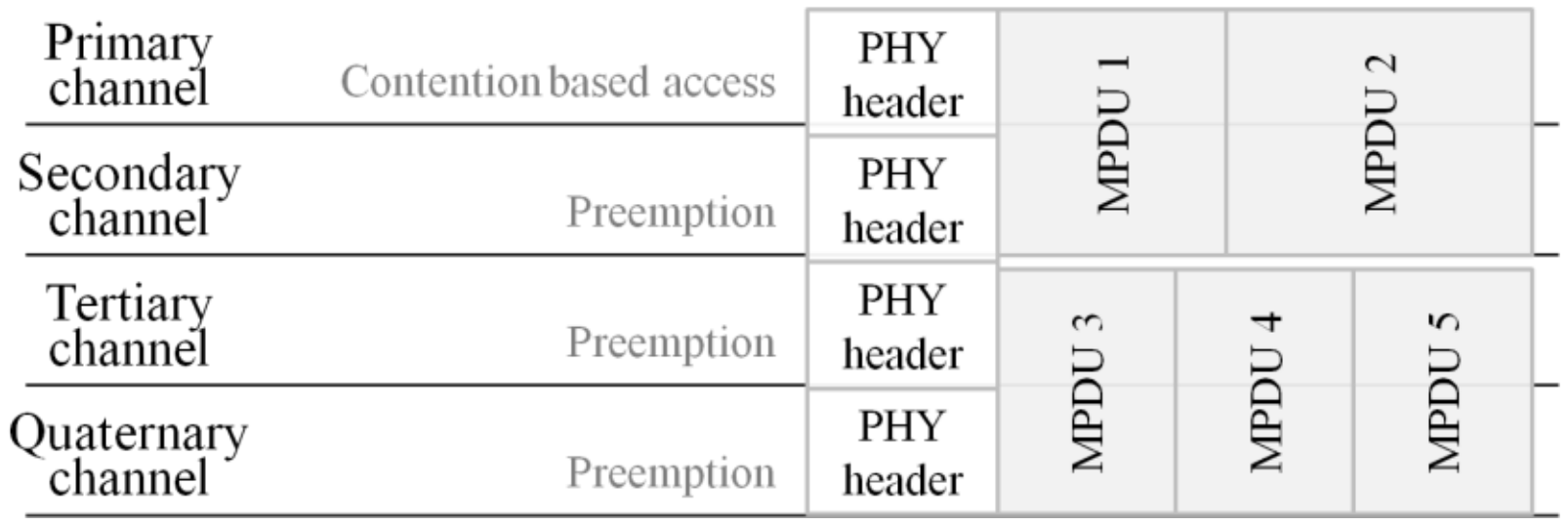

**Figure 3.4 Example of 'horizontal' aggregation over 4 channels**

The global transmission time remains almost unchanged compared to the case illustrated in Figure 3.2 if the A-MPDUs can be arranged to be of equal size. The difference is that the MDPUs are now arranged horizontally, explaining the 'horizontal aggregation' name that was given to this transmission scheme. The PHY header remains globally the same, except that the used partitioning should be signaled if it is not the default transmission mode. However, the PHY layer also needs to be modified so as to be 'horizontal aggregation'-aware, in order for this solution to properly function.

### 3.2.1.2 Modified Transmit and Receive Chain Structures

We can consider that there are $B$ parallel groups of streams to transmit over $B$ different subgroups of channels. The groups of streams have to be independent in order to dissociate the effects of the subgroups of channels. They should hence be scrambled, coded, interleaved, and mapped independently. The structure that can be directly deduced from these criteria is given in Figure 3.5. In this ‘tree structure’, there are as many ‘branches’ as there are A-MPDUs.

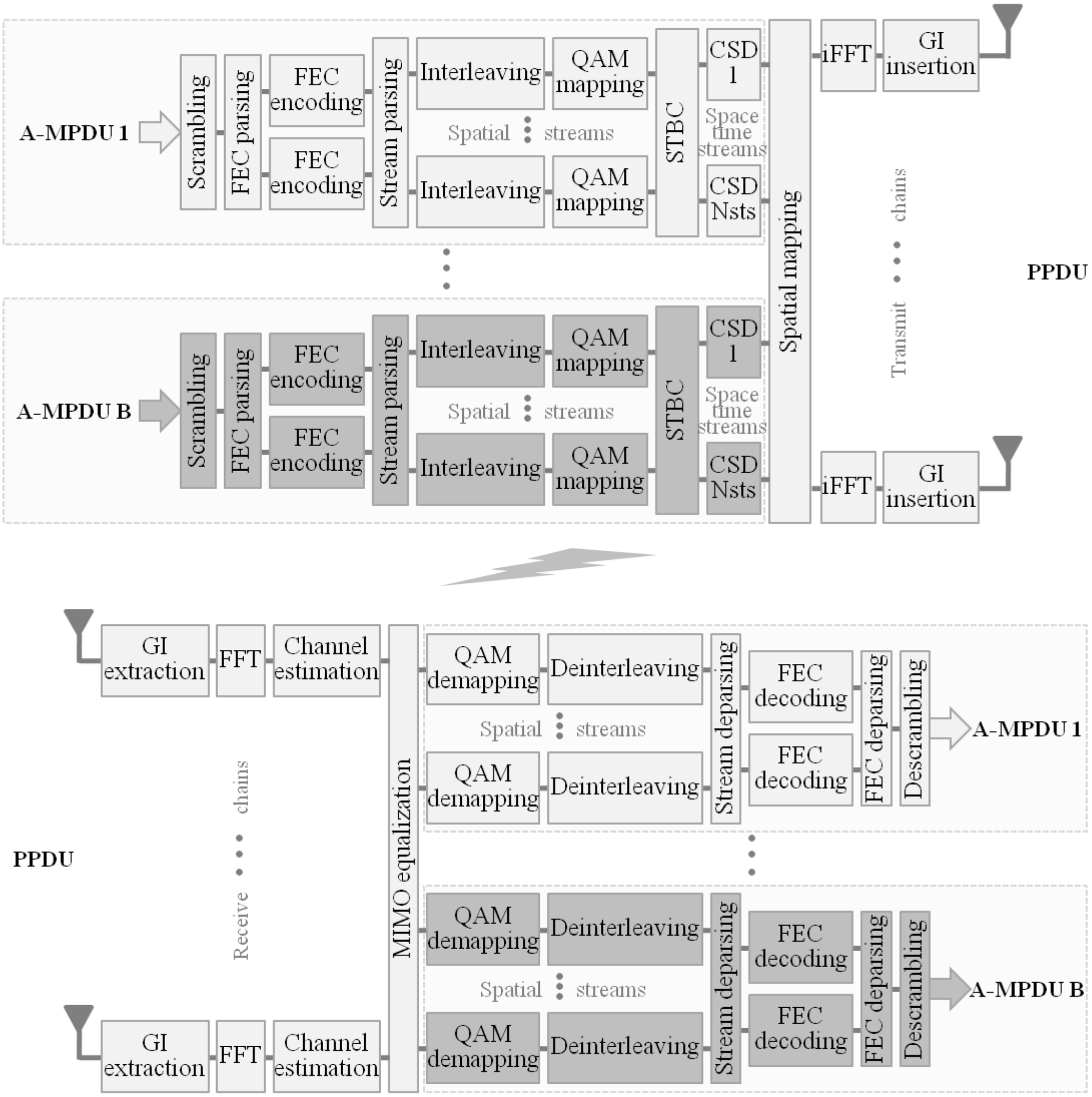

**Figure 3.5 Horizontal aggregation adapted IEEE 802.11n transmission block diagram with adapted reception**

The new blocks, relatively to the classical chain (given in Figure 3.3), are darkened. Each branch takes into account the application of the MIMO scheme seen above. We also point out that all of these branches contain functions that are less complex than the original ones. Indeed the interleavers, scramblers, and coding/decoding data block sizes are $B$ times shorter than the ones in the classical solution. This is due to the fact that the computing complexity needed for data transmission over the whole bandwidth is shared among the branches. Therefore, the global order of complexity should not be that much increased in the proposed aggregation scheme.

However, some functions (such as bit to symbol mapping or cyclic shift diversity inclusion) need not be modified intrinsically by the scheme. Their redundant use can thus be optimized, so that the structure stays very close to the classical one, especially when $B$ equals 2. Therefore, at first sight, the complexity of implementation will not be a big problem compared to other throughput increasing schemes (e.g. spatial division multiplexing which also requires parallel processing, see section 2.4.2).

## 3.2.2 Comparison with Vertical Aggregation

With all these properties of the horizontal aggregation in mind, we can carry out a comparison with the classical vertical aggregation.

### 3.2.2.1 Improved Robustness to Collisions

We have considered up to this point that collisions, when they occur, have serious implications. It is not always the case because the closer the interferer, the more consequences the collision has. Signal-to-interference ratio (SIR) should thus be a good parameter to characterize the intensity of collisions.

If the SIR is very low (close interferer), collisions will greatly corrupt the signal. The partitioning of the horizontal aggregation should then be of great use to confine the effect of the collision(s) to some A-MPDUs. It should thus be the best scheme in this situation. If on the other hand the SIR is very high (far away interferer), we would expect the decoder to easily recover the useful signal. Vertical aggregation, with its A-MPDU spread out over the whole bandwidth, should better benefit from frequency diversity. It should thus be the best scheme in this dual situation.

### 3.2.2.2 Slight PHY Layer Performance Degradation

Frequency diversity is indeed the main advantage of the vertical aggregation scheme. Horizontal aggregation, which spreads MPDUs over less channels, should therefore have lower PHY layer performance. This degradation of performance needs to be taken into account in the cross-layer compromise.

## 3.2.3 Analog to Digital Converter Saturation

Until now, we have focused on the digital part of the transmission chain. If we now consider the front-end segment, the effects of collisions become much more serious, justifying the multiplicity of protection mechanisms in IEEE 802.11 standards (see section 1.2.3.4 for one of these mechanisms).

In order to maximize efficiency, the ADC calibrates its range to the level of the incoming sequence during the PHY preamble (automatic gain control, see section 1.1.1.5). The rest of the signal is supposed to be at the same level. Therefore there is no risk of saturation. However, if a collision arises once the automatic gain control has been done, an interferer is added onto the existing signal and the output could then be saturated. In this case, no information at all can be recovered. Considering its consequences, such a scenario is worth an analysis.

### 3.2.3.1 Contiguous and Non-contiguous Channels

In the case where the used channels are all contiguous, there is only need for one front-end segment. A powerful enough collision will thus prevent recovery of the useful information. The ADC could saturate regardless of the aggregation scheme. Therefore it is not a good comparison basis.

On the other hand, if the used channels are non-contiguous, there are at least two front-end segments. These two parallel reception modules can naturally fit the horizontal aggregation scheme and are transparent to the vertical one. In such a configuration, a collision will only affect one ADC and one segment of the aggregation. The advantage of having independent chains can thus be exploited to correctly receive at least a part of the A-MPDUs in both aggregation schemes (whereas all would have been surely lost in the contiguous case). That is why we took this channel configuration to carry out the simulations developed hereafter.

### 3.2.3.2 Analog to Digital Converter Scaling Offset

Here we consider the case were the interferer occurs before automatic gain control (see section 1.1.1.5). The calibration process thus takes into account the interference signal level. The opposite is not worth simulating here since the collision-affected ADC will saturate. The vertical aggregation will thus be too disadvantaged and comparison would not have been valid.

Let us express the effects of this scaling shift due to addition of the interference. If $\beta/\alpha$ is the band occupied by the interferer relative to that of the signal,

$$P_{\max} = S + \frac{\beta}{\alpha} \cdot I \tag{3.2}$$

With $S$, $I$, and $P_{\max}$ being the power of the signal, the interferer, and the saturation power of the

ADC (resp.). If we divide (3.2) by noise power, we have:

$$SNR = SNR_{\max} - \frac{\beta}{\alpha} \cdot \frac{SNR}{SIR} \tag{3.3}$$

With $SNR$ and $SNR_{\max}$ being the real signal-to-noise ratio (SNR) and the maximum ADC-allowed SNR (resp.). Therefore the real SNR can be ultimately written as:

$$SNR = \frac{1}{1 + \frac{\beta}{\alpha \cdot SIR}} \cdot SNR_{\max} \tag{3.4}$$

This expression of SNR will be used to take into account ADC scaling offset in what follows.

## 3.3 Simulation Scenarios and Parameters

In order to simulate the horizontal aggregation scheme and soundly compare it to the vertical one, we have built a modified IEEE 802.11n transmission chain, using a custom Ptolemy II-like [54] PHY tool (C code encapsulated in C++ environment for block manipulation) to which we added some MAC functions. The high throughput task group (TGn) [74] channel model is used (see section 4.2.1.4.1) in all six considered scenarios. Per scenario simulation parameters and common simulation parameters are given in Table 3.1 and Table 3.2 (resp.).

The same probability of collision of 0, 5, 10, or 15% is assumed for each 20 MHz channel. Furthermore, we consider that collisions always affect the PHY header (without impacting information retrieval from the signaling field) and thus the effect of the interferer is modeled by additive white Gaussian noise (AWGN), in the frequency domain, upon the part of the frame transiting through the collision-affected channel(s). Modulation and coding schemes (MCSs) are identified using the classification of the 802.11n standard. An easy equivalence can be drawn with IEEE 802.11ac's[1]. Indeed, scenario 1.1's MCS 8 corresponds to index 0 with 2 spatial streams (SSs) of Table D.1 in Appendix D (13 Mbps data rate). Scenario 1.2's MCS 13 corresponds to index 5 with 2 SSs of Table D.1 (104 Mbps).

[1] For supported bandwidths, IEEE 802.11n's MCS index can be obtained by taking IEEE 802.11ac's MCS index (from 0 to 7) and adding $7 \cdot (\text{number of SS} - 1)$ to it. Refer to Appendix D for 802.11ac MCSs.

**Table 3.1 Per scenario simulation parameters**

| Scenario | $P_{collision}$ (%) | Rate adaptation algorithm | # aggregation blocks $B$ | Antenna configuration | # SS | SNR (dB) | ADC scaling offset |
|---|---|---|---|---|---|---|---|
| 1.1 | 0, 5, 10, 15 | Fixed (MCS 8) | 1, 2, 4 | $2\times2$ | 2 | 18 | No |
| 1.2 | 0, 5, 10, 15 | Fixed (MCS 13) | | | | | |
| 2.1 | 0, 5 | AMRR | 1, 2, 4 | $4\times4$ | 4 | 20 | No |
| 2.2 | 0, 10 | | | | | | |
| 3.1 | 0, 5 | AMRR | 1, 2, 4 | $4\times4$ | 4 | 20 | Yes |
| 3.2 | 0, 10 | | | | | | |

**Table 3.2 Common simulation parameters**

| | |
|---|---|
| **MAC layer parameters** | |
| Access category | Best effort |
| MPDU size | 1530 octets (typical 1500 octet video packet with a 30 octet header) |
| Aggregation size | 20 MPDUs per 20 MHz band |
| Total number of TxOPs | 200 per simulation point |
| Validation of received frame | Through received FCS |
| Acknowledgment validation | Always consider as correctly returned |
| **PHY layer parameters** | |
| SIR | -10 dB to 30 dB |
| Interference type | AWGN in frequency domain |
| Channel coding | Binary convolutional coding |
| Guard interval | Long |
| Channel estimation | Perfect knowledge at receiver |
| **TGn channel parameters** | |
| Channel model | B (residential) |
| Bandwidth | 80 MHz (4 non-contiguous 20 MHz channels, 20 MHz subcarrier mapping) |
| Central carrier frequency | 5.2 GHz (channel n° 40) |
| Channel seeds | Randomly chosen at the end of each TxOP |

The adaptive multi-rate retry (AMRR) rate control algorithm [33] (see Appendix B) was also modified to suit the multichannel case. Statistics are computed for MPDUs instead of A-MPDUs.

Counts (which are the numbers of retries per rate) are thus set to $c_0 = 2$, $c_1 = 1$, $c_2 = 1$, and $c_3 = 1$ and the update period (of the four rates) set to 25 ms for better reactivity. The algorithm starts at MCS 27 (index 3 with 4 SSs of Table D.1). We shall note that, for each scenario, SNR is fixed but SIR ranges from -10 dB to 30 dB. The vertical aggregation, denoted as '$1 \times 80$ MHz' (with $B = 1$), will be compared to the horizontal ones, '$2 \times 40$ MHz' (with $B = 2$) and '$4 \times 20$ MHz' (with $B = 4$). We will also note that the FCS is used to check frame integrity, instead of comparing sent and received bits, so as to enable quasi-realistic performance estimation.

# 3.4 Simulation Results and Discussion

## 3.4.1 Results without ADC Scaling Offset

### 3.4.1.1 Fixed Data Rates

The PHY layer notion of packet error rate (number of bad packets over the number of received packets) is not adapted for a quasi-realistic aggregation context. Indeed the PHY notion of 'packet' is usually quite vague and rather corresponds to the PHY service data unit (see section 1.2.5). All MPDUs of an A-MPDU are thus considered as a whole 'packet'. In addition, packet error rate is computed by comparing the sent and received bits.

However, in Wi-Fi devices, the correctness of each MAC frame in the A-MPDU is verified through the associated FCSs. The received FCS is the only means of verifying the integrity of each MPDU, because sent bits are normally not available at the receiver. The number of incorrect FCSs thus corresponds to the number of retries, if acknowledgments are considered correctly received. Consequently the MPDU retry ratio should rather faithfully characterize system performance considering PHY and MAC layers.

We thus used this ratio to compare the vertical aggregation scheme with the two horizontal ones for different interference levels and probabilities of collision, and for a fixed MCS. The results obtained for scenario 1.1 are given in Figure 3.6. The higher the probability of collision and the relative interference power (i.e. low SIRs), the more the horizontal scheme becomes attractive. Indeed partitioning minimizes loss in harsh (but realistic) conditions. On the other hand, for low interference-level or collision-free transmissions, vertical aggregation outperforms the proposed scheme. Frequency diversity gain allows the coder to recover errors, leading to better performance.

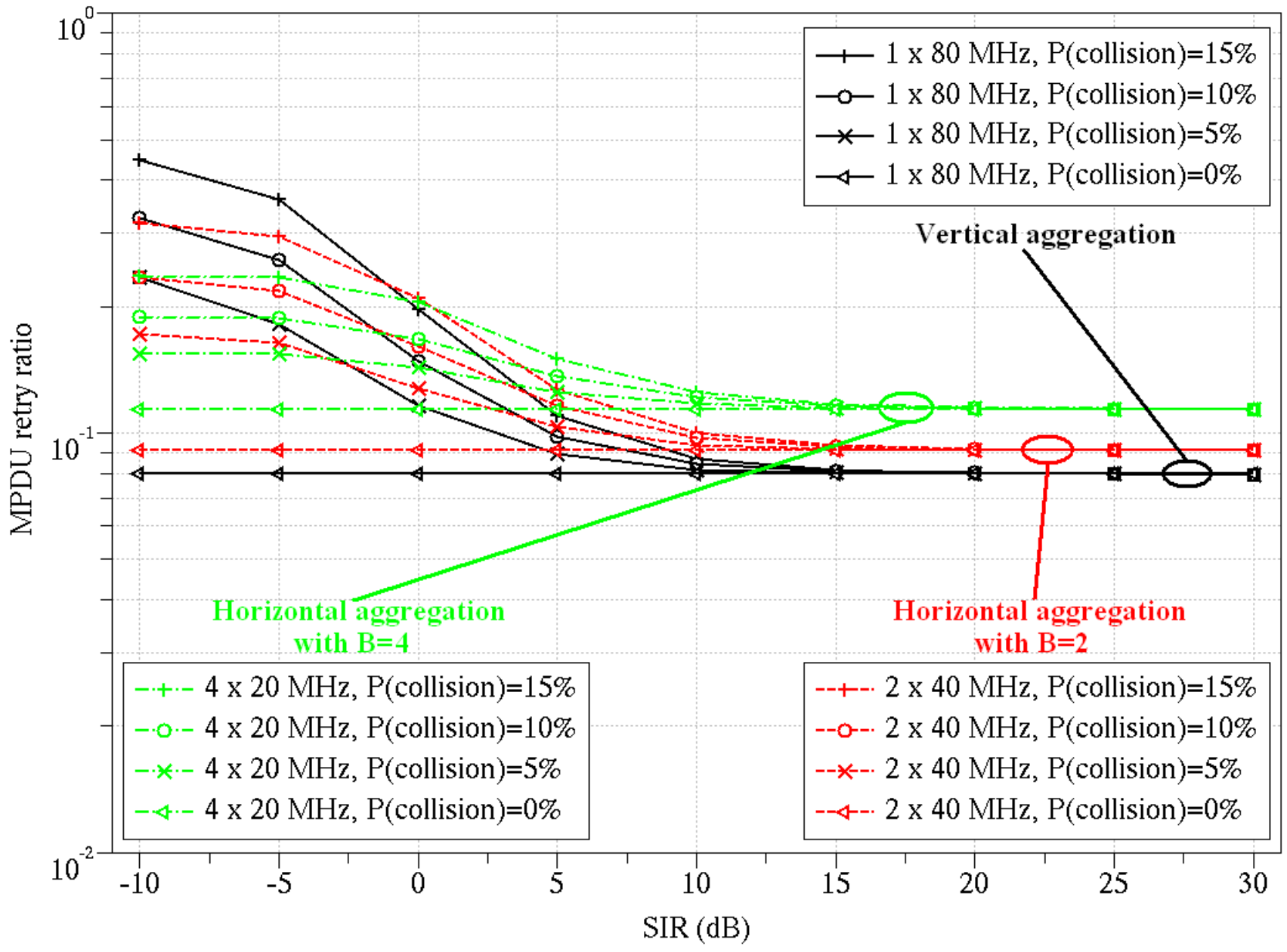

**Figure 3.6 MPDU retry ratio as a function of SIR for scenario 1.1: MCS 8 (13 Mbps data rate)**

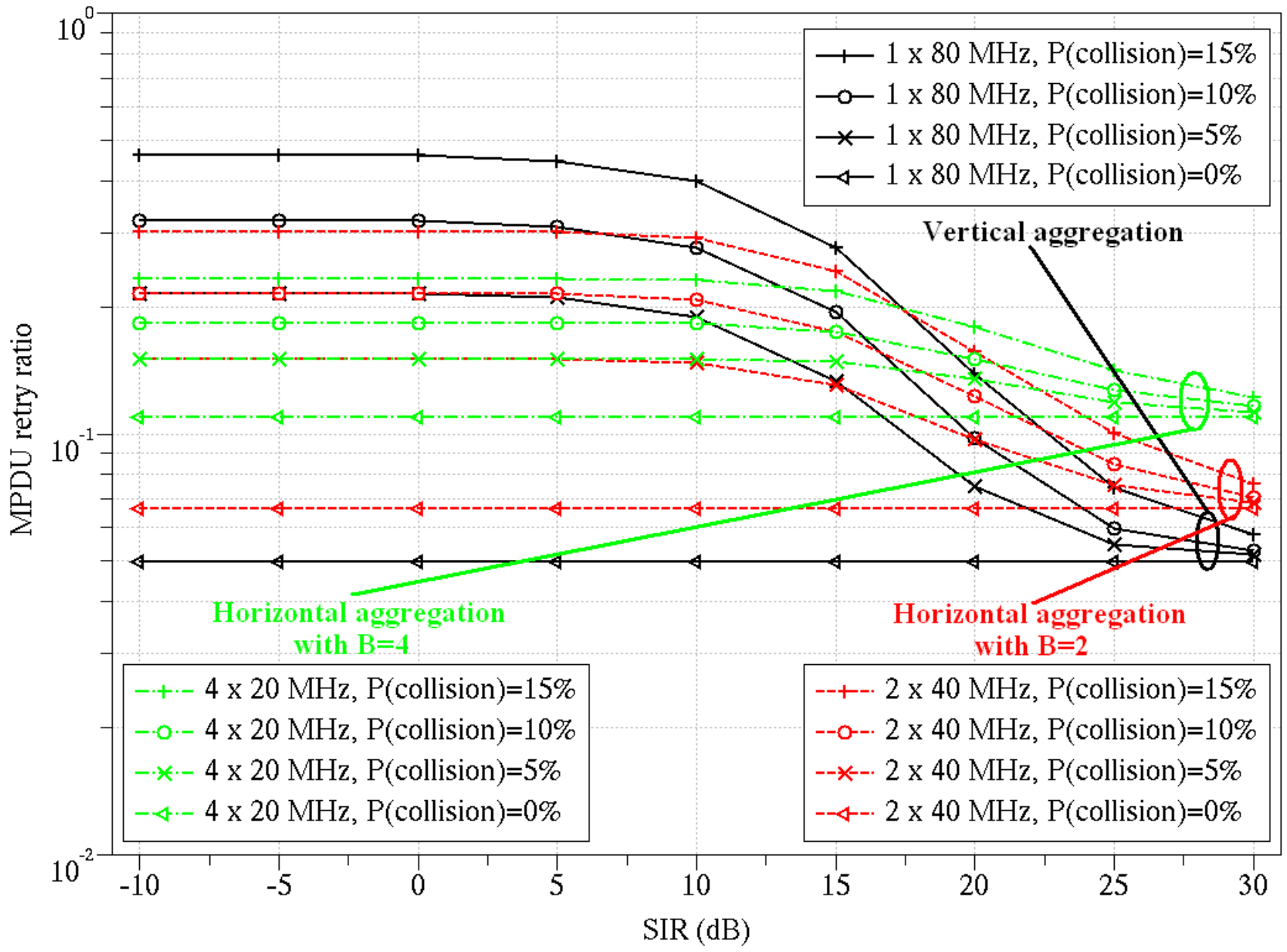

**Figure 3.7 MPDU retry ratio as a function of SIR for scenario 1.2: MCS 13 (104 Mbps data rate)**

We have considered non-contiguous channels for reasons explained previously (see section 3.2.3.1). In the case of contiguous channels, vertical aggregation would have shown better performance because of the greater number of data subcarriers (see section 2.2.1 for the difference between 40+40 MHz and 80 MHz).

The MCS that was used in scenario 1.1 applies a robust modulation (binary phase-shift keying with 1/2 coding rate) but offers a rather low data rate (13 Mbps). The results obtained with the higher data rate (104 Mbps), offered by the MCS used in scenario 1.2, are given in Figure 3.7. In this scenario, the horizontal aggregation schemes improve performance for SIRs of up to 18 dB (-2 dB in Figure 3.6). Considering these two situations, the $2\times 40$ MHz scheme seems to be a good compromise with regards not only to performance but also to complexity of implementation. The latter is very important to account for. Indeed, implementation cost constraints often lead to system simplification. This can be seen with the 80+80 MHz channel transmission mode. Despite the increase in the number of possible 160 MHz channels it offers, this non-contiguous mode was defined as optional because of the additional front-end segment it implies. Therefore, the $2\times 40$ MHz scheme has more chances of being normalized than the more efficient but more complex $4\times 20$ MHz scheme.

### 3.4.1.2 With AMRR Rate Adaptation

For a more pragmatic analysis, a rate adaptation algorithm has to be used. Indeed fixed MCS schemes are never implemented. Considering the transmitted loads, the algorithm needs to be very reactive to channel condition changes in order to minimize losses due to rate calibration. That is why we chose the AMRR algorithm [33] with parameters modified as exposed in section 3.3.

The use of the AMRR algorithm will not favor an aggregation over the other. The first count ($c_0$) has the highest of the four transmission rates associated to it ($r_0$) and is used at the beginning of a new transmission. This count is expressly set to 2 so as to avert reacting to collisions (which are spurious by definition) as with channel conditions. The comparison criterion between the two schemes is also modified. The algorithm is designed to maximize throughput by adapting the rate in order to keep the retry ratio to a certain interval (between 10% and 33%). We will therefore take the effective throughput (correctly received useful data averaged over the global transmission time) as the comparison criterion.

The obtained effective throughputs for different SIRs are plotted in Figure 3.8 and Figure 3.9, for scenarios 2.1 and 2.2 (resp.). The probabilities of collision of 5% (scenario 2.1) and 10% (scenario 2.2) are considered, the 15% probability being deemed a little too high to be pragmatic.

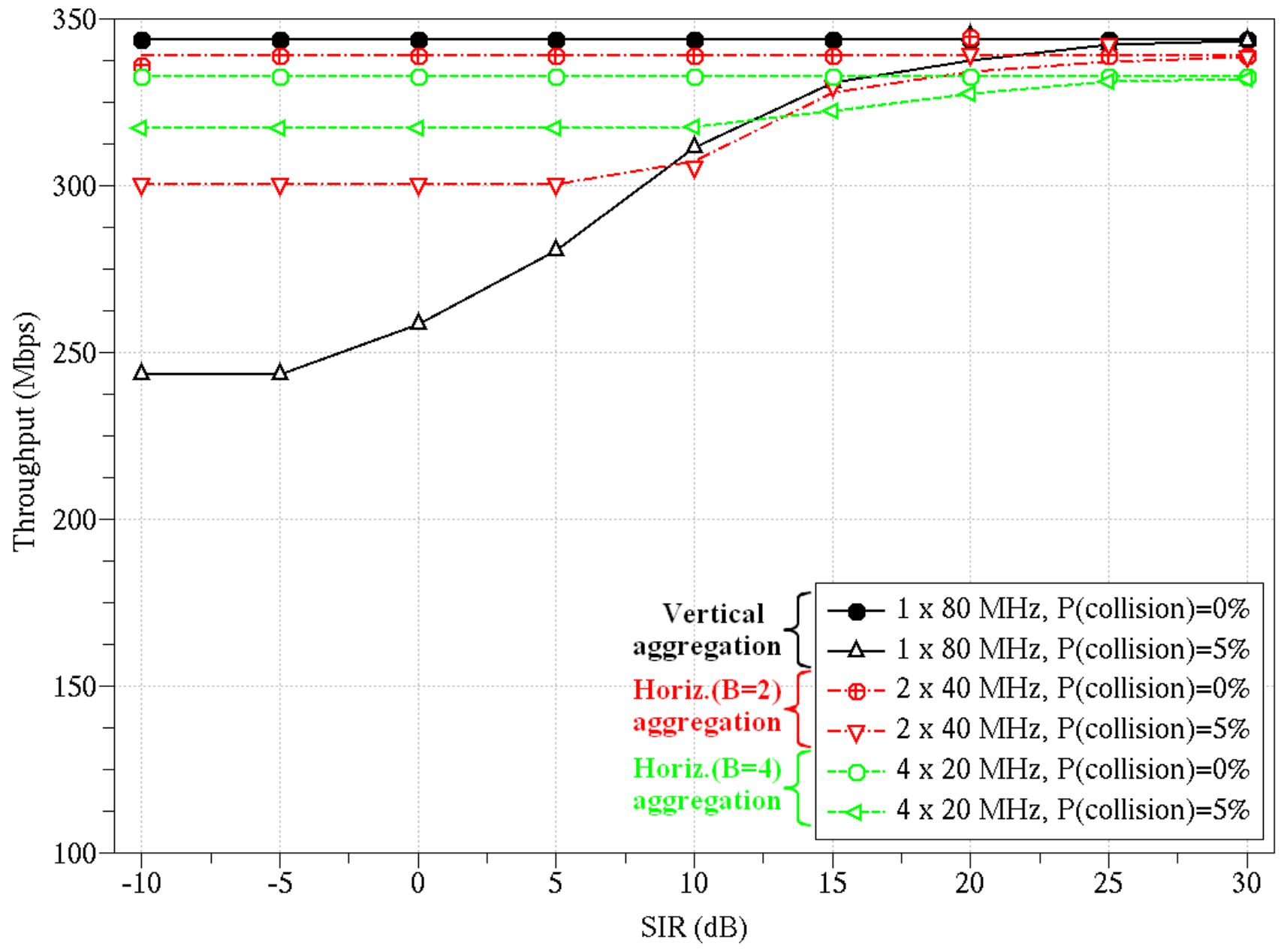

**Figure 3.8 Effective MAC throughput as a function of SIR for scenario 2.1: $P_{collision}$ = 0% and 5%**

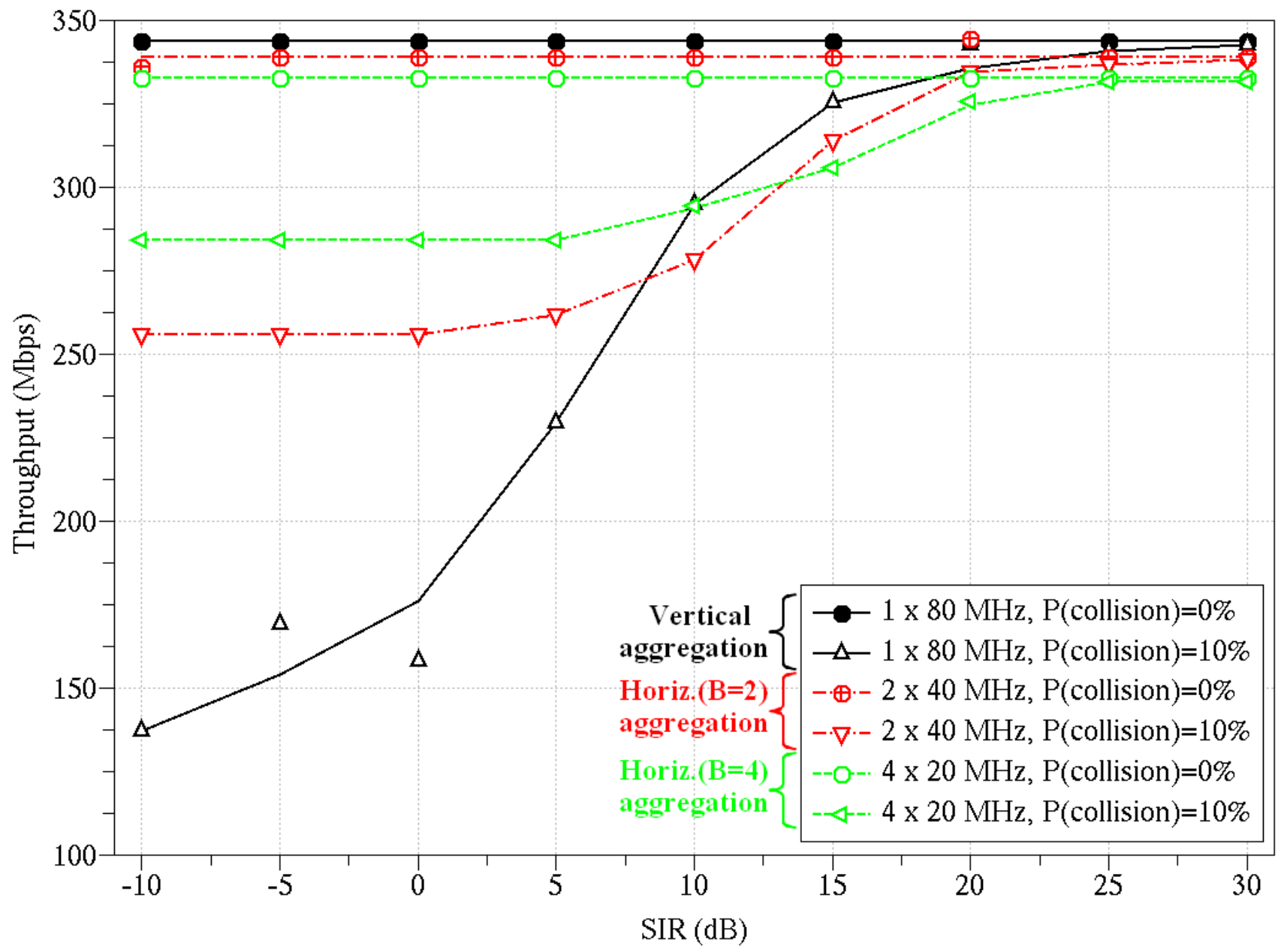

**Figure 3.9 Effective MAC throughput as a function of SIR for scenario 2.2: $P_{collision}$ = 0% and 10%**

Apart for very low interference levels, the improvements introduced through the use of horizontal aggregation are very significant:

- An increase in throughput as high as 24% for $P_{collision} = 5\%$ and 86% for $P_{collision} = 10\%$, when using the $2 \times 40$ MHz scheme (relative to the vertical aggregation scheme). When using the $4 \times 20$ MHz scheme, there is even more increase in throughput. It can get as high as 30% for $P_{collision} = 5\%$ and 107% for $P_{collision} = 10\%$ (relative to the vertical aggregation scheme still);
- A betterment of system performance for SIRs up to 8 dB or 9 dB (for $P_{collision} = 5\%$ and 10% resp.), i.e. where the interfering station is 6 to 8 times farther than the useful signal transmitting station, when using the $2 \times 40$ MHz scheme. With the $4 \times 20$ MHz scheme, the betterment of system performance can be obtained for SIRs up to 10 dB or 12 dB (for $P_{collision} = 10\%$ and 5% resp.), i.e. where the interfering station is 10 to 16 times farther than the useful signal transmitting station;
- Performance loss in collision-free transmissions is minimized through the use of a horizontal aggregation scheme with a reasonably small number of blocks $B$, i.e. $2 \times 40$ MHz (2% loss in throughput) instead of $4 \times 20$ MHz (3% loss in throughput).

Hence horizontal aggregation turns out to be the best transmission scheme for rate adaptive algorithms (even more than it did for fixed MCS transmissions) in collision-prone environments.

### 3.4.2 Results with ADC Scaling Offset

If we now modify the reception chain so as to comprehend the consequences of the ADC scaling offset following (3.4), we obtain the effective throughputs given in Figure 3.10 and Figure 3.11 for scenario 3.1 ($P_{collision} = 5\%$) and scenario 3.2 ($P_{collision} = 10\%$) respectively.

Performance, contrarily to what would be expected, does not change that much except that horizontal aggregation is even more profitable for $P_{collision} = 10\%$. The transition SIR is now 9.5 dB (i.e. interferer 9 times farther than useful signal emitting station) and the increase in throughput gets as high as 93%, when using the $2 \times 40$ MHz scheme. For the $4 \times 20$ MHz scheme, the transition SIR is now 12.5 dB (i.e. interferer 18 times farther than useful signal emitting station) and the increase in throughput gets as high as 117%.

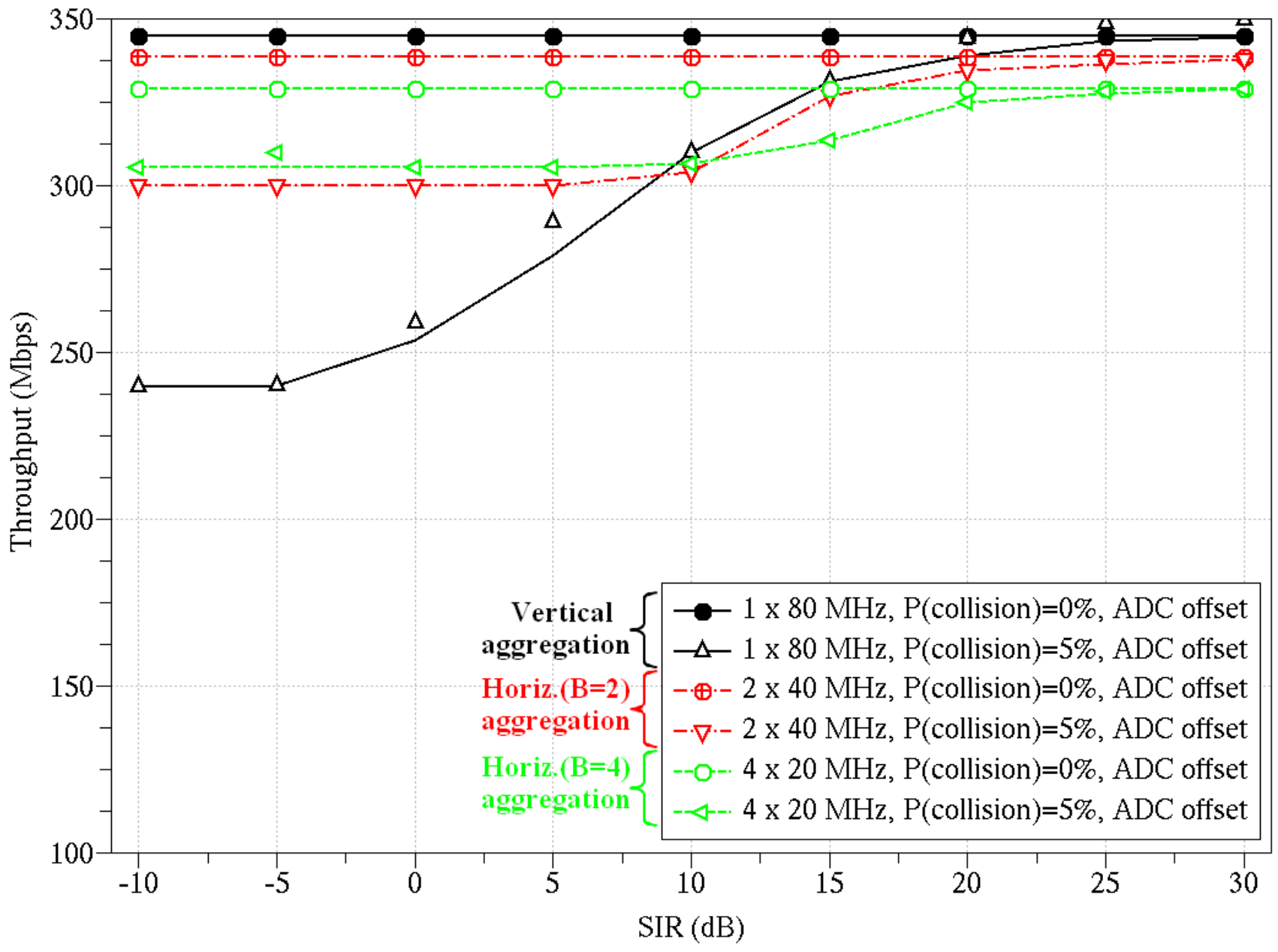

**Figure 3.10 Effective MAC throughput as a function of SIR for scenario 3.1: $P_{collision}$ = 0% and 5% with ADC scaling offset**

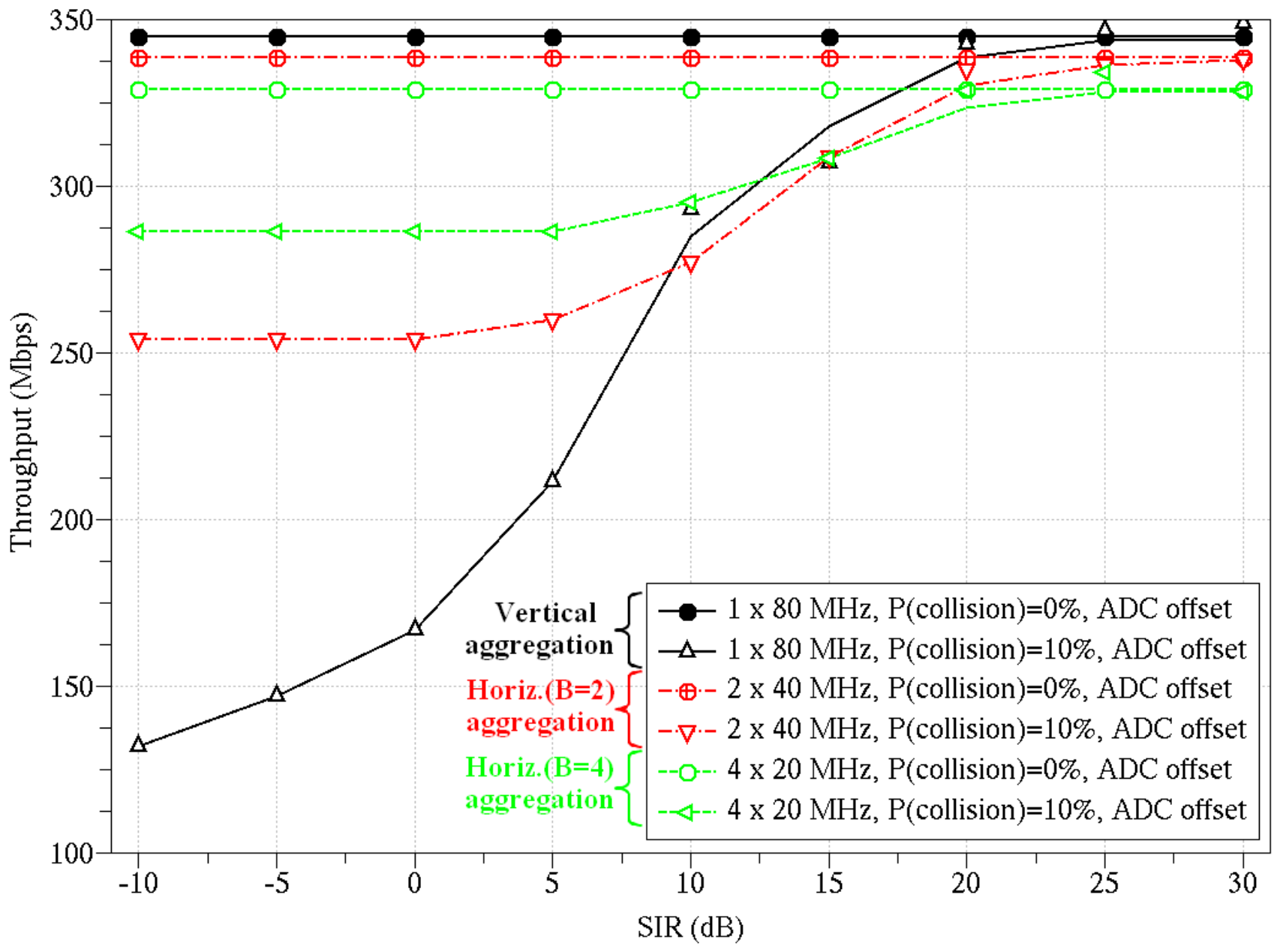

**Figure 3.11 Effective MAC throughput as a function of SIR for scenario 3.2: $P_{collision}$ = 0% and 10% with ADC scaling offset**

## 3.5 Summary

The classical vertical aggregation solution optimizes data distribution by prioritizing PHY layer data rate. We have shown that global system performance of multichannel transmissions can be improved by considering a cross-layer criterion. Collisions occurring in any of the used channels can indeed corrupt the whole transmission if no partitioning measure is taken. The $2\times40$ MHz horizontal aggregation thus stands out as the best transmission scheme when taking simultaneously into account PHY layer (complexity of implementation, SNR degradation in case of collisions, and PHY data rate) and MAC layer (throughput and robustness to collisions) criteria. Despite the increase in the complexity of implementation (due to partitioning), we have seen through simulation results that there can be an increase in throughput as high as 93% in a realistic collision context. If better performance is desired, the $4\times20$ MHz horizontal aggregation can be used, but at a higher implementation cost. An increase in throughput as high as 117% can thus be obtained in the same realistic collision context. Furthermore, through partitioning, channels can be known with greater granularity, especially concerning the occurrence of collisions. The rate adaptation algorithm could therefore be modified so as to take this extra information into account.

We have evaluated the gains of the proposed scheme using detailed PHY chains and a fine grained channel modeling, to which we added some MAC functions. Such a system is sufficient to evaluate tendencies. Nonetheless, if a more realistic analysis is desired (e.g. modeling contention access or acknowledgement errors), a cross-layer simulator faithfully modeling both PHY and MAC layers becomes necessary.

Chapter 4

# PHY+MAC Simulation Platform for IEEE 802.11n/ac

In the wireless local access network (WLAN) context, physical (PHY) layer or medium access control (MAC) layer performance evaluation can either use link-level or system-level simulations, depending on the layer to study in depth. Elaborated PHY layer simulators, notably including very fine channel models, simplify MAC layer mechanisms (see section 1.5.1 for details). On the other hand, complex MAC layer simulators allow fine simulation of contention access and queue refilling, which is done according to application layer needs, while using a simplified PHY model (see section 1.5.2 for details). Therefore in both types of simulators, i.e. either ‘PHY-centric’ or ‘MAC-centric’ (resp.), one of the layers has been reduced in complexity, somewhat biasing the behavior of the global system.

That is why we propose a new multiple-user simulation platform composed of two parts working in a symbiotic manner. Each part finely characterizes the corresponding layer. The incorporation of an all inclusive IEEE 802.11n/ac [6,7] PHY layer, containing a realistic channel model, and an elaborated IEEE 802.11n/ac MAC layer results in an accurate modeling of reality. PHY layer simulations are very time consuming because of channel tap computations. Consequently the simulation platform performs these simulations only when necessary. Such a knowledgeable mix of detailed PHY simulations and lookup tables (LUTs) enables optimization with regards to time and computational resource consumption. The reader can refer to section 1.5.3 where other cross-layer simulators are exposed. The associated works are very interesting but do not encompass PHY techniques, MAC mechanisms, and precise channel modeling specific to 802.11n/ac systems.

We firstly present the proposed simulator structure and detail the characteristics of the PHY chain,

dynamic channel model, and MAC section. Then we validate our simulation model through a well-known MAC functionality, data aggregation (see section 1.2.7). A modified version of network simulator 2 (ns-2) [59] using LUTs serves as reference. Finally, the advantages of using such a model are shown through a hidden node [73] scenario (see section 1.2.1.4).

## 4.1 Global Architecture

### 4.1.1 Presentation

The principle that served as a basis for the elaboration of the simulator is the following: develop a MAC simulation module which interfaces with a PHY simulation module in a symbiotic manner. Indeed, in order to have a complete PHY+MAC simulator, it might be tempting to expand a PHY-exhaustive simulator to somehow faithfully comprehend the MAC layer, or the other way around. But this first sight solution implies a considerable increase in complexity. One of the layers' tailor-made simulator architecture has to be distorted so as to accommodate for the other layer. Therefore we propose to have two separate functions that interact only when necessary and exchange pre-processed information. We have a 'layered' structure, but with enhanced and completely dynamic interaction.

### 4.1.2 PHY Section with Realistic Channel Modeling

As implied before, we have used a fine grained PHY simulator as a basis of our cross-layer simulator. This custom Ptolemy II-like [54] PHY tool uses C++ environment, which eases functional block manipulation. It has allowed us to build IEEE 802.11n [6] and IEEE 802.11ac [7] compliant transmit (Tx) and receive (Rx) chains (see sections 3.1.3 and 2.2.4 respectively). High throughput task group (TGn) [74] and very high throughput task group (TGac) [75] channel model blocks, which are used in the standardization process, represent the most substantial portion of the PHY section. These blocks enable a faithful representation of channel variations through time for 802.11n and 802.11ac systems (resp.) in a multiple-input, multiple-output (MIMO) context.

### 4.1.3 MAC Section

The MAC section, based on network simulator 3 (ns-3) [61] architecture, deals with contention access, basic service set (BSS) management, and control frame exchanges. Each frame is acknowledged if it is correctly received (see section 1.2.2). The classical IEEE 802.11 contention access (exposed in section 1.2.3) has also been extended to the IEEE 802.11e [30] quality of service (QoS) contention access[1] (see section 1.2.4) and to the IEEE 802.11n packet aggregation (see section 1.3.2.2). Multiple-user MIMO (MU-MIMO), as detailed in IEEE 802.11ac (see section 2.4), is also implemented. Furthermore, rate adaptation algorithms, which are not specified in standards but are indispensable in any commercial product, are taken into account (see section 1.2.8). It is also the case of MU-MIMO station selection algorithms, which will certainly be used in 802.11ac devices. In short, we try to be as close as possible to actual implementation and carry out realistic simulations.

### 4.1.4 Interactions between the PHY and MAC Sections

Once the transmitting station has been selected by the MAC section, information regarding the transmission (mainly modulation and coding scheme - MCS - and data size) is handed over to the PHY section. The latter simulates transmission of data bits using the received parameters. Channel coding and modulation is done accordingly, followed by interaction with a fine grained channel, and a demodulation and decoding process to finish with. The PHY section then forwards the decoded bits, which will be checked through the included forward check sum (FCS), to the MAC section. Channel state information (CSI) can also be handed over, if necessary, to improve rate adaptation and MU-MIMO station selection. The two sections thus work sequentially and in an interdependent manner, so as to be sure to compass each layer’s characteristics.

## 4.2 Detailed Structure of each Section

### 4.2.1 PHY Section

#### 4.2.1.1 Global Structure

The most time and computational resource consuming block in WLAN link-level simulators is the channel block. This is due to channel tap generation and matrix manipulations. But a fine grained channel block is critical when modeling channel evolutions. Therefore for an efficient use of resources, we have two PHY layers. One consists of a complete Tx chain, channel, and Rx chain. This model is used for the transmission of data frames. The precise effects (benefits as well as handicaps) of the physical medium and Tx/Rx chains upon data transmission can thus be faithfully captured. The problem is that this process is time consuming.

[1] However we have used Wi-Fi multimedia (see section 1.3.2.2) parameters for all the simulations of this thesis.

That is why we use another PHY model for control and management frames. This model is a simplified version of the first one. The success or failure of a frame is probabilistically determined based on LUTs (just like in some modified versions of ns-2 and ns-3). Control and management frames are normally sent using robust modulations. Therefore, compared to less robust IEEE 802.11n or IEEE 802.11ac data frames, there is greater margin for correct reception. The particularities of the PHY chain have less impact on the outcome than for data frame reception, because of this inherent robustness.

The resulting optimized architecture will allow a simulation that is as close as possible to reality while minimizing global computational complexity (see Appendix E).

#### 4.2.1.2 Parameters

Our chain is designed so as to be compliant to IEEE 802.11n and IEEE 802.11ac standards. It is also shared by all stations of the currently simulated scenario. Consequently 802.11n and 802.11ac parameters can be tuned during the scenario setup (see sections 1.3.2 and 1.3.3 respectively for these parameters). In addition, all transmission parameters (number of antennas, MCS, etc.) are dynamically reconfigurable according to the characteristics of the selected pair of stations and the current transmission rate. This modular structure enables to compass a wider scope of scenarios.

#### 4.2.1.3 Dynamic Reconfiguration

The PHY layer chain is common to all stations therefore it needs to be dynamically reconfigurable so as to meet the characteristics of each (e.g. number of antennas and data size). These are given over by the MAC section every time a station has won access to the channel.

However, most PHY layer simulators (e.g. Ptolemy II and COSSAP [55]) have functional blocks with static in/out (I/O) first-in, first-out (FIFO) buffer sizes. The sizes of I/O FIFO buffers are set at the beginning of the simulation so as to fit a particular MCS. This is understandable considering that PHY layer simulations usually evaluate the performance of a single link, i.e. between a transmitter and a receiver. Having the same bridle in our fine grained PHY simulator while simulating multiple links, we built a ‘water pipe’-like (data + padding padding) structure for I/Os to workaround this flaw. At the beginning of the simulation, I/O FIFO sizes are set large enough so as to support the current simulation's maximum data size. However this bypass structure should not be much resource consuming because processing is only limited to the data, padding being neglected. Dynamic reconfiguration is enabled despite static I/O FIFO sizes.

#### 4.2.1.4 Dynamic Channel Modeling

##### 4.2.1.4.1 Channel Model

As indicated before, one of the elements that renders our PHY layer simulation faithful to reality is the TGn [74] (or TGac [75]) channel modeling block. The latter is a 3GPP SCM-like [76] geometric model based on stochastic modeling of scatterers, also called cluster model [77,39]. Fast fading and shadowing phenomena are also taken into account in the TGn and TGac models.

##### 4.2.1.4.2 Channel Tap Handling

Channel taps characterize channel conditions between a pair of conversing stations (see section 1.1.2.2). The way they are handled determines the correspondence of the simulated scenario to reality. In addition, the channel modeling block being the most computational resource consuming element of the PHY layer chain, optimization can be done through wise channel management.

###### *4.2.1.4.2.1 Temporal*

Most PHY-centric simulations look after channel capacity. Therefore different channel conditions have to be considered so as to obtain an ergodic fading process, if possible. In that case, the capacity of the channel is reached. To do so, new channel taps are generated for every transmission. As a matter of fact, the notion of frame segmentation is often vague in PHY-centric simulations (as already mentioned in section 3.4.1.1). The proposed simulation platform uses frames as in MAC-centric simulators.

Channel taps are estimated by the receiver during the training sequence at the beginning of every frame. CSI is available at the receiver through this estimation process. On the other hand, CSI cannot be continually available at the transmitter for beamforming purposes, as assumed in classical PHY-centric simulators. In the 'real world', as in the proposed simulation platform, the transmitter has to send a sounding frame and wait for an estimate of channel taps (at the given time) to be returned through a response frame (see section 1.1.3.3). In addition, channel taps evolve through time (coherently to simulation time) while still remaining correlated. The returned estimates can thus be used in following transmissions.

##### *4.2.1.4.2.2 Multiple Station Support*

Another advantage is facilitated support for multiple stations. Channel taps between an oriented pair of stations can be stored away so that the chain can be reused to simulate transmission between another pair of stations, while having the possibility to recover, later on, the stored channel context. Space division multiple access (or MU-MIMO) is also rendered possible for IEEE 802.11ac implementation, by using stored taps to model crosstalk interference.

#### 4.2.1.5 Gains Compared to a MAC-centric Approach

We can see from what precedes that compared to a MAC-centric approach (see section 1.5.2) which oversimplifies the PHY layer, our simulator structure, through realistic channel modeling and complete Tx/Rx chains, allows a more reliable and more flexible PHY section. Channel adaptive PHY transmissions can thus be faithfully characterized.

### 4.2.2 MAC Section

#### 4.2.2.1 Global Structure

The MAC section contains the main MAC functions of the IEEE 802.11n (and IEEE 802.11ac) in conjunction with an application layer which generates packets. This is where the advantage of using C++ programming language can be most clearly seen: we can generate as many applications per station as desired, and create also as many stations as necessary for the simulation thanks to the object concept in C++. In the actual state of things, we have considered the infrastructure mode, i.e. with an access point (AP) assuming the management of the BSS. However, this can be easily extended to an ad hoc mode. In addition, we can also use our MAC section to generate multiple APs (operating on at least one common 20 MHz channel) and see how the system reacts in an overlapping BSS context (see section 1.2.1.3). Another advantage is that stations supporting different bandwidths can also be associated to the same AP and one can easily study system behavior in such a scenario. We will note that ns-2 and ns-3 frameworks were used to develop this MAC section. Hence, this kind of structure offers a lot of possibilities for modeling different scenarios corresponding to every day use cases.

#### 4.2.2.2 Function Presentation

As indicated above, we can define as many applications as desired per node. The latter centralizes topology information (see Figure 4.1). Applications are managed by an application function which models higher layers and where the traffic category, rate, and duration are defined. Each application generates packets periodically or in an adaptive manner depending on the transport layer protocol, but with a random jitter for arrival fairness. The obtained traffic is then handed over to a network interface. The latter can either be AP-specific or station-specific and handles traffic it relays to the contention and queueing function. In IEEE 802.11n, this corresponds to the enhanced distributed contention access (EDCA, see section 1.3.2.2). The access category (AC) having won access of the channel gives over its packet to the function handling data/acknowledgement transactions. Afterwards the PHY section takes over.

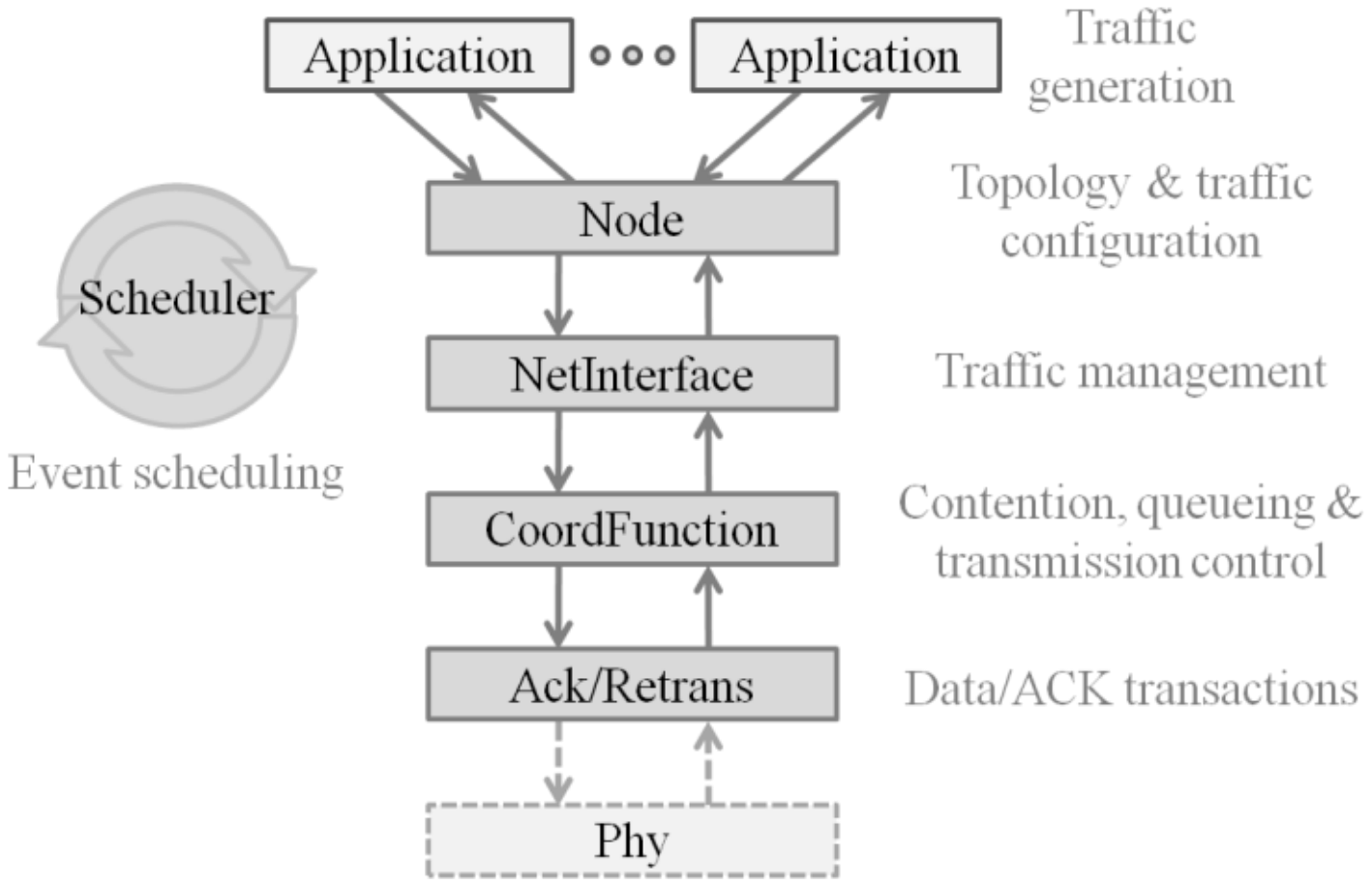

**Figure 4.1 MAC section structure**

We note that there is an event scheduler which takes care of the sequencing of channel access requests through callbacks. This way data, control, and management frames, and even collisions, are accounted for meticulously and in a timely manner.

### 4.2.2.3 Main MAC Functions

MAC particularities are considered mostly in NetInterface, CoordFunction and Ack/Retrans (as defined in Figure 4.1). Indeed, carrier sense multiple access with collision avoidance (CSMA/CA) with QoS, which is a schematic definition of EDCA, is an essential part of IEEE 802.11n. Every AC of every station can contend for the channel and the outcome is decided in the CoordFunction function, which also builds the frames to transmit. Positive acknowledgment (ACK) is also a particularity of IEEE 802.11 systems (see section 1.2.2). Ack/Retrans ensures that simple data MAC protocol data units (MPDUs) or even aggregate MPDUs (A-MPDUs) are correctly acknowledged.

Still some MPDUs do not need acknowledgment but are essential in 802.11 systems. Beacons are among such frames. They are generated by the AP's NetInterfaces, which consider them as one of the many control, management or data flows that a NetInterface must manage.

#### 4.2.2.4 Gains Compared to a PHY-centric Approach

As can be seen, traffic generation, queueing, and channel access, as well as acknowledgment, are all taken into account in this model. It would not have been the case in a PHY-centric approach (see section 1.5.1) where all of the previous MAC and upper layer functions are simplified. The system behavior would diverge from reality. This is especially true in a multiple-user context.

## 4.3 Simulation Scenarios and Parameters

### 4.3.1 Model Validation through Aggregation using ns-2

Before showing any of the improvements that this new simulation platform enables (see section 4.3.2), the first thing to do is to validate the model. We propose to do this through a cut-and-dried concept of IEEE 802.11n systems, MPDU aggregation (see section 1.2.7), using ns-2 simulator with LUTs as reference. Associated simulation results are given in section 4.4.1.

#### 4.3.1.1 Reference Structure: ns-2 System Simulator with LUT Channel Abstraction

In the modified version of ns-2 we have used, PHY layer performance is taken into account through LUTs, which are computed off-line through link-level simulations. This way the general particularities of the Tx chain, channel, and Rx chain can be accounted for in a statistical manner. The success or failure of reception is established by randomly selecting a packet error rate (PER) value and comparing it with the reference LUT PER value, for a given signal-to-noise ratio (SNR). It will be our reference structure for the validation of the proposed simulation platform.

#### 4.3.1.2 Simulation Parameters

The TGn channel, IEEE 802.11n PHY, and IEEE 802.11n MAC simulation parameters are given in Table 4.1. The scenario is summarized in Figure 4.2, where the station is tagged STA.

We shall note that the maximum transmit opportunity (TxOP) is set to the maximum video TxOP duration (see section 1.3.2.2) so as to have realistic durations.

The AP and the station are placed close to one another so that the rate adaptation algorithm can use the maximum PHY service data unit (PSDU) transmission rate (i.e. 130 Mbps). The CoordFunction can aggregate as many MPDUs as allowed by the maximum TxOP, using the current maximum rate $r_0$ to compute the frame's possible duration. We can thus send A-MPDUs with a high number of aggregates (as much as 30). The proof of concept, where A-MPDU_20s are to be used, is applicable. In addition, the chosen application rate corresponds to the maximum PSDU data rate. We thus insure that saturation could only be at MAC layer and/or PHY layer.

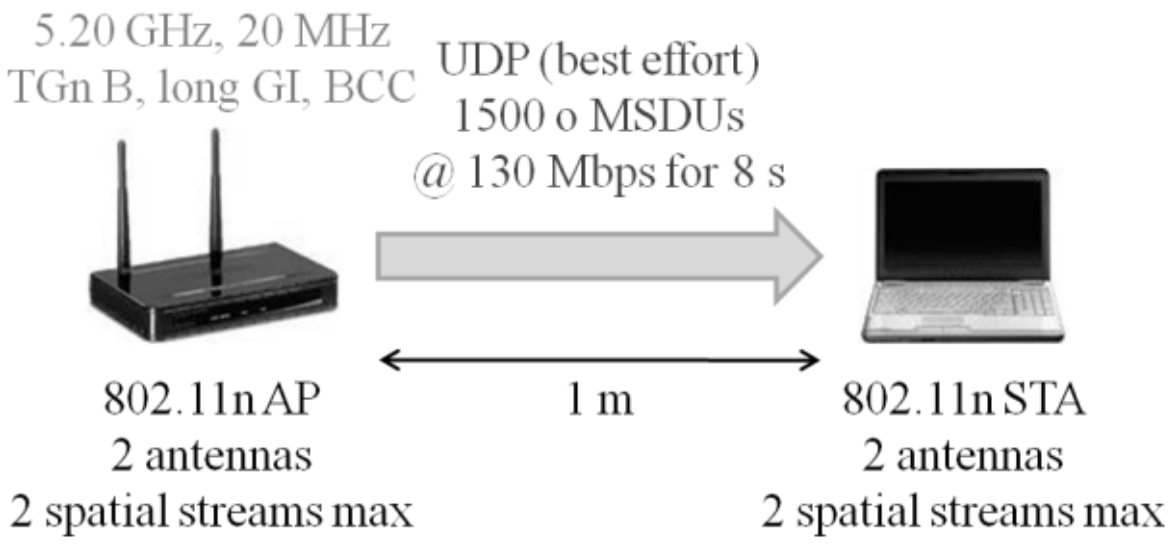

**Figure 4.2 Simulation scenario and parameters for model validation**

We will also note that all listed simulation parameters except the last PHY layer parameter and the last TGn channel parameter, which are specific to the PHY section, are also used by the ns-2 simulator. The two simulators are initialized with identical conditions.

### 4.3.2 Improvements in Modeling in a Hidden Station Scenario

A station establishes whether there is a signal through its carrier sense function. However if two or more stations are out of range of each other but in range of another station, carrier sense is eluded and collisions often occur. This is the hidden nodes problem [73] (see section 1.2.1.4). The proposed simulator can be used to better characterize the consequences of this phenomenon. Associated simulation results are given in section 4.4.2.

**Table 4.1 Main simulation parameters for model validation**

| | | | |
|---|---|---|---|
| **Global simulation parameters** | | | |
| Application rate (duration) | | 130 Mbps (8 seconds) | |
| Transport layer protocol | | User datagram protocol (UDP) | |
| BSS configuration | | 1 AP and 1 station | |
| AP-station distance | | 1 meter | |
| **MAC layer parameters** | | | |
| Access category | | Best effort | |
| MAC service data unit (MSDU) size | | 1500 octets (typical payload format) | |
| Maximum allowed TxOP | | 3008 µs | |
| Maximum aggregation size | | 1, 2, 8 or 20 depending on simulation (tagged no aggregation, A-MPDU_2, A-MPDU_8, and A-MPDU_20 resp.) | |
| Rate adaptation algorithm | Adaptive multi-rate retry (AMRR) algorithm | Initial rates | $r_0 = r_1 = r_2 = r_3 = 6.5\ Mbps$ |
| | | Counts | $c_0 = 3, c_1 = 3, c_2 = 1, c_3 = 3$ |
| Activated modulation and coding schemes | | MCS 0 to 7 for 1 spatial stream[1] and MCS 8 to 15 for 2 spatial streams[2] | |
| **PHY layer parameters** | | | |
| Number of antennas | | 2 at both AP and station | |
| Maximum number of spatial streams | | 2 | |
| Transmit power / system loss / antenna gain | | 17 dBm / 8.5 dB / 0 dB | |
| Additive white Gaussian noise (AWGN) level | | 7 dB | |
| Channel coding | | Binary convolutional coding (BCC) | |
| Guard interval (GI) | | Long | |
| Channel estimation | | Done once at the beginning of the received frame. Affected by AWGN | |
| **TGn channel parameters** | | | |
| Channel model | | B (residential) | |
| Bandwidth | | 20 MHz | |
| Central carrier frequency | | 5.2 GHz (channel n° 40) | |
| TGn channel seeds | | 50 for all simulations of this scenario | |

[1] Correspond to indexes 0 to 7 for 1 spatial stream in Table D.1 (Appendix D).
[2] Correspond to indexes 0 to 7 for 2 spatial streams in Table D.1 (Appendix D).

### 4.3.2.1 Classical Structures

#### 4.3.2.1.1 Reference Structure: ns-2 System Simulator with LUT Channel Abstraction

The ns-2 (MAC-centric) simulator presented above can finely model contention access. However, if there is a collision detected by the event scheduler, colliding frames are automatically considered as corrupt, not acknowledged and must be retransmitted. This is done whatever the collision power, duration and location within the received frame. Clearly this way of doing things seems harsh. But considering that ns-2 uses an abstracted PHY layer, it has no way of determining whether the collision leads to decoding errors or not. This simulator serves as our reference structure.

#### 4.3.2.1.2 Contribution of Fine Channel Modeling

A PHY-centric simulator cannot, as such, model the complex channel access procedure of CSMA/CA. It cannot be used as a reference structure. However this approach can be interesting in that the interference caused by collisions can be simulated [78] (see work exposed in Chapter 3). We can thus verify whether the PHY layer Rx chain can recover from the induced signal distortions. The interference caused by the collision is modeled by white Gaussian noise having the same power as the collision causing frame. It is added to the received signal on the concerned OFDM (orthogonal frequency division multiplexing) symbols. We have chosen to do so if the collision occurs in the PHY payload and if the collision-causing frame has a much lower SNR than the locked-on frame. Therefore the analog to digital converter (ADC) will not saturate and the Rx chain has chances of recovering signal distortions. If the PHY header is affected, we use LUTs to determine whether it has been correctly received. Indeed the PHY header is much more robust to errors than the PHY payload. Therefore, for low SNR collisions, we have chosen not to use the PHY layer Rx chain for the reasons exposed in section 4.2.1.1.

That is why in our simulation platform the MAC section hands over data frames having undergone collision during PHY payload transmission to the fine grained PHY simulation chain, in addition to collision-free data frames. Information on the power of the collision-inducing signal and on the relative position of collision-affected OFDM symbols is also handed over.

### 4.3.2.2 Simulation Parameters

In this scenario, two stations placed diametrically with regards to the AP are considered. Simulation parameters given in Table 4.1 are maintained except concerning the points given in Table 4.2. In addition, the considered streams are now uplink. The scenario is summarized in Figure 4.3, where station 1 and station 2 are tagged as STA1 and STA2 respectively.

**Table 4.2 Changes to bring to previous simulation parameters to expose model contributions**

| Concerned station | Initial distance from AP | Mobility (away from AP) | MSDU size | Max. aggregation size | TGn channel seed |
|---|---|---|---|---|---|
| Station 1 | 25 m | 5 m every 2 s | 100 octets | No aggregation | 50 |
| Station 2 | 10 m | None | 1500 octets | 2 | 52 |

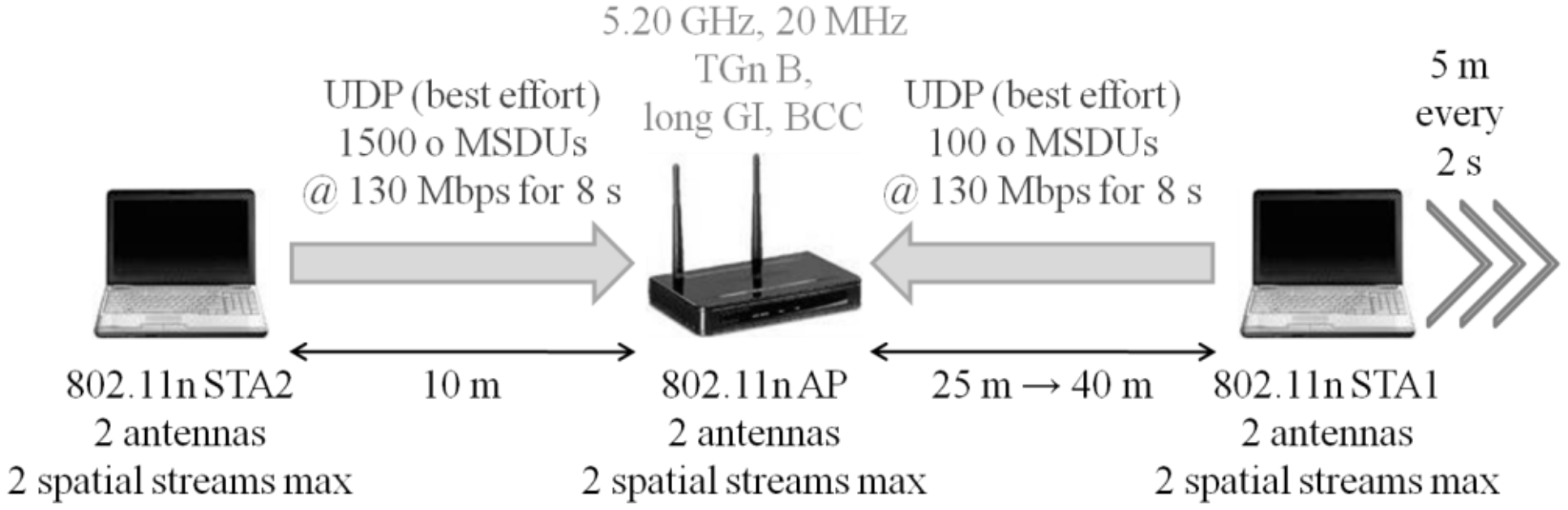

**Figure 4.3 Simulation scenario and parameters exposing model contributions**

Here also, the new simulation parameters, except the TGn related parameter (i.e. channel seed), are also used in the ns-2 reference simulator.

# 4.4 Simulation Results and Discussion

## 4.4.1 Model Validation through Aggregation using ns-2

The downlink MAC throughputs obtained through the reference ns-2 simulator and the new cross-layer simulator (denoted as XLS) for simple MPDU and A-MPDU_2 transmissions, using parameters given in 4.3.1, are illustrated in Figure 4.4. The reference ns-2 simulator uses LUTs to decide whether a frame is correctly received or not. These correspondence tables are obtained by averaging PERs using at least 500 different channel seeds for each SNR value. Particular evolutions of channel taps are thus smoothed out and a consistent PER-SNR curve is obtained for each MCS. XLS throughput curves are plotted for a specific channel (TGn with seed 50). They are thus much affected by the evolutions of this channel (fast fading). In addition, XLS throughput is at times lower than ns-2's because of shadowing effect (slow fading). Still the obtained average throughputs are roughly the same for the two simulators, for simple MPDUs and A-MPDU_2s. In addition, by performing a sufficient number of simulations using different channels (thus different TGn seeds) and averaging, the results obtained by XLS will much more stable and should match those of ns-2 in this scenario.

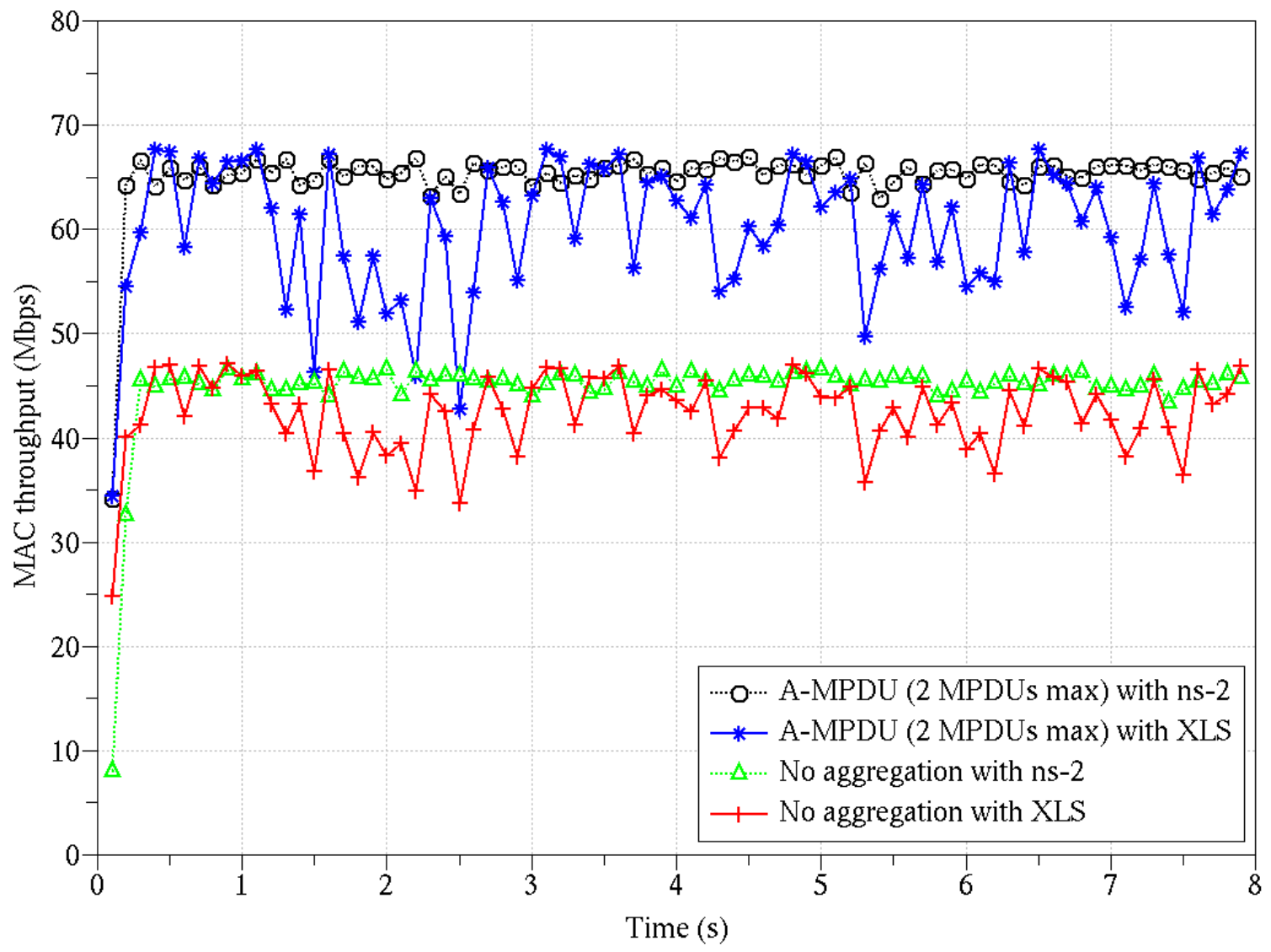

**Figure 4.4 Downlink throughput as a function of simulation time for transmissions without MPDU aggregation and with A-MPDU_2s using the reference ns-2 simulator and the proposed cross-layer simulator**

An easy way to reach higher throughput, and see the interest of the new XLS platform, without modifying PHY parameters would be to use A-MPDUs with more MPDUs (for greater MAC efficiency, see Appendix C for more details). The results of such simulations are given in Figure 4.5 (a), where A-MPDU_8s and A-MPDU_20s are used. When using the XLS platform, performance of A-MPDU_20 transmissions is even worse than that of simple MPDU transmissions (compare Figure 4.5 (a) and Figure 4.4). The latter have an average rate of 45 Mbps whereas the former have 40 Mbps.

Having designed the XLS platform to be as close as possible to real systems, channel taps are estimated at the beginning of the incoming frame. Rx channel estimation is simulated through the PHY header's long training frame sequences [8]. These estimates are then used for channel equalization of the rest of the frame. However if channel taps have notably changed in between, correct equalization cannot be done. The aging of these estimates can induce errors.

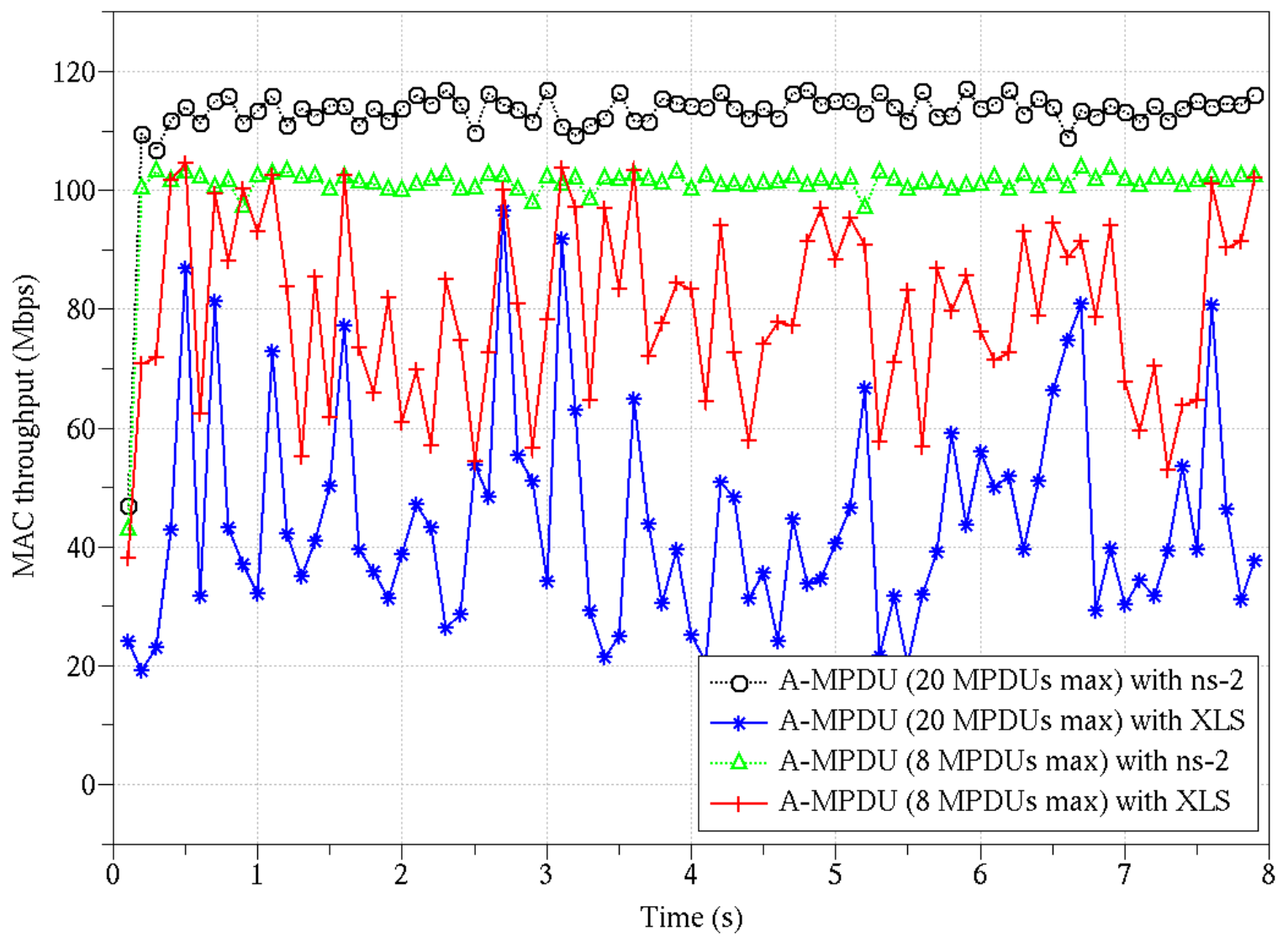

(a)

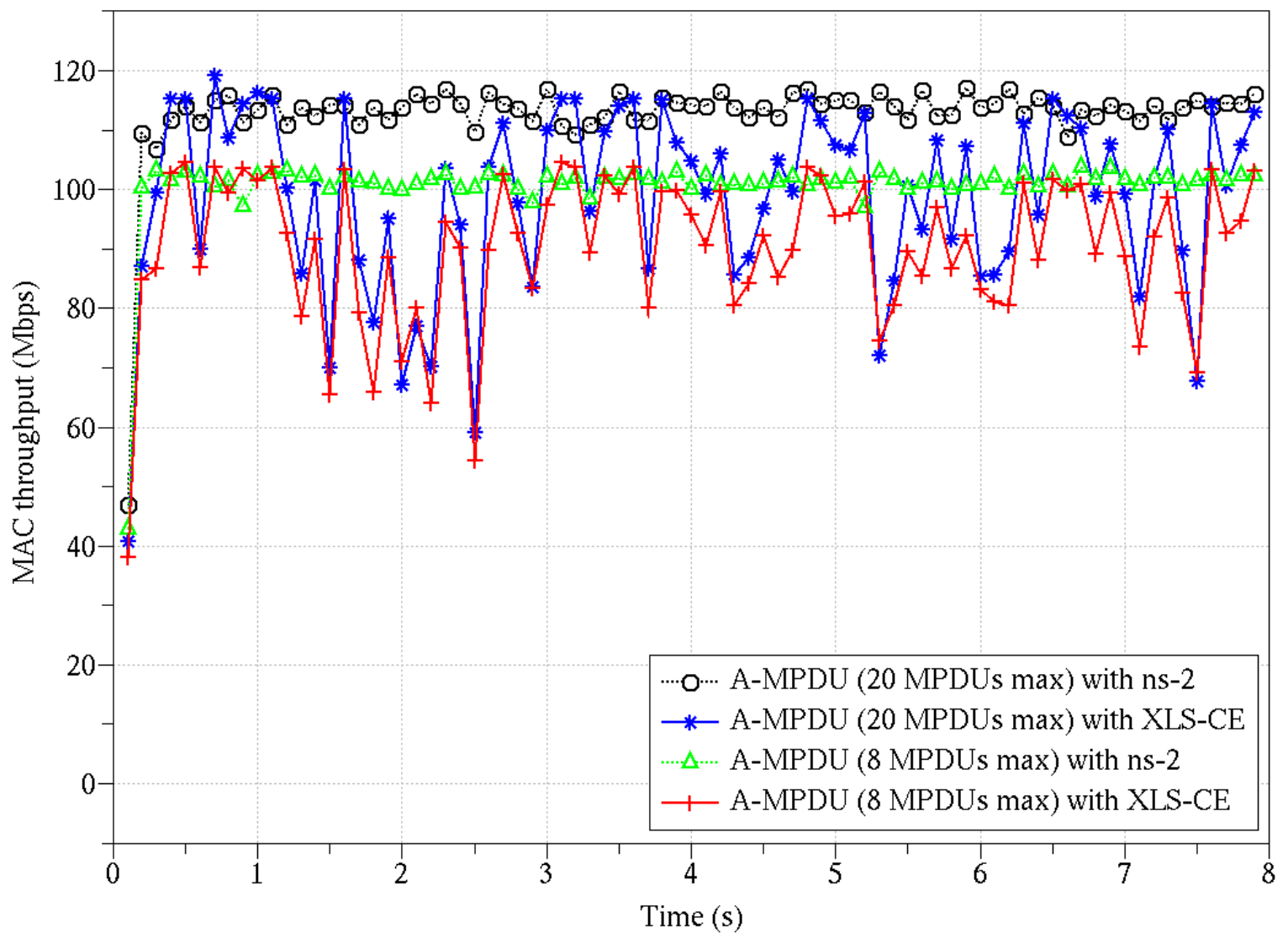

(b)

**Figure 4.5 Downlink throughput as a function of simulation time for transmissions with A-MPDU_8s and A-MPDU_20s using ns-2 and XLS. Channel estimation is done either once at the beginning of each frame (a), or for every OFDM symbol of each frame (b)**

There is nonetheless an amplifying factor: as indicated above, the number of MPDUs to be aggregated is determined using the highest rate $r_0$ of the AMRR algorithm. For this rate and a maximum duration of 3008 µs, the TGn channel does not globally change much. But if consecutive errors induce rate decrease (through AMRR rate fallback), the frame will last longer[1]. The odds of having important channel tap changes increase, eventually leading to errors. This chain reaction causes the associated throughput to drop notably. To verify this assertion, we have slightly modified the XLS PHY section so as to have OFDM symbol by OFDM symbol continuous estimation (XLS-CE). Estimates are updated throughout the whole frame reception process. The results obtained with this alteration are illustrated in Figure 4.5 (b). These results are on average similar to those obtained by ns-2.

Therefore the difference in throughput for A-MPDU_8 and A-MPDU_20 transmissions between ns-2 and XLS is due to the aging of channel estimates. Taking into account this phenomenon can be a strong advantage in some studies (e.g. CSI aging in MU-MIMO). It is only with this kind of simulator that channel evolution (PHY) and MCS adaptation (MAC) can be simultaneously accounted for.

As stated above, the effects of the rate adaptation algorithm on throughput are quite important. On all previous downlink throughput graphs obtained using XLS, there is a rate decrease between 1 s and 1.5 s. The evolution of PHY payload data rates is given in Figure 4.6 for this period.

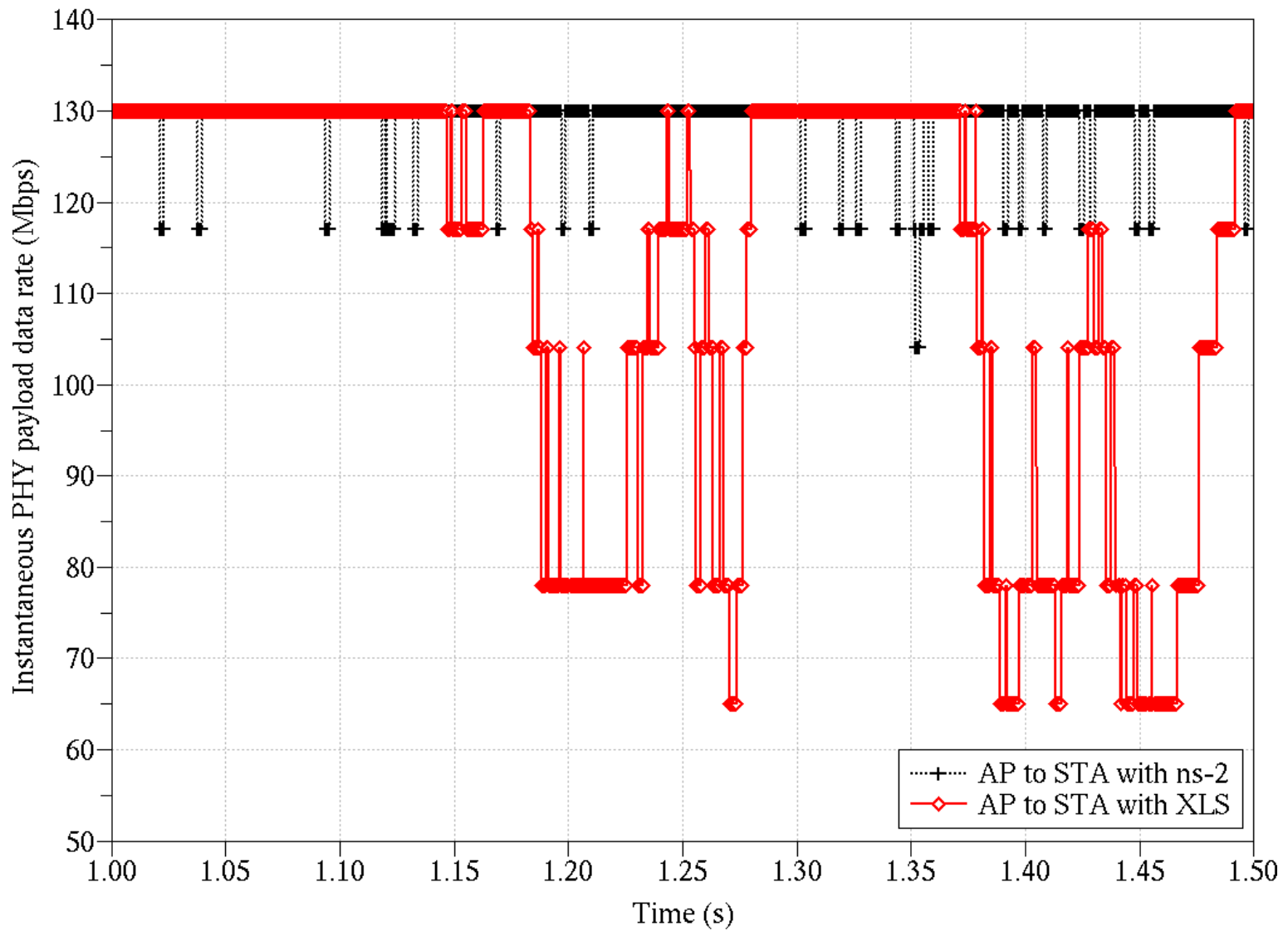

**Figure 4.6 Instantaneous PHY payload data rates as a function of simulation time for A-MPDU_2 transmissions using ns-2 and XLS**

[1] In this version of the simulator, the number of aggregated MPDUs does not change during retrial.

It can be seen that after 1.17 s the rate adaptation algorithm falls back to more robust modulations. The fallbacks could be due to ACK loss. As a control frame, the success or failure in receiving an ACK frame in XLS is determined through the use of LUTs. However, a uniformly distributed random variable is used for this purpose. Thus the odds of having bursts of error on ACKs only is very rare (if it were so, PHY payload data rates of ns-2 simulations, based also on LUTs, would have been less stable). The fallbacks could also be due to a deep fade of the TGn channel at that moment, due to shadowing. These evolutions induce bursts of error. Knowing that the same channel (seed 50) is used for all the simulations of this scenario, the second assumption explains the rate decrease.

All the same, the coherence in throughput results between ns-2 and XLS, in Figure 4.4 and Figure 4.5 (b), show that the proposed model is valid. It can thus be used to fully model an IEEE 802.11n/ac environment. Some additional options, like support for channel estimate aging, can come in handy for finer analysis. Studies on the aging of CSI feedback available at the transmitter, for beamforming purposes, can benefit from this PHY+MAC simulator.

### 4.4.2 Improvements in Modeling in a Hidden Station Scenario

In this scenario (see section 4.3.2), STA1 and STA2 are placed diametrically with regards to the AP. With STA1 moving away from the AP by 5 m every 2 s, the consequent gap growing between STA1 and STA2 favors hidden node scenario. Indeed by using the minimum receiver sensitivity (-82 dBm for binary phase-shift keying, with 1/2 coding rate, modulations [6]) we can compute the range of each station's transmission. Figure 4.7 (a), Figure 4.7 (b), Figure 4.7 (c), and Figure 4.7 (d) show the positions of the devices during the first 2 s, between 2 s and 4 s, between 4 s and 6 s, and finally during the last 2 s (resp.). The maximum coverage ranges of the AP, STA1, and STA2, computed using power parameters given in Table 4.1, are also illustrated.

We can see from Figure 4.7 (a) that when STA1 is 25 m away from the AP, the two stations can sense each other (see section 1.2.3.1). Collisions should thus be very rare. Because of CSMA/CA, there is collision only if both stations have backoffs ending at the same time. In this case both station's frames start at the same time and collision occurs during the PHY header (see section 4.3.2.1). When STA1 is 30 m and 35 m away from the AP though (see Figure 4.7 (b) and Figure 4.7 (c)), STA1 and STA2 become hidden nodes. Both stations will transmit without taking the other into account, not being able to sense its transmissions. However ACKs coming from the AP are received by both. When STA1 is 40 m away from the AP (Figure 4.7 (d)), it is also out of range of the AP. Therefore STA2 is alone in the BSS during the last two seconds of the simulation.

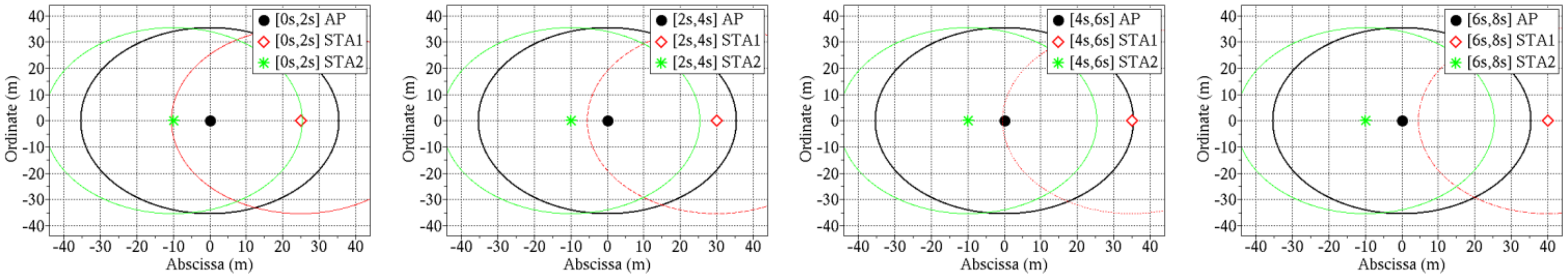

(a) (b) (c) (d)

**Figure 4.7 Positions of the AP, STA1, and STA2 through time on an AP-centered grid displaying maximum range between 0 s and 2 s (a), 2 s and 4 s (b), 4 s and 6s (c), and finally 6 s and 8 s (d)**

During the third interval (i.e. between 4 s and 6s), STA1 and STA2 are in hidden node scenario, with STA1's SNR being much smaller than STA2's. Therefore the ADC does not saturate and the AP's receive PHY chain could recover from errors. We check the validity of this affirmation by analyzing frame exchanges between 4.230 s and 4.260 s. Signal power of frames from STA1 and STA2, as perceived by the AP, are given in Figure 4.8 (a). The difference in payload size can be seen. Figure 4.8 (b) illustrates the power and duration of collisions, relative to signal power, as perceived by the AP. Figure 4.8 (c) illustrates the per-MPDU FCS error rate of frames received from STA2. The reader shall note that errors are detected by comparing the transmitted FCS and receiver computed FCS (as exposed in section 1.2.5). It is based on the outcome of this test that the receiving station decides whether or not to acknowledge the concerned MPDU. FCS error rate is thus a concise indication of the impact of errors on each MPDU. The PSDU rates of ACK frames sent towards each station are given in Figure 4.8 (d). The four graphs are aligned so that by combining them we can see whether the AP is receiving frames from STA1 or STA2 (in Figure 4.8 (a)), whether there was a collision (Figure 4.8 (b)) or not, if the received MDPU was correct (null FCS error rate in Figure 4.8 (c)) or not, and if the data frame has been acknowledged (strictly positive rate in Figure 4.8 (d)) or not. In addition, all expected ACKs and block ACKs (BA) being registered, a '0 Mbps ACK rate' is equivalent to a missed ACK or a missed BA.

During interval i, STA1 has sent 3 frames which have interfered STA2's frame (Figure 4.8 (a)) and caused three collisions, as shown in Figure 4.8 (b). We shall note that STA1's frames are not detected as IEEE 802.11 frames because the AP has locked on STA2's frame and is processing it. Once all OFDM symbols of STA2's frame are received, decoded bits are transferred to the MAC section which verifies for each MPDU whether the FCS is correct or not. Using this information we can see that STA2’s A-MPDU has been correctly received (null FCS error rate in Figure 4.8 (c)).

However, STA2’s A-MPDU is not acknowledged (Figure 4.8 (d)) despite its correct reception. This is because STA1’s third frame of the interval occupies the channel even after the end of STA2’s frame reception. The channel not being idle, the AP cannot acknowledge as expected. A similar phenomenon can be observed during interval ii. Again STA2’s frame is correctly received despite a collision due to STA1’s frame. But the latter activating the AP’s carrier sense, acknowledgment cannot be done. During interval iii though, collision with STA1’s frame is contained within STA2’s PSDU. Decoding being successful (Figure 4.8 (c)), the AP acknowledges STA2’s frame (Figure 4.8 (d)). Therefore, using the XLS platform thus enables to encompass the Rx chain’s error correction capacity.

During interval iv, STA1 manages to send its frame before STA2 (Figure 4.8 (a)). The AP thus locks on STA1’s frame and starts decoding the information within. But STA2’s being right after, an important collision occurs (19.8 dB as illustrated in Figure 4.8 (b)). Here, the ADC is saturated and STA1’s frame cannot be recovered (FCS error rate of 0.44 in Figure 4.8 (c)). It is thus not acknowledged (Figure 4.8 (d)). In addition, STA2’s frame has not been detected as an 802.11 frame by the AP and is consequently not acknowledged. Therefore a simulator finely modeling both PHY (to compass the Rx chain) and MAC (to simulate CSMA/CA-driven access) layers is needed to study the effect of collisions.

Figure 4.9 (a) illustrates the total uplink throughput for this scenario using XLS and ns-2. We can see that during the first and the last 2 s the same average throughput is obtained with both simulators. During the first 2 s, the two stations are in range and CSMA/CA is effective. Figure 4.9 (b) and Figure 4.9 (c), which illustrate STA1’s and STA2’s uplink throughputs (resp.), confirm that both stations have access to the channel. However during the last 2 s, STA2 monopolizes the channel. This is because STA1 is out of range of the AP, and is thus no longer part of the BSS (see Figure 4.7 (d)).

In between (i.e. from 2 s to 6 s of simulation time), throughput results obtained with XLS and ns-2 differ. ns-2 considers that all collisions corrupt the frames they affect. Most stations’ frames are considered lost in the hidden node situation. STA2 having much longer frames than STA1, the odds of having a collision during STA2’s transmissions are much higher than during STA1’s transmissions. In addition, the rate adaptation algorithm falls back and the CSMA/CA mechanism increases contention windows. This explains the low uplink throughput of STA2 (Figure 4.9 (c)). STA1 can thus benefit from this situation and send its frames when possible, explaining its relatively high uplink throughput peaks (Figure 4.9 (b)). Performance is not steady though because STA2 periodically tries to access the channel.

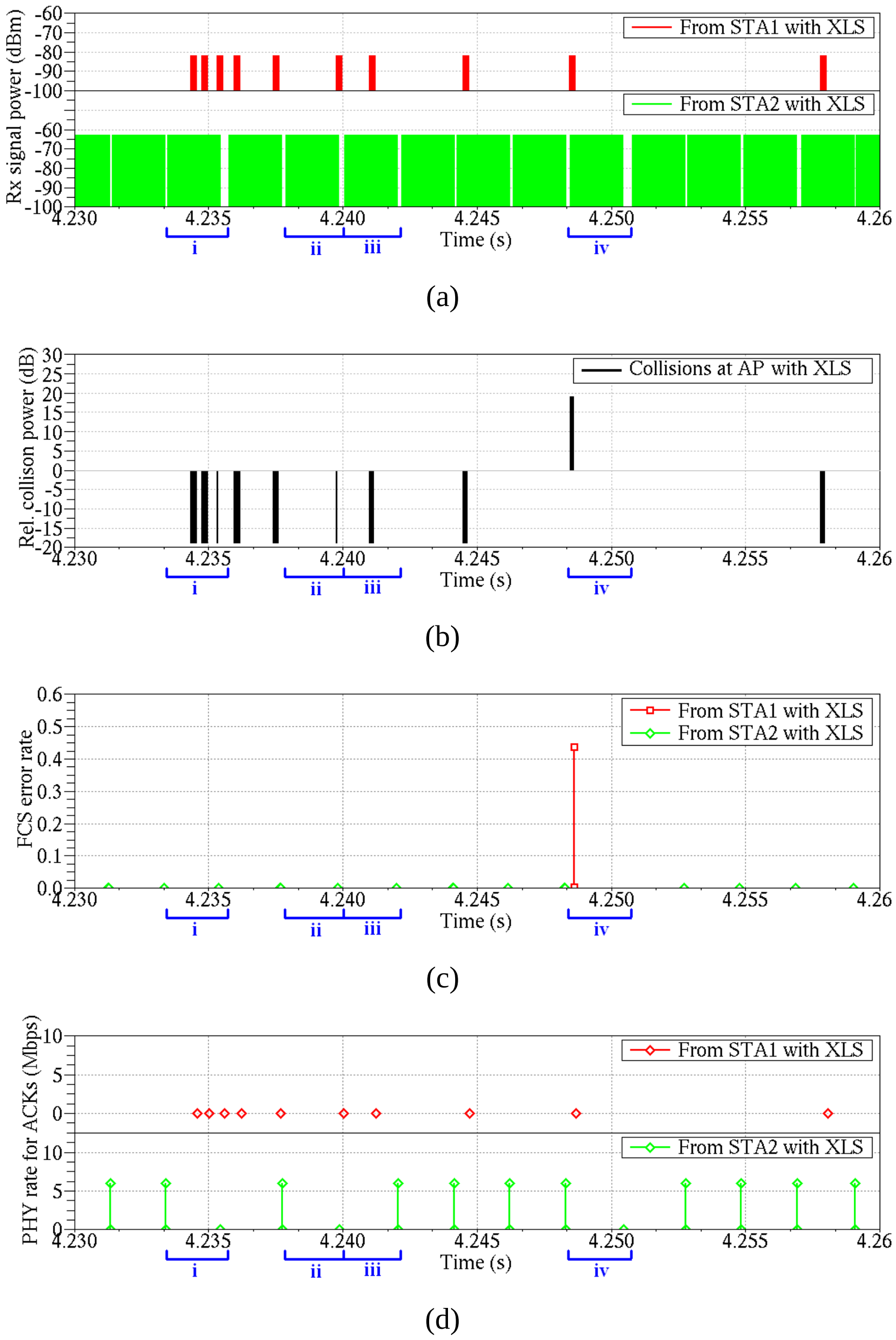

**Figure 4.8 Received signal power at AP for frames sent by STA1 and STA2 (a), relative power of collisions handled by the AP's PHY layer (b), FCS error rate for frames received from STA1 and STA2 (c), and PHY data rate for ACK frames sent by the AP towards STA1 and STA2 (d)**

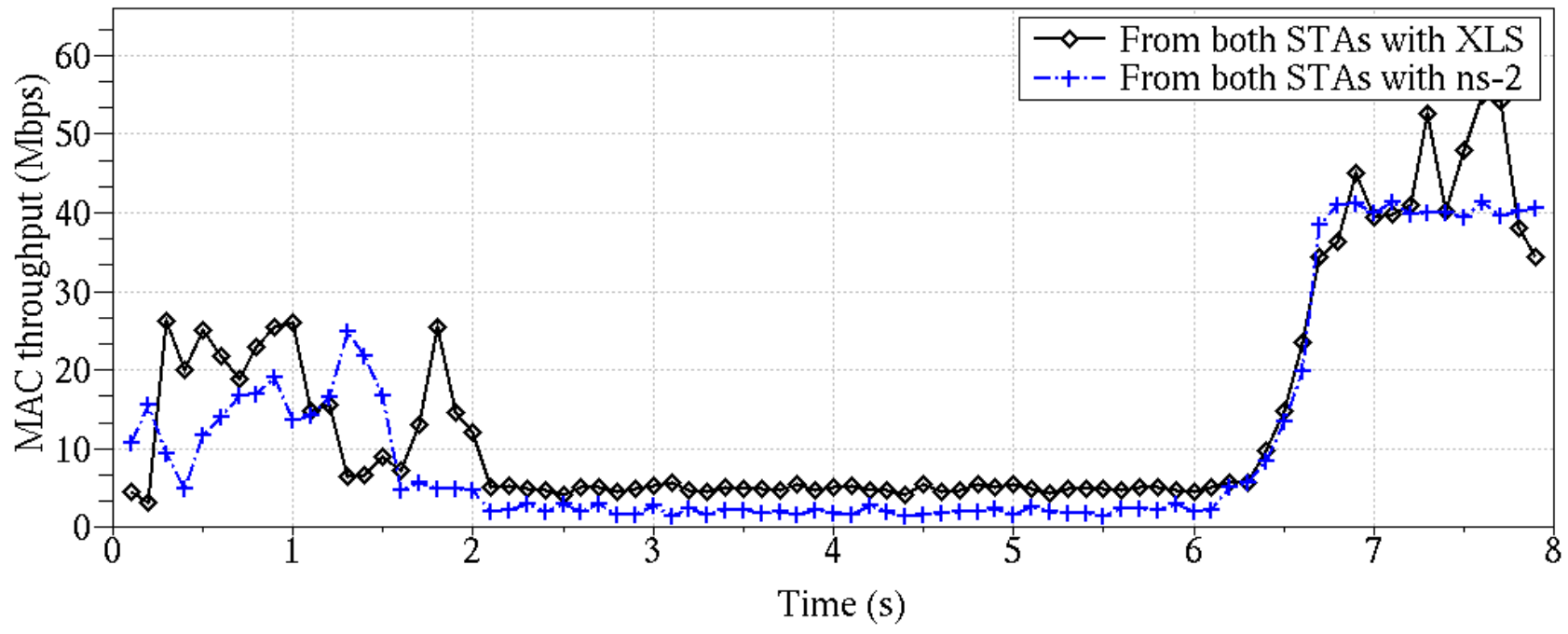

(a)

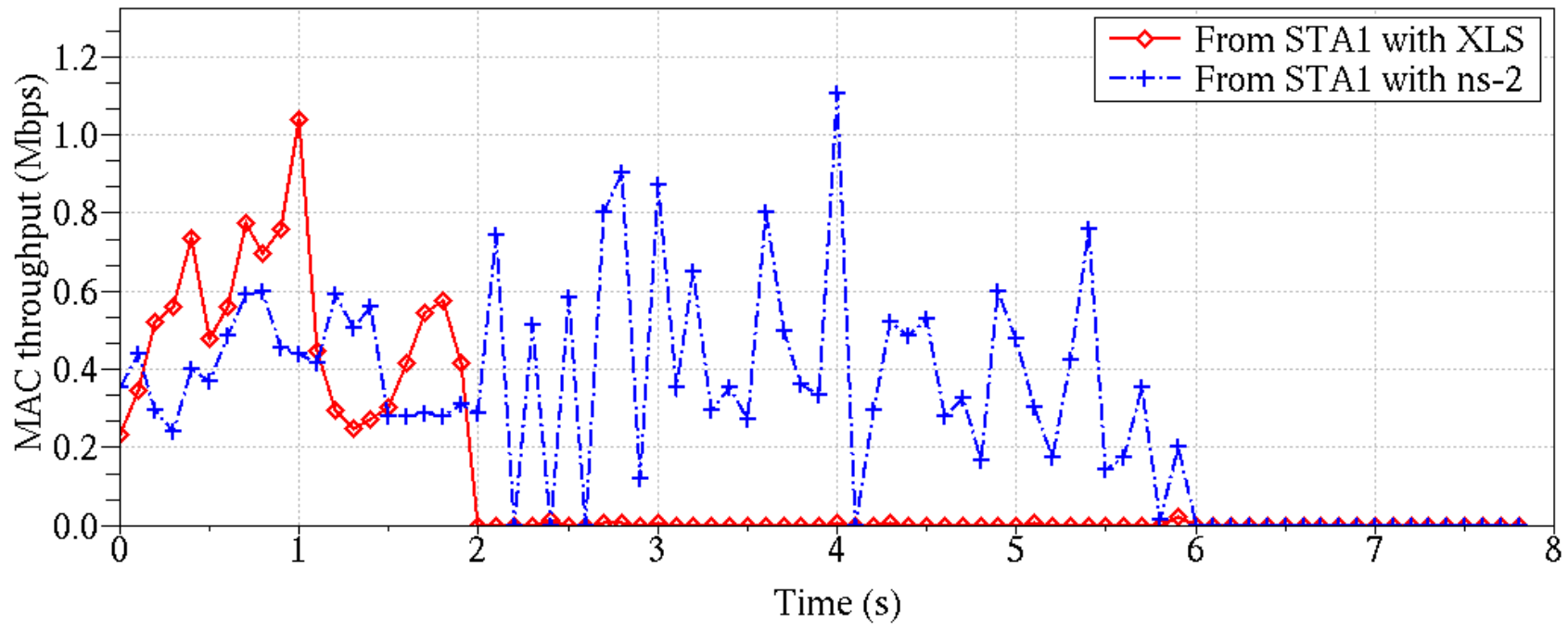

(b)

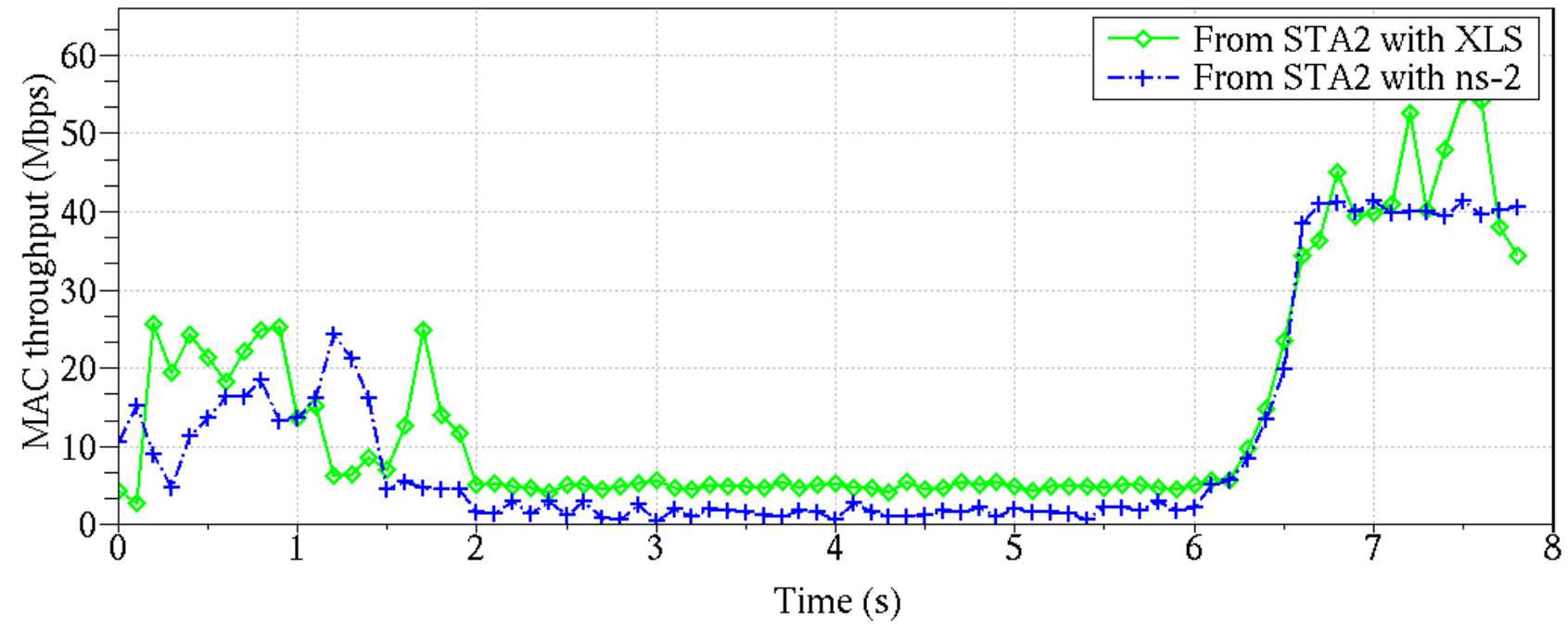

(c)

**Figure 4.9 Uplink throughput as a function of simulation time for A-MPDU_2 transmissions using ns-2 and the proposed cross-layer simulator for both STAs (a), for STA1 only (b), and for STA2 only (c)**

With XLS, the tendency is reversed because of the PHY layer can correct a part of the collisions. The AP can thus better recover from collisions when decoding STA2's frames, which are longer and have better SNR. Some ACKs are lost though, as seen above, therefore the rate adaptation algorithm falls back to more robust rates, thus the rather low, but steady, uplink throughput for STA2. STA1, on the other hand, rarely manages to effectively send its frames to the AP. It thus has very low throughput. Therefore there is clearly an advantage in having fine models of PHY and MAC layers when considering some phenomena.

## 4.5 Summary and Additional Features

In this chapter, we presented a novel IEEE 802.11n/ac simulation platform composed of a MAC simulation module which interfaces with a PHY simulation module in a symbiotic manner. This way, channel variations, Tx/Rx chain specificities, and channel access mechanisms are faithfully taken into account, while minimizing computational resource consumption.

Through MPDU aggregation performance evaluation, we have validated our fine-grained PHY+MAC simulator in that similar performances are obtained as with a ns-2 simulator using LUT channel abstraction. We have also shown in this chapter that the proposed PHY+MAC simulation platform provides a more realistic modeling of some phenomena. The impact of channel estimate aging and deep fades can be accounted for, while finely modeling contention access. In addition, the collision correction ability of the PHY chain was shown in a hidden nodes context.

The proposed 802.11n/ac simulator structure is thus a very interesting platform for modeling phenomena and testing optimizations simultaneously involving PHY and MAC layers. MU-MIMO performance evaluation studies can fully benefit from this platform.

Chapter 5

# PHY+MAC Performance Comparison of SU-MIMO and MU-MIMO in IEEE 802.11ac

Most of studies dealing with single-user multiple-input, multiple-output (SU-MIMO) and multiple-user MIMO (MU-MIMO) rely on physical (PHY) layer only or medium access control (MAC) layer only simulations. In the first case, the apparent considerable performance gains enabled by space division multiple access (SDMA) in downlink transmissions are often obtained without taking into account sounding and channel access induced overhead. In the second case, crosstalk interference (CTI, also called multiple access interference in literature) between stations served by the SDMA transmission is not always accounted for. Consequently, in order to take into account all the parameters and characteristics of these two transmission schemes, precise PHY+MAC modeling is particularly needed. A realistic and fair comparison may thus be performed. The new approach presented in this chapter is based on the multiple-user IEEE 802.11ac [7] simulation platform exposed in Chapter 4. In order to discriminate the characteristics of each transmission scheme, the single-user (SU) versus multiple-user (MU) comparison is done for single-antenna stations firstly, channel inversion being used as the precoding scheme. It is then extended to multiple-antenna stations using block diagonalization as the precoding scheme. In addition, the influence of CTI in SDMA transmissions is studied, so as to compare the real gains of MU-MIMO with the generally accepted gains.

In this chapter, we firstly detail the reasons for such a PHY+MAC performance comparison of SU-MIMO and MU-MIMO in 802.11ac. We then present the studied schemes. This is followed by an exposition of simulation scenarios and parameters. Finally, the corresponding simulation results are given and discussed.

# 5.1 Motivations for a Cross-layer Analysis of Single-user and multiple-user MIMO

## 5.1.1 Single-user MIMO and Multiple-user MIMO in IEEE 802.11

### 5.1.1.1 Single-user MIMO

SU-MIMO, or ‘plain’ MIMO, has been introduced in the IEEE 802.11 standards by the IEEE 802.11n amendment [6] (refer to sections 1.1.2 and 1.3.2 for details). The multiple antennas at the transmitter and receiver enable the use of spatial diversity or spatial multiplexing to improve performance (see sections 1.1.2.3 and 1.1.2.4 respectively). Channel knowledge, or channel state information (CSI), is an important parameter when using MIMO (see section 1.1.2.5). With the use of channel knowledge, the transmitter can use beamforming to improve even more performance (see section 1.1.3). Most of beamforming schemes use singular value decomposition (SVD) and precode the signal with the right singular vectors of the channel matrix so as to obtain an inter-antenna interference-free equivalent channel. CSI feedback can be done through explicit or implicit feedback (see section 1.3.2.2).

MAC layer efficiency is not affected by the implementation of SU-MIMO if CSI is not requested, and thus no beamforming is done, except for the fact that more long training fields and other new fields are needed (see section 1.3.2.1.2). Indeed SU-MIMO is a PHY layer technique offering robustness and throughput increase. However, when beamforming is applied for greater throughput and/or robustness, special control and management frames should be used. The overhead induced by these sounding or calibration procedures directly impacts MAC layer performance.

### 5.1.1.2 Multiple-user MIMO

MU-MIMO is introduced by the IEEE 802.11ac standard [7], where downlink SDMA has been chosen as the MU transmission scheme (refer to sections 1.1.4, 1.3.3, and 2.4 for details). The access point (AP) can make profit of the spatial distribution of stations to simultaneously transmit independent MIMO transmissions to selected stations. There is no explicit indication on which precoding technique to use for SDMA transmissions in the 802.11ac [7]. However CTI between served stations should be minimized. Non-linear, i.e. dirty paper coding-like, precoding techniques can utterly cancel CTI but at the expense of increased complexity (see section 1.1.4.1). That is why linear precoding is often considered (as exposed in section 1.1.4.2). Considering that we have one antenna at the stations in the first part of this study, and as no particular technique is standardized, we have chosen the channel inversion precoding technique because of the good performance-complexity tradeoff it offers. This scheme tries to cancel CTI through zero-forcing (see section 1.1.4.2.1). Its simplicity makes it the precoding technique that is often mentioned by chip makers and used as reference in standardization. In the second part of this study, multiple-antenna stations are considered. Block diagonalization is used so as to also have reduced receiver complexity (see section 1.1.4.2.3).

MU-MIMO can enable an increase in throughput through transmission parallelization, and thus by sharing contention periods and PHY preambles. However there needs to be CSI of all the concerned stations at the AP in order to operate the best beamforming matrices (see section 1.1.4). The overhead induced by this feedback can be more or less important depending on the number of served stations, number of antennas, and bandwidth (see section 2.4.4).

### 5.1.2 Pros and Cons of a PHY-centric Analysis

MIMO as well as SDMA are clearly PHY layer techniques. That is why quite a few studies comparing SU-MIMO and MU-MIMO rely on PHY layer only simulations. Indeed, the comparison criterion is often packet error rate (PER), as in [79,80], or symbol error rate, as in [81]. The former can even be exploited to extract PHY payload data rate, which is much more explicit with regards to higher layers, as in [80]. These link-level, or PHY-centric (see section 1.5.1), simulations show very often that SDMA enables considerable performance gains in downlink transmissions.

However, most of these results are obtained neither considering sounding and channel access induced overhead, nor using a rate adaptation algorithm. As exposed above, MU-MIMO transmissions require prior knowledge of all downlink channels of the stations which are to be served by the SDMA transmission. That is why CSI feedback, and thus sounding, is a central point of such techniques and should be accounted for.

### 5.1.3 Pros and Cons of a MAC-centric Analysis

The other tendency is to favor system-level simulations [82,83]. In these MAC-centric simulations, the channel and PHY layer models are abstracted (see section 1.5.2). At best, the last two are modeled by a lookup table, giving a correspondence between signal-to-noise ratio (SNR) and PER.

However such a simplification could have important implications in a SDMA context, where the decoding performance is intricately related to the used precoding scheme. As exposed above, the AP must have precise knowledge of the downlink channels of the served stations. In addition, CTI between stations served by the AP may seriously impact performance.

### 5.1.4 Advantages of a PHY+MAC Analysis

The new approach presented in this chapter offers detailed PHY and MAC layers modeling IEEE 802.11ac techniques and mechanisms. We use the 802.11ac multiple-user simulation platform exposed in Chapter 4. This simulation platform contains an all inclusive PHY layer combined with an elaborated MAC layer and working in a symbiotic manner [84]. Taking into account a detailed PHY layer, along with a realistic channel model, is necessary to precisely model channel evolutions and system response.

This approach thus enables a thorough cross-layer analysis of SU-MIMO and MU-MIMO techniques for an adequate comparison of these two schemes in 802.11ac. The relevance of accounting for CTI between served stations is also studied, thus evaluating the pertinence of a MAC-centric approach.

## 5.2 Studied Schemes

In this analysis three different schemes are compared: SU-MIMO with beamforming (Figure 5.1 (a)), MU-MIMO with CTI (Figure 5.1 (b)), and MU-MIMO without CTI (Figure 5.1 (c)).

SU-MIMO is used with beamforming based on SVD, and using right singular vectors of the channel matrix for precoding, so as to put it on equal footing with the MU-MIMO schemes. In the third scheme, MU-MIMO is considered as if it were composed of two (or more) completely independent beams towards each station. The precoders are thus completely independent. If framing overhead is not considered, this would be equivalent to having as many SU beamforming transmissions as there are SDMA-served stations. The reader should note that MAC-centric simulators use a similar approach. A comparison between the last two schemes can thus enable to evaluate the impact of accounting for CTI.

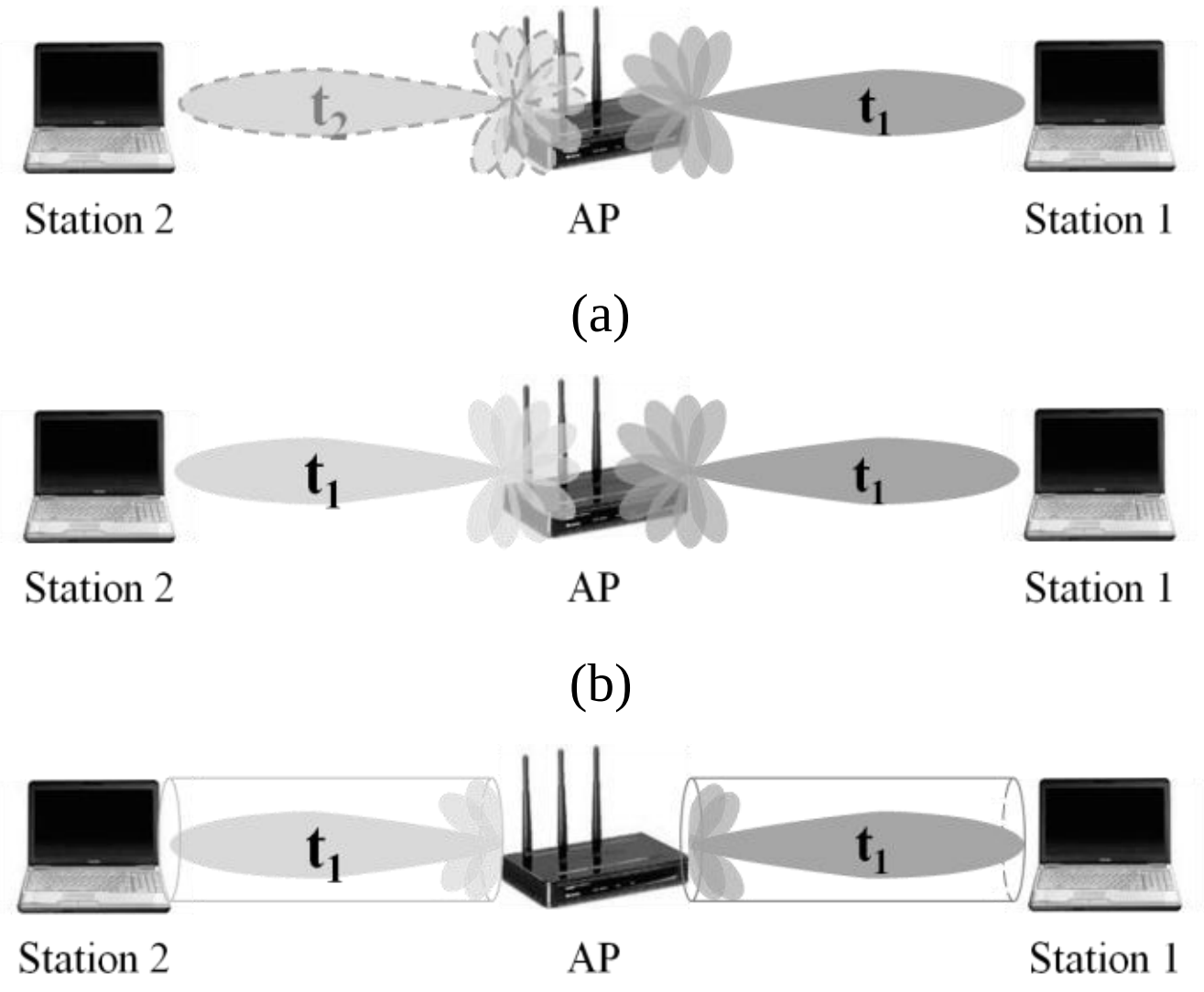

(c)

**Figure 5.1 Compared schemes: SU-MIMO (a), MU-MIMO (b), and MU-MIMO without crosstalk interference (c)**

# 5.3 Simulation Scenarios and Parameters

The very high throughput task group (TGac) channel model [74], IEEE 802.11ac PHY, and IEEE 802.11ac MAC simulation parameters of scenarios 1 to 5 are given in Table 5.1 (and illustrated in Figure 5.2). Parameters common to all 5 scenarios are given in Table 5.2, as defined in the 802.11ac standard [7].

The AP is equipped with 3 antennas in the first 4 scenarios, so as to have one degree of freedom in the null space search (see Appendix A for details). Indeed if the AP has more antennas than the sum of served stations' antennas, the null space search will be facilitated and thus reasonable performance can be achieved [19]. In the last scenario, the AP is equipped with 4 antennas and stations with 2 antennas. There is no degree of freedom in the null space search but spatial diversity is available.

We shall note that the maximum transmit opportunity (TxOP) is set to the maximum video TxOP duration (see section 1.3.2.2) so as to have realistic transmission durations. The maximum aggregation size is deduced from this maximum TxOP. In addition, one access category queue per served station is reserved so as to prevent starvation when using MU-MIMO (see sections 1.2.4.1 and 1.3.2.2 for details on access categories). Indeed if only one buffer is used for all packets of the same access category, packets intended for a station with bad channel conditions can monopolize the unique buffer, thus starving applications sending same access category packets to other stations.

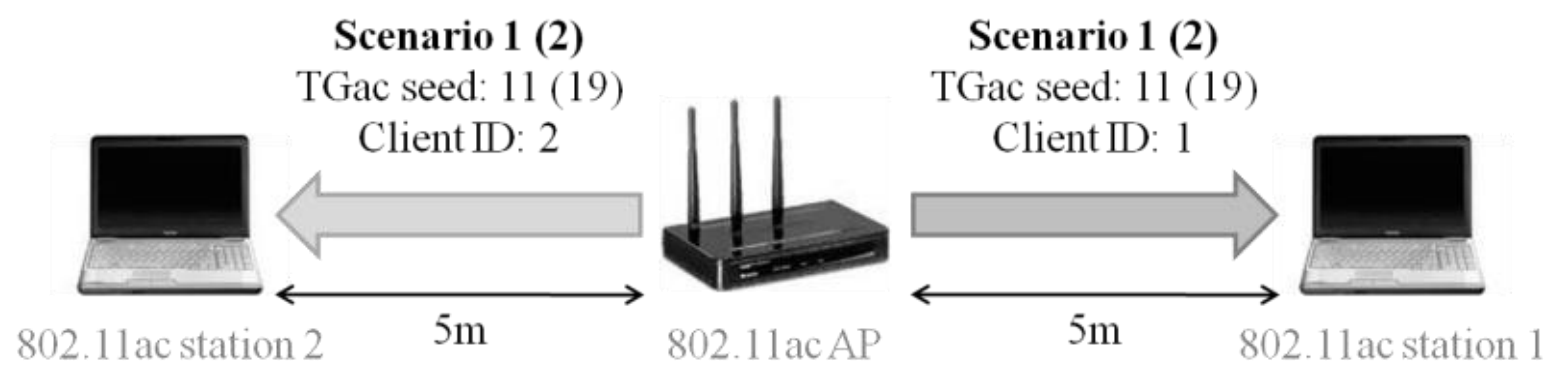

(a)

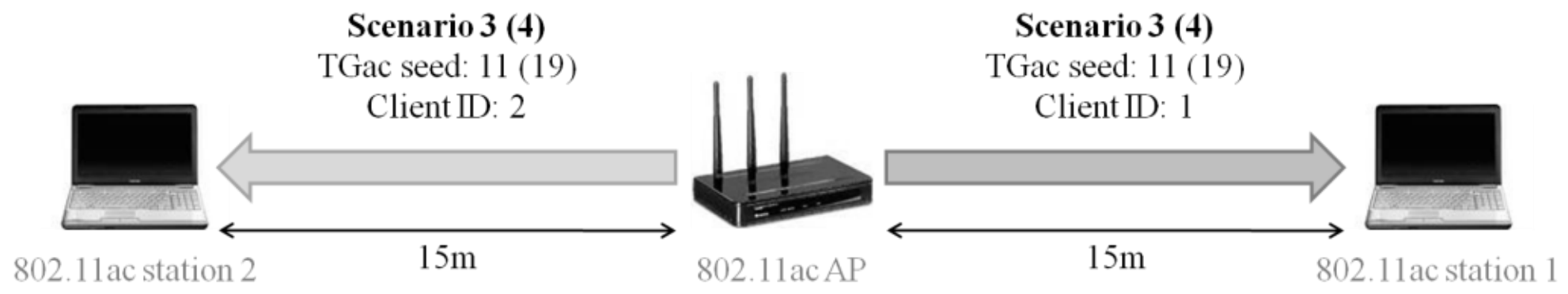

(b)

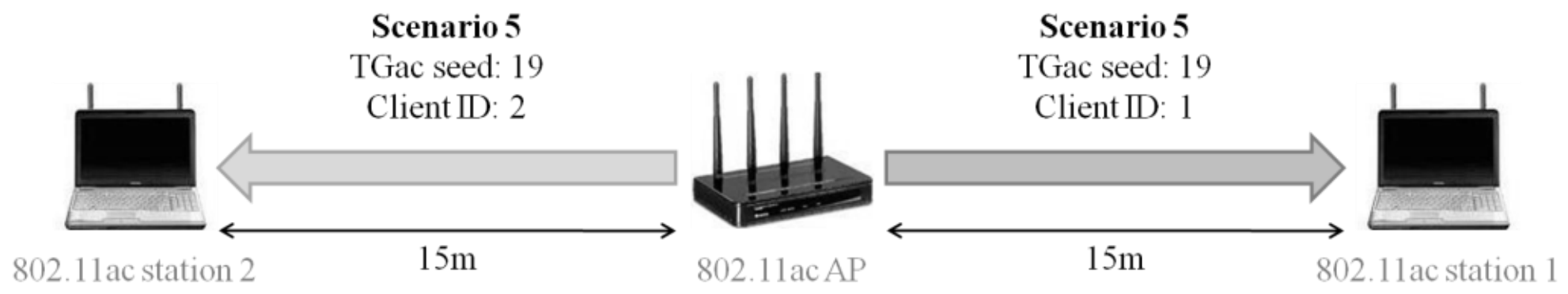

(c)

**Figure 5.2 Simulation scenarios 1 and 2 (a), 3 and 4 (b), and 5 (c)**

**Table 5.1 Per scenario simulation parameters**

| Scenario | Number of antennas | | AP-station distance (m) | TGac channel seed |
|---|---|---|---|---|
| | At AP | At each station | | |
| 1 | 3 | 1 | 5 | 11 |
| 2 | 3 | 1 | 5 | 19 |
| 3 | 3 | 1 | 15 | 11 |
| 4 | 3 | 1 | 15 | 19 |
| 5 | 4 | 2 | 15 | 19 |

**Table 5.2 Common simulation parameters**

| | |
|---|---|
| **Global simulation parameters** | |
| Application rate / duration | 80 Mbps / 2 seconds |
| Transport layer protocol | User datagram protocol (UDP) |
| Number of stations | 2 |
| **MAC layer parameters** | |
| Access category | Best effort |
| MAC service data unit size | 1500 octets (typical payload format) |
| Max. allowed transmit opportunity | 3008 µs |
| Max. aggregation size | 18 |
| Queue management | One access category queue per station |
| Rate adaptation algorithm — Adaptive multi rate retry algorithm | Initial rates: $r_0 = r_1 = r_2 = r_3 = 6.5\ Mbps$<br>Counts: $c_0 = 3, c_1 = 3, c_2 = 1, c_3 = 3$ |
| Activated modulation and coding schemes (MCSs) | MCS 0 to 8 for 1 spatial stream (see Table D.1) |
| Sounding and feedback | Sounding frequency: Every 20 ms<br>Subcarrier grouping: None<br>Ψ angle quantization: 2 bits (SU) and 5 bits (MU)<br>Φ angle quantization: 4 bits (SU) and 7 bits (MU) |
| **PHY layer parameters** | |
| Number of spatial streams | 1 |
| Transmit power / system loss / antenna gain | 17 dBm / 8.5 dB / 0 dB |
| Additive white Gaussian noise (AWGN) level | 7 dB |
| Channel coding | Binary convolutional coding |
| Guard interval | Long |
| Channel estimation | Done once at the beginning of the received frame. Affected by AWGN |
| **TGac channel parameters** | |
| Channel model | B (residential) |
| Bandwidth | 20 MHz |
| Central carrier frequency | 5.2 GHz (channel n° 40) |
| TGac client identification index | 1 and 2 for station 1 and 2 respectively |

SU-MIMO transmissions also use the same TGac client identification indexes as MU-MIMO, following TGac recommendations [75]. The two schemes are thus compared on equal footing.

## 5.4 Simulation Results and Discussion

### 5.4.1 Scenario 1: Favorable Downlink Channels

The throughput results obtained for station 2 (STA2) using scenario 1 are given in Figure 5.3 (a) for the three studied transmission schemes. Both MU-MIMO schemes' throughputs have almost doubled relative to SU-MIMO. The inclusion of CTI, and consequently the use of channel inversion precoding, does not affect performance with scenario 1's parameters. Indeed the PHY data rates of frames sent to STA2 using MU-MIMO with CTI (Figure 5.3 (b)) confirm throughput stability. The AP's adaptive multi-rate retry (AMRR) algorithm (see Appendix B) adapts the PHY data rate according to channel conditions, and thus received signal strength, through statistics on packet acknowledgements. After a warm-up phase of 200 ms, the data rate remains at 78 Mbps, which is the maximum data rate for 1 spatial stream (see Table D.1 in Appendix D).

The solid and dashed curves in Figure 5.3 (c) give the received signal powers' evolution over time, for MU-MIMO with and without CTI (resp.). The displayed power levels are obtained during the long training fields' reception, the latter being precoded with STA2's beamforming vector (see section 2.4.1). The observed evolutions are due to environment mobility. The displayed power levels are normalized over the noise-free and deep fade-free received signal power. Firstly, the reader can notice that the use of channel inversion precoding reduces received signal strength. Indeed, in channel inversion, precoding vectors have to be chosen in the null space of other stations' channel matrices. This constraint lowers the odds of choosing the best (i.e. power maximizing) vector, as in classical SVD beamforming.

The received interference levels for MU-MIMO with and without CTI (circular markers and triangular markers resp.) are also given in Figure 5.3 (c). These interference levels are determined using the long training fields precoded with the other stations' beamforming vectors and are normalized over useful signal power. Naturally CTI power is null for the second scheme. For the first scheme however, the reader can notice that the interference level is highest between 0.3 s and 0.7 s. However this corresponds to the period during which the CTI-including scheme's received signal power is closest to that of the CTI-free scheme's. STA2 can thus easily decode its information, despite CTI.

Therefore in this favorable scenario, there is no advantage in accounting for CTI in the performance evaluation of MU-MIMO. A MAC-centric simulation could have given similar results.

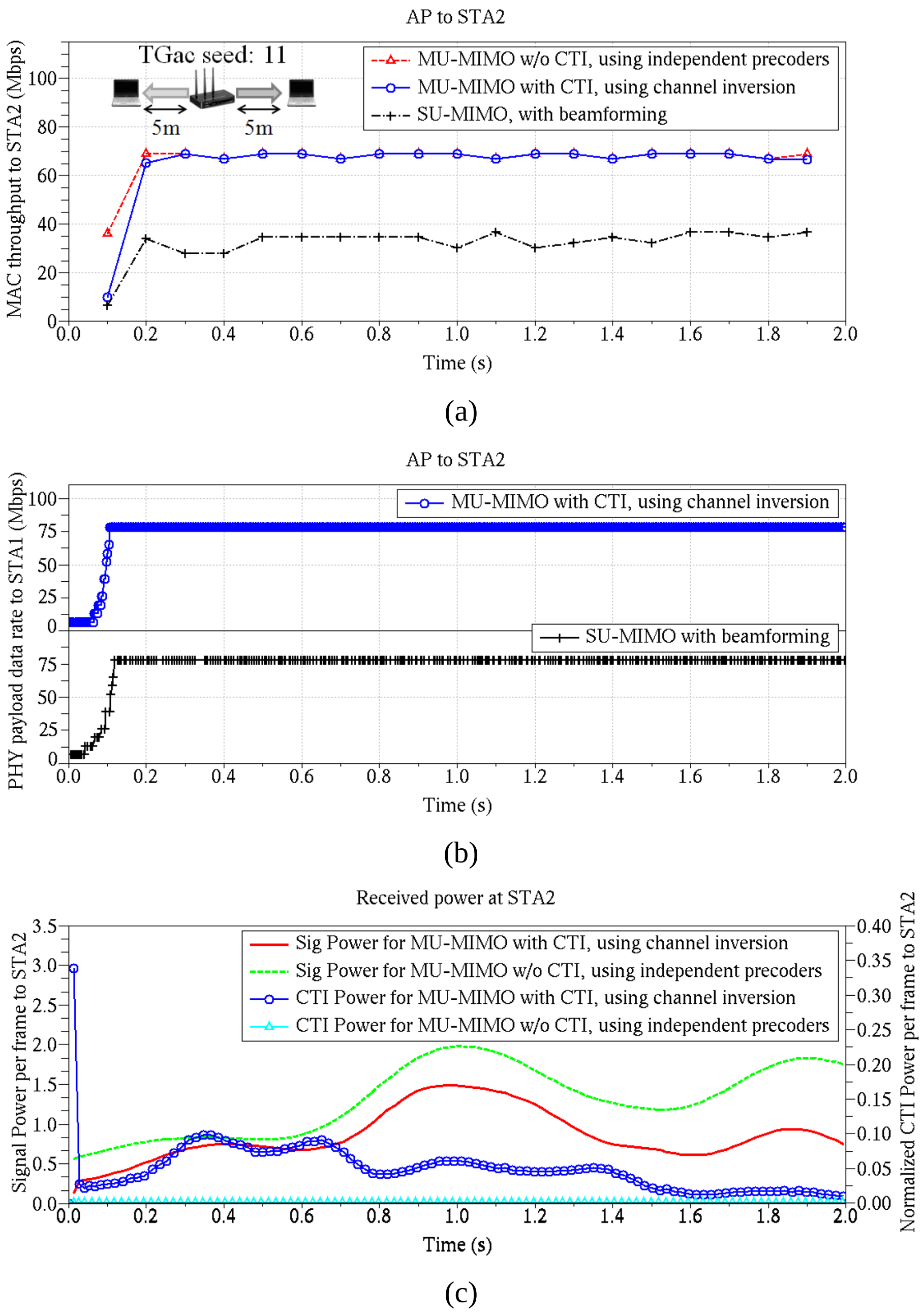

**Figure 5.3 Scenario 1: favorable downlink channels: MAC throughput towards STA2 for the three schemes (a), PHY data rate of each transmitted data MAC payload frame for the CTI-including MU-MIMO scheme and SU-MIMO scheme (b), and received useful signal and normalized interfering signal powers for the compared MU-MIMO schemes (c)**

### 5.4.2 Scenario 2: Correlated Downlink Channels

The throughput results obtained for station 1 (STA1) using scenario 2 are given in Figure 5.4 (a), again for the three studied transmission schemes. The performance of SU-MIMO and CTI-free MU-MIMO schemes are almost the same as with the first scenario. This, along with received signal power levels for CTI-free MU-MIMO scheme in Figure 5.4 (c), lead to the conclusion that channel conditions are roughly the same as before. A MAC-centric simulation would have treated both scenarios the same way and obtained similar performance (no seed differentiation).

However, the inclusion of CTI considerably changes things in scenario 2. In this case, the throughput, far from being stable as previously, can even be worse than SU-MIMO's over some periods. Indeed the data rates (Figure 5.4 (b)) confirm the throughput falls observed between 0.4 s and 1.1 s, and between 1.4 s and 2.0 s. MU-MIMO is thus not always better than SU-MIMO, either in a throughput-oriented or a stability-oriented analysis. The use of channel inversion also seems to have greater impact on the received signal strength than in the previous scenario (solid curve in Figure 5.4 (c)). Another noticeable fact is that the local minima coincide with the previous throughput falls. The received power being lower than during other time intervals, the AMRR algorithm chooses slower but more robust modulations (Figure 5.4 (b)). We can also notice that despite channel inversion precoding, CTI is not decreased (curve with circular markers in Figure 5.4 (c)). The received interference levels are twice as high compared to the other scenario's levels.

Therefore, the SU-MIMO view of channel conditions, which would have classified these channels as favorable, is not suited for MU-MIMO schemes including CTI. Indeed MAC-centric simulators do not account for the correlation between downlink channels. The performance of MU-MIMO schemes should thus be evaluated with PHY+MAC simulators considering CTI.

### 5.4.3 Scenarios 3 and 4: Farther Stations

In scenarios 1 and 2, where stations are placed 5 meters from the AP, the maximum PHY layer data rate is reached. If stations are 3 times farther away from the AP as in scenarios 3 and 4, the SNR received by each station decreases from 42 dB to 25 dB for SU-MIMO, and from 39 dB to 22 dB for MU-MIMO. Indeed an AP using MU-MIMO towards two stations has the same total transmitted power constraints as when using SU-MIMO, thus the 3 dB gap. This gap may have more implications when stations are 15 meters away because different data rates may be chosen for MU-MIMO and SU-MIMO, and performance comparison is less skewed towards MU-MIMO.

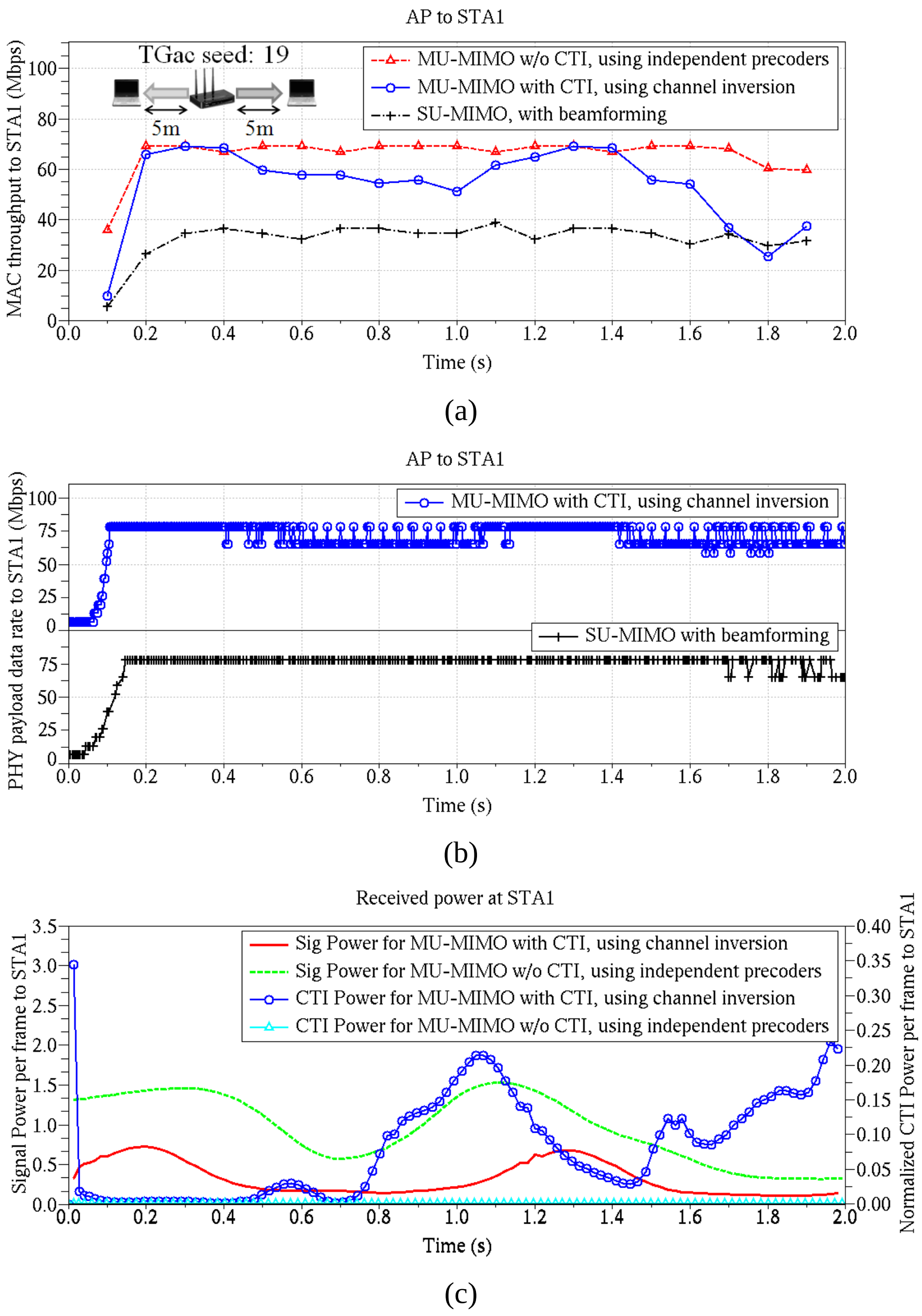

**Figure 5.4 Scenario 2: correlated downlink channels: MAC throughput towards STA1 for the three schemes (a), PHY data rate of each transmitted data MAC payload frame for the CTI-including MU-MIMO scheme and SU-MIMO scheme (b), and received useful signal and normalized interfering signal powers for the compared MU-MIMO schemes (c)**

Total downlink throughput results obtained using scenarios 3 and 4 are given in Figure 5.5 (a) and Figure 5.5 (b) respectively. In scenario 3, MU-MIMO with CTI roughly performs twice as better as SU-MIMO. With favorable downlink channels, the inclusion of CTI brings no additional information to performance evaluation, even when AP-station distance is increased.

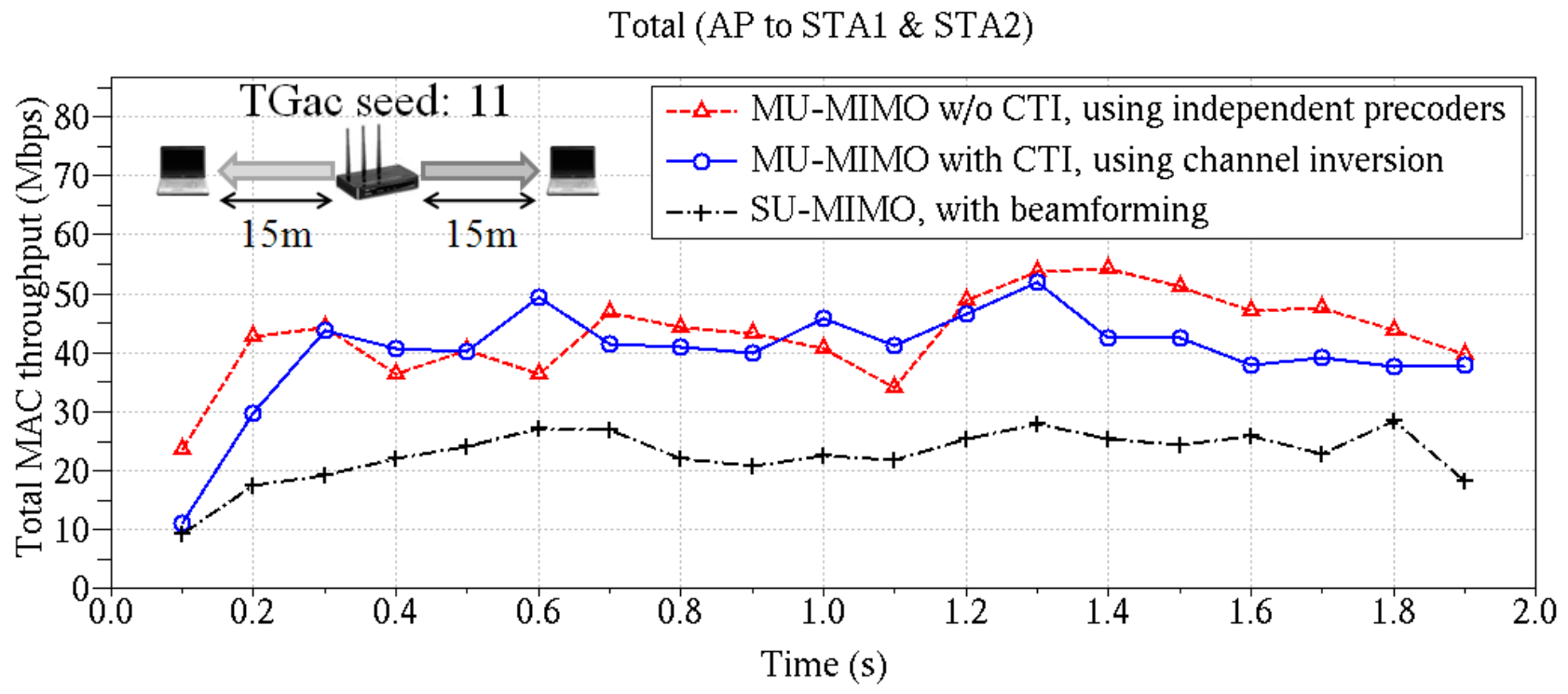

(a)

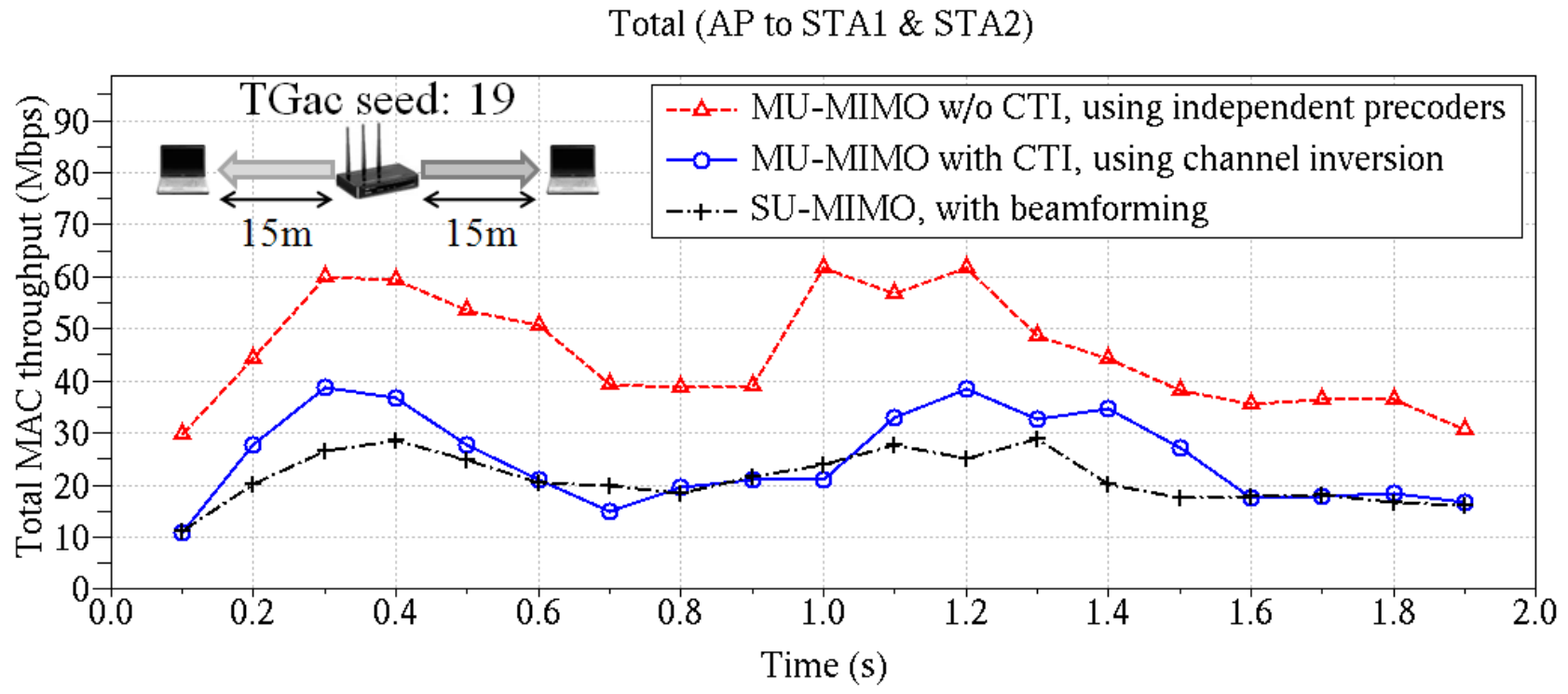

(b)

**Figure 5.5 Total downlink MAC throughput for the three schemes using scenario 3: favorable downlink channels with farther stations (a), and scenario 4: correlated downlink channels with farther stations (b)**

However, in scenario 4 where the correlated downlink channels are used, MU-MIMO with CTI and SU-MIMO show similar performance. When AP-station distance increases, the performance degradation of MU-MIMO, in presence of correlated downlink channels, becomes important. The MAC-centric view is even less suited than in scenario 2. In this kind of scenario, it would be much more profitable to completely switch from MU-MIMO to SU-MIMO so as to gain sounding overhead and processing.

### 5.4.4 Scenario 5: Greater Diversity

Spatial diversity is a means of improving reception and decoding. In scenario 5, one antenna is added per device. The total downlink throughput obtained with the resulting four AP-antennas and two per-station antennas (while still activating 1 spatial stream only) is illustrated in Figure 5.6.

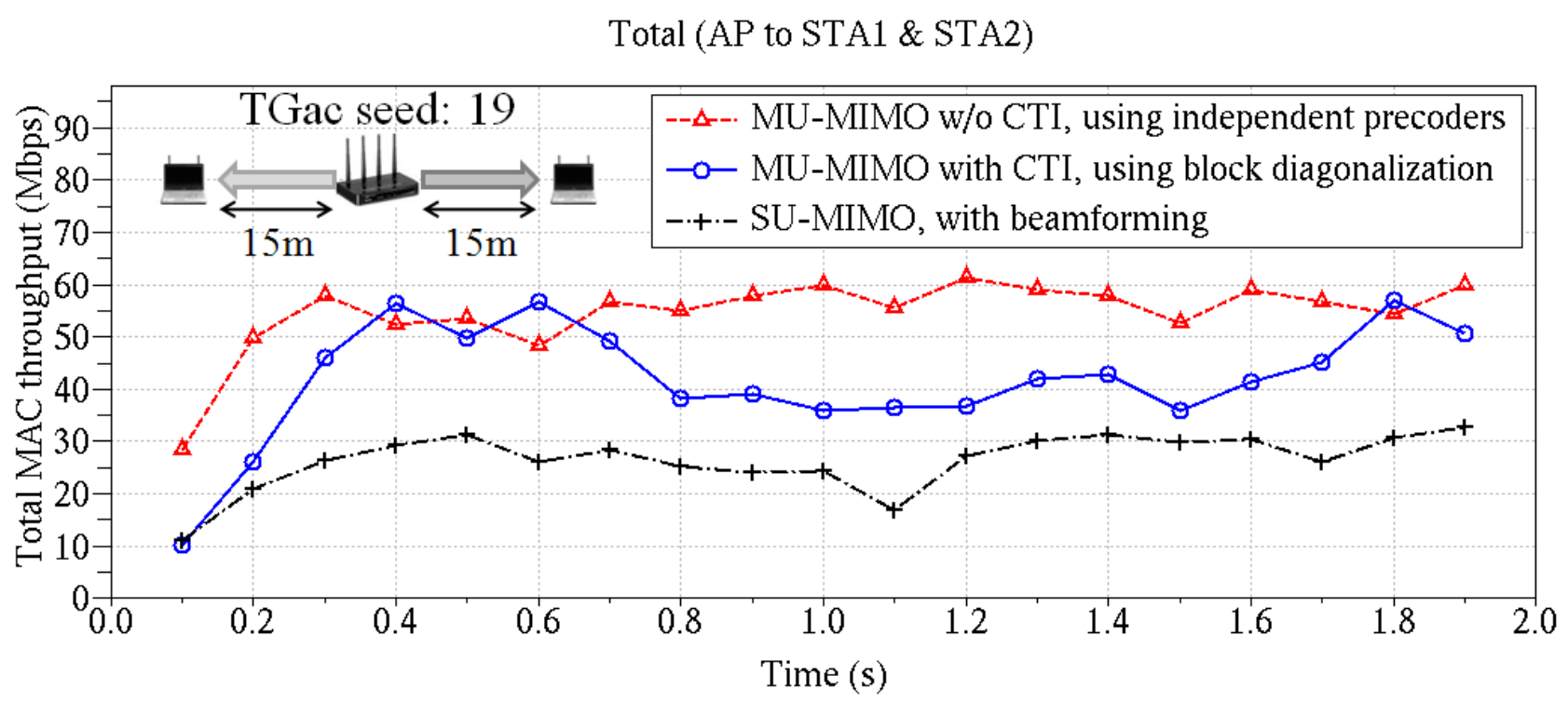

**Figure 5.6 Total downlink MAC throughput for the three schemes using scenario 5: correlated downlink channels with farther stations and greater diversity**

Despite the fact that the used downlink channels are not exactly the same as in scenarios 2 and 4[1], channel conditions here are rather good in this scenario. Indeed, even when accounting for CTI, the MU-MIMO scheme is better than SU-MIMO. However, its performance should not be as important as assumed by MAC-centric simulators. Therefore this confirms that a PHY+MAC analysis is needed to evaluate MU-MIMO's performance.

## 5.5 Summary

In this chapter a new approach based on a detailed PHY+MAC simulation platform is used. A realistic and fair comparison of SU-MIMO and MU-MIMO has been done for single-antenna stations and multiple-antenna stations benefitting from spatial diversity. It has thus been shown that it is important to consider CTI when evaluating the performance of MU-MIMO transmission schemes. The performance of MU-MIMO when CTI is not accounted for may be too optimistic. Indeed, MU-

[1] The phase shifts and correlation matrices, used in the TGac channel model [75] (see [74] for full equations), are modified by the addition of antennas at the AP and the stations. The differentiation between stations served by the MU-MIMO transmission is based on the client identifier and the phase shift between antennas. Therefore the more the antennas, the greater the dissimilarity between the downlink channels over time.

MIMO transmissions can have less throughput gain and be much less stable than SU-MIMO transmissions, when downlink channels are correlated. Thus time sensitive applications may suffer from jitter and higher packet loss when using MU-MIMO. When AP-station distance is increased, MU-MIMO's performance is degraded and the advantage of pooling channel access is lost. In these conditions, using SU-MIMO would enable similar performance. On the other hand, when downlink channels are favorable or when there is greater antenna diversity, MU-MIMO enables the expected throughput increase. This remains true even when AP-station distance is increased. Therefore using a PHY+MAC performance evaluation is useful for evaluating the real performance of MU-MIMO under different channel conditions.

The easily implementable channel inversion scheme being used in this chapter, other more complex precoding schemes, e.g. minimum mean square error precoding (as exposed in section 1.1.4.2.2), are to be tested. An intelligent switching algorithm between MU-MIMO and SU-MIMO schemes based on CSI and received SNR can also be considered in future studies.

Another possible PHY+MAC performance test, which does not require additional modification of the used configurations, would be to increase the number of stations. As exposed above, the overhead induced by the CSI feedback can be more or less important depending on the number of served stations. Therefore, obtained performance may depend on the period of the related sounding procedures.

Chapter 6

# PHY+MAC Analysis of IEEE 802.11ac MU-MIMO Channel Sounding Interval

We have seen that the IEEE 802.11ac standard [7] enables medium access control (MAC) throughputs of up to 1 Gbps [42] by incorporating a multiple-user dimension to the classical multiple-input, multiple-output (MIMO) configuration. Indeed the access point (AP) can simultaneously transmit independent streams to multiple stations by applying crosstalk interference (CTI) minimizing precoding vectors. However, the AP has to have precise knowledge of the channels of all the stations which are to be served by multiple-user MIMO (MU-MIMO) transmissions. To that end, a channel sounding procedure is regularly engaged so that the concerned stations can return a feedback of their respective channel estimates. The frequency of this sounding procedure, and thus the aging of the obtained estimates, has an important impact on performance. However, there is no indication, in the standard, on which interval to take.

In this chapter, we present the impact that the channel sounding interval can have on throughput. To do so, we use the 802.11ac simulation platform, containing an all inclusive physical (PHY) layer combined with an elaborated MAC layer and working in a symbiotic manner [84], exposed in Chapter 4. The aging of channel estimates is therefore precisely accounted for while finely simulating contention mechanisms. A similar study has been done in [85]. This interesting work was written during the first phases of the 802.11ac standardization process, thus not being compliant to the current version of the standard. In addition, the resulting analysis was done firstly from a PHY perspective then from a MAC perspective. In the present chapter however, we present a combined PHY+MAC analysis using 802.11ac compliant chains, thus leading to recommendations taking into account both PHY and MAC considerations [86].

We firstly recall PHY and MAC implications of MU-MIMO, with a special focus on 802.11ac's channel sounding and feedback procedure. Simulation scenarios and parameters are then given. Finally the corresponding simulation results are presented and discussed. These throughput results are given firstly using a statistical MAC analysis, then using a MAC-centric simulator, and finally using a PHY+MAC simulator. The impact of increasing the granularity of simulations can thus be evaluated.

## 6.1 Motivations for a Cross-layer Analysis of Channel Sounding Interval for MU-MIMO

Just as MIMO, aggregation, and quality of service management (see sections 1.1.2, 1.2.7, and 1.2.4 resp.) are the main features of the IEEE 802.11n standard [6] (see section 1.3.2), MU-MIMO is that of IEEE 802.11ac (see section 1.3.3). In 802.11n, features could either be classified as 'PHY-related' or 'MAC-related'. MU-MIMO though has strong PHY and MAC implications. The reader can refer to section 2.4 for a more detailed presentation of these implications.

### 6.1.1 PHY Implications of MU-MIMO

With MU-MIMO, an AP can simultaneously transmit independent groups of streams to multiple stations. It can thus make use of one channel access to transmit specific data to stations belonging to the same group. Antennas available at the AP can therefore be used to increase system efficiency. As mentioned in 5.1.1.2, there is no explicit indication on which precoding technique to use for MU-MIMO transmissions in IEEE 802.11ac. However CTI between the served stations should be minimized [7]. We have chosen the block diagonalization precoding technique [19] because of the good performance-complexity tradeoff it offers (see section 1.1.4.2.3). This scheme tries to cancel CTI through zero-forcing. It simplifies to channel inversion precoding for single-antenna stations (see section 1.1.4.2.1).

### 6.1.2 MAC Implications of MU-MIMO

Using MU-MIMO implies precoding each station's streams, and thus, having precise knowledge of their channels (see section 1.1.4). This is done through explicit channel feedback. Special frames, which are overhead with regards to data, are exchanged so as to regularly return a feedback of the channel state information (CSI) to the AP. Another implication is that groups of stations are to be defined by the AP. It is based on the resulting group identifier that stations will retrieve their data from received MU-MIMO frames (see section 1.4.2). Finally, the acknowledgment procedure is also adapted to enable destination stations to acknowledge their received frames.

### 6.1.3 Channel Sounding and Feedback in IEEE 802.11ac

Let us take a closer look at the channel sounding and feedback protocol in IEEE 802.11ac. In the IEEE 802.11n standard, the multiplicity of options for the sounding protocol has made things difficult for interoperability when using beamforming techniques [41] (see section 1.3.2.2). Consequently, 802.11ac uses a unique protocol based on the use of a null data packet for channel sounding and compressed beamforming matrices for feedback (see section 2.3.3).

The duration of the channel sounding procedure depends on the parameters given in Table 2.1 of section 2.3.3. Clearly the main parameters are the number of beamformees and the number of spatial streams. Another overhead-adding parameter is the channel sounding interval. As explicated in [41], MU-MIMO is much more sensitive to feedback errors and aging than classical single-user MIMO (SU-MIMO) beamforming (i.e. using singular value decomposition – SVD – see section 5.1.1.1 for more information on SU-MIMO with beamforming). This implies that channel sounding has to be done more frequently than for SU-MIMO. There is no indication, in the standard, on which interval to take but studies show that it will be smaller than SU-MIMO's 100 ms [85].

## 6.2 Simulation Scenarios and Parameters

### 6.2.1 Simulation Parameters

The very high throughput task group (TGac) channel model [75], IEEE 802.11ac PHY, and IEEE 802.11ac MAC simulation parameters common to all scenarios are given in Table 6.1, as defined in the 802.11ac standard [7]. Simulations are undertaken using three levels of simulation granularity, i.e. a MAC statistical analysis (see Appendix EAppendix C for details), MAC-centric simulations (network simulator 2 [59]), and PHY+MAC simulations (see Chapter 4). This way the performance of multiple-user (MU) transmissions expected by each level of granularity can be compared. Single-user (SU) transmissions are taken as reference. We shall note that simulation parameters that do not belong to the first two levels (i.e. MAC statistical analysis and MAC-centric simulations) are tagged as '[1]' and '[2]' (see note under Table 6.1). In addition, precisions given in section 5.3 still apply.

**Table 6.1 Common simulation parameters**

| | | | |
|---|---|---|---|
| **Global simulation parameters** | | | |
| Application rate / duration | | 80 Mbps / 2 seconds | |
| Transport layer protocol | | User datagram protocol (UDP) | |
| [1]AP-station distance | | 5 meters | |
| **MAC layer parameters** | | | |
| Access category | | Best effort | |
| MAC service data unit size | | 1500 octets (typical payload format) | |
| Max. allowed transmit opportunity (TxOP) | | 3008 μs | |
| Max. aggregation size | | 17 for MU and 18 for SU | |
| Queue management | | One access category queue per station | |
| Activated modulation and coding schemes (MCSs) | | MCS 0 to 8 for 1 spatial stream (see Table D.1) | |
| Sounding and feedback | | Sounding frequency | 10/20/30/40 ms for MU, 100/140 ms for SU |
| | | Subcarrier grouping | None |
| | | Ψ angle quantization | 2 bits (SU) and 5 bits (MU) |
| | | Φ angle quantization | 4 bits (SU) and 7 bits (MU) |
| [1]Rate adaptation algorithm | Adaptive multi-rate retry algorithm | Initial rates | $r_0 = r_1 = r_2 = r_3 = 6.5\ Mbps$ |
| | | Counts | $c_0 = 3, c_1 = 3, c_2 = 1, c_3 = 3$ |
| **PHY layer parameters** | | | |
| Number of spatial streams | | 1 | |
| Guard interval | | Long | |
| [1]Transmit power / system loss / antenna gain | | 17 dBm / 8.5 dB / 0 dB | |
| [1]Additive white Gaussian noise (AWGN) level | | 7 dB | |
| [1]Channel coding | | Binary convolutional coding | |
| [1,2]Channel estimation | | Done once at the beginning of the received frame. Affected by AWGN | |
| **TGac channel parameters** | | | |
| Bandwidth | | 20 MHz | |
| [1]Channel model | | B (residential) | |
| [1]Central carrier frequency | | 5.2 GHz (channel n° 40) | |
| [1,2]Channel seeds | | 19 (and 11 for second group) | |
| [1,2]Client identification index | | 1 and 2 (and 3) for station 1, 2 (and 3) respectively | |

[1] and [2] indicate parameters that are not used by the statistical analysis and the MAC-centric analysis (resp.)

### 6.2.2 Studied Scenarios

In this chapter, the impact that the channel sounding interval may have on MU-MIMO is evaluated. To this end, the four scenarios presented in Table 6.2 are studied using each level of simulation granularity. Through these scenarios, we can evaluate how the system reacts to an increase in the number of stations and to an increase in the number of antennas.

**Table 6.2 Per scenario simulation parameters**

| Scenario | Number of stations | Number of antennas | |
|---|---|---|---|
| | | At AP | At each station |
| 1 | 1 group of 2 stations | 3 | 1 |
| 2 | 1 group of 3 stations | 3 | 1 |
| 3 | 2 groups of 2 stations each | 3 | 1 |
| 4 | 1 group of 2 stations | 4 | 2 |

## 6.3 Simulation Results and Discussion

### 6.3.1 Statistical Saturation Throughput Results

By taking average channel access durations and the highest available PHY data rate (i.e. 78 Mbps because 1 spatial stream), the maximum average saturation throughput can be obtained. The statistically obtained results for MU-MIMO and SU-MIMO with different channel sounding intervals are given in Table 6.3. We can see that channel sounding overhead has more impact on MU-MIMO than on SU-MIMO. Indeed, for MU-MIMO, the more frequent the channel sounding, the lower the obtained throughput. This is natural considering the implicated overhead. For SU-MIMO, the sounding overhead is rather small and the intervals are important (relative to MU-MIMO's). This explains why throughput results are roughly the same for all four scenarios when using SU-MIMO.

When comparing scenarios 1 and 4, which both have 2 stations, it might seem surprising, at first sight, that MU-MIMO in scenario 4 would offer lower throughput than scenario 1 despite the increase in antenna diversity. However, these results are obtained for one spatial stream and using the maximum available PHY data rate. Therefore the increase in sounding-related overhead explains the difference. There are indeed more channel estimates to return. Based on these results, we could recommend using only one of the two antennas (i.e. the one having the best channel). But these throughput results being statistically obtained, neither pathloss nor spatial diversity can be accounted for. This analysis and the resulting recommendation need to be revisited based on PHY and MAC simulations.

When comparing scenarios 1 and 2, the benefit of using MU-MIMO can be seen. Adding one more station in the group increases the throughput by almost 50%. However, sensitivity with regards to the channel sounding interval is increased. An 11 Mbps (or 6%) difference can be seen between a 10 ms interval and a 40 ms interval for scenario 2, whereas the difference is only of 5 Mbps (4%) for scenario 1. When the number of groups is increased as in scenario 3, the sounding interval has even greater relative difference in throughput. Thus the 10 Mbps increase in throughput between a 10 ms interval and a 40 ms interval corresponds to an 8% increase.

**Table 6.3  Statistically obtained maximum average saturation throughputs**

| Used scheme | Sounding interval (ms) | Total downlink saturation throughput (Mbps) | | | |
|---|---|---|---|---|---|
| | | Scenario 1 | Scenario 2 | Scenario 3 | Scenario 4 |
| MU-MIMO | 10 | 130.25 | 183.74 | 123.36 | 127.56 |
| | 20 | 133.70 | 190.85 | 130.25 | 132.35 |
| | 30 | 134.85 | 193.22 | 132.55 | 133.95 |
| | 40 | 135.42 | 194.41 | 133.70 | 134.75 |
| SU-MIMO | 100 | 71.17 | 70.98 | 70.79 | 71.09 |
| | 140 | 71.28 | 71.14 | 71.01 | 71.22 |

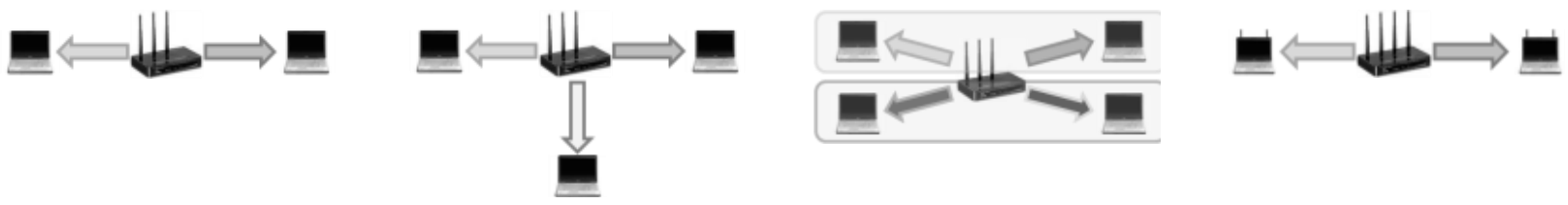

Therefore statistically obtained results show that the best solution for increasing throughput, when using MU-MIMO, would be to increase the number of stations per group. Increasing the overhead would be worth it considering the resulting gains. On the other hand, sounding related overhead should be minimized. Thus this analysis shows that a 40 ms sounding interval would offer best performance.

### 6.3.2 System Simulations Using Lookup Tables

Previous results give the maximum attainable average throughputs. However neither power related parameters (e.g. pathloss), nor channel access mechanisms (e.g. contention, deferral), nor errors, nor rate adaptation algorithms are considered in the analysis. All the latter are incorporated in MAC-centric simulations using lookup tables (LUTs, see section 1.5.2 for details). Finer system characterization is thus enabled. We shall note that the LUTs are computed offline using TGac channels and employing beamforming with SVD (see section 1.1.3.2).

The thus obtained total downlink throughput results for scenarios 1, 2, 3, and 4 are illustrated in Figure 6.1, Figure 6.2, Figure 6.3, and Figure 6.4 (resp.). In each of these figures, a warm-up phase of approximately 200 ms can be observed. During this initialization period, the rate adaptation algorithm chooses the 'best' data rate to use in order to transmit downlink frames towards each of the stations (see section 1.2.8 and Appendix B). Thus the following analysis does not account for this period.

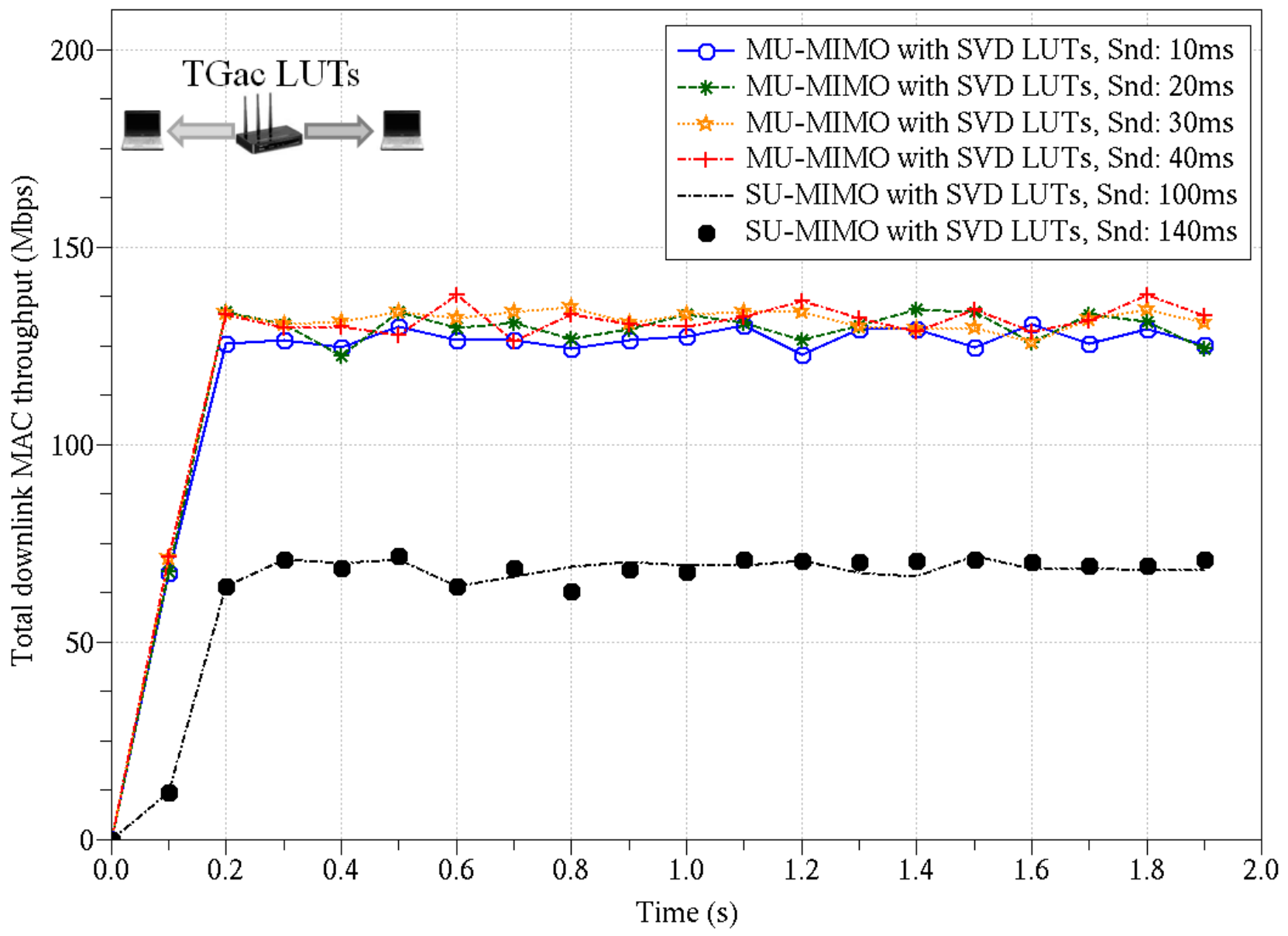

**Figure 6.1 Total downlink throughput of scenario 1 using a MAC-centric simulator**

We can see that the results obtained for all four scenarios are, on average, almost identical to the previously exposed statistical results (though slightly lower because of transmission errors and retransmission overhead).

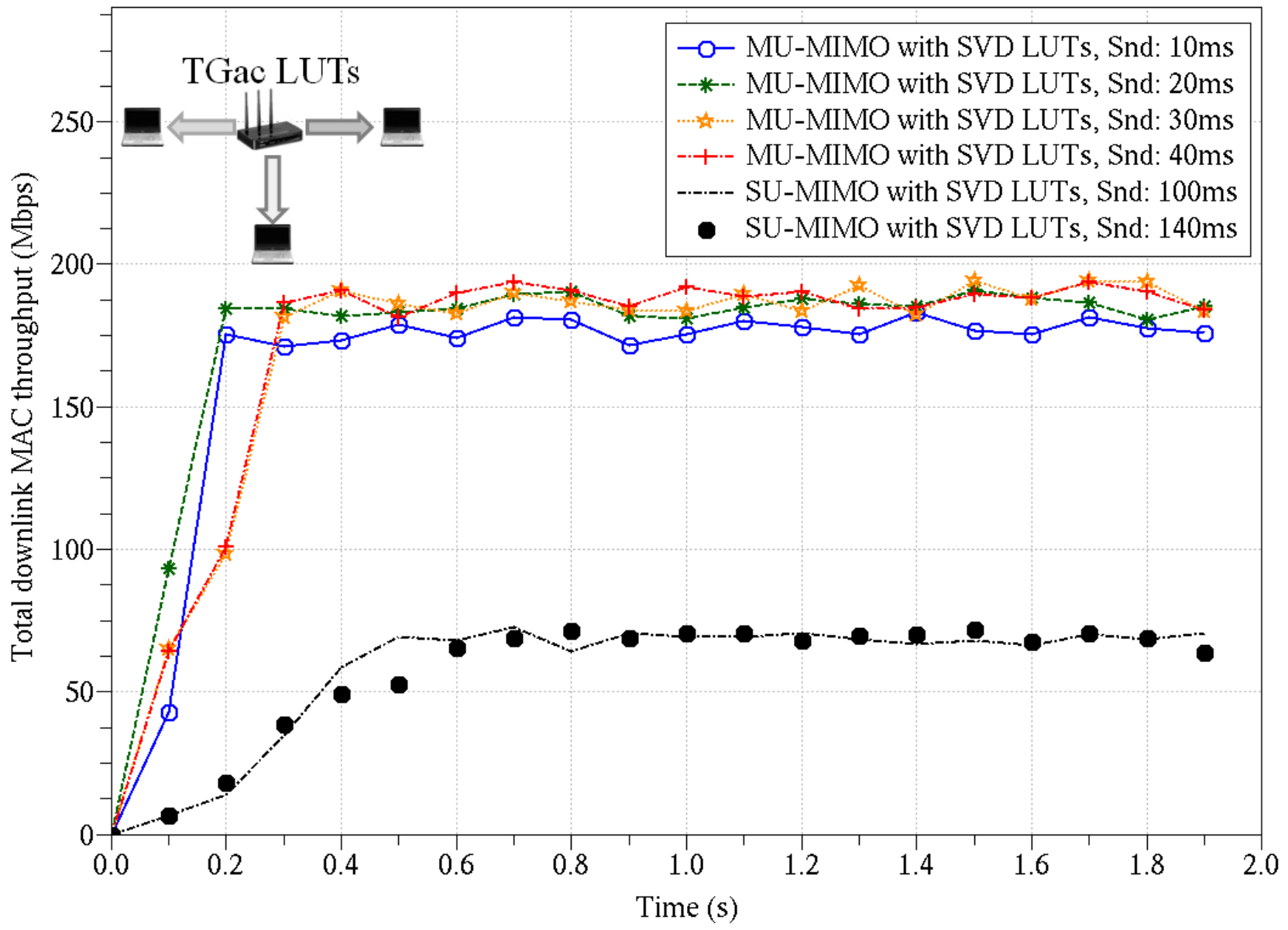

**Figure 6.2 Total downlink throughput of scenario 2 using a MAC-centric simulator**

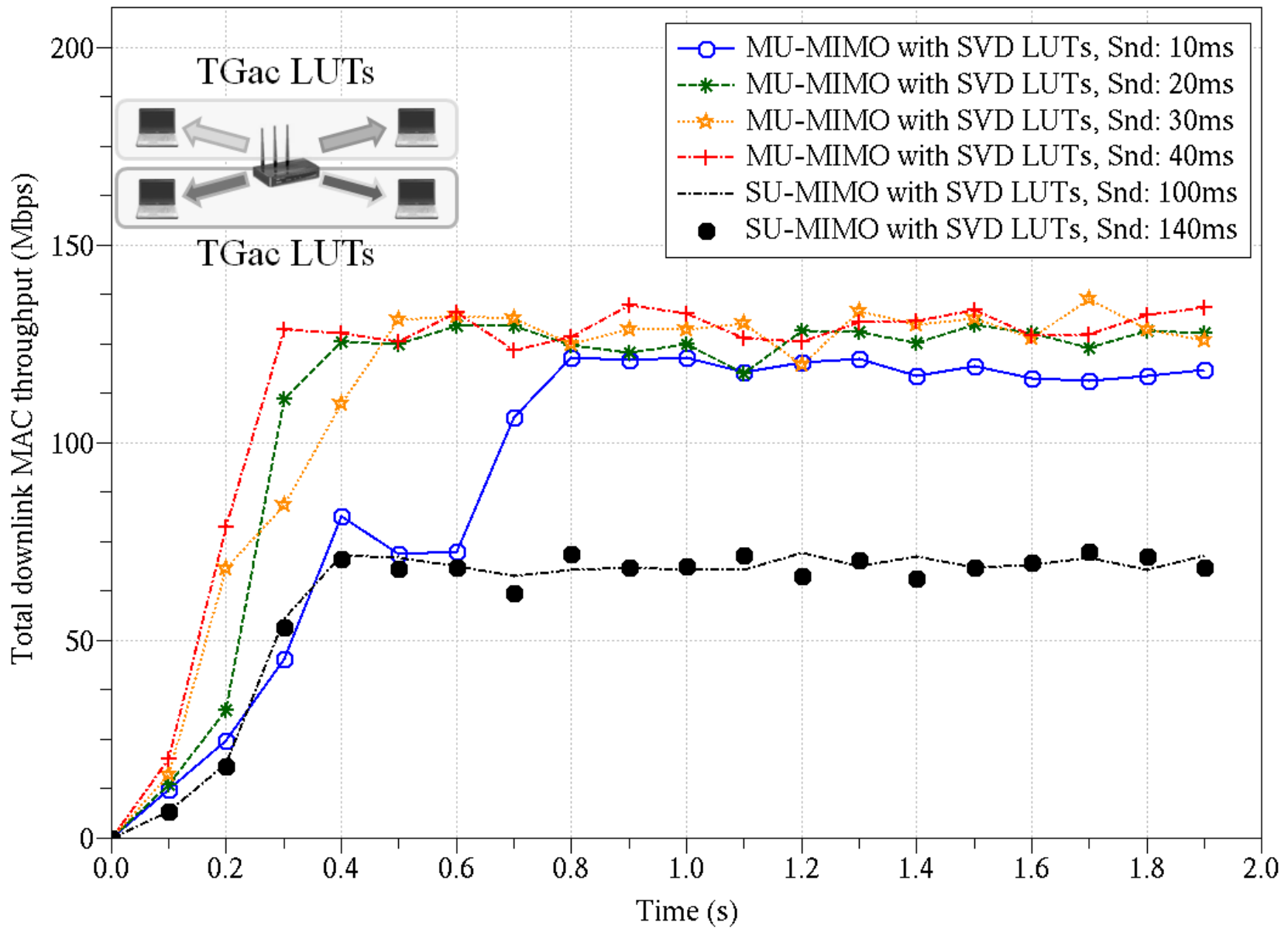

**Figure 6.3 Total downlink throughput of scenario 3 using a MAC-centric simulator**

However, in scenario 3 throughput curves take some time before stabilizing (e.g. 800 ms for a 10 ms sounding interval). This is due to channel access contention. Having chosen to take one access category queue per station (see section 5.3), packets destined to different groups internally contend for channel access as would packets belonging to different access categories (see section 1.2.4.1 for the contention process). It is the same contention phenomenon causing SU throughput curves, in scenarios 2 and 3, to stabilize after a longer period than in scenarios 1 and 4. Indeed in the latter case, there are fewer stations therefore less chances of drawing the same backoff. To avoid such contention, it would be beneficial to create only one group of 4 users and schedule only 3 out of 4 at each TxOP. The AP can thus perform scheduling over contention channel access.

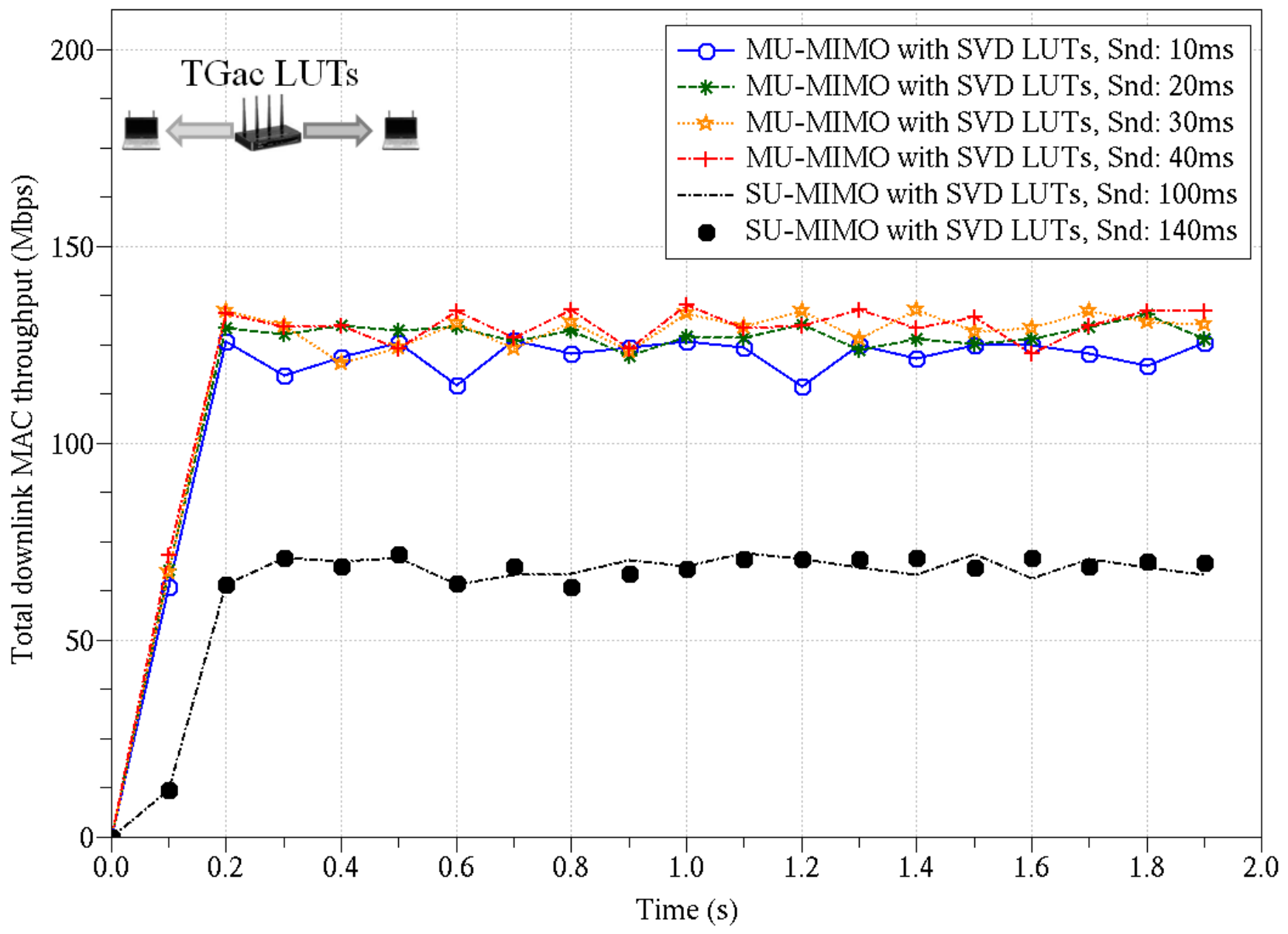

**Figure 6.4 Total downlink throughput of scenario 4 using a MAC-centric simulator**

Through the first two levels of simulation granularity, we come to the conclusion that the channel sounding interval should be increased to 40 ms (at least). Overhead can thus be reduced and performance increased. Indeed if more stations are considered, this overhead reduction can result in even greater throughput. However, such a conclusion has been drawn without considering channel estimate aging and CTI. Indeed, as mentioned in section 6.1, the effects of these two phenomena on MU-MIMO performance may be quite important.

### 6.3.3 Simulations Using the PHY+MAC Cross-layer Simulator

This leads to using the PHY+MAC simulator, presented in Chapter 4, to obtain realistic throughput results, i.e. jointly accounting for PHY and MAC parameters. The analyses underwent in Chapter 4 and Chapter 5 show the interest of using such a simulator.

#### 6.3.3.1 Scenario 1: Triple-antenna AP and 2 Single-antenna Stations

The total downlink throughputs for scenario 1 using the PHY+MAC simulator are given in Figure 6.5 (a). Although the throughputs for SU-MIMO roughly correspond to the previously obtained values, those of MU-MIMO are quite different. The throughput gap between MU-MIMO with 10 ms and 40 ms channel sounding interval is not only inversed but multiplied by more than 5. Indeed the average throughputs obtained with the PHY+MAC simulator for 10 ms and 40 ms sounding intervals are 110 Mbps and 83 Mbps (resp.), whereas those obtained with the MAC-centric simulator are 127 Mbps and 132 Mbps (resp.).

The stability of SU-MIMO curves can be partly explained by the difference in channel states at a given time. The AP can take advantage of this diversity. When using MU-MIMO, the AP transmits to both stations as long as there is data available for each, because they belong to the same group. Considering that saturated throughputs are evaluated here, the AP transmits to both stations every channel access, thus benefiting less from this diversity. The other, and main reason why SU-MIMO throughputs are so stable is that, considering that stations are only 5 meters away from the AP, channel conditions are good. However, for MU-MIMO, the correlation between these channels becomes an important factor. The TGac channel 19 used here is a channel with important correlation, thus leading to quite poor performance [87] (see Chapter 5).

The solid-line and cross-marker curves in Figure 6.5 (b) give the received signal powers' evolution over time for MU-MIMO and SU-MIMO (resp.) for both stations. The displayed power levels are normalized over the noise-free and deep fade-free received signal power. Firstly, the reader can notice that the use of channel inversion reduces received signal strength. Indeed, in channel inversion precoding, precoding vectors have to be chosen in the null space of other stations' channel matrices (see Appendix A), lowering the odds of choosing the power maximizing vector. The received interference levels for MU-MIMO (circular markers) are also given in Figure 6.5 (b) for both stations. These interference levels are determined using the long training fields that are precoded with the other station's beamforming vector (see 1.3.3.1.2) and are normalized over useful signal power.

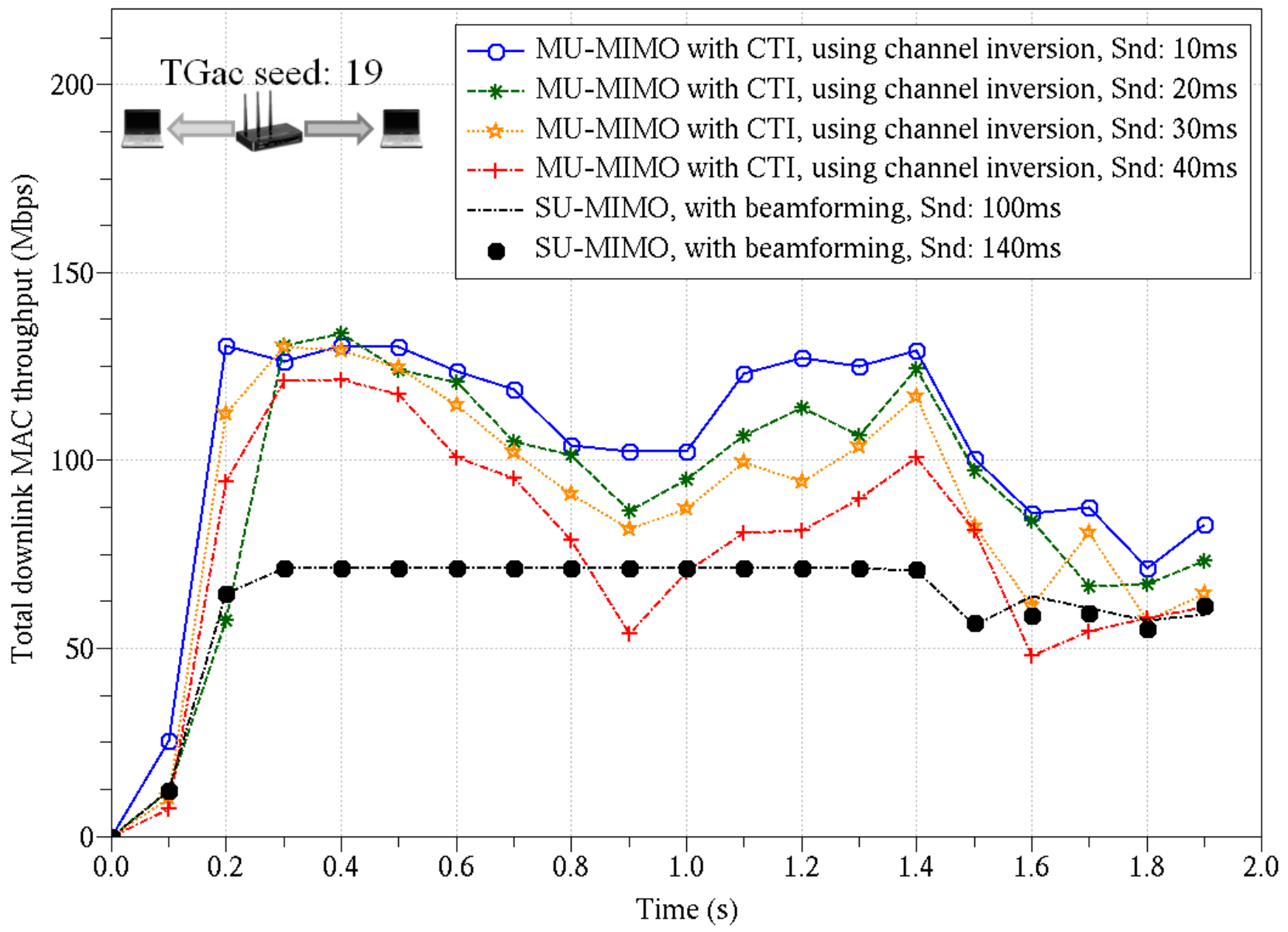

(a)

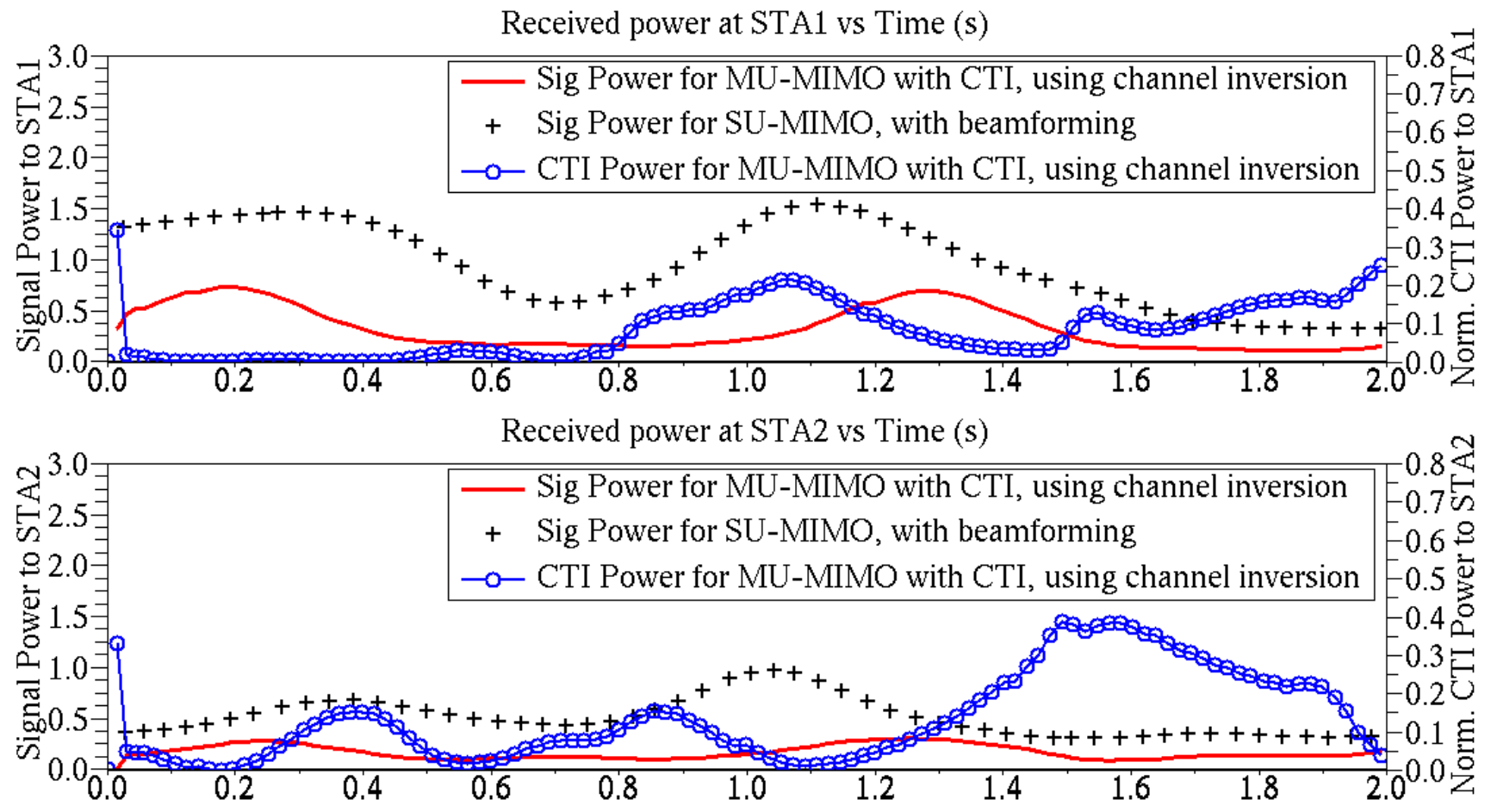

(b)

**Figure 6.5 Scenario 1 using the PHY+MAC simulator: total downlink throughput (a), and received useful signal power and normalized crosstalk interference power for both stations (STA1 and STA2) (b)**

By corresponding Figure 6.5 (a) and Figure 6.5 (b), it can be seen that the received power level has a direct influence on throughput. Indeed, estimation errors have greater impact when signal powers are low. During these low power phases (e.g. between 0.6 s and 0.8 s or between 1.6 s and 2 s), the CTI levels rise considerably. This leads the rate adaptation algorithm to decrease the PHY data rate. Thus, channel estimates have to be refreshed more frequently during these phases to minimize the impact that estimation errors and aging can have on performance. Through this dynamic dimensioning of the sounding interval, important gains could be observed despite an increase in overhead. In this particular scenario however using a static 10 ms sounding interval can enable important gains.

#### 6.3.3.2 Scenario 2: Triple-antenna AP and 3 Single-antenna Stations

In scenario 2, the AP having three antennas and stations supporting only one spatial stream, the former can support MU-MIMO transmissions towards a group of three such stations. Figure 6.6 (a) illustrates the total downlink throughputs for this scenario.

The analysis done in scenario 1 is still valid here. However the impact on performance is much worse in scenario 2. Contrary to MAC-centric results where a 40% increase in throughput is expected when moving to scenario 2, a 60% decrease can be observed. The MU scheme shows poorer performance than the SU scheme, especially when using a 40 ms sounding interval. This is mainly due to two phenomena. The first is that the IEEE 802.11 standards limit the maximum energy that a station can transmit. Accordingly, the AP has to satisfy this constraint when engaging MU-MIMO transmissions.

Therefore there is less available energy per station in scenario 2 than in scenario 1, leading to the use of more robust, but slower, PHY data rates. The second reason is that the AP has no longer the degree of freedom in the null space search it had in the previous scenario. Having three independent streams to transmit on three antennas complicates things for channel inversion precoding. The obtained precoding vectors are quite low. This can be seen in Figure 6.6 (b) through the useful signal power received by station 1. The interference power levels can thus be very high, with regards to those of scenario 1. Therefore it is again preferable to sound the channels more frequently. A static 10 ms interval is well suited for this particular scenario with correlated downlink channels.

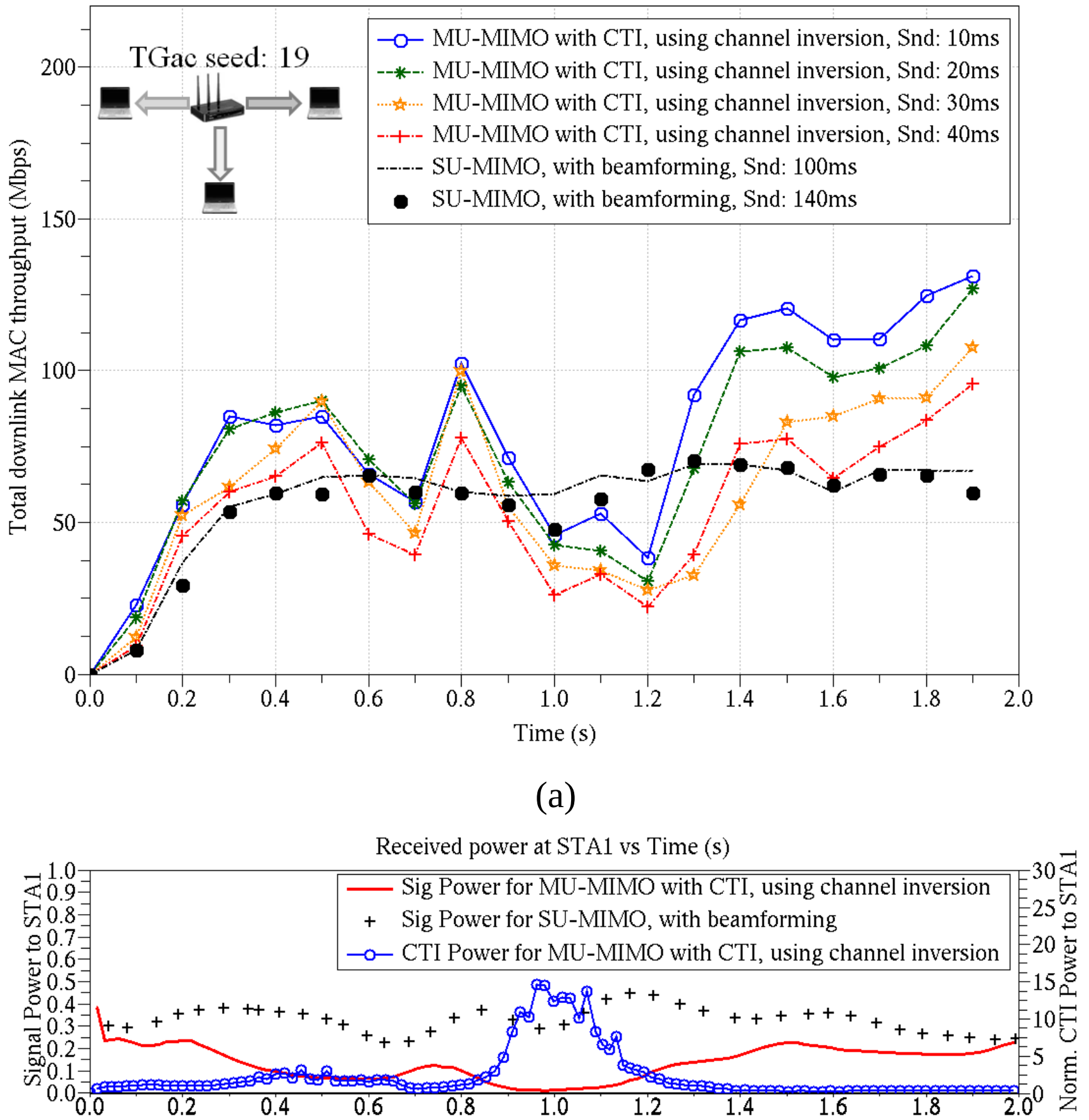

(a)

(b)

**Figure 6.6 Scenario 2 using the PHY+MAC simulator: total downlink throughput (a), and received useful signal power and normalized crosstalk interference power for station 1 (STA1) (b)**

### 6.3.3.3 Scenario 3: Triple-antenna AP and Two Groups of 2 Single-antenna Stations Each

Total downlink throughput results for scenario 3, where two groups are considered, are illustrated in Figure 6.7. Again, the results obtained with the PHY+MAC simulations, taking into account channel aging and CTI, are quite different from results obtained with the MAC-centric simulator. MU transmissions shows best overall performance when a sounding interval of 10 ms is used. If optimal performance is desired, a dynamic switching between a 10 ms and a 20 ms interval, based on the current channel states and received signal-to-noise ratios, can be envisioned. When the sounding interval is less frequent (i.e. for 30 ms and 40 ms), scenario 1's performance is achieved. Indeed with less sounding overhead, having two groups of stations leads to MU diversity. The latter thus compensate for the doubling in sounding overhead between scenario 1 and scenario 3. We shall note that MU and SU throughput curves are not stable because of contention (see section 6.3.2).

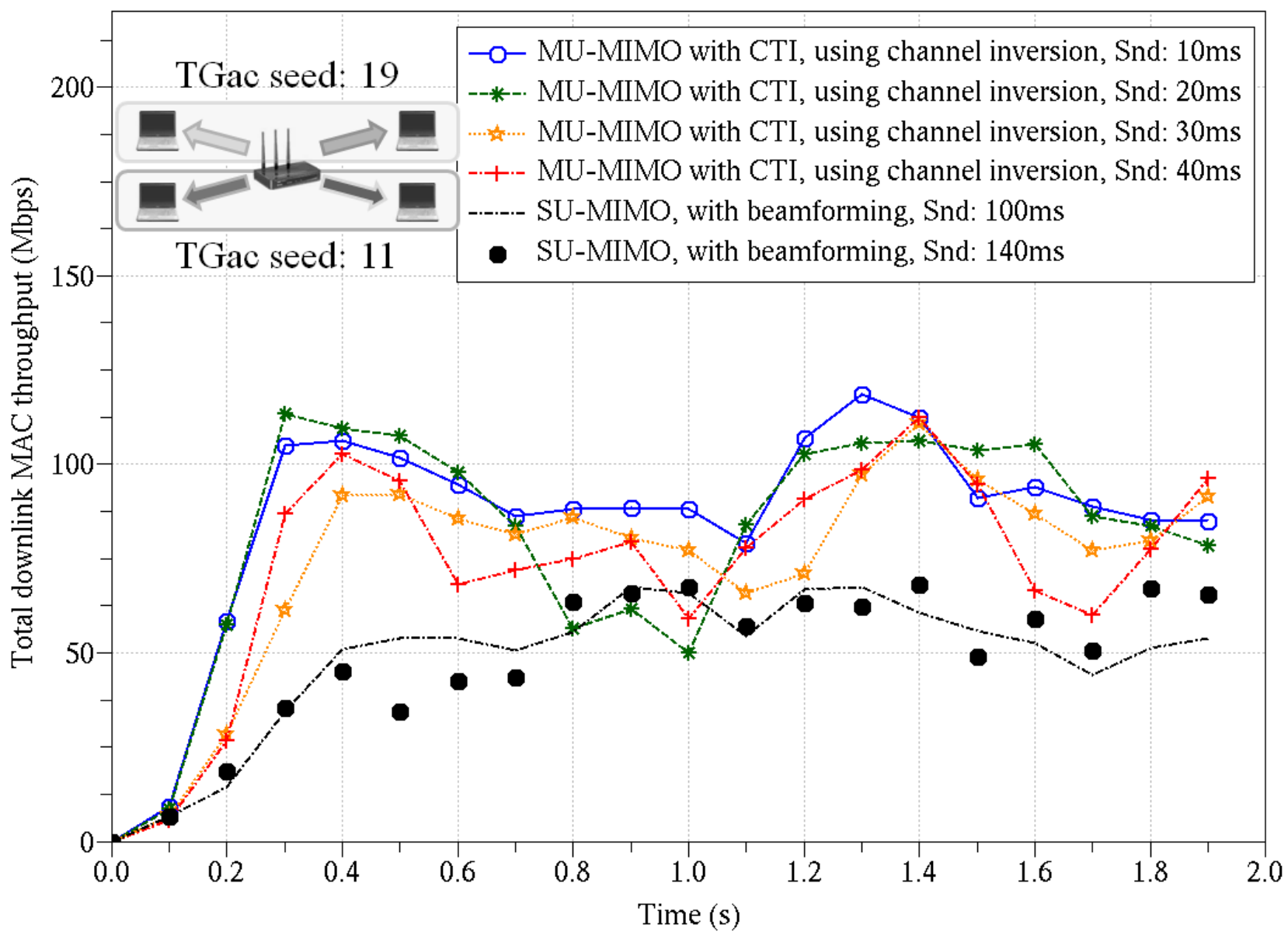

**Figure 6.7 Total downlink throughput of scenario 3 using the PHY+MAC simulator**

### 6.3.3.4 Scenario 4: Quadruple-antenna AP and 2 Double-antenna Stations

In scenario 4 the effect of increasing antennas, both at AP and stations, is studied. Figure 6.8 illustrates the total downlink throughputs towards the two available stations. The reader shall note that the number of spatial streams is still set to 1. The stations can thus make profit of spatial diversity to improve their respective receptions. This explains the overall good performance. Results obtained here are similar to those obtained with the MAC-centric simulator. In addition, there is little throughput difference between sounding intervals. The AP could therefore dynamically dimension the channel sounding interval based on the received signal-to-noise ratio that has been returned by the stations. Performance can thus be maintained while reducing channel occupation to accommodate for other transmissions. A simpler solution in this scenario would be to choose a 40 ms sounding interval.

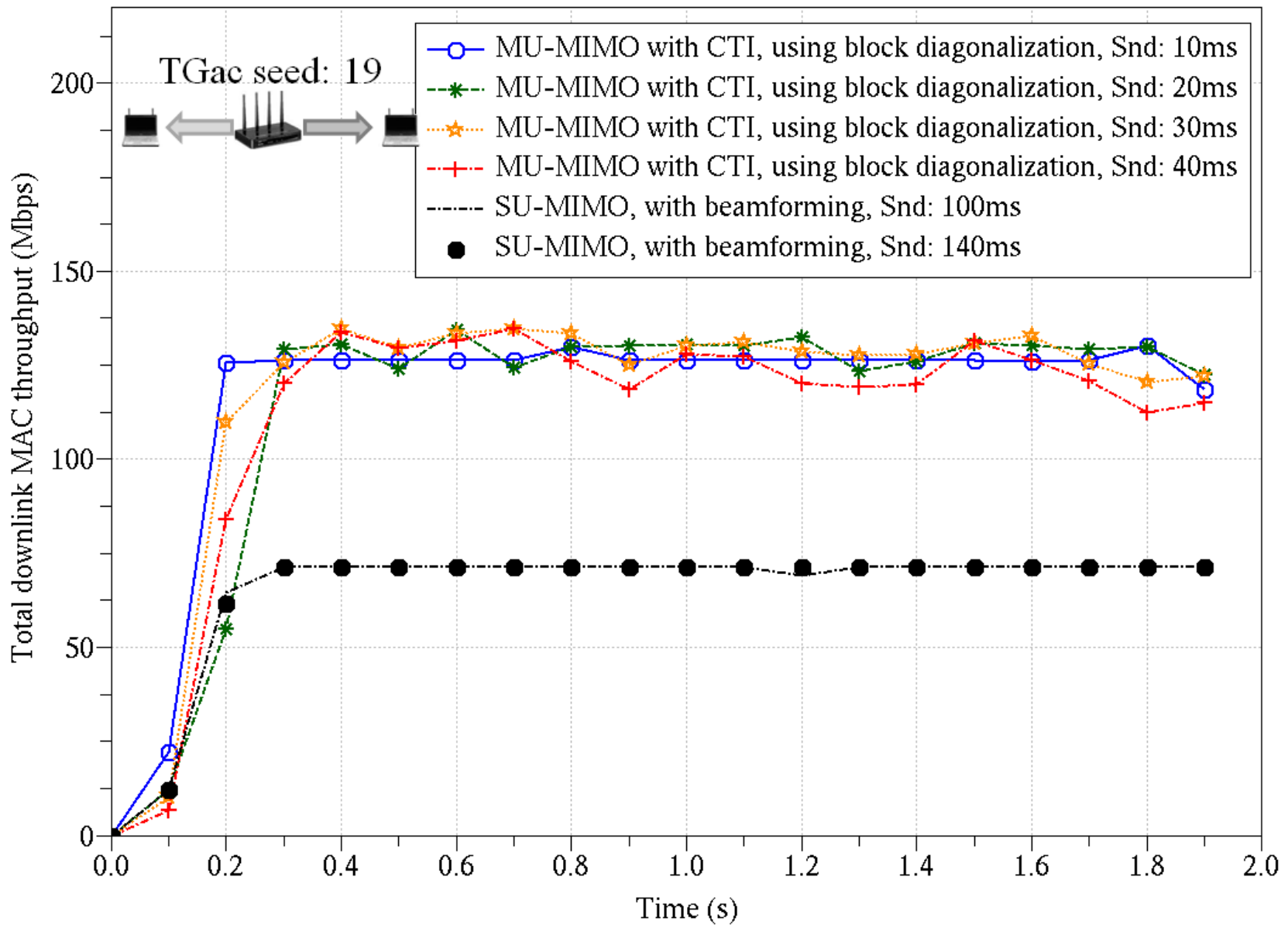

**Figure 6.8 Total downlink throughput of scenario 4 using the PHY+MAC simulator**

Therefore when using the PHY+MAC simulator to evaluate the impact of the sounding interval on MU-MIMO, the effect of channel estimate aging and CTI between stations can clearly be seen. In addition, total transmitted power constraints are also taken into account in the analysis. Through the four scenarios, it can be seen that frequent sounding enables best performance despite the induced overhead. However spatial diversity can compensate for degradations due to estimate aging. Longer sounding intervals can thus be used.

### 6.3.4 Summarizing Statistics

The throughput results obtained using the three simulators for all four scenarios are summarized through the statistics given in Figure 6.9. These statistics do not include the warm-up phase.

Statistically obtained saturation throughput results and those obtained with a MAC-centric simulator are almost identical. Indeed we do not consider many stations in the studied scenarios. Therefore channel access contention does not have much impact on results (though a disparity of throughput can be observed even in scenarios 2 and 3).

Results obtained with the PHY+MAC simulator show that statistical and MAC-centric results are optimistic, except when spatial diversity improves channel conditions in scenario 4.

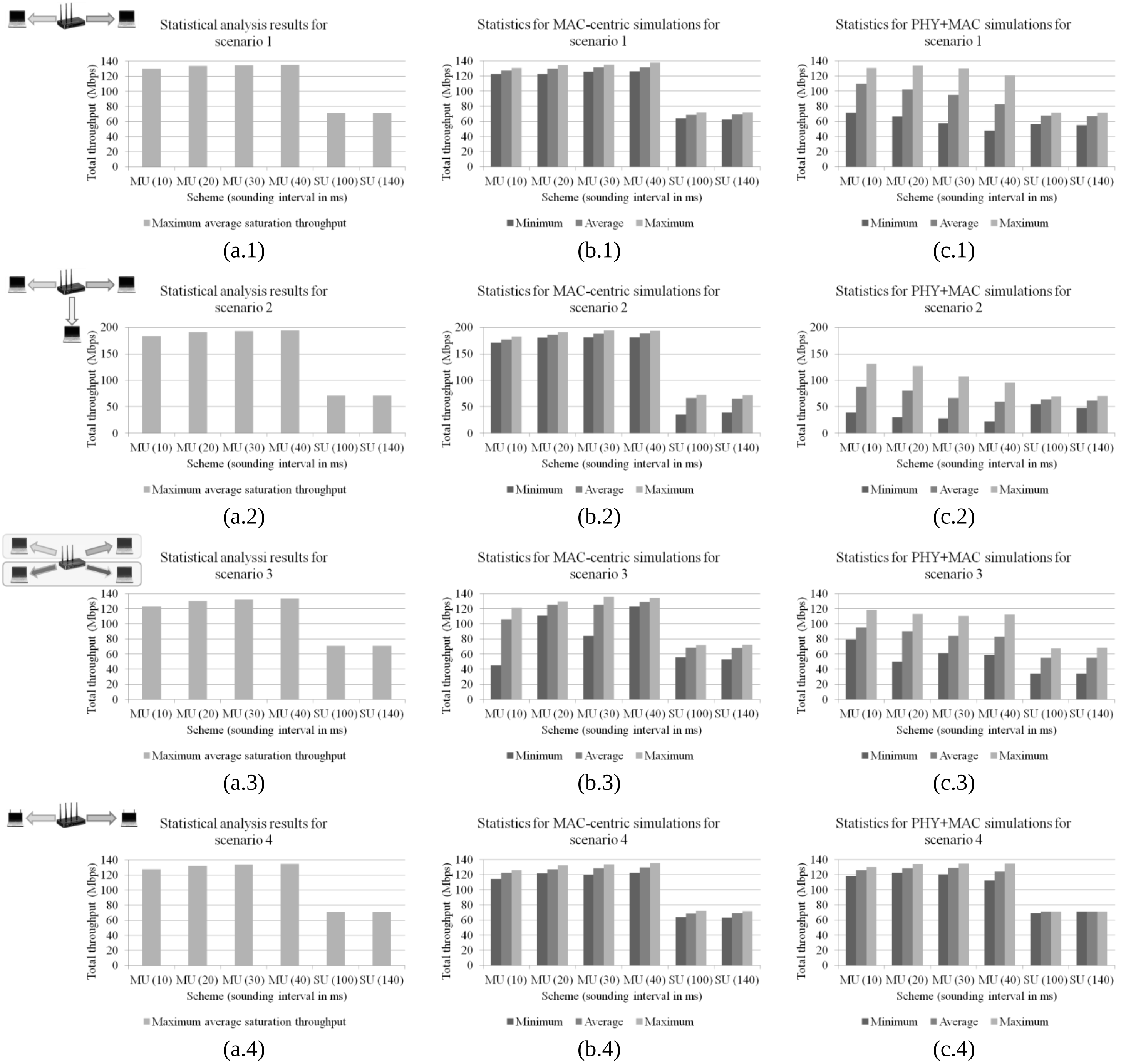

**Figure 6.9 Summary of simulation results obtained with the 3 levels of simulation (i.e. statistical MAC (a.#), MAC-only (b.#), and PHY+MAC (c.#)) for simulation 1 (x.1), simulation 2 (x.2), simulation 3 (x.3), and simulation 4 (x.4)**

## 6.4 Summary

In this chapter, MU-MIMO channel sounding interval is analyzed from a PHY and MAC layer perspective and results are compared to a MAC layer only perspective (be it statistical or simulated). Results obtained with a MAC layer only perspective show that a 40 ms sounding interval would enable best performance in the considered scenarios. Throughput is increased when sounding overhead is minimized. However, the aging of channel estimates and the CTI due to MU precoding should be accounted for, because MU-MIMO is much more sensitive to feedback errors than SU-MIMO. The joint PHY+MAC analysis of the four scenarios using quite correlated channels show that frequent sounding can lead to an increase in average throughput despite the associated overhead. In the considered scenarios, a 10 ms sounding interval enables best performance. Indeed, channel estimates at AP should be refreshed frequently, especially when received signal strength is relatively low, so as to enable precise knowledge of channel conditions. Nevertheless the interval can be increased when antenna diversity is used.

A dynamic dimensioning of the interval based on the current channel states can be considered either to increase throughput or reduce overhead. Indeed, when the AP has more stations to serve using MU-MIMO, the previous channel sounding interval may no longer be valid. In addition, when channel conditions change, the optimal sounding interval also changes. Thus a dynamic dimensioning can be very interesting with regards to channel condition changes and the increase in the number of stations.

Chapter 7

# Low Power Cross-layer Optimization for Future WLANs

After having focused on the IEEE 802.11ac standard [7] in the previous five chapters, we now turn to a current evolution of the 802.11ac standard. As exposed in section 1.3.4, the IEEE 802.11ah standard [44] extends IEEE 802.11 to below 1 GHz by using the ISM (industrial, scientific, and medical) bands for license-exempt wireless machine to machine communications. Applications such as smart grid, sensor networks, industrial process automation, and healthcare are to be addressed [48]. The main challenges are thus enhanced signal robustness, increased range, and increased number of associated stations. In such a context, overhead reduction becomes an important issue because of the great number of users exchanging small packets. That is why we propose to use an ultra short acknowledgment (ACK) frame [88]. Through this cross-layer technique, system performance can be much enhanced. More stations can be served while enhancing robustness of the acknowledgment procedure. The 802.11ah working group proposed a similar solution, the short ACK [89]. We shall note that the latter work was exposed after the ultra short ACK principle had been proposed.

We firstly provide more details on the physical (PHY) and medium access control (MAC) layer specificities of the 802.11ah standard. In this first section we focus on the short ACK solution. We then present the ultra short ACK solution and compare its performance with those of the classical ACK and the short ACK. To do so, we use the analytical MAC simulator used in Chapter 6 (and exposed in Appendix C), while adapting it to 802.11ah specifications.

# 7.1 IEEE 802.11ah

As mentioned above, the IEEE 802.11ah standard builds upon the IEEE 802.11ac standard. A lot of functionalities of the latter standard (see Chapter 2) are thus incorporated in the former. In this section, we expose the new features and specificities with regards to the 802.11ac standard. A detailed presentation of channelization, PHY layer, and MAC layer specificities, exposed in section 1.3.4, is done here.

## 7.1.1 Channelization

In IEEE 802.11, orthogonal frequency division multiplexing (OFDM) based specifications (see section 1.3) use 20 MHz-wide channels [4,38,6,7]. Wider channels (i.e. 40 MHz, 80 MHz, and 160 MHz) can be obtained by bonding such channels (see sections 1.3.2.1.1 and 1.3.3.1.1). In IEEE 802.11ah, the reference bandwidth becomes 1 MHz. Indeed basic service set (BSS) range and capacity in number of stations are more an issue than throughput. BSSs covering greater space, overlapping of collocated BSSs (see section 1.2.1.3) can be reduced by choosing channels with smaller bandwidth. However, just as with previous standards, greater throughput is obtained through channel bonding. Channels of 2 MHz, 4 MHz, 8 MHz, and 16 MHz are thus defined. The resulting channelization, as defined by American and European regulation authorities, is illustrated in Figure 7.1. Japan, South Korea, China, and Singapore have their own regulations [44].

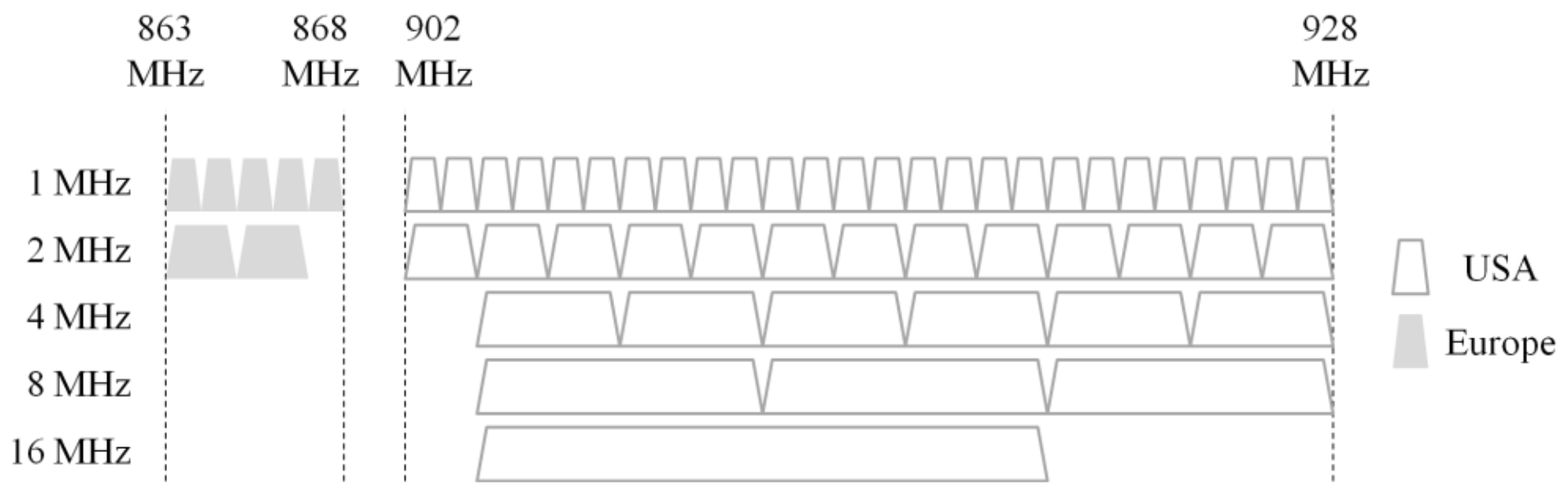

**Figure 7.1 802.11ah channelization in USA and Europe**

## 7.1.2 PHY Layer

The changes brought on the PHY layer concern mainly subcarrier design, for the physical medium dependent (PMD) sub-layer, and beamforming dependent PHY preamble, for the PHY layer convergence procedure (PLCP) sub-layer.

### 7.1.2.1 PMD Sub-layer

Transmissions using 2 MHz bandwidth use a total of 64 subcarriers. The IEEE 802.11n [6] 20 MHz subcarrier design (see section 1.3.2.1.1) is used for 2 MHz transmissions in IEEE 802.11ah. This implies a subcarrier spacing of 31.25 kHz. 4 MHz, 8 MHz, and 16 MHz transmissions use the same spacing. Such spacing can be obtained through a simple downclocking[1] of previous chipsets. IEEE 802.11ac's subcarrier designs for 40 MHz, 80 MHz, and 160 MHz (respectively) are thus used [44]. This correspondence with 802.11ac enables to maintain the same nomenclature for modulation and coding schemes (MCSs). 802.11ah's MCS tables for block convolutional coding are the same as the corresponding tables for 802.11ac before downclocking. Table D.1, Table D.2, Table D.3, and Table D.4 of Appendix D can be used for above 2 MHz 802.11ah transmissions. Tabulated data rates should however be divided by 10.

1 MHz transmissions also use the same subcarrier spacing. This leads to the definition of the new subcarrier design illustrated in Figure 7.2. 24 subcarriers, out of a total of 32 subcarriers, are used for data transmission. MCSs for 1 MHz (0 to 9) are limited to 1 spatial stream (SS). A special robust scheme, the MCS0 Rep2, is also defined. It consists in using MCS 0 after duplication of 12 coded symbols. We shall note that beamforming and multiple-user (MU) transmissions are supported (but optional) and that the feedback protocol is the same as 802.11ac's. In addition, the maximum number of space-time streams, across all users, is limited to 4. Therefore if no space-time block coding is used, up to 4 SSs can be transmitted.

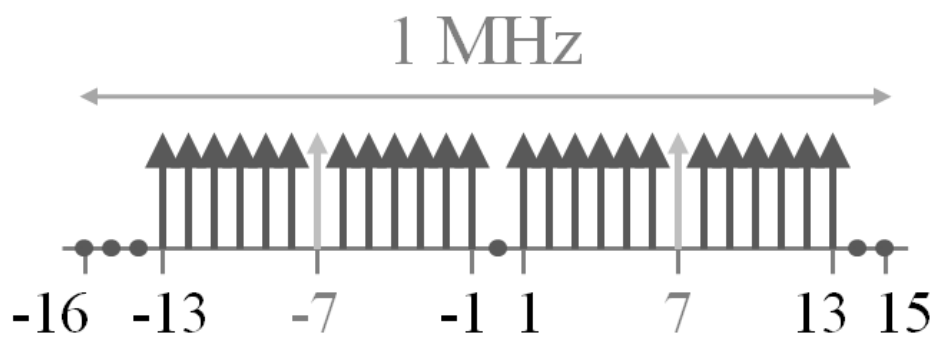

**Figure 7.2 1 MHz subcarrier design for IEEE 802.11ah**

[1] Having maintained the same number of subcarriers over a bandwidth which is 10 times smaller relative to IEEE 802.11ac, the IEEE 802.11ah standard uses a subcarrier spacing which is also 10 times smaller. This results in 10 times longer symbol durations. Information processing rate at PMD sub-layer is thus slower.

### 7.1.2.2 PLCP Sub-layer

Beamforming (single-user – SU – or MU) being optional two different PHY preambles are defined. The short PHY preamble, illustrated in Figure 7.3 (a), is used for SU transmissions without any precoding. The long PHY preamble, illustrated in Figure 7.3 (b), is used for SU beamforming and MU transmissions [44]. The latter preamble is composed of an 'omni' (or non-precoded, as in section 2.2.3) portion followed by a precoded portion. The 'omni' portion, containing the short training frame (STF), the first long training frame ($LTF_1$), and the first signaling field (SIG-A), ensures compatibility with non-beamforming stations, which also have an STF, $LTF_1$, and a signaling (SIG) field.

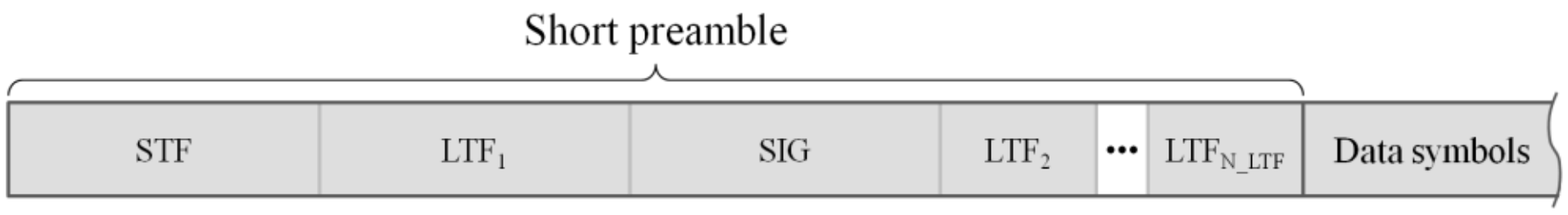

(a)

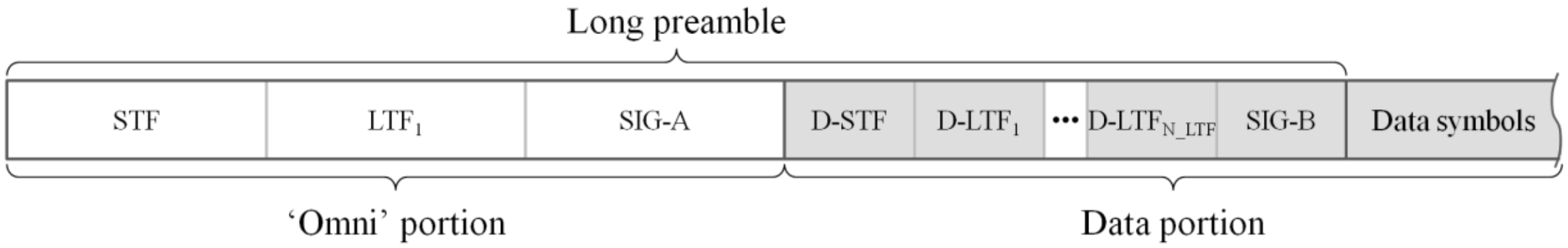

(b)

**Figure 7.3 PLCP framing in IEEE 802.11ah when using short PHY preamble (a) and long PHY preamble (b)**

An interesting cross-layer optimization is that the acknowledgment policy (ACK, block ACK, or no ACK) of the carried frame is notified in the signaling field. In older standards, this information resides in the quality of service control field of the MAC header (see section 1.3.2.2). The receiving station can determine how long the data-ACK exchange should last, without even having to decode the PHY service data unit (PSDU, see section 1.2.5) and extract the information from the MAC header.

## 7.1.3 MAC Layer

IEEE 802.11ah is to be used for applications involving a great number of stations (e.g. for outdoor applications the amendment should support more than 2007 associations [48]) with the latter often having stringent power consumption constraints. Accordingly the MAC layer is modified to accommodate for such constraints. Long power save periods, spreading of uplink transmissions to mitigate the hidden node problem (see section 1.2.1.4), and intelligent grouping are some of the proposed enhancements [44]. Other solutions involve overhead reduction of classical frames (see Appendix C for the impact of classical frames' overhead). These solutions are presented in sections 7.1.3.1, 7.1.3.2, and 7.1.3.3.

### 7.1.3.1 MAC Header Compression

The MAC header (see section 1.3.1.2) is shortened by using the association identifier (AID, see section 2.2.3) instead of the full MAC addresses of the receiver (Rx) and transmitter (Tx) in uplink and downlink transmissions (resp.). The corresponding optional MAC framing are illustrated in Figure 7.4 (a) and Figure 7.4 (b), respectively, for the latter.

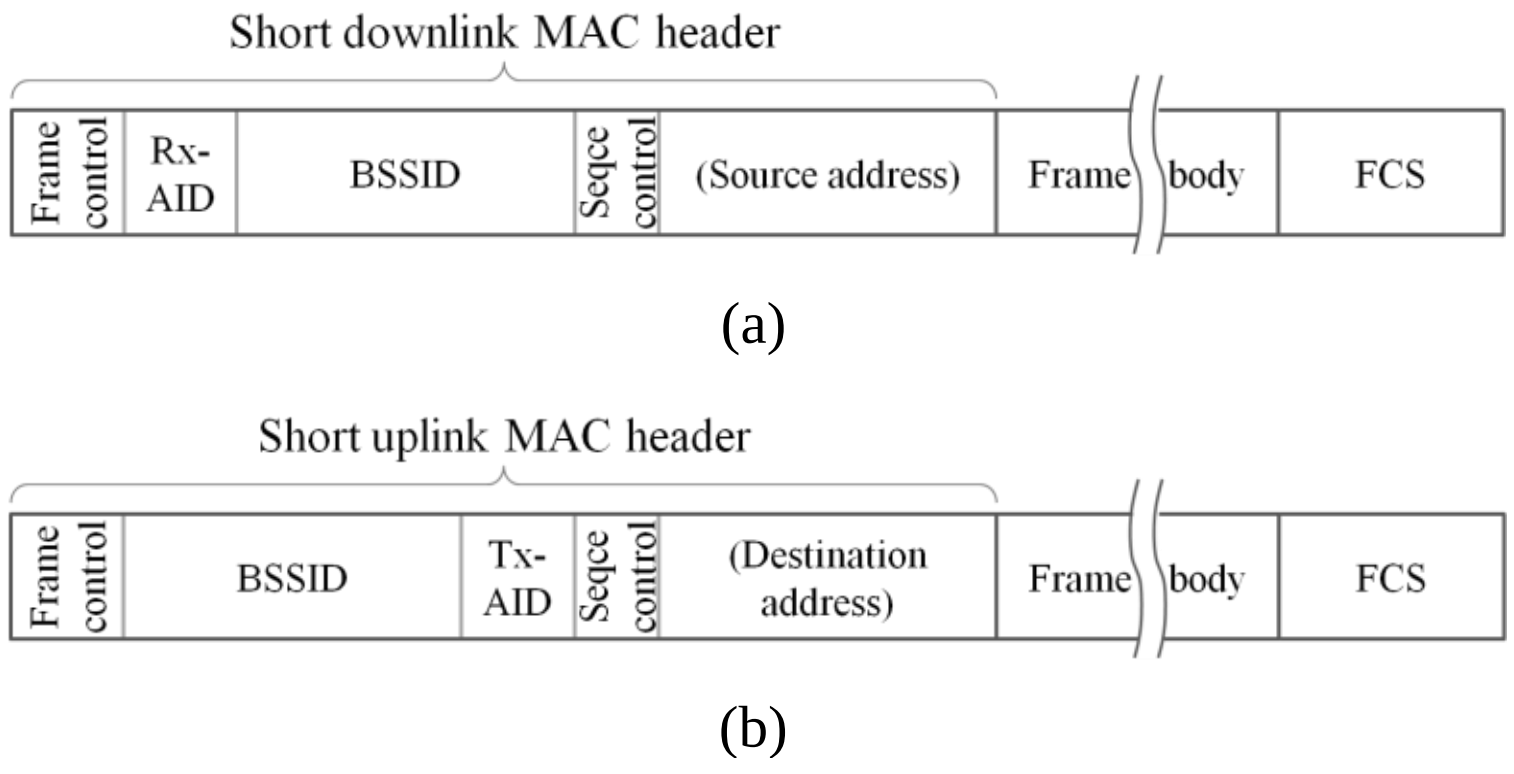

**Figure 7.4 MAC framing in IEEE 802.11ah when using short downlink MAC header (a) and short uplink MAC header (b)**

We shall recall that the BSS identifier (BSSID) is the MAC address of the access point (AP) and that it should be notified in all exchanged MAC frames when using infrastructure mode. The inclusion of source and destination addresses (i.e. the originator and recipient) of the frame is optional.

### 7.1.3.2 Short ACK

We underlined in section 7.1.2.2 that the inclusion of the acknowledgment policy is an important step towards cross-layer solutions. The IEEE 802.11ah standard exploits this optimization further by defining a short ACK frame format with cross-layer design [89].

#### 7.1.3.2.1 Frame Format

The increase in range and number of stations in IEEE 802.11ah implies using robust, thus low data rate, transmissions. In addition, applications such as smart grid, sensor networks, and industrial process automation result in the transmission of relatively small frames. Therefore the overhead induced by the acknowledgment procedure is quite important. For example, when transmitting a 100 octet MAC protocol data unit at 600 Kbps using 2 MHz bandwidth, the 480 µs-long ACK takes up almost 17 % of the total 2,828 µs transmit opportunity [89].

That is why a short ACK frame comprised of only an STF, an $LTF_1$, and a special SIG field has been proposed, as illustrated in Figure 7.5 (a). A reserved MCS value is used to indicate that the current frame is a short ACK and that, consequently, there is no PSDU. The SIG field also contains an ACK identifier (ACK ID) field. The latter is computed from the partial frame check sequence (FCS, see section 1.2.5) and the scrambling seed of the frame to acknowledge [44]. This way the short ACK can be recognized by the station expecting it. The 'normal' (or classical) ACK frame format is given in Figure 7.5 (b) for comparison. Frame type and receiver address are contained in the MAC header.

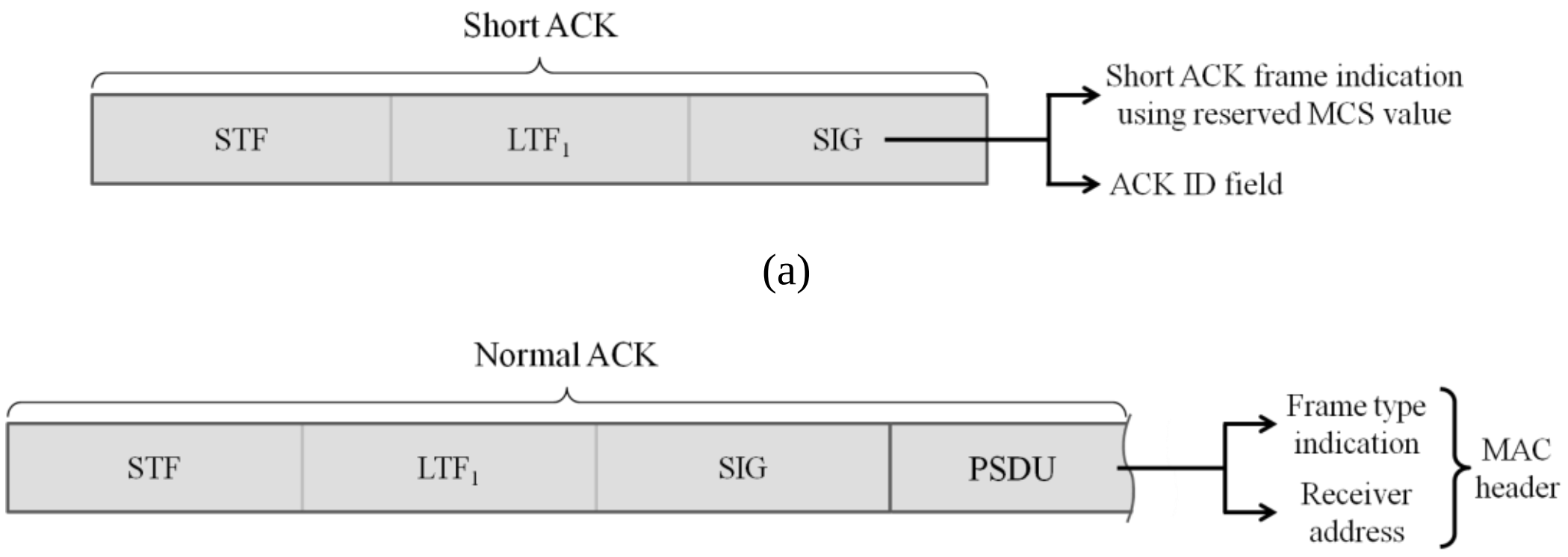

(b)

**Figure 7.5 Frame format for short ACK (a) and normal ACK (b)**

#### 7.1.3.2.2 False Positive Case

There is clearly a risk of confusing ACKs if no receiver identifier information is carried in the short ACK. The ACK ID has been incorporated in the SIG field for this reason. Indeed, even if acknowledgment is expected a short inter-frame spacing (SIFS, see section 1.3.1.2) after the end of the frame to acknowledge, different stations could expect a short ACK at the same moment. This can happen in overlapping BSSs (OBSSs) in a hidden node configuration as illustrated in Figure 7.6 (a) [89]. A short ACK from another BSS can be perceived as the response to a frame which has not been correctly received by the intended receiver, if no identification information is carried.

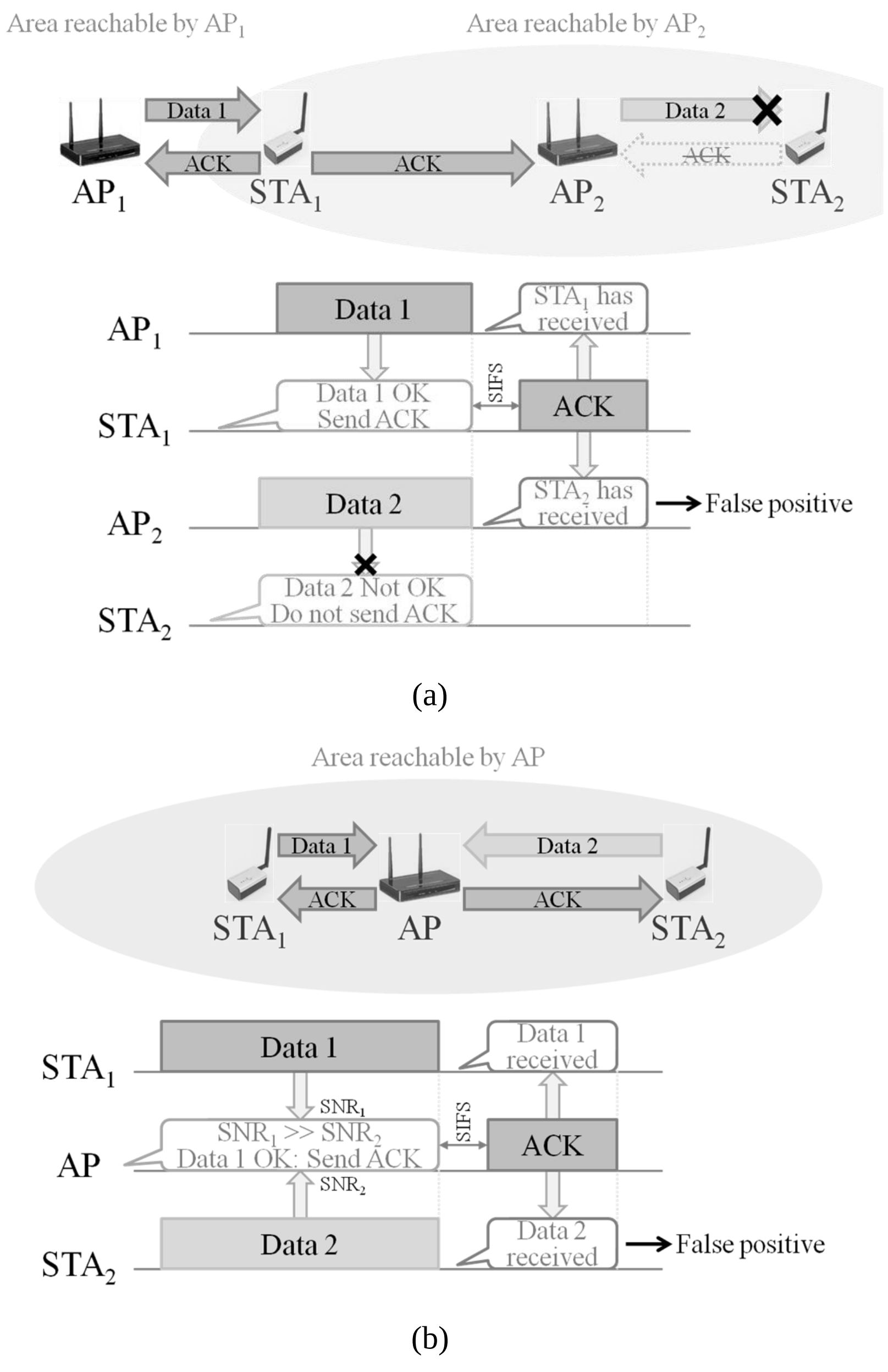

**Figure 7.6 False positive case with short ACK in an OBSS hidden node scenario (a) and an intra-BSS scenario (b)**

This false positive case can even occur in an intra-BSS scenario with stations in range of one another, as illustrated in Figure 7.6 (b). The two stations having drawn the same backoff (see section 1.2.3.1), carrier sense is eluded. If the AP manages to decode the frame with the highest signal-to-noise ratio (SNR), it acknowledges it. Both stations could thus consider their frames as correctly received.

We have to admit that such cases may seem rare, but considering the great number of users and the importance of some of the conveyed information (e.g. in healthcare), a false positive can have serious implications. Its occurrence should thus be minimized through identification of ACK frames.

#### 7.1.3.3 Other Short Frames

Beacon frames convey important information on the BSS. However they are quite long. Therefore a short beacon format has also been proposed [44]. Full beacon frames could be periodically transmitted though. Protection frames also benefit from IEEE 802.11ah's frame shortening. Clear to send (CTS, see section 1.2.3.4) frames can thus be shortened to the PHY preamble. A special SIG field is consequently used to notify the type of frame [44], just as with short ACK.

## 7.2 Ultra Short Acknowledgment for IEEE 802.11ah

As seen above, the short ACK concept is very interesting for reducing overhead. However, the concept can be pushed further[1]. An ultra short ACK can be used for greater system capacity but also for longer battery life [88]. This is obtained in the latter by shorter awake periods when power save mode is activated.

### 7.2.1 Ultra Short Acknowledgment Design

#### 7.2.1.1 Basic Design

When a station sends a frame with an immediate ACK policy, it expects an ACK frame SIFS seconds after having finished its transmission. In addition, start-of-frame detection is done through the STF sequence in IEEE 802.11 standards (see sections 1.1.1.5 and 1.3.1.1.2). Therefore sending the STF as a standalone frame can serve as acknowledgment. Indeed in other frames, the $LTF_1$ sequence and SIG field always follow. The receive chain having been synchronized and gain control done through the STF, following fields (especially $LTF_1$) are detected if present. Therefore, the ultra short ACK can be easily recognized. The principle of this ultra short ACK is illustrated in Figure 7.7.

[1] We shall note that the ultra short ACK concept was presented in the IEEE 802.11ah work group in Jan. 2012 [88], whereas the short ACK in Mar. 2012 [89]. Only the latter was accepted and thus incorporated.

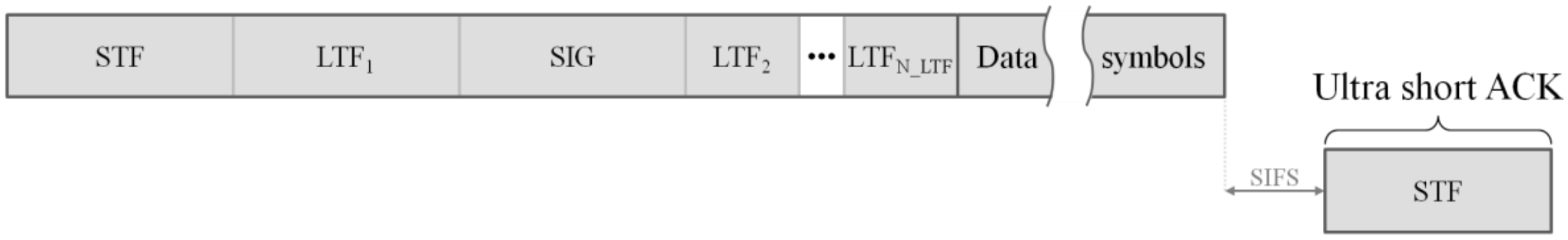

**Figure 7.7 Ultra short ACK principle**

Through this technique, acknowledgment overhead can be reduced to two OFDM symbol durations (i.e. 40 µs when using long guard interval) instead of five with the short ACK. Moreover, the ACK information is more robust than when using classical ACK or short ACK. With the classical ACK structure, the station having sent the data frame has to correctly decode the ACK frame's PSDU before establishing whether its frame has been correctly received or not. With the short ACK structure, the ACK-expecting station has to decode the SIG field before deciding that an ACK frame has indeed been received. In both cases, the decision depends on correct channel estimation and demodulation process. When using the ultra short ACK, the decision depends on the detection of a standalone STF sequence. Because of its implications on the rest of the frame, this process is very robust. Therefore the ultra short ACK procedure is much more robust to noise and interference. However, with this basic design of the ultra short ACK frame, false alarms (as described in 7.1.3.2.2) cannot be mitigated if two (or more) ACKs are expected within the same period.

### 7.2.1.2 Robust Design

The impediments of the basic ultra short ACK design can be minimized by using the time-reversal technique [90] as shown in the example illustrated in Figure 7.8.

Whenever a frame goes through a channel, the latter acts as a filter and the delay profile of the signal is modified. This profile is specific to the location of the transmitting and receiving stations (i.e. AP and station, resp., in the illustrated example). The STF sequence contained in the frame to be acknowledged can be used to extract the response of the channel. If the obtained channel response is reversed in time and used to filter the ultra short ACK, this ACK frame is focalized (temporally and spatially) at the initial transmitter (i.e. the AP in the example). Upon receiving a relatively focalized STF, the initial transmitter establishes that its frame has been acknowledged.

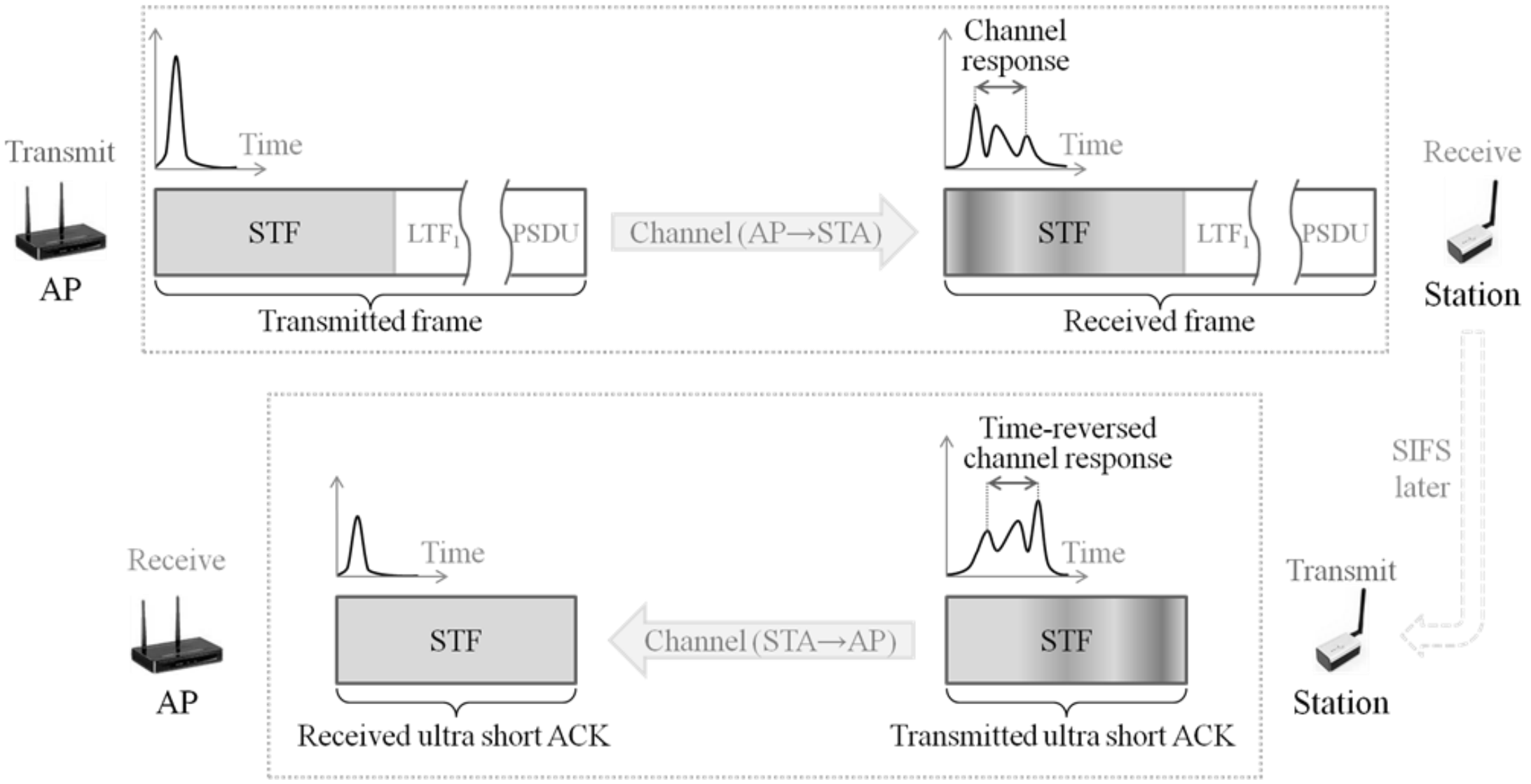

**Figure 7.8 Ultra short ACK principle with time-reversal**

This design for the ultra short ACK is robust to false alarms. Figure 7.9 (a) and Figure 7.9 (b) illustrate data transmission and robust ultra short ACK transmission (resp.) in the intra-BSS scenario presented previously.

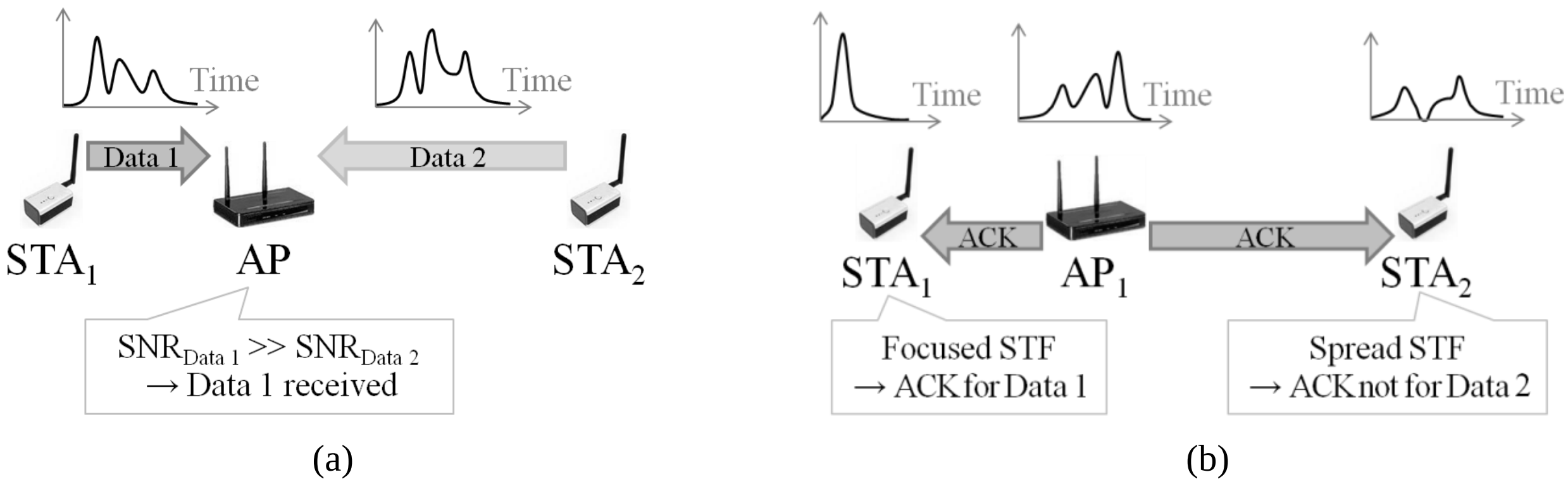

**Figure 7.9 Data-ACK exchange with the robust time-reversal version of the ultra short ACK in an intra-BSS collision scenario. Data transmission (a) and ACK transmission (b)**

The AP locks on $STA_1$'s frame and correctly decodes it. It then uses the time-reversed version of the STF sequence received from $STA_1$ to acknowledge Data 1. The channel being reciprocal during the data+ACK exchange, $STA_1$ receives a focused STF whereas $STA_2$ receives a spread one. The former can thus assert that its frame has been acknowledged whereas the latter engages a retransmission procedure. We shall add that the decision window (enabling to determine whether the received STF is focused or not) shall be long enough to absorb differences between uplink and downlink channels, but short enough to enable good discrimination.

Therefore, ultra short ACK with time-reversal should mitigate false alarms, just as short ACK with ACK ID, while inducing less overhead.

### 7.2.2 Simulation Scenarios and Parameters

The PHY layer, MAC layer, and application layer simulation parameters of the considered four scenarios are given in Table 7.1. Parameters common to these scenarios are given in Table 7.2. The impact of the ACK procedure on sensor network and industrial process automation applications is tested. Hence the classical ACK with a compressed MAC header, the short ACK, and the ultra short ACK are compared in each scenario.

**Table 7.1 Per scenario simulation parameters**

| Scenario | Bandwidth | Application | MSDU size | Refreshment cycle |
|---|---|---|---|---|
| 1.1 | 1 MHz | Industrial process automation | 64 octets | Every 5 s |
| 1.2 | 2 MHz | | | |
| 2.1 | 1 MHz | Sensor networks | 256 octets | Every 35 s |
| 2.2 | 2 MHz | | | |

**Table 7.2 Common simulation parameters**

| MAC layer parameters | |
|---|---|
| Access category | Best effort |
| MAC header type | Compressed |
| Aggregation | None |
| **PHY layer parameters** | |
| Number of spatial streams | 1 |
| Preamble type | Long |
| Beacon type | Short beacon (every 100 ms, with long beacon every 10 s) |
| Guard interval | Long |

The reference refreshment cycles of sensor networks and industrial process automation are 10 to 60 s and 0.1 to 10 s (resp.) [48]. We have thus taken average values in simulations. The MAC service data unit (MSDU) sizes correspond to typical MAC payload for the concerned applications. We shall note that we have taken 1 MHz and 2 MHz bandwidth so as to evaluate performance in a configuration where BSS overlapping is reduced (see section 7.1.1). Furthermore, IEEE 802.11ac's inter-frame spacing and contention window sizes (see sections 1.3.1.2 and 1.3.2.2) are still applicable, as no change is specified in [44].

### 7.2.3 Simulation Results and Discussion

#### 7.2.3.1 Industrial Process Automation Application

The duration of a data+ACK exchange for 64 octet MSDUs using 1 MHz bandwidth (scenario 1.1) and 2 MHz bandwidth (scenario 1.2) for the available MCSs are given in Figure 7.10 and Figure 7.11 (resp.). In Figure 7.10, MCS0 Rep2 (see section 7.1.2.1) is represented with a '-1' index. We take the same notation for following 1 MHz curves.

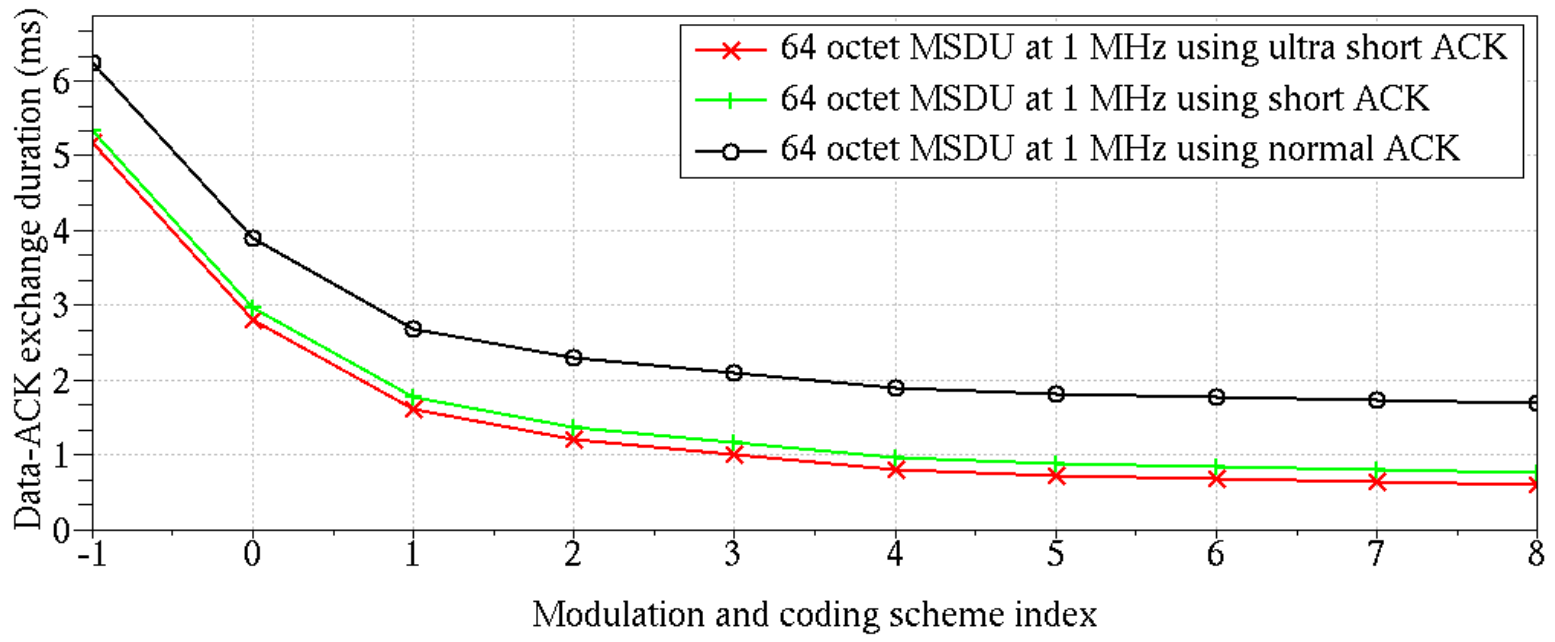

**Figure 7.10 Data+ACK exchange duration as a function of MCS index for the three ACK procedures using scenario 1.1**

We can see in scenario 1.1 that the short ACK procedure enables to reduce the data+ACK exchange duration by 15% using the lowest MCS and 54% using the highest MCS, with regards to the normal ACK. Using ultra short ACK enables to reduce an additional 3% for the lowest MCS and 21% for the highest, with regards to the short ACK. Considering this small gain in duration, ultra short ACK does not seem worth using in this scenario. Removing the PSDU of the ACK frame is sufficient to have interesting performance in overhead reduction.

In scenario 1.2 however, the ultra short ACK offers 10% reduction for the lowest MCS, with regards to the short ACK, making it more interesting.

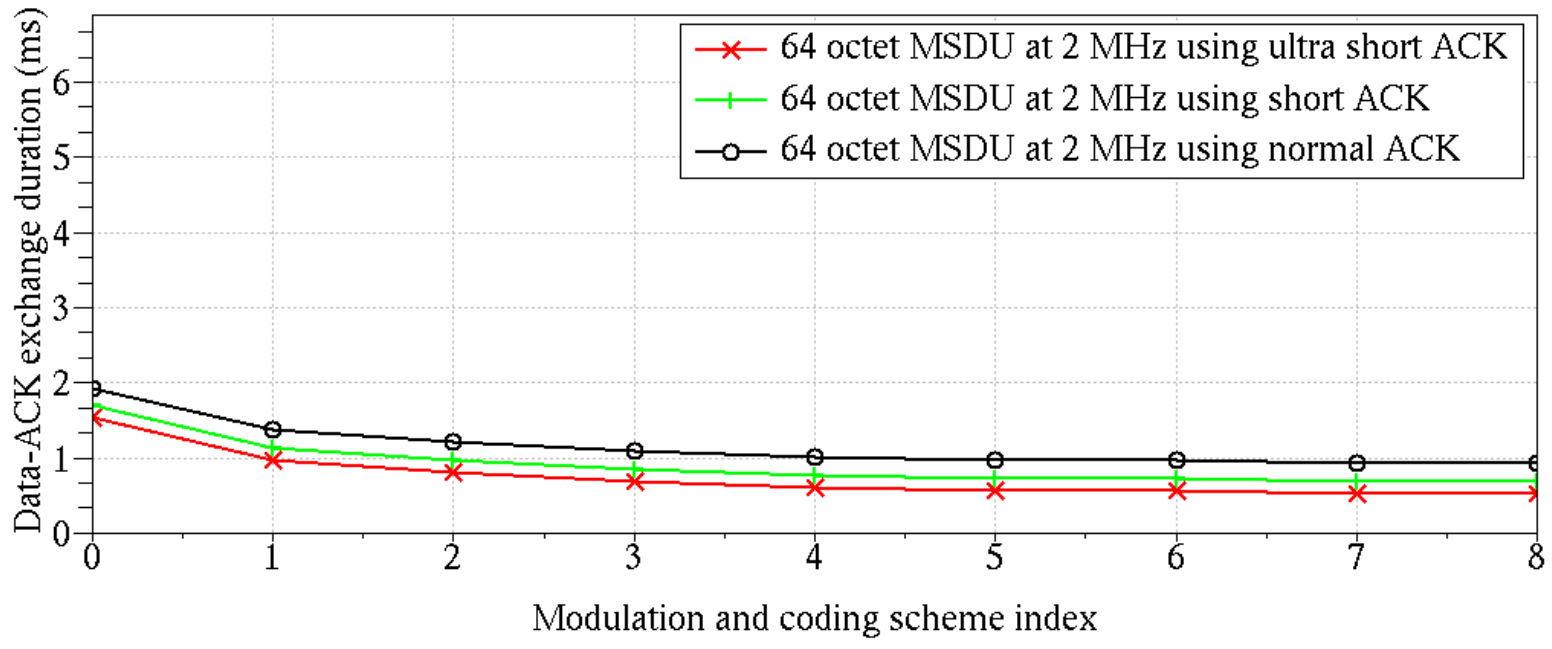

**Figure 7.11 Data+ACK exchange duration as a function of MCS index for the three ACK procedures using scenario 1.2**

Another way of evaluating performance is to analyze the maximum number of active stations (i.e. contending for the medium) per second. System capacity is indeed an important issue in IEEE 802.11ah. Figure 7.12 and Figure 7.13 give this number for scenario 1.1 and 1.2 (resp.). The aim of these curves is to give optimal performance. We assume that there are no collisions and that each station has one frame to transmit every 5 s [48].

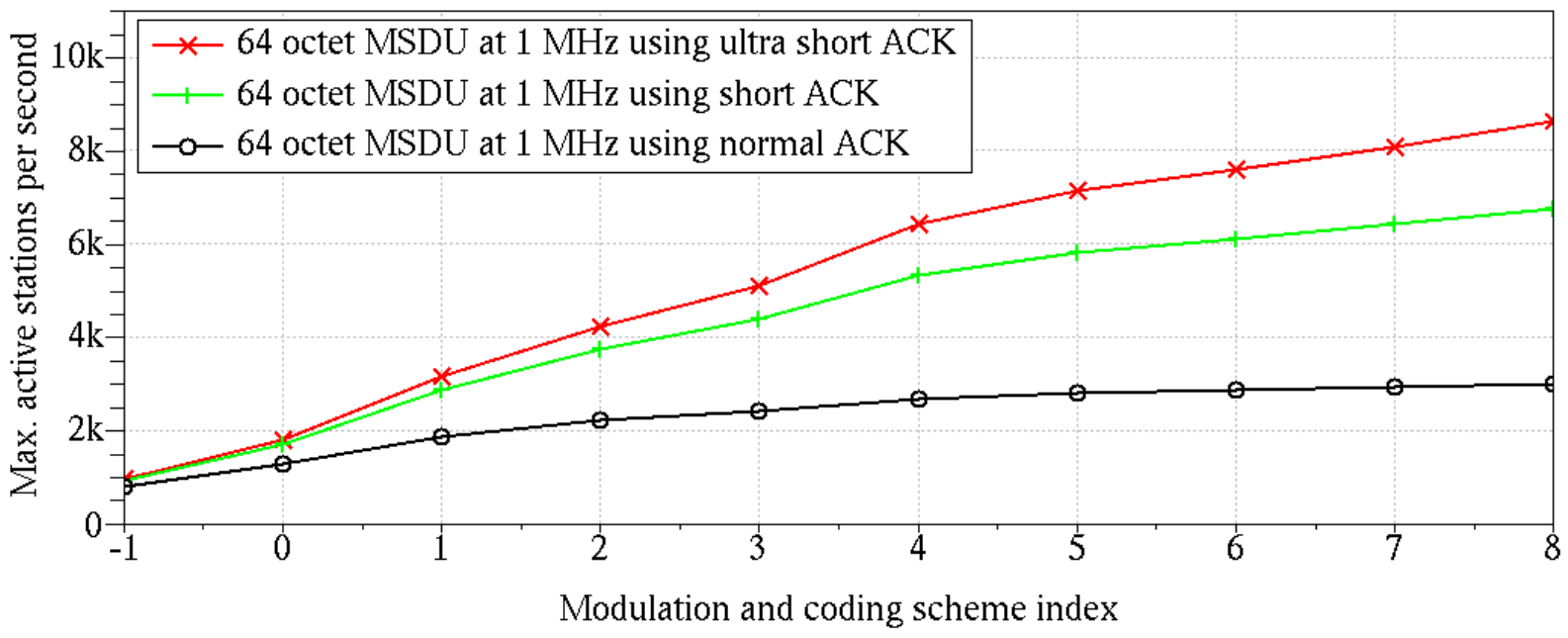

**Figure 7.12 Maximum number of active stations per second as a function of MCS index for the three ACK procedures using scenario 1.1**

With the proposed ultra short ACK, up to 8,635 transmitting stations could ideally be fitted within 1 second in scenario 1.1. This is almost double the maximum number of stations obtained using normal ACK. If the bandwidth is doubled (i.e. scenario 1.2), a 32% increase can be expected when using the ultra short ACK technique instead of the short ACK technique.

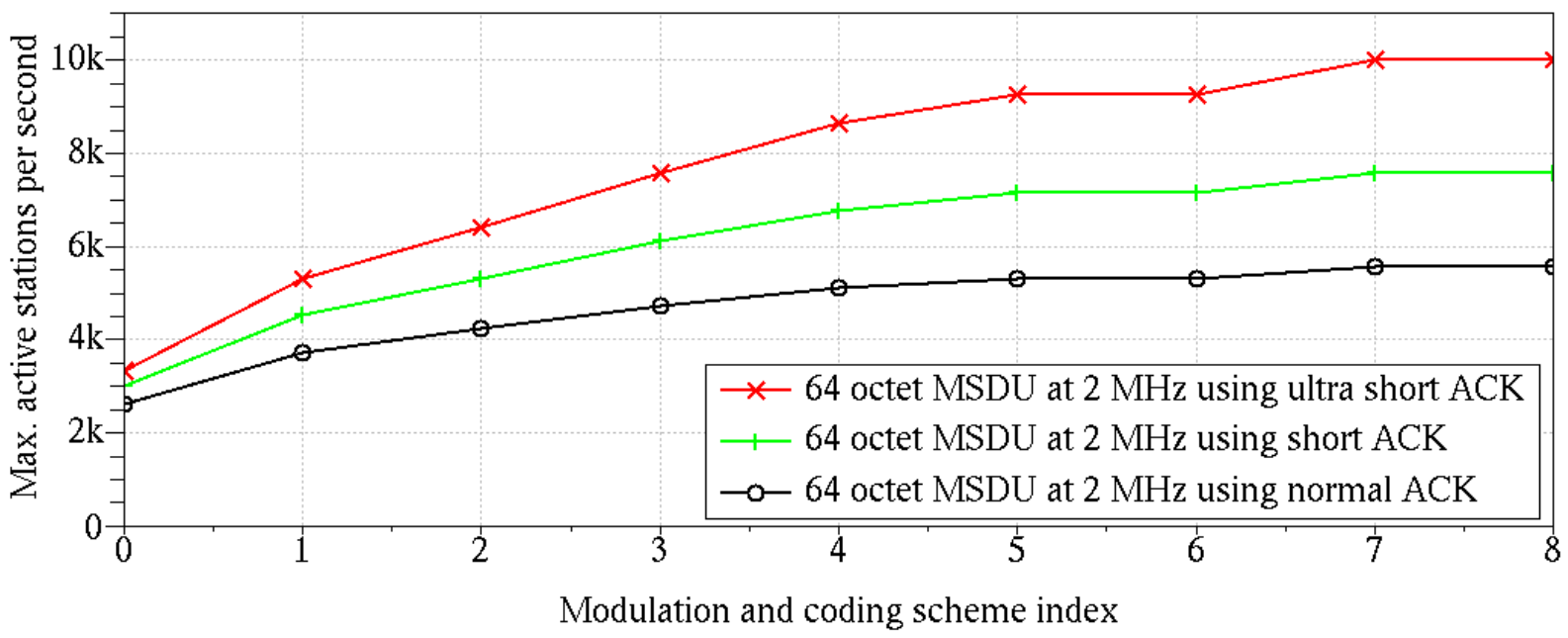

**Figure 7.13 Maximum number of active stations per second as a function of MCS index for the three ACK procedures using scenario 1.2**

Therefore, despite the apparently small gains when considering data+ACK exchange duration, the ultra short ACK technique can be very interesting to increase system capacity. Indeed it offers significantly better performance than the standard's short ACK technique for industrial process automation applications.

### 7.2.3.2 Sensor Network Application

The maximum number of active stations per second for 256 octet MSDUs using 1 MHz bandwidth (scenario 2.1) and 2 MHz bandwidth (scenario 2.2) are given in Figure 7.14 and Figure 7.15 (resp.). Again, we assume that there are no collisions. We have considered that each station has one frame to transmit every 35 s [48].

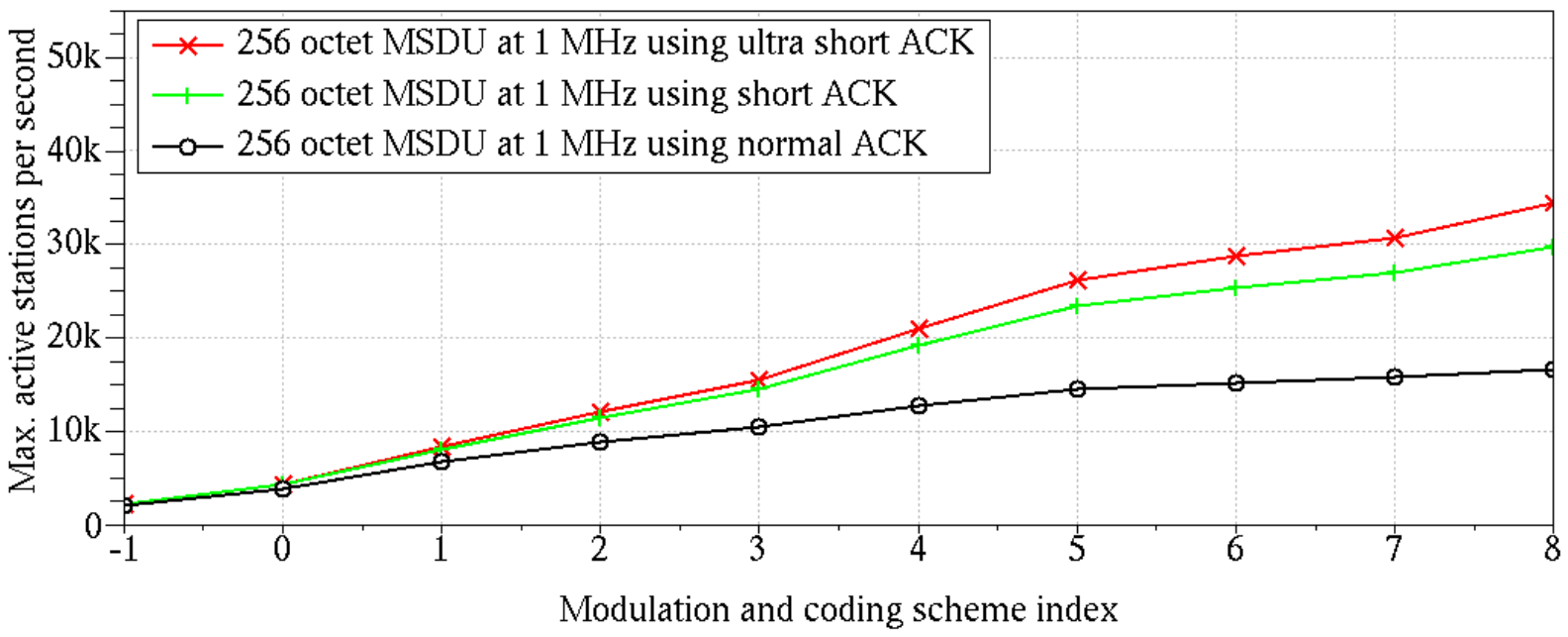

**Figure 7.14 Maximum number of active stations per second as a function of MCS index for the three ACK procedures using scenario 2.1**

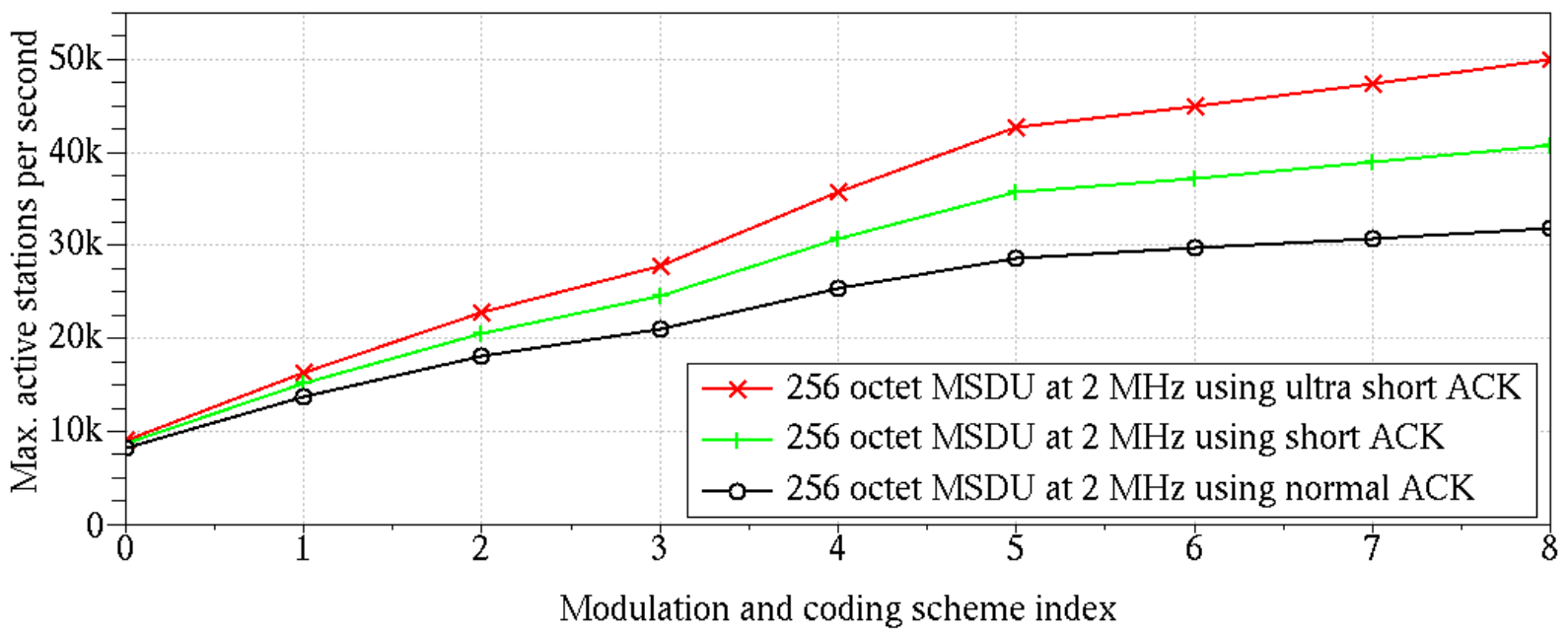

**Figure 7.15 Maximum number of active stations per second as a function of MCS index for the three ACK procedures using scenario 2.2**

Here also the ultra short ACK technique offers significant gain. In scenario 2.1, using it instead of the short ACK enables to fit up to 4,655 additional stations. This number leaps to 17,600 if ultra short ACK is used instead of normal ACK. In scenario 2.2, using the ultra short ACK procedure enables to fit up to 50,050 stations, which is 9,300 more than when using short ACK. The ultra short ACK is thus also adapted to sensor network applications.

# 7.3 Summary

In this chapter, IEEE 802.11ah is exposed in detail. This current evolution of IEEE 802.11ac implements interesting cross-layer solutions. The short ACK procedure enables to reduce overhead by acknowledging frames at PHY layer by putting the ACK information in the SIG field. We propose to use an ultra short ACK procedure, which better benefits from cross-layer optimization. A standalone STF field is sent to acknowledge frames in a more robust manner than in classical ACK and short ACK procedures. False alarms can be mitigated by using the time-reversal technique. The station receiving the ACK can thus precisely determine whether its frame has been acknowledged.

The exposed industrial process automation and sensor network scenarios show that this ultra short ACK results in performance gain with regards to normal ACK and short ACK. By reducing the ACK overhead, more stations can thus transmit within the same band. This is one of the important challenges 802.11ah. Signal robustness being another important issue, the impact of using time-reversal should be investigated in future studies.

# Conclusions and Future Work

In this thesis we have focused on studying and proposing cross-layer optimization solutions for the physical (PHY) and medium access control (MAC) layers of the IEEE 802.11ac standard. Our approach was to revisit standardized layering so as to improve performance without completely modifying the global architecture.

A detailed presentation of the 802.11ac standard, and standards it builds upon, has first been done. In 802.11ac, more than two channels can be bonded to increase throughput. But in multichannel transmissions collisions become a serious issue. Aggregated MAC protocol data units (MPDUs) being spread over the whole bandwidth, a single collision on one of the bonded channels may corrupt the whole data. We have shown that much better robustness to collisions can be achieved by considering PHY and MAC layer criteria, rather than just PHY data rate increase. Horizontal aggregation schemes have thus been proposed. The MAC layer aggregates available MPDUs per blocks of sub-channels, and the PHY layer processes these aggregates independently for robustness to collisions. With these schemes, an increase in throughput as high as 93%, or even 117%, can be achieved for 80 MHz transmissions in a realistic collision context by using different levels of partitioning. A trade-off can thus be established between robustness to collisions and partitioning-induced implementation complexity.

While evaluating the gains of the latter solution, we came to the conclusion that a cross-layer simulator was needed to perform in-depth analysis of 802.11ac's PHY and MAC layers. Studies involving the channel adaptive multiple-user multiple-input, multiple-output (MU-MIMO) scheme, which simultaneously involves PHY and MAC layers, can benefit from such a simulator. Consequently we have implemented a novel IEEE 802.11n/ac PHY+MAC simulation platform composed of a MAC simulation module which interfaces with a PHY simulation module in a symbiotic manner. Channel variations, transmit/receive chain specificities, and channel access mechanisms are faithfully taken into account, while minimizing computational resource consumption. The simulation platform has been validated using an IEEE 802.11n implementation of network simulator 2 (ns-2). It was shown that the proposed platform outperforms this system-level simulator in that channel state evolutions, channel estimate aging, and even the collision-correction ability of receive chains can be faithfully accounted for.

The first study that was carried out using this 802.11ac simulation platform concerned performance comparison of single-user MIMO (SU-MIMO) with beamforming and MU-MIMO. This was done for single-antenna stations and multiple-antenna stations benefitting from spatial diversity. Channel state information (CSI) feedback, enabling precoding, is a central point of such techniques. Taking into account crosstalk interference (CTI) between precoded multiple-user (MU) streams, the PHY+MAC simulator has proven very beneficial for a realistic and fair comparison of these two schemes. The performance of MU-MIMO when CTI is not accounted for was too optimistic. Indeed, it has been shown that, because of CTI, MU-MIMO transmission can sometimes have less throughput gain and be much less stable than SU-MIMO. Thus time sensitive applications may suffer from jitter and higher packet loss when using MU-MIMO. In addition, when destination stations with correlated downlink channels are relatively far from the access point (AP), MU-MIMO and SU-MIMO with beamforming have shown similar results. An intelligent switching algorithm between MU-MIMO and SU-MIMO schemes based on CSI and received SNR can be a good solution for optimum performance.

The second study concerned channel sounding interval for 802.11ac's MU-MIMO. When using MU-MIMO, the AP has to have precise knowledge of the channels of all destination stations. To that end, a channel sounding procedure is regularly engaged. The frequency of this sounding procedure, and thus the aging of the obtained estimates, has an important impact on performance. The joint PHY+MAC analysis of scenarios using quite correlated channels has showed that frequent sounding can lead to an increase in throughput, despite the associated overhead. In the considered scenarios, a 10 ms sounding interval has enabled much better performance than a 40 ms sounding interval. This is contrary to results that have been obtained with a MAC layer only perspective (be they statistical or simulated). In addition these MAC-centric results have proven to be very optimistic for MU-MIMO. The aging of channel estimates, and the CTI due to MU precoding, should thus be accounted for. This is because MU-MIMO is much more sensitive to feedback errors than SU-MIMO. Here also a dynamic dimensioning of the interval, channel conditions and the number of stations, can be very interesting for optimum performance.

This thesis has mainly focused on 802.11ac features. However, we have also studied the IEEE 802.11ah standard. This current evolution of 802.11ac standard addresses machine to machine communications. Power consumption and support for a great number of users are its main challenges. This standard has defined a short acknowledgment (ACK) procedure which enables to reduce overhead by acknowledging frames at PHY layer. In this interesting cross-layer solution the ACK information is contained in the signaling field. We had proposed to use an ultra short ACK procedure, which better benefits from cross-layer optimization. A standalone short training field (STF) is sent to acknowledge frames in a more robust manner than in classical ACK and short ACK procedures. False alarms should be mitigated by using the time-reversal technique. We have shown that a 23% and 57% increase in the number of stations can be obtained by using the ultra short ACK instead of the short ACK and normal ACK (resp.) for sensor networks. For industrial process automation, the increase is even greater, i.e. 32% and 80% (resp.), thus showing the interest for such a solution in 802.11ah systems.

# Research Perspectives

The results of this thesis have given some interesting insight on the performance of MU-MIMO in IEEE 802.11ac, especially, through PHY+MAC simulations. Interesting research directions include, but are not limited to, the following three aspects:

- Evaluate the performance of MU-MIMO for a greater number of stations. However, this implies developing grouping algorithms enabling to exploit multiuser diversity (for better quality of service or higher throughput) or to optimize power consumption. With the same logic of studying more use cases, the influence of the increase of the number of available spatial streams and bandwidth could be studied. Furthermore, other precoding techniques could also be evaluated;
- Implement dynamic switching algorithms between SU-MIMO and MU-MIMO and dynamic dimensioning of MU-MIMO’s channel sounding interval for optimum performance;
- Extend the PHY+MAC 802.11ac simulation platform to IEEE 802.11ah systems in order to analyze PHY layer implications of using an STF with time-reversal technique.

The different cross-layer optimization principles, which have been studied and modeled through the PHY+MAC simulator for 802.11ac systems, can be applied to future generations of Wi-Fi systems. The new thesis on evolutions of the 802.11ac, which will start within the same laboratory, will surely benefit from the simulation platform and the studies exposed in this thesis.

# Appendices and References

# Appendix A

# Block Diagonalization Algorithm for Throughput Maximization

Block diagonalization (BD), introduced in section 1.1.4.2.3, is a spatial division multiple access (SDMA) technique. This linear precoding solution for multiple antenna receivers optimizes power transfer to a group of antennas belonging to the same receiver, while zero-forcing inter-user interference [15]. The channel is thus block diagonalized, with interference inside each block being tackled by a per-user processing [91].

When BD is used, the precoding matrix $\mathbf{W}_c$ of each user $c$ is chosen such that $\mathbf{H}_e \cdot \mathbf{W}_c = \mathbf{0}$, for all $e \neq c$ of the selected set $C$, with $\mathbf{H}_e$ being the channel to user $e$. The received signal vector $\mathbf{y}_c$ for user $c$ is thus given by:

$$\mathbf{y}_c = \mathbf{H}_c \cdot \mathbf{W}_c \cdot \mathbf{d}_c + \sum_{j \in C, j \neq c} \mathbf{H}_{\not{c}} \cdot \mathbf{W}_{\not{c}} \cdot \mathbf{d}_{\not{c}} + \mathbf{z}_c \qquad \text{(A.1)}$$

With $\mathbf{d}_c$ being the vector of data symbols transmitted to user $c$ and $\mathbf{z}_c$ the noise perceived by the same user. This means that, for every user of $C$, the zero crosstalk interference (CTI) constraint forces $\mathbf{W}_c$ to lie in the null space of $\tilde{\mathbf{H}}_c$. This matrix is obtained by removing the rows belonging to $\mathbf{H}_c$ from the downlink channel $\mathbf{H}$ towards $N$ users:

$$\tilde{\mathbf{H}}_c = \begin{bmatrix} \mathbf{H}_1^T & \cdots & \mathbf{H}_{c-1}^T & \mathbf{H}_{c+1}^T & \cdots & \mathbf{H}_N^T \end{bmatrix}^T \qquad \text{(A.2)}$$

To achieve this, we introduce matrix $\mathbf{F}_c$ which holds the orthogonal basis for the null space of $\tilde{\mathbf{H}}_c$. This matrix ensures block diagonalization. Another matrix $\mathbf{E}_c$ can then be introduced, per user, so as to maximize throughput. The precoding matrix $\mathbf{W}_c$ is equal to $\mathbf{F}_c \cdot \mathbf{E}_c$. Each receiver can thus process its CTI-free symbols with matrix $\mathbf{G}_c$. The obtained BD system is elegantly illustrated in Figure A.1 [91]. The algorithm enabling to find the precoding matrix, given in [19], is exposed here.

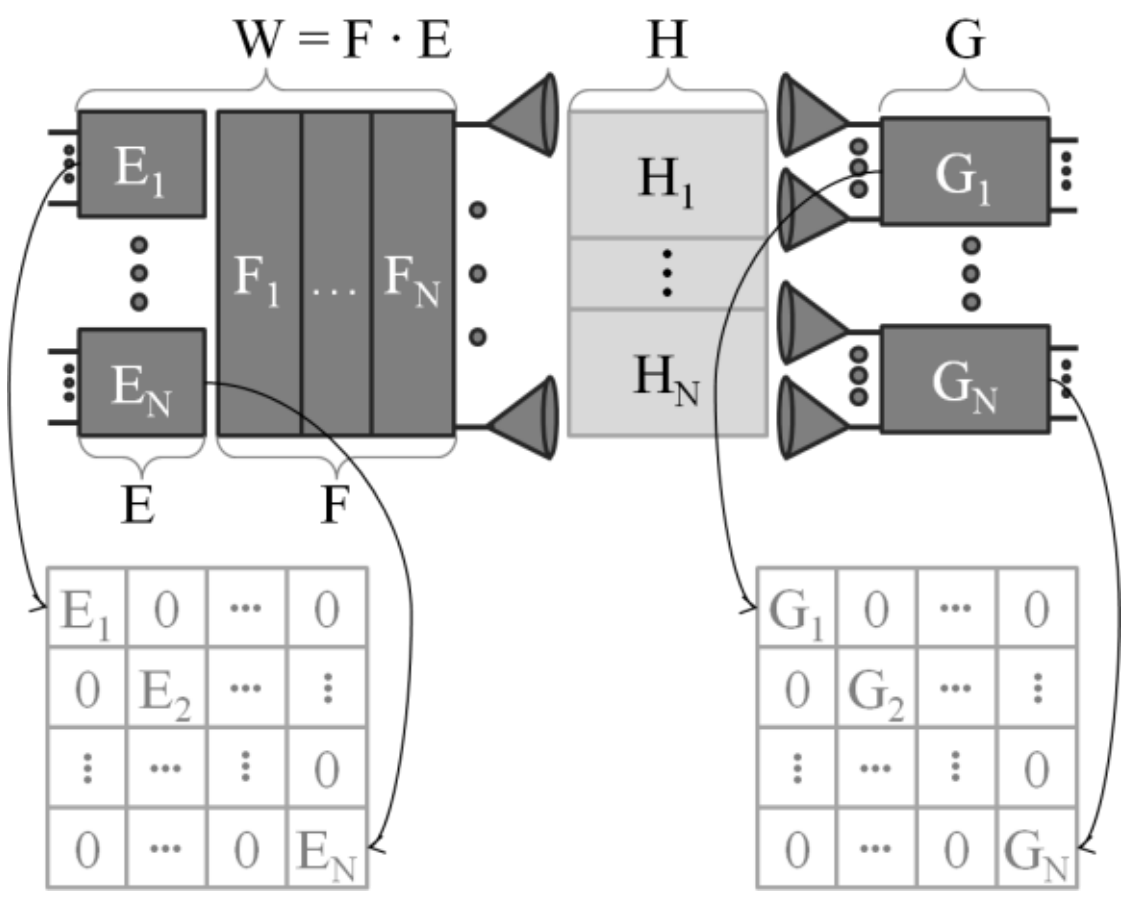

**Figure A.1 SDMA system and coding matrices when using block diagonalization**

Firstly, finding a matrix $\mathbf{F}_c$ implies that the null space of $\tilde{\mathbf{H}}_c$ has a dimension greater than 0, i.e. $\mathrm{rank}\left(\tilde{\mathbf{H}}_c\right) < n_T$. Let the singular value decomposition (SVD, see section 1.1.3.1) of $\tilde{\mathbf{H}}_c$ be:

$$\tilde{\mathbf{H}}_c = \tilde{\mathbf{U}}_c \cdot \tilde{\mathbf{\Sigma}}_c \cdot \left[\tilde{\mathbf{V}}_c^{(1)} \quad \tilde{\mathbf{V}}_c^{(0)}\right]^* \tag{A.3}$$

If we note $\tilde{r}_c = \mathrm{rank}\left(\tilde{\mathbf{H}}_c\right)$, $\tilde{\mathbf{V}}_c^{(1)}$ holds the first $\tilde{r}_c$ right singular vectors and $\tilde{\mathbf{V}}_c^{(0)}$ the remaining $n_T - \tilde{r}_c$ right singular vectors. The columns of $\tilde{\mathbf{V}}_c^{(0)}$ are thus candidates for $\mathbf{F}_c$. However, in order to have a CTI-free transmission to user $c$, the rank $\bar{r}_c$ of the product $\overline{\mathbf{H}}_c = \mathbf{H}_c \cdot \tilde{\mathbf{V}}_c^{(0)}$ should be at least equal to one. To satisfy this condition, the users in the selected set $C$ should be carefully selected so as not to have highly correlated channel matrices. If so, the desired block diagonal structure is obtained for the channel matrix $\overline{\mathbf{H}} = \mathbf{H} \cdot \mathbf{F}$. SVD can thus be used on a per-user basis to find the throughput maximizing matrix $\mathbf{E}_c$. The SVD of $\overline{\mathbf{H}}_c$ gives:

$$\overline{\mathbf{H}}_c = \overline{\mathbf{U}}_c \cdot \begin{bmatrix} \overline{\mathbf{\Sigma}}_c & \mathbf{0} \\ \mathbf{0} & \mathbf{0} \end{bmatrix} \cdot \left[\overline{\mathbf{V}}_c^{(1)} \quad \overline{\mathbf{V}}_c^{(0)}\right]^* \tag{A.4}$$

$\overline{\mathbf{V}}_c^{(1)}$ holds the first $\bar{r}_c$ singular vectors. The columns of $\overline{\mathbf{V}}_c^{(1)}$ are thus candidates for $\mathbf{E}_c$. The BD precoding matrix which maximizes throughput for user $c$ subject to zero inter-user interference is:

$$\mathbf{W}_c = \tilde{\mathbf{V}}_c^{(0)} \cdot \overline{\mathbf{V}}_c^{(1)} \tag{A.5}$$

Power loading can then be applied to scale the power transmitted into each precoder. We thus have given the details needed to obtain equation (1.13) in section 1.1.4.2.3.

# Appendix B

# Adaptive Multi-rate Retry Algorithm

The adaptive multi-rate retry algorithm (AMRR) offers short-term adaptation at medium access control (MAC) layer for high-latency systems (see section 1.2.8). To this aim, its adaptation algorithm is composed of two evolutions. The 'macro-evolution' enables long-term adaptation. We will note that this process would be sufficient for low-latency systems. Short-term adaptation is enabled for high-latency systems through the 'micro-evolution' process. Therefore the AMRR algorithm is well suited for both types of systems [33].

In the macro-evolution process, decisions are taken considering statistics on failed and successful transmissions of MAC protocol data units (MPDUs, see section 1.2.5). If transmissions, using a certain data rate, are globally successful, it means that channel conditions may be good enough to support higher data rate transmissions. The data rate is thus increased for following transmissions, considering that enough frames have been received to establish reliable statistics. Indeed, the given group of transmissions is considered as globally successful if the ratio of failed frames over successful frames is below a success ratio. A minimum number of successful frames have to be obtained for this ratio to be reliable.

On the other hand, the system has to be very reactive to transmission failures. Indeed, if channel conditions are not good enough to support a certain data rate, it is better to fall back to a lower but more robust transmission rate. In AMRR, there is fallback if the ratio of failed frames over successful frames is above a failure ratio. In order to avoid penalizing the station for sporadic bad channel conditions (e.g. collisions), a probing frame[1] is sent at every rate increase. The acknowledgment of this frame meaning that channel conditions are stable, the rate is validated if so. If not, the initial data rate is restored and the minimum success threshold is doubled. The use of this binary exponential backoff results in a stable algorithm when in presence of long-term channel evolutions.

[1] The probing frame is the first frame being sent after a data rate increase so as to 'probe' whether the receiver can correctly decode the sent information, considering the current channel state. A normal data frame can be used for this purpose. Correct reception is detected through classical acknowledgment procedures.

If the ratio of failed frames over successful frames is between the previous intervals, the data rate is maintained. The state machine of AMRR's macro-evolution is illustrated in Figure B.1.

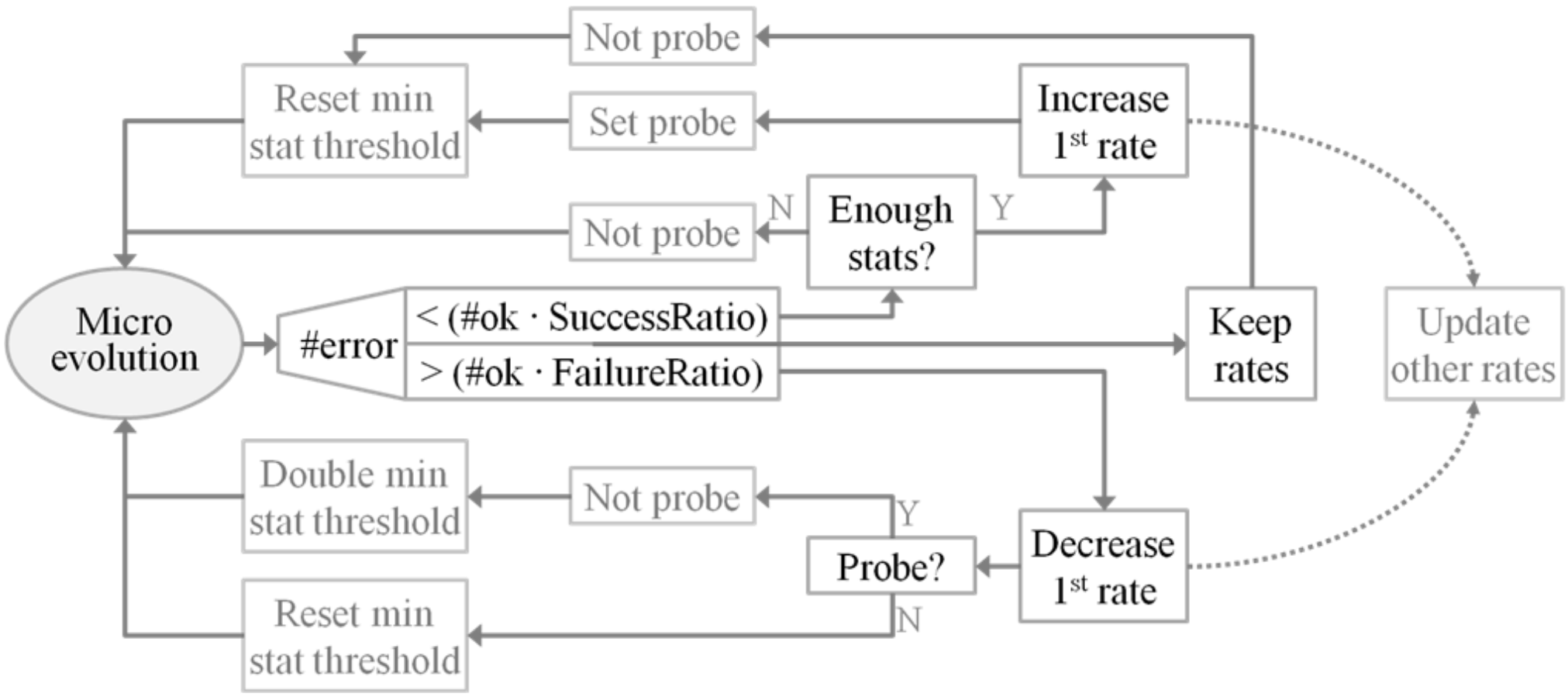

**Figure B.1 State machine of the macro-evolutions of AMRR**

As mentioned above, some chipsets, using software drivers, may not be fast enough to perform per-packet decisions (see [33] for more details). Therefore the AMRR algorithm uses micro-evolutions for faster response to short-term channel variations. The state machine of AMRR's micro-evolution is illustrated in Figure B.2.

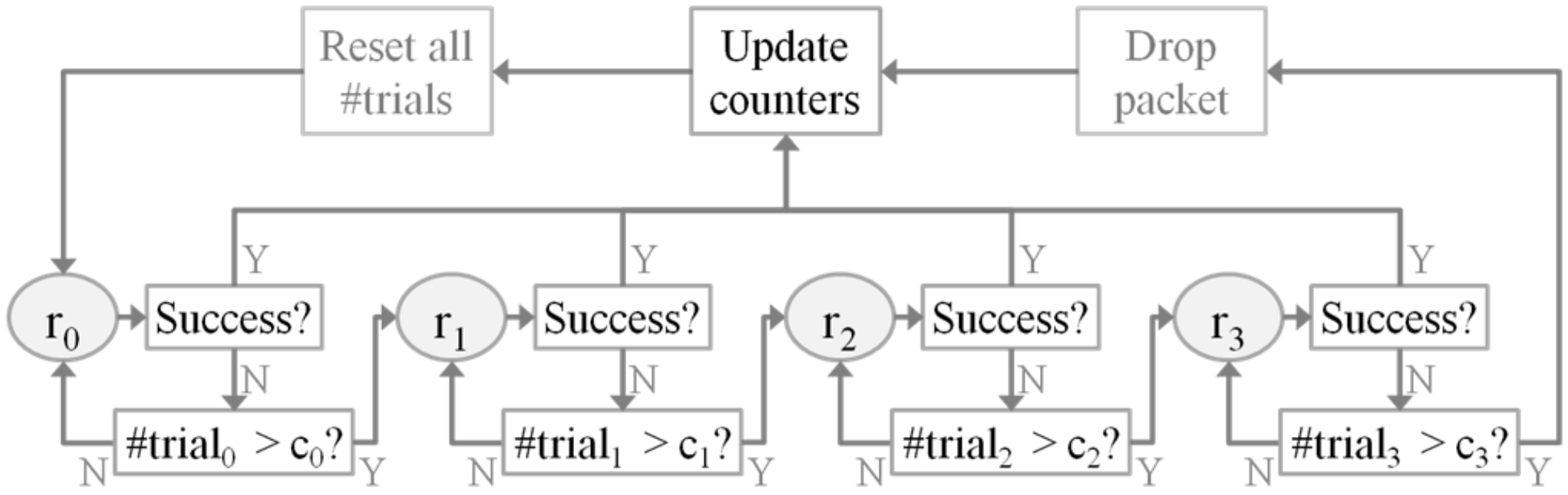

**Figure B.2 State machine of the micro-evolutions of AMRR**

An ordered set of 4 pairs of data rates ($r_0$, $r_1$, $r_2$, and $r_3$) and transmission counts ($c_0$, $c_1$, $c_2$, and $c_3$) are loaded in the hardware's local memory. Whenever the medium is available for transmission, the first data rate ($r_0$) is used. If the frame is correctly acknowledged, counters are updated for the macro-evolution mechanism. If not, data rate $r_0$ is used again, up to $c_0 - 1$ times. If all these transmissions fails, the second rate ($r_1$) is tried up to $c_1$ times, and so on. If, even with the last rate ($r_3$) transmission fails, i.e. $c_0 + c_1 + c_2 + c_3$ failures, the packet is dropped and counters are updated. Rates are updated according to AMRR's macro-evolution decisions (see Figure B.1). An increase, or decrease, directly affects the rate $r_0$. The following two rates, based on the first one, are consequently updated. Indeed, $r_1 = r_0 - 1$ and $r_2 = r_0 - 2$, if minimum rate is not reached. The last rate is always equal to minimum rate, as a final effort to correctly transmit data.

For reasons exposed in section 1.2.8, we have used this implementation of AMRR[1] when using aggregation (see sections 1.2.7 and 1.3.2.2) and multiple-user multiple-input, multiple-output (MU-MIMO, see sections 1.1.4 and 2.4). Thus, success and failure statistics are still established using MPDUs, when using aggregation. For MU-MIMO, statistics are also established using MPDUs. However, one instantiation of the algorithm is used per served station.

We hope that, through this explanation, the reader has acquired better understanding of the AMRR algorithm introduced in section 1.2.8. For more information on the used source code, please refer to the multiband Atheros driver for Wi-Fi project [92].

[1] Except in Chapter 3, where some modifications are proposed for collision-prone environments.

Appendix C

# MAC Efficiency

As indicated in 1.3.2.2, medium access control (MAC) efficiency is obtained by dividing the average throughput by the average physical (PHY) data rate. Therefore efficiency is an important parameter to take into account when making technical choices. Indeed, the set of possible PHY data rates can be easily computed according to transmitter and receiver characteristics (bandwidth, number of antennas, guard interval size) and the maximum data rate is often clearly displayed on devices' packaging. However, PHY data rates only account for the transfer rate of the PHY service data unit (PSDU). There is no notion of time nor PHY and MAC overhead. The latter are accounted for in the MAC throughput. Therefore if there is an *a priori* notion of the achievable efficiency, MAC throughput can be roughly estimated from the maximum PHY data rate, thus orienting technical choices.

In IEEE 802.11e [30] and IEEE 802.11n [6] new MAC functionalities were introduced to improve this efficiency. MAC protocol data unit (MDPU) aggregation is one of these functionalities and is much used because of the complexity/efficiency tradeoff it offers [8,31] (see section 1.2.7 for more details on aggregation).

Let us consider a data+acknowledgment (ACK) exchange at different data rates with different payloads for IEEE 802.11a/g and IEEE 802.11n devices. The parameters of these exchanges are given in Table C.1. A 120 octet MAC service data unit (MSDU) generated by voice traffic (see section 1.3.2.2) and a 1500 octet MSDU are considered. Three data rates are used for 802.11a/g transmissions (highest, lowest and middle range). Likewise, three data rates are also used for 802.11n transmissions. The former range from the lowest 20 MHz and 1 spatial stream (SS) data rate, to the highest 40 MHz and 4 SS data rate, with 130 Mbps corresponding to the highest 20 MHz and 2 SS data rate. In 802.11n, a PSDU can consist of a simple MDPU or an aggregate MPDU (A-MPDU). An A-MPDU of 4 MPDUs, each composed of a 1500 octet MSDU, and an A-MPDU of 20 MPDUs, each composed also of a 1500 octet MSDU, are considered to evaluate the impact of aggregation. When A-MPDUs are sent by a transmitter, the concerned receiver replies with a block ACK (BA).

**Table C.1 Parameters of different data+ACK exchanges for an illustration of MAC efficiency**

| Standard | Payload type | Used data rates (Mbps) 6 | 24 | 54 | 6.5 | 130 | 540 |
|---|---|---|---|---|---|---|---|
| 802.11a/g | 120 o MSDU | × | × | × | | | |
| | 1500 o MSDU | × | × | × | | | |
| 802.11n | 120 o MSDU | | | | × | × | × |
| | 1500 o MSDU | | | | × | × | × |
| | Aggregate of 4 MDPUs (1500 o MSDU each) | | | | | × | × |
| | Aggregate of 20 MDPUs (1500 o MSDU each) | | | | | × | × |

Figure C.1 illustrates the detailed durations of the considered 16 data+ACK exchanges.

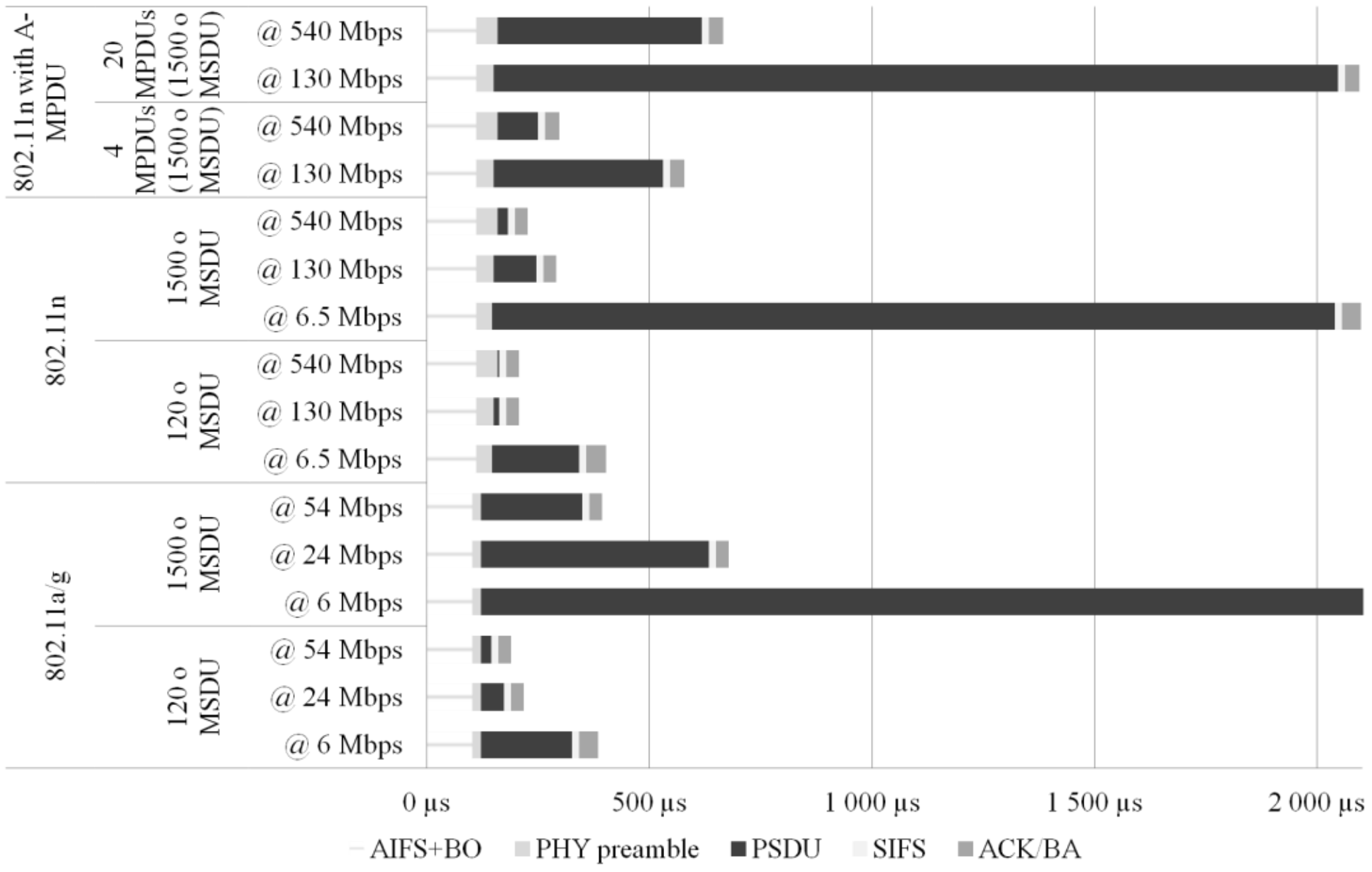

**Figure C.1 Detailed durations of the considered data+ACK exchanges**

The channel access phase, consisting of the arbitration inter-frame space (AIFS)[1] and an average backoff duration (BO)[2], precedes the data transmission phase (with preamble and PSDU). After a short inter-frame spacing (SIFS), the receiver replies with an ACK or a BA frame. The impact of data rate on total duration can be clearly seen. One can also notice that an increase in data rate is not always the optimal solution. It takes exactly the same time, i.e. 206.5 µs, to transmit a 120 octet payload at 130 Mbps than at 540 Mbps. But for this particular payload, it is more interesting to use the legacy 54 Mbps data rate, with which the exchange takes 189.5 µs, because of a shorter PHY preamble. 802.11a/g is thus more adapted to short payload transmission than 802.11n (17 µs difference at lowest rate for 120 octet MSDUs).

On the other hand, when aggregation is used transmissions last longer. Indeed there is more data to transmit (4 or 20 MPDUs) therefore the duration alone cannot be a good criterion. MAC throughput, i.e. payload over occupied time, needs to be considered. The throughputs, along with the PHY data rates, are given in Figure C.2. MAC efficiencies, the ratio between the two, are given in Figure C.3.

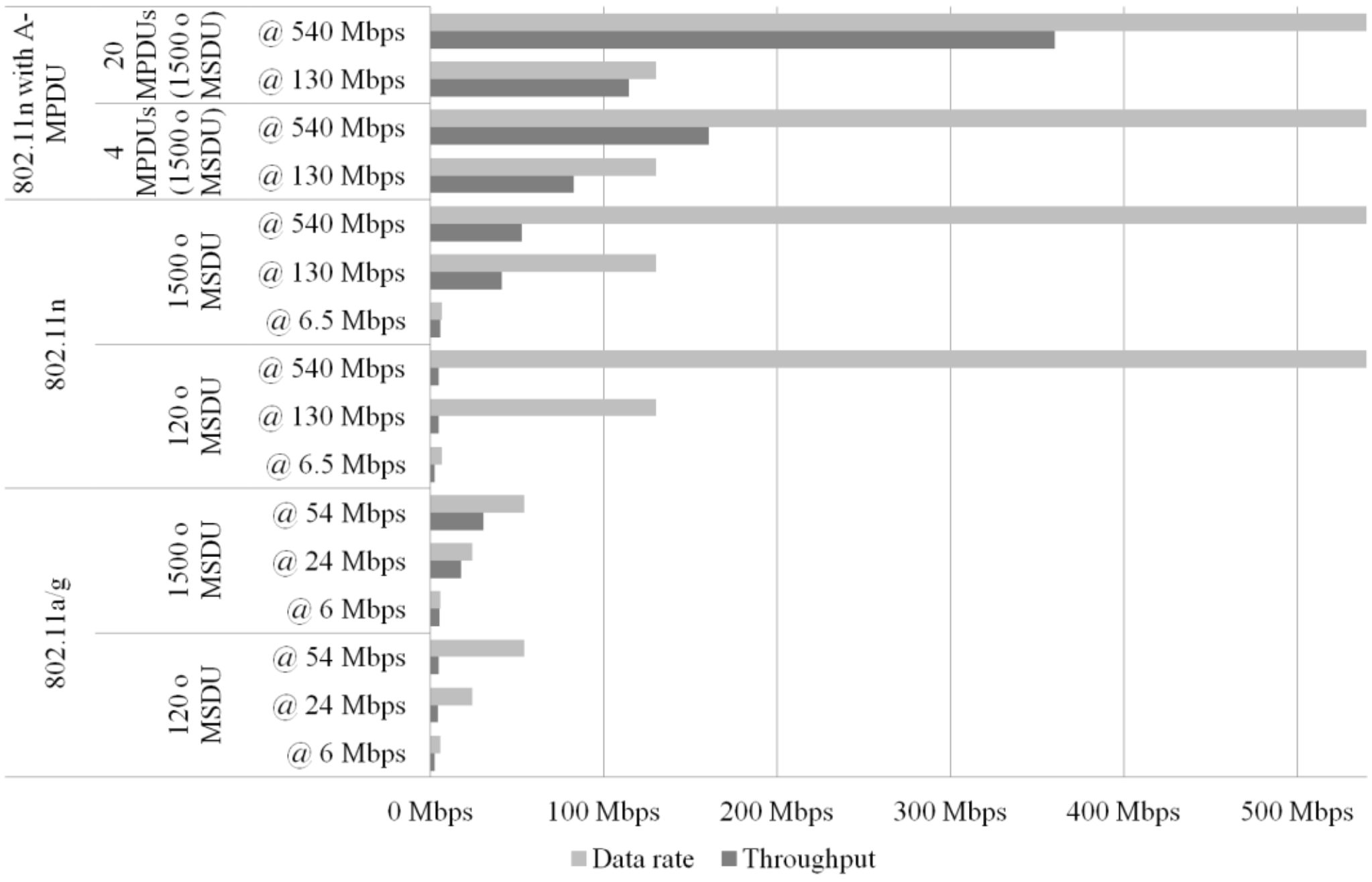

**Figure C.2 Throughputs and data rates of the considered data+ACK exchanges**

[1] Best effort frames are considered here (with legacy inter-frame spacing for IEEE 802.11a/g exchanges).
[2] Half the minimum contention window is taken as average.

Naturally, throughput cannot exceed PHY data rate, because of overhead (preamble, header channel access, spacing, and ACK/BA), but it increases as higher data rates are used. Indeed the time occupied by the data+ACK exchanges is generally smaller as data rates increase (considering that overhead has not changed, which is not always true as stated above). However, the MAC efficiency decreases as the difference between data rate and throughput increases. The transmission time occupied by the MAC payload consists in an important part of the total exchange if the payload is important or if the data rate is small.

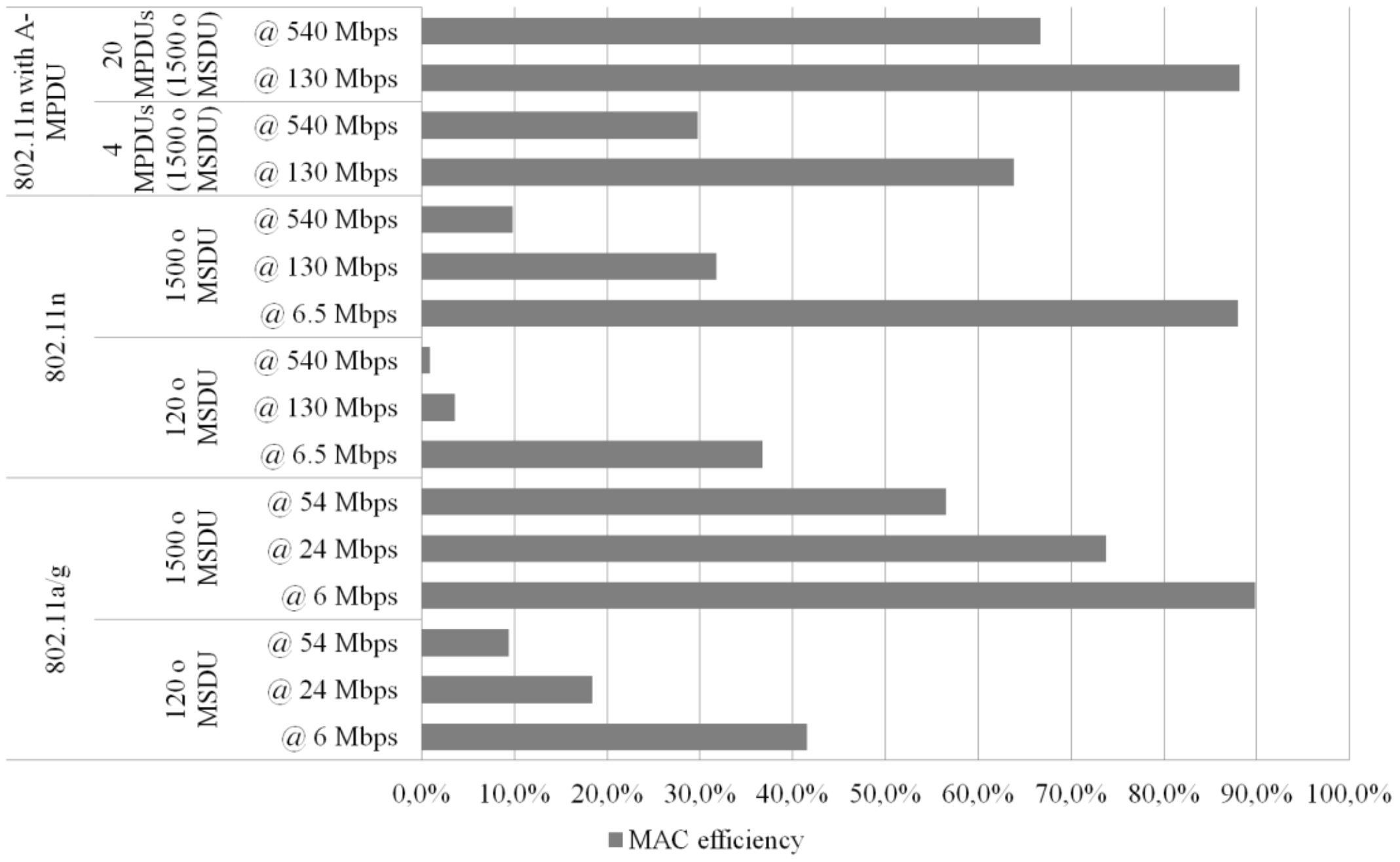

**Figure C.3 MAC efficiencies of the considered data+ACK exchanges**

We can indeed see that transmitting a 1500 octet MSDU at 6 Mbps offers more efficiency (i.e. an additional 35%) than transmitting twenty 1500 octet MSDUs at 540 Mbps. On the other hand, the latter exchange has a throughput which is 66 times greater than the former. Therefore good efficiency does not always imply high throughput. Both should be considered as in Figure C.4.

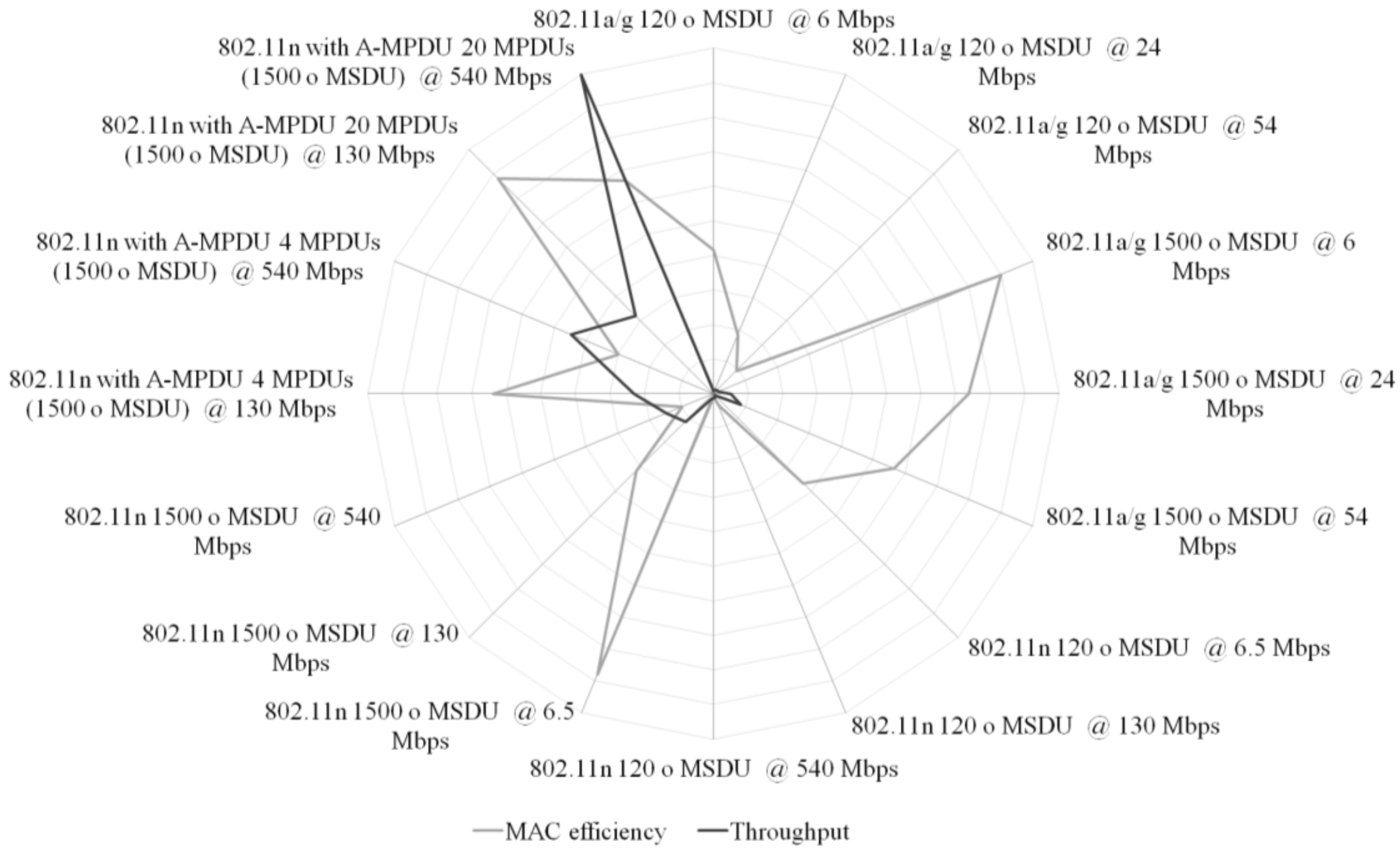

**Figure C.4 MAC efficiencies and throughputs of the considered data+ACK exchanges**

Clearly, the best throughput/efficiency compromise is obtained through aggregation. In other words, MPDU aggregation offers high throughput while offering high efficiency at high data rates. This explains its introduction in the 802.11n amendment to improve efficiency as indicated in 1.3.2.2.

We shall note that in this analysis, we have used analytical results. To this aim, durations induced by PHY and MAC framing, in conjunction with average channel access durations, are used, according to 802.11n specifications [6] and the considered system configuration. Total frame exchange duration (i.e. channel access + data frame + SIFS + ACK/BA) is thus obtained. Throughput is then computed by dividing the MAC payload of the data frame by the echange duration. This gives a good indication of the ideal MAC capacity of the medium (i.e. no pathloss, no noise, no interferer, no collisions, and no retries). Nevertheless, overhead induced by beacon frames and channel sounding procedure is considered.

A similar approach is used in Chapter 6 and Chapter 7 by using IEEE 802.11ac [7] and IEEE 802.11ah [44] specifications (resp.).

Appendix D

# Very High Throughput Modulation and Coding Schemes

The physical (PHY) layer data rates for the IEEE 802.11ac standard [7], labeled very high throughput (VHT), are given in Table D.1, Table D.2, Table D.3, and Table D.4 for 20 MHz, 40 MHz, 80 MHz, and 160 MHz (or two non-contiguous 80 MHz band transmission, called 80+80MHz) bandwidth transmission respectively. PHY data rates are sorted by modulation and coding scheme (MCS) index, grouped by the used number of spatial streams (SSs), and given for long guard intervals (GIs). PHY data rates for short GI can be obtained by multiplying the tabulated data rates by 10/9.

**Table D.1 VHT MCSs for 20 MHz bandwidth with long GI**

| MCS index | Data rate (Mbps) | | | | | | | | MCS index |
|---|---|---|---|---|---|---|---|---|---|
| | 1 SS | 2 SS | 3 SS | 4 SS | 5 SS | 6 SS | 7 SS | 8 SS | |
| 0 | 6.5 | 13 | 19.5 | 26 | 32.5 | 39 | 45.5 | 52 | 0 |
| 1 | 13 | 26 | 39 | 52 | 65 | 78 | 91 | 104 | 1 |
| 2 | 19.5 | 39 | 58.5 | 78 | 97.5 | 117 | 136.5 | 156 | 2 |
| 3 | 26 | 52 | 78 | 104 | 130 | 156 | 182 | 208 | 3 |
| 4 | 39 | 78 | 117 | 156 | 195 | 234 | 273 | 312 | 4 |
| 5 | 52 | 104 | 156 | 208 | 260 | 312 | 364 | 416 | 5 |
| 6 | 58.5 | 117 | 175.5 | 234 | 292.5 | 351 | 409.5 | 468 | 6 |
| 7 | 65 | 130 | 195 | 260 | 325 | 390 | 455 | 520 | 7 |
| 8 | 78 | 156 | 234 | 312 | 390 | 468 | 546 | 624 | 8 |
| 9 | - | - | 260 | - | - | 520 | - | - | 9 |

Support for 20 MHz, 40 MHz, and 80 MHz with 1 SS is mandatory. In addition all VHT devices shall support MCSs 0 through 7 for all supported bandwidth. All other configurations are optional.

**Table D.2 VHT MCSs for 40 MHz bandwidth with long GI**

| MCS index | Data rate (Mbps) | | | | | | | | MCS index |
|---|---|---|---|---|---|---|---|---|---|
| | 1 SS | 2 SS | 3 SS | 4 SS | 5 SS | 6 SS | 7 SS | 8 SS | |
| 0 | 13.5 | 27 | 40.5 | 54 | 67.5 | 81 | 94.5 | 108 | 0 |
| 1 | 27 | 54 | 81 | 108 | 135 | 162 | 189 | 216 | 1 |
| 2 | 40.5 | 81 | 121.5 | 162 | 202.5 | 243 | 283.5 | 324 | 2 |
| 3 | 54 | 108 | 162 | 216 | 270 | 324 | 378 | 432 | 3 |
| 4 | 81 | 162 | 243 | 324 | 405 | 486 | 567 | 648 | 4 |
| 5 | 108 | 216 | 324 | 432 | 540 | 648 | 756 | 864 | 5 |
| 6 | 121.5 | 243 | 364.5 | 486 | 607.5 | 729 | 850.5 | 972 | 6 |
| 7 | 135 | 270 | 405 | 540 | 675 | 810 | 945 | 1080 | 7 |
| 8 | 162 | 324 | 486 | 648 | 810 | 972 | 1134 | 1296 | 8 |
| 9 | 180 | 360 | 540 | 720 | 900 | 1080 | 1260 | 1440 | 9 |

**Table D.3 VHT MCSs for 80 MHz bandwidth with long GI**

| MCS index | Data rate (Mbps) | | | | | | | | MCS index |
|---|---|---|---|---|---|---|---|---|---|
| | 1 SS | 2 SS | 3 SS | 4 SS | 5 SS | 6 SS | 7 SS | 8 SS | |
| 0 | 29.25 | 58.5 | 87.75 | 117 | 146.25 | 175.5 | 204.75 | 234 | 0 |
| 1 | 58.5 | 117 | 175.5 | 234 | 292.5 | 351 | 409.5 | 468 | 1 |
| 2 | 87.75 | 175.5 | 263.25 | 351 | 438.75 | 526.5 | 614.25 | 702 | 2 |
| 3 | 117 | 234 | 351 | 468 | 585 | 702 | 819 | 936 | 3 |
| 4 | 175.5 | 351 | 526.5 | 702 | 877.5 | 1053 | 1228.5 | 1404 | 4 |
| 5 | 234 | 468 | 702 | 936 | 1170 | 1404 | 1638 | 1872 | 5 |
| 6 | 263.25 | 526.5 | - | 1053 | 1316.25 | 1579.5 | - | 2106 | 6 |
| 7 | 292.5 | 585 | 877.5 | 1170 | 1462.5 | 1755 | 2047.5 | 2340 | 7 |
| 8 | 351 | 702 | 1053 | 1404 | 1755 | 2106 | 2457 | 2808 | 8 |
| 9 | 390 | 780 | 1170 | 1560 | 1950 | - | 2730 | 3120 | 9 |

**Table D.4 VHT MCSs for 160 MHz and 80 + 80 MHz bandwidth with long GI**

| MCS index | Data rate (Mbps) | | | | | | | | MCS index |
|---|---|---|---|---|---|---|---|---|---|
| | 1 SS | 2 SS | 3 SS | 4 SS | 5 SS | 6 SS | 7 SS | 8 SS | |
| 0 | 58.5 | 117 | 175.5 | 234 | 292.5 | 351 | 409.5 | 468 | 0 |
| 1 | 117 | 234 | 351 | 468 | 585 | 702 | 819 | 936 | 1 |
| 2 | 175.5 | 351 | 526.5 | 702 | 877.5 | 1053 | 1228.5 | 1404 | 2 |
| 3 | 234 | 468 | 702 | 936 | 1170 | 1404 | 1638 | 1872 | 3 |
| 4 | 351 | 702 | 1053 | 1404 | 1755 | 2106 | 2457 | 2808 | 4 |
| 5 | 468 | 936 | 1404 | 1872 | 2340 | 2808 | 3276 | 3744 | 5 |
| 6 | 526.5 | 1053 | 1579.5 | 2106 | 2632.5 | 3159 | 3685.5 | 4212 | 6 |
| 7 | 585 | 1170 | 1755 | 2340 | 2925 | 3510 | 4095 | 4680 | 7 |
| 8 | 702 | 1404 | 2106 | 2808 | 3510 | 4212 | 4914 | 5616 | 8 |
| 9 | 780 | 1560 | - | 3120 | 3900 | 4680 | 5460 | 6240 | 9 |

Appendix E

# Computational Complexity Evaluation of the Cross-layer Simulation Platform

As stated in section 4.2.1.1, precisely modeling channel evolutions implies having a fine grained channel block. The latter however is very resource consuming. A physical (PHY) layer model consisting of a complete transmit chain, channel, and receive chain can faithfully capture the precise effects and specificities of the physical medium and transmit/receive chains (see section 1.5.1). On the other hand a PHY layer model consisting of lookup tables (LUTs) is very interesting with regards to computational complexity. The global (or ergodic) behavior of the physical medium and transmit/receive chains can be modeled (see section 1.5.2). However specificities are smoothed out and fine accurate PHY layer modeling cannot be done. This is often the case with simulators detailing medium access control (MAC) layer mechanisms.

The difference in computational resource consumption between a simulation platform for IEEE 802.11n [6] and IEEE 802.11ac [7] systems using LUTs only and one using detailed PHY simulations only is illustrated in Figure E.1. The time factor indicates the increase in duration of simulation compared to the first simulation (LUTs only, one MAC protocol data unit – MPDU) taken as reference. A thousand-fold increase in simulation time can be observed. It is the price to pay for fine channel modeling. A compromise between complexity and faithfulness is thus needed.

A knowledgeable mix of these two PHY layer models has been used in the implemented cross-layer simulation platform for 802.11n and 802.11ac systems (see Chapter 4 for a detailed description of the simulation platform). Control and management frames being sent using much more robust modulation schemes than data frames, the former have much more chances of being correctly received. The issue is rather on the more sensitive data frames. There is great interest in using the detailed PHY layer model for data frames. Control and management frames can be modeled using LUTs to lower the overall simulation time. Computational resource use can thus be reduced while faithfully modeling system behavior. In Figure E.2, a system using the proposed knowledgeable mix of PHY layer models is compared to one using the detailed PHY layer model only, using simulation parameters exposed in section 4.3.1.2.

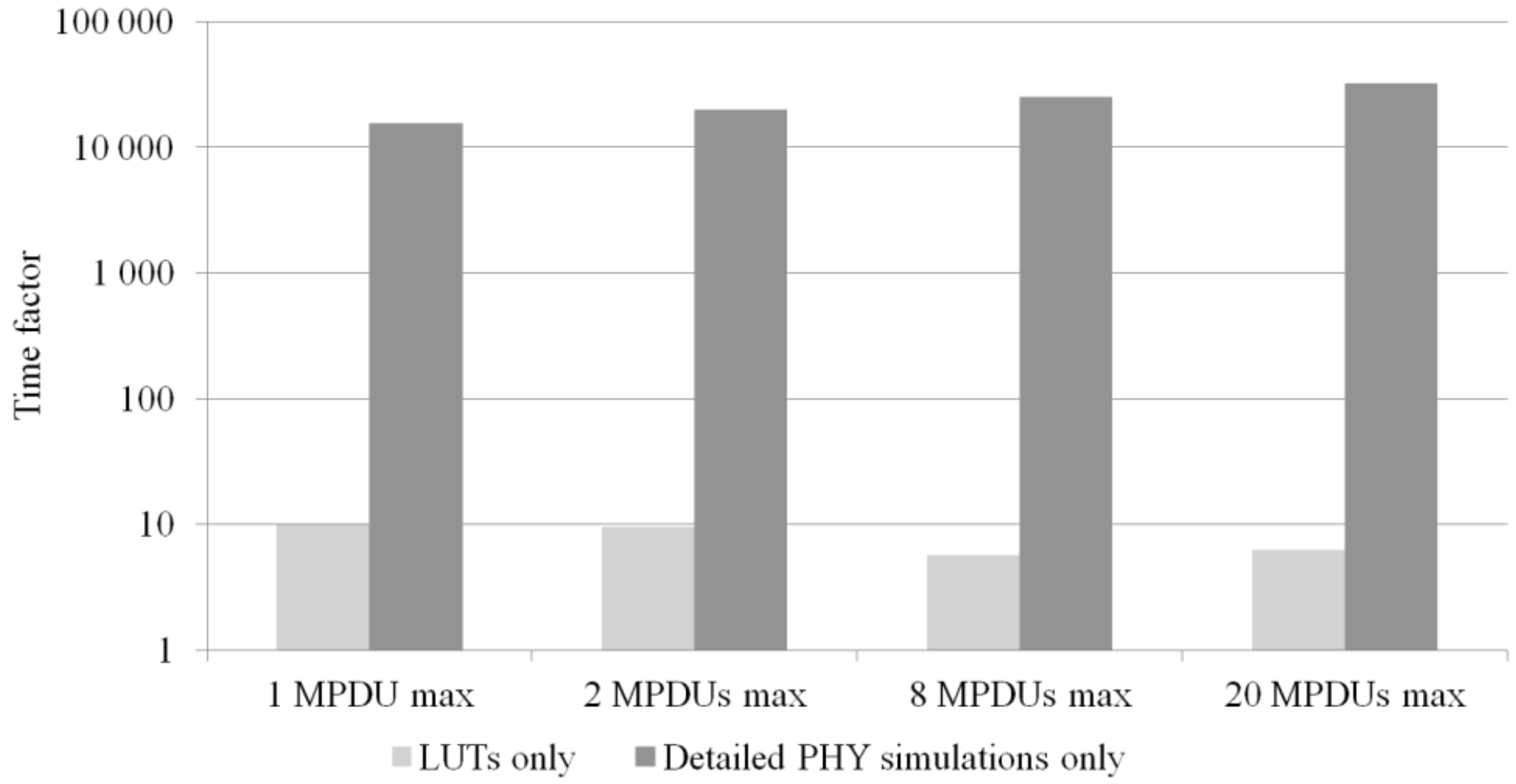

**Figure E.1 Computational complexity of an IEEE 802.11n/ac simulation platform, when using LUTs only or detailed PHY simulations only, for different maximum aggregation sizes**

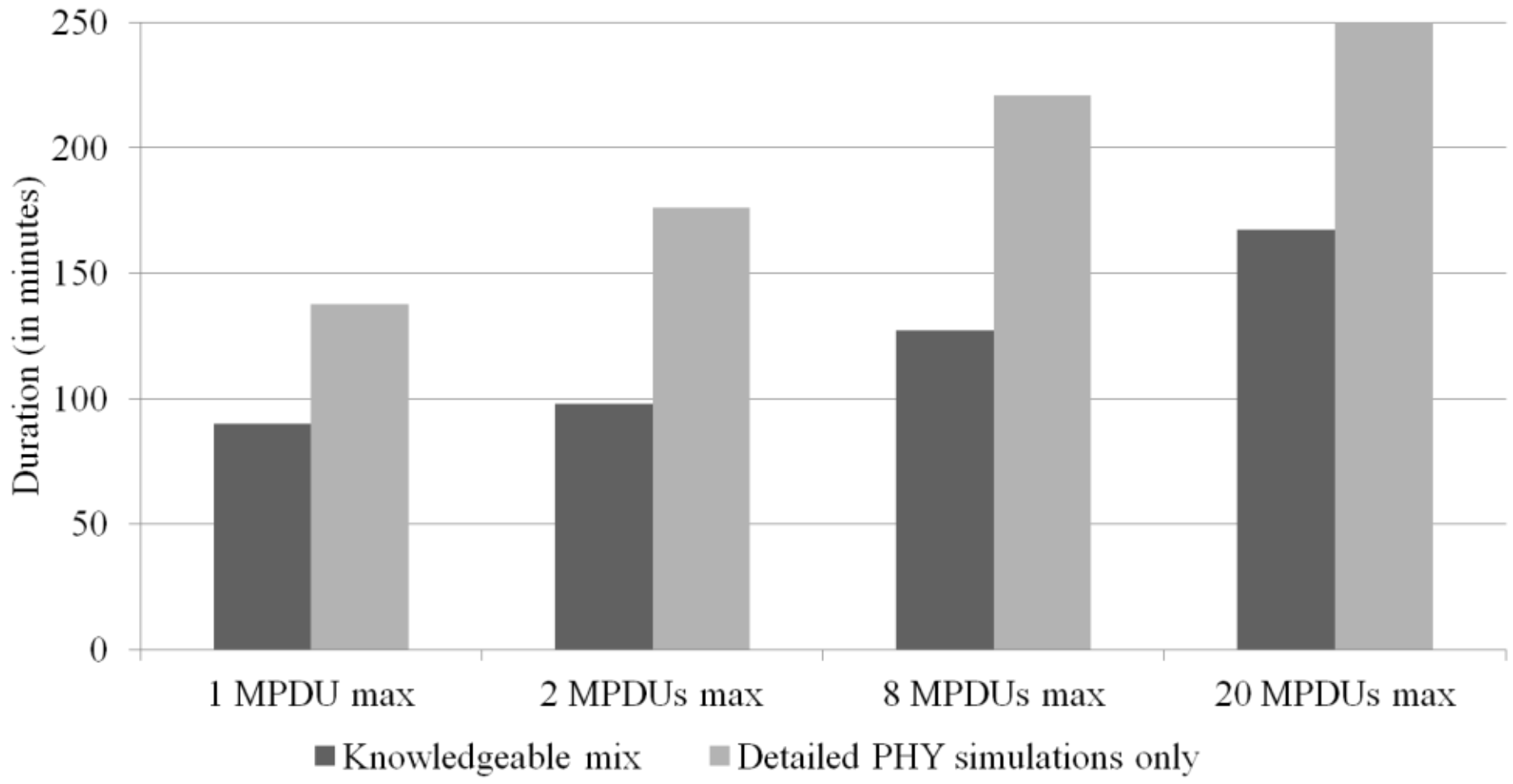

**Figure E.2 Duration of simulation for different aggregation sizes, when using the chosen compromise or detailing all simulations**

It can be seen that simulation duration is reduced by almost 50% with the proposed architecture. Realistic simulations can thus be done while minimizing global computational complexity.

# List of Figures

## Figures dans Résumé Etendu

# Figures in Thesis

# Figures in Appendices

# List of Tables

## Tables in Thesis

# Tables in Appendices